\documentclass[11pt]{article}
\usepackage{graphicx}
\usepackage[utf8]{inputenc}
\usepackage{fullpage}
\usepackage{amsmath,amssymb,amsfonts,amsthm,stmaryrd}
\usepackage{thmtools,thm-restate}
\usepackage[colorlinks=true,linkcolor=blue,allcolors=blue]{hyperref}
\usepackage{todonotes}
\usepackage{physics}
\usepackage[ruled,vlined,linesnumbered]{algorithm2e}
\usepackage[capitalize,nameinlink]{cleveref}
\usepackage{tikz}
\usetikzlibrary{decorations.pathreplacing}

\newcommand{\F}{\mathbb{F}}
\newcommand{\opn}[1]{\operatorname{#1}}
\newcommand{\im}{\opn{im}}
\newcommand{\spn}{\opn{span}}
\newcommand{\gX}{\opn{X}}
\newcommand{\gZ}{\opn{Z}}
\newcommand{\gCX}{\opn{CX}}
\newcommand{\gCCX}{\opn{CCX}}
\newcommand{\gCCZ}{\opn{CCZ}}
\newcommand{\gH}{\opn{H}}

\newcommand{\gInit}{\opn{Init}}
\newcommand{\gCO}{\opn{CF}}
\newcommand{\gSwap}{\opn{Swap}}
\newcommand{\Enc}{\opn{Enc}}
\newcommand{\Dec}{\opn{Dec}}
\newcommand{\supp}{\opn{supp}}
\newcommand{\evl}{\opn{ev}}
\newcommand{\poly}{\opn{poly}}
\newcommand{\polylog}{\opn{polylog}}
\newcommand{\titi}[1]{\tilde{\tilde{#1}}}
\newcommand{\concat}{\diamond}

\newcommand{\cB}{\mathcal{B}}
\newcommand{\cC}{\mathcal{C}}\newcommand{\cD}{\mathcal{D}}
\newcommand{\cE}{\mathcal{E}}\newcommand{\cF}{\mathcal{F}}
\newcommand{\cG}{\mathcal{G}}

\newcommand{\cO}{\mathcal{O}}
\newcommand{\cQ}{\mathcal{Q}}

\newcommand{\bC}{\mathbb{C}}
\newcommand{\bF}{\mathbb{F}}

\newcommand{\bN}{\mathbb{N}}

\newcommand{\bR}{\mathbb{R}}

\newcommand{\bZ}{\mathbb{Z}}

\DeclareMathOperator*{\argmin}{arg\,min}

\newtheorem{theorem}{Theorem}[section]
\newtheorem{lemma}[theorem]{Lemma}
\newtheorem{claim}[theorem]{Claim}
\newtheorem{definition}[theorem]{Definition}
\newtheorem{corollary}[theorem]{Corollary}
\newtheorem{proposition}[theorem]{Proposition}
\newtheorem{fact}[theorem]{Fact}
\theoremstyle{remark} 
\newtheorem{remark}[theorem]{Remark}

\newcommand{\lzg}[1]{}
\newcommand{\npb}[1]{}

\title{Fault-Tolerant Quantum Computation with Adversarial Errors\thanks{To appear in 67th IEEE Symposium On Foundations Of Computer Science (FOCS) 2026.}}
\author{
  Nikolas P.~Breuckmann \\
  University of Bristol, UK \\
  \href{mailto:niko.breuckmann@bristol.ac.uk}{\texttt{niko.breuckmann@bristol.ac.uk}}
  \and
  Louis Golowich \\
  UC Berkeley, USA \\
  \href{mailto:lgolowich@berkeley.edu}{\texttt{lgolowich@berkeley.edu}}
  \and
  Umesh Vazirani \\
  UC Berkeley, USA \\
  \href{mailto:vazirani@eecs.berkeley.edu}{\texttt{vazirani@eecs.berkeley.edu}}
}

\begin{document}

\maketitle

\begin{abstract}

  We prove a fault-tolerance theorem for quantum computation against adversarial noise. 
  For every quantum circuit on $\bar{N}$ logical qudits of depth $\bar{T}$, we construct a fault-tolerant circuit on $N=\poly(\bar{N})$ physical qudits of depth $\bar{T}\cdot\bar{N}^{o(1)}$, which is robust against an adversary who may arbitrarily choose and corrupt an almost-linear number $N^{1-o(1)}$ of physical qudits at each time step.
  This robustness significantly improves upon prior fault-tolerance theorems, which assumed corruptions were either local and stochastic, or else only act on a
  polynomially vanishing fraction of qudits.

  Our fault-tolerance scheme addresses a key bottleneck towards constructing quantum PCPs via the circuit-to-Hamiltonian mapping of Anshu, Breuckmann, and Nguyen (STOC'24).
  More fundamentally, our result demonstrates that fault-tolerant quantum computation remains possible under noise models that are global, worst-case, and non-Markovian over the full duration of the computation, directly countering concerns that correlated noise could fundamentally undermine quantum fault tolerance.

  Our construction is based on a new family of subsystem product codes we develop, which have large dimension and distance along with low-weight parity-checks, and which support transversal non-Clifford gates. 
  We show how to perform single-shot fault-tolerant error correction on these codes using a Floquet-like procedure based on the local testability of classical tensor codes. 
  We then obtain a universal fault-tolerance scheme using repeated code switching in a hypercubic qudit architecture.
  Finally, we recursively compose our scheme with itself to reduce an initially exponential qudit dimension down to a constant.
\end{abstract}

\newpage

\tableofcontents

\newpage

\section{Introduction}

The feasibility of quantum computation relies on our ability to control the effects of noise.
A major achievement in this regard is the \emph{quantum fault-tolerance theorem} \cite{aharonov_fault-tolerant_1997,kitaev_quantum_1997,knill_resilient_1998} which shows that a quantum device made from faulty components can perform any quantum computation with arbitrarily high fidelity.
The theorem relies on a number of assumptions, one of which is that errors are probabilistic and localized.
More precisely, the fault-tolerance theorem allows for errors that are correlated in space (between qudits) and time (between execution steps), as long as these correlations decay exponentially.

Subsequent work has refined and extended the fault-tolerance theorem in several directions.
Aliferis, Gottesman and Preskill \cite{aliferis_accuracy_2006} gave a rigorous quantitative treatment.
A separate line of work has focused on reducing the resource overhead of fault-tolerant protocols: 
Gottesman \cite{gottesman_fault-tolerant_2014} showed that constant space overhead suffices assuming good quantum LDPC codes, and Yamasaki and Koashi \cite{yamasaki_time-efficient_2024} subsequently achieved constant space and polylogarithmic time overhead simultaneously.
Most recently, Nguyen and Pattison~\cite{nguyen_quantum_2025} improved the time overhead to almost logarithmic, matching classical fault tolerance up to sub-polylogarithmic factors, while He, Nguyen and Pattison~\cite{he_composable_2025} introduced a composable framework that allows threshold proofs to be built modularly from independently analyzed gadgets.
All of these results, however, still operate under the assumption that noise is local and stochastic, with correlations decaying exponentially in space and time.

Fault-tolerance is more difficult to achieve when the noise is \emph{adversarial}. In this model, in each time step an adversary may choose an arbitrary set of qudits of some bounded size, and then corrupt these qudits by applying an arbitrary channel, which may introduce long-range correlated errors. Tan et al.~\cite{tan_single-shot_2025} presented a fault-tolerance scheme against such adversarial noise that is based on single-shot code switching, building upon a related scheme of Bomb\'{i}n \cite{bombin_dimensional_2016}.
However, for a circuit on $N$ physical qudits, these works can only handle at most $O(N^{1/3})$ adversarial errors per time step, which has remained the state-of-the-art.\footnote{To the best of our knowledge.}

Our main result is a fault-tolerance scheme with significantly improved robustness to adversarial noise. Specifically, our scheme protects against adversarial corruptions to an \emph{almost-linear} number~$N^{1-o(1)}$ of physical qudits at every time step.

\begin{theorem}[Main result; Informal statement of \Cref{thm:main}]
  \label{thm:maininf}
  For every fixed prime power $q$ and every $\epsilon>0$, there exists a fault-tolerance scheme over $q$-dimensional qudits with the following guarantee: Every quantum circuit using $\bar{N}$ logical qudits and time (i.e.~depth) $\bar{T}$ is compiled to a fault-tolerant physical circuit using $N\leq\bar{N}^{5+\epsilon}$ physical qudits and time $T\leq\bar{T}\cdot 2^{O(\sqrt{\log\bar{N}})}$, which is robust to adversarial corruptions on $N/2^{O(\sqrt{\log\bar{N}})}$ qudits per timestep.\footnote{Recall that $2^{O(\sqrt{\log N})}=N^{o(1)}$ grows subpolynomially.}
\end{theorem}

We prove \Cref{thm:maininf} for quantum circuits using a fixed universal\footnote{This gate set was shown to be universal over $\bF_q$ for every prime $q\geq 5$ in \cite{aharonov_fault-tolerant_1997}. Note that to obtain a universal scheme over qubits, one could simulate each $5$-dimensional qudit using three qubits. This modification would change the gate set; we do not pursue it for simplicity.} gate set---namely, Paulis, Hadamard, CNOT, Toffoli, and initialization/reset---that also have access to free noiseless classical computation. We emphasize that we allow our circuits to have arbitrary quantum inputs and quantum outputs, which may be chosen after compilation of the physical circuit. In this setting, fault-tolerance inherently requires physical quantum circuits, as even an all-powerful classical computer cannot process quantum information. As such, like many other fault-tolerance theorems (e.g.~\cite{gottesman_fault-tolerant_2014,he_composable_2025,tan_single-shot_2025}), we primarily use side classical computation in our decoder. See \Cref{sec:faulttolinf} for details.

\Cref{thm:maininf} implies that quantum computation can be made robust against noise patterns that are global, worst-case and non-Markovian over the full duration of the computation. Quantitatively, we allow corruptions on $N^{-o(1)}$-fraction of the physical qudits in each timestep, which is a super-polynomial improvement over the $O(N^{-2/3})$-fraction of corrupted qudits allowed by prior constructions. Qualitatively, we emphasize the global nature of the adversarial corruptions we protect against. 
Because any inverse-polynomial fraction $N^{-\alpha}$ of qudits can be entirely corrupted, we cannot use traditional approaches to fault-tolerance that encode each logical qudit in its own code block (e.g.~\cite{neumann_probabilistic_1956,dobrushin_upper_1977,pippenger_networks_1985,aharonov_fault-tolerant_1997}). 
Indeed, such schemes are vulnerable to adversarial corruptions on $1/\bar{N}$-fraction\footnote{It is helpful to consider the regime where the number of physical qudits $N\leq\bar{N}^{O(1)}$ grows at most polynomially in the number of logical qudits $\bar{N}$, as is the case in all our results.} of the physical qudits, which can wipe out an entire code block, making fault-tolerance impossible. In contrast, we prove \Cref{thm:maininf} by jointly encoding our qudits in a code of high rate and distance throughout the computation, so that there is never a small code block presenting an easy target for an adversary to corrupt.

From a foundational perspective, our strong robustness addresses a concern raised by skeptics of quantum computing, who have argued against its feasibility by pointing out that correlated noise could fundamentally undermine fault tolerance~\cite{levin2003tale, kalai2011how, kalai2016puzzle}. 
The concern is that physical noise need not satisfy the locality and independence assumptions underlying existing fault-tolerance theorems, and that uncontrolled degrees of freedom could place the environment in ``devilish states capable of conspiracies which defy imagination''~\cite{levin_qc_essay}. 
Our result directly addresses this concern: 
we show that fault-tolerant quantum computation is possible even if nature conspires against us.
An adversary with full knowledge of the circuit and the ability to apply coherent, correlated errors to an almost-linear number of qudits at every time step cannot prevent the computation from succeeding.


Furthermore, our result addresses a key bottleneck towards constructing quantum PCPs (qPCPs) via adversarial fault-tolerance. 
The qPCP conjecture posits the existence of a constant \emph{soundness gap} $\delta>0$ along with a family of local Hamiltonians\footnote{We consider Hamiltonians $H=\sum_{i=1}^m H_i$ acting on $N$ qudits with $m=\poly(N)$ terms $H_i$, each of which acts on $O(1)$ of the qudits and has bounded norm $\|H_i\| \leq O(1)$.
The ground energy density $\min_{\ket{\psi}}\bra{\psi}H\ket{\psi}/m$ simply equals the ground energy divided by the number of terms.} for which it is \textsf{QMA}-hard to determine if the ground energy density is $\leq 0$ vs $\geq\delta$.
This conjecture has remained a major open question for which relatively little remains known.
In contrast, the classical analogue, namely the celebrated PCP theorem \cite{arora_probabilistic_1998,arora_proof_1998}, has seen multiple different proofs (see e.g.~\cite{dinur_pcp_2007,meir_combinatorial_2012}), with numerous implications throughout complexity theory and cryptography.
However, all of these approaches to PCPs have resisted quantization due to various constraints imposed by quantum mechanics~\cite{aharonov_quantum_2013}.

Anshu, Breuckmann, and Nguyen \cite{anshu_circuit--hamiltonian_2024} recently proposed a new approach to constructing quantum PCPs that evades these issues and which is based on fault-tolerance. 
The idea is to compile a logarithmic-depth \textsf{QMA}-verifier circuit into a fault-tolerant version, which can then be mapped to a local Hamiltonian. \cite{anshu_circuit--hamiltonian_2024} gave a circuit-to-Hamiltonian mapping for which they conjectured that low-energy states correspond to noisy executions of the circuit, and proved some results towards this conjecture. The noise here may consist of large adversarial corruptions---precisely the error model against which our fault-tolerance scheme is the first to protect! Our results hence provide a key ingredient towards quantum PCPs with soundness gap $\delta\geq 1/N^{o(1)}$, due to our tolerance of errors on $1/N^{o(1)}$-fraction of the qudits; the main remaining missing piece is the conjecture of \cite{anshu_circuit--hamiltonian_2024}. In contrast, the best previously known soundness gap for \textsf{QMA} was $\delta=1/\poly(N)$ from the Feynman--Kitaev construction~\cite{kitaev2002classical}.

In fact, the exact reasoning above carries through for the classical case. In \cite{anshu2026classical}, a classical fault-tolerance theorem against adversarial noise is used to prove a `$\polylog$-weaker' classical PCP theorem with inverse-polylogarithmic soundness.
The proof follows precisely the strategy outlined above: 
a circuit is compiled into a fault-detecting circuit via adversarial fault tolerance and then using the Cook--Levin mapping turned into a constraint satisfaction problem whose soundness gap is inherited from the error threshold of the fault-tolerance scheme.
The classical construction achieves fault detection against a $N/\polylog(N)$-fraction of adversarial errors per layer, which translates directly into a PCP with polylogarithmic query complexity.
This provides strong evidence that the quantum analogue of this strategy is a viable path toward the quantum PCP conjecture.

As a technical remark, as described above our fault-tolerance scheme in \Cref{thm:maininf} relies on noiseless classical computation, which may be difficult to handle under the circuit-to-Hamiltonian mapping. However, \cite{anshu2026classical} faced a related challenge in the classical setting, which they overcame by relaxing the fault-tolerance requirement to a \emph{fault-detection} requirement. That is, because the PCP setting inherently concerns \emph{verifying} computation, it is sufficient to detect errors, rather than correct them. Such error-detection can be performed without decoding, which as mentioned above is our primary use of noiseless classical computation. Using this idea, we believe our fault-tolerance scheme can be adapted to the PCP setting, though the ultimate required circuit model will depend on the details of any resolution to the conjecture of \cite{anshu_circuit--hamiltonian_2024}.

\section{Technical Overview}

We now provide a more detailed overview of our fault-tolerance scheme. 
Like nearly all prior fault-tolerance schemes, we assume our input is encoded in a quantum error-correcting code, and we perform computation on the encoded data to compute the desired encoded output. However, our strong adversarial error model places stringent requirements on the codes that we can use. For instance, as described above, we cannot encode each logical qudit into its own code block, as then the adversary could corrupt an entire such block. Hence we must jointly encode our qudits in a code of large (i.e.~almost-linear) distance. But then to perform the desired logical computation (which may apply arbitrary gates in parallel to different logical qudits) with just $N^{o(1)}$ time overhead as in \Cref{thm:maininf}, we need to perform:
\begin{enumerate}
\item Error-correction in the presence of adversarial noise, which corrects more errors than it introduces, and
\item Parallel targeted gates on the logical qudits, with limited error propagation.
\end{enumerate}
The first item above can be achieved using good quantum LDPC codes, which have low-weight parity-checks that can be measured by low-depth circuits to perform error-correction. The low circuit depth ensures that only a small number of new errors occur during error-correction. Meanwhile, the second item above can be achieved using codes supporting \emph{transversal gates}, meaning that a constant-depth physical application of the gate induces the logical action of the gate on the encoded message. We will specifically need transversal gates that are non-Clifford, such as Toffoli, as Clifford gates alone cannot achieve universal computation.

A recent breakthrough line of works culminated in the construction of linear-distance quantum LDPC codes, using a specific ``balanced/lifted product'' of classical LDPC codes \cite{breuckmann_balanced_2021,panteleev_asymptotically_2022}. Separately, linear-distance codes supporting transversal non-Clifford gates were constructed using algebraic codes such as Reed-Solomon and algebraic-geometry \cite{wills_constant-overhead_2024,golowich_asymptotically_2025,nguyen_quantum_2025}. However, these algebraic codes are inherently non-LDPC, as they possess linear-weight parity-checks. Prior attempts to combine the LDPC and transversal non-Clifford properties have yielded length-$N$ codes with either distance $\leq O(\sqrt{N})$ \cite{bombin_exact_2007,bombin_topological_2007,bombin_gauge_2015,zhu_non-clifford_2023,scruby_quantum_2024,zhu_topological_2025,breuckmann_cups_2024,lin_transversal_2024,golowich_quantum_2025-1}, or else check weight $\geq\Omega(N^{1/3})$ \cite{golowich_near-asymptotically-good_2025}. These parameters are insufficient to prove \Cref{thm:maininf}, which instead would require distance $\geq N^{1-o(1)}$ and check weight $\leq N^{o(1)}$.

The independent and concurrent work \cite{li_transversal_2026-1} recently obtained quantum LDPC codes of nearly linear distance $\tilde{\Omega}(N)$ with a transversal non-Clifford gate. However, the transversal gate is only shown to induce a constant number of logical non-Clifford gates on the encoded qudits, analogously to how the classical repetition code allows for transversal multiplication on a single logical bit. This construction is therefore also insufficient to prove \Cref{thm:maininf}, which requires performing (on average) at least a polynomial number of logical non-Clifford gates per timestep.

Furthermore, while some level of addressability has been achieved for transversal non-Clifford gates (see e.g.~\cite{he_asymptotically_2025}), existing constructions still for instance do not support parallel fault-tolerant Toffoli gates on arbitrary triples of logical qudits. Hence known constructions of addressable gates remain insufficient to prove \Cref{thm:maininf}.

\subsection{Key Ingredient: New Quantum Subsystem Product Codes}
\label{sec:codesinf}
We overcome these challenges by introducing a new family of quantum codes that simultaneously have high distance and dimension, low check weight, support transversal non-Clifford gates, and are flexible enough to support logical qudit permutations that facilitate parallel targeted gates. As described below, we specifically obtain all of these properties for a new family of subsystem codes over exponentially large alphabets. Error-correction on these subsystem codes requires measuring non-commuting parity-checks (called \emph{gauge operators}), which we nevertheless show how to perform in a single-shot fault-tolerant manner by applying classical local testability. Meanwhile, to reduce the alphabet size from exponential down to constant, we recursively compose our fault-tolerance scheme with itself, to simulate high-dimensional qudits using lower-dimensional qudits.

\subsubsection{Reed-Solomon Codes}
We construct our subsystem codes from products of Reed-Solomon codes, building upon the techniques of \cite{golowich_near-asymptotically-good_2025}.
Recall that classical Reed-Solomon codes consist of evaluations of bounded-degree univariate polynomials over a finite field $\bF_q$ (see \Cref{def:polyeval}). Reed-Solomon codes have linear dimension and distance, along with a \emph{transversal multiplication property}: 
the component-wise product\footnote{The component-wise product of $a,b\in\bF_q^n$ is $a*b=(a_1b_1,\dots,a_nb_n)\in\bF_q^n$.} of two Reed-Solomon codewords lies in another, higher-dimensional Reed-Solomon code. It was shown by \cite{wills_constant-overhead_2024,golowich_asymptotically_2025,nguyen_quantum_2025} that a quantum CSS code constructed from two Reed-Solomon codes exhibits a transversal $\gCCX$ (i.e.~Toffoli) gate, which is non-Clifford.\footnote{\label{footnote:ccz} Strictly speaking \cite{wills_constant-overhead_2024,golowich_asymptotically_2025,nguyen_quantum_2025} provided a transversal $\gCCZ$ gate across three code states. However, we can instead obtain transversal $\gCCX$ by replacing the third code $C=(C_X,C_Z)$ with its dualized version $(C_Z,C_X)$, as $\gCCX$ is equivalent up to $\gCCZ$ up to conjugation of the third qudit by Hadamard gates.} 
This result is natural given that the $\gCCX$ gate, defined by $\gCCX\ket{x_1,x_2,x_3}=\ket{x_1,\;x_2,\;x_3+x_1x_2}$, simply performs multiplication in the standard basis. In fact, a similar idea was used decades ago by~\cite{aharonov_fault-tolerant_1997} to prove the fault-tolerance theorem against local noise.

However, the duals of Reed-Solomon codes are themselves Reed-Solomon codes, meaning that the parity-checks have linear weight. \cite{golowich_near-asymptotically-good_2025} proposed a way to reduce the parity-check weight while preserving transversal $\gCCX$ by taking a \emph{subsystem product} of Reed-Solomon codes \cite{zeng_minimal_2020}, as defined in \Cref{def:subsysteminf} below.

\subsubsection{Subsystem Product Construction with Transversal Non-Clifford Gates}
\label{sec:subprodinf}
The subsystem products we use can be viewed as a quantum generalization of classical tensor products. Recall that for classical codes $C^1,\dots,C^u\subseteq\bF_q^n$, the tensor product code $C=\bigotimes_{i\in[u]}C^i$ consists of all $c\in\bF_q^{[n]^u}$ such that the restriction of $c$ to every column in every direction $i$ lies inside $C^i$. 
Here a direction-$i$ column simply consists of the $n$ points $(j_1,\dots,j_u)\in[n]^u$ for every $j_i\in[n]$, with $j_{i'}$ fixed for $i'\neq i$. 
Naturally, tensor codes have weight-$n$ parity-checks supported inside these columns.
Tensor products of Reed-Solomon codes also retain the transversal multiplication property, which was used by \cite{anshu2026classical} to construct a classical fault-tolerance scheme. 
We will similarly use transversal $\gCCX$ gates on subsystem products of quantum Reed-Solomon codes to obtain our quantum fault-tolerance scheme.

We say a pair $C=(C_X,C_Z)$ of classical codes $C_X,C_Z\subseteq\bF_q^n$ forms a \emph{CSS non-subsystem code} if they satisfy the CSS orthogonality condition $C_X^\perp\subseteq C_Z$; otherwise, $C$ is a \emph{CSS subsystem code}. For both the subsystem and non-subsystem cases, the code has dimension
\begin{equation*}
  k = \dim(C_Z+C_X^\perp)-\dim(C_X^\perp) = \dim(C_Z)-\dim(C_Z\cap C_X^\perp)
\end{equation*}
and distance $d=\min\{d_X,d_Z\}$, where we define the
\begin{equation*}
  \text{$X$-distance }\; d_X = \min_{c\in(C_X+C_Z^\perp)\setminus C_Z^\perp}|c| \hspace{2em} \text{ and } \hspace{2em} \text{$Z$-distance }\; d_Z = \min_{c\in(C_Z+C_X^\perp)\setminus C_X^\perp}|c|.
\end{equation*}

\begin{definition}
  \label{def:subsysteminf}
  Let $u\in\bN$ and for $i\in[u]$ let $C^i=(C^i_X,C^i_Z)$ be a CSS non-subsystem code of length $n$ and dimension $k_i$. The \emph{subsystem product} code $C=(C_X,C_Z)=\bigotimes_{i\in[u]}C^i$ is the CSS subsystem code given by
  \begin{align*}
    C_X &= \bigotimes_{i\in[u]}C^i_X \hspace{1em} \text{ and } \hspace{1em} C_Z = \bigotimes_{i\in[u]}C^i_Z.
  \end{align*}
\end{definition}

\begin{figure}
  \centering
  \begin{tikzpicture}[
    scale=0.7,
    font=\large,
    every path/.style={thick},
    brace/.style={decorate, decoration={brace, amplitude=8pt}},
    mirror/.style={decorate, decoration={brace, mirror, amplitude=8pt}}
    ]

    \def\W{13}    
    \def\H{11}    

    \def\a{3}
    \def\b{8}

    \def\c{3.5}
    \def\d{10}

    \draw (0,0) rectangle (\W,\H);

    %
    %
    %
    %
    \draw (\c, \a) -- (\W, \a);          
    \draw (\c, \a) -- (\c, \H);          
    \draw (\d, \a) -- (\d, \b);          
    \draw (\c, \b) -- (\d, \b);          

    \pgfmathsetmacro{\logx}{(\c+\d)/2}
    \pgfmathsetmacro{\logy}{(\a+\b)/2}
    \node[align=center] at (\logx, \logy) {
      $X$ Logical operators \\[4pt]
      $C_Z / (C_Z \cap C_X^\perp)$
    };

    \pgfmathsetmacro{\gaux}{(\c+\d)/2}
    \pgfmathsetmacro{\gauy}{\a/2}
    \node[align=center] at (\gaux, \gauy) {
      $X$ Gauge operators \\[4pt]
      $C_X^\perp = (C_X^1)^\perp \otimes \mathbb{F}_q^n + \mathbb{F}_q^n \otimes (C_X^2)^\perp$
    };

    \draw[brace] (-0.6, 0) -- (-0.6, \a)
    node[pos=0.5, left=10pt] {$(C_X^1)^\perp$};

    \draw[brace] (-3.2, 0) -- (-3.2, \b)
    node[pos=0.5, left=10pt] {$C_Z^1$};

    \draw[brace] (-5.4, 0) -- (-5.4, \H)
    node[pos=0.5, left=10pt] {$\mathbb{F}_q^n$};

    \draw[mirror] (0, -0.6) -- (\c, -0.6)
    node[pos=0.5, below=10pt] {$(C_X^2)^\perp$};

    \draw[mirror] (0, -2.2) -- (\d, -2.2)
    node[pos=0.5, below=10pt] {$C_Z^2$};

    \draw[mirror] (0, -3.8) -- (\W, -3.8)
    node[pos=0.5, below=10pt] {$\mathbb{F}_q^n$};

  \end{tikzpicture}
  \caption{\label{fig:subprod} An illustration of the subsystem product $C=C^1\otimes C^2$ of $u=2$ codes $C^1=(C^1_X,C^1_Z)$ and $C^2=(C^2_X,C^2_Z)$. Recall that $C_X=C^1_X\otimes C^2_X$ and $C_Z=C^1_Z\otimes C^2_Z$. The entire square in the figure represents an $n\times n$ grid of basis elements of $\bF_q^{n\times n}$, so that each axis contains $n$ basis elements of $\bF_q^n$. The space $C_X^\perp\subseteq\bF_q^{n\times n}$ represents $X$ parity-checks, i.e.~gauge operators. Meanwhile, $C_Z/(C_Z\cap C_X^\perp)$ represents the space of $X$ logical operators.}
\end{figure}
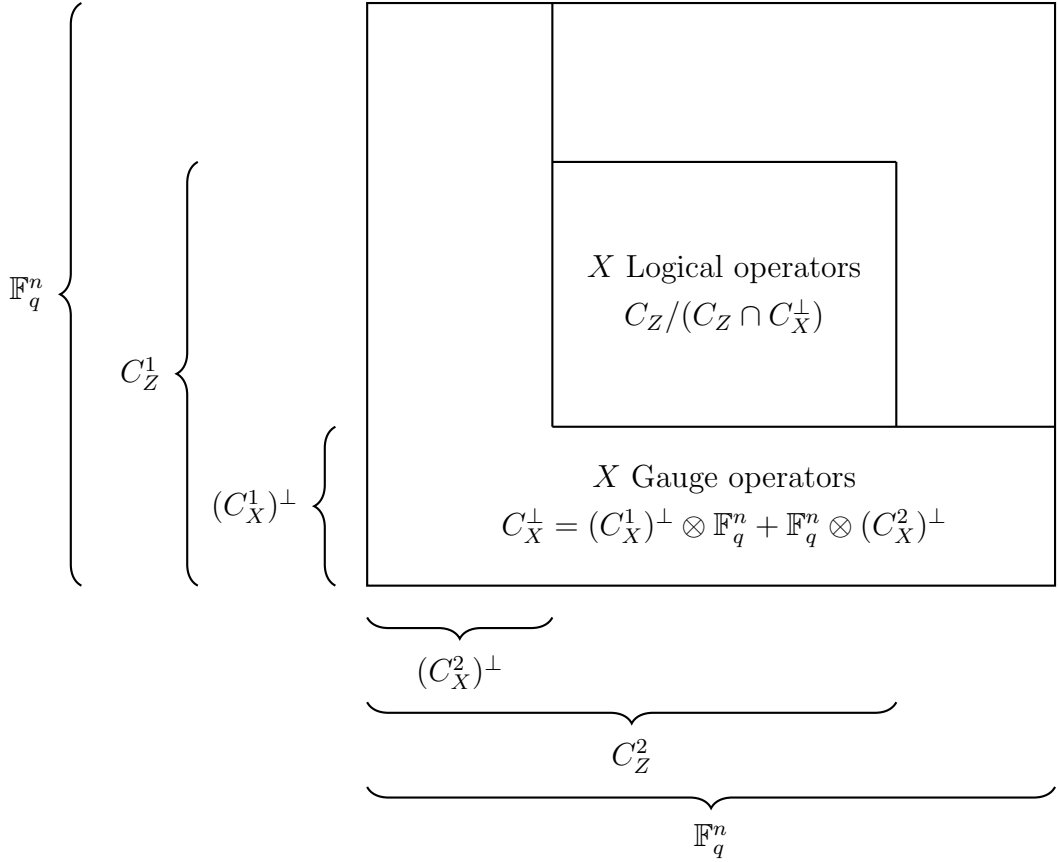

\Cref{fig:subprod} provides an illustration of the subsystem product of $u=2$ factor codes.

For general $u\in\bN$, taking each $C^i_X$ and $C^i_Z$ for $i\in[u]$ to be an appropriate Reed-Solomon code, then the transversal multiplication property of these factor codes translates to a transversal $\gCCX$ gate on the subsystem product code $C$. Indeed, a similar result was shown by \cite{golowich_near-asymptotically-good_2025} (who considered $\gCCZ$ instead of $\gCCX$; see \Cref{footnote:ccz}).

\subsubsection{Fault-Tolerant Error-Correction}
\label{sec:errcorrinf}
Performing error-correction on these subsystem product codes in \Cref{def:subsysteminf} is more involved. In the language of stabilizer subsystem codes, the product code $C$ has weight-$n$ \emph{gauge operators} given by $X^c$ for $c\in C_X^\perp$ and $Z^c$ for $c\in C_Z^\perp$. Each $c$ here corresponds to a weight-$n$ parity-check of the tensor code $C_X$ or $C_Z$ (see above), where the product code $C$ has length $n^u\gg n$ assuming $u\gg 1$ (we will take $u=\log n$). Hence the gauge-operators are low-weight, and can be measured by low-depth circuits with limited error propagation.

However, because $C$ is a subsystem code, these gauge operators do not commute. Thus if we try to correct $X$-errors by measuring $Z$-gauge operators, we collapse our state to a $Z$-gauge eigenstates, which are not $X$-gauge eigenstates. We can then try to correct $Z$-errors by measuring $X$-gauge operators, but will collapse the state to $X$-gauge eigenstates, which are not $Z$-gauge eigenstates. Alternating between measuring low-weight $X$ and $Z$ gauge operators in this way, we obtain a dynamic error-correction protocol reminiscent of floquet codes \cite{hastings_dynamically_2021}.

In particular, after measuring $Z$-gauge operators, we apply a correction to move the code state into a simultaneous $+1$-eigenspace of all $Z$-gauge operators. It is on this gauge-fixed code state that we are able to apply transversal $\gCCX$ (see \Cref{sec:toffoli} for details).

However, we must perform the error-correction described above in the presence of measurement noise. For non-subsystem LDPC codes, commuting measurements can be repeated to identify measurement errors (see e.g.~\cite{gottesman_fault-tolerant_2014}). In contrast, errors in the non-commuting measurements of subsystem codes are more difficult to handle, as repeating the measurements will collapse the state into different eigenspaces at each step.

Nevertheless, we are able to perform fault-tolerant error-correction on subsystem product codes using a single-shot strategy suggested in \cite[Appendix~C]{golowich_near-asymptotically-good_2025}. Specifically, the noisy gauge operator measurements in a given basis ($X$ or $Z$) give a noisy syndrome $s$ consisting of parity-checks for the corresponding tensor code ($C_X$ or $C_Z$). We then find the $s'$ that is closest to $s$ in Hamming weight such that $s'$ is the true syndrome of some error $e'$ (i.e.~$s'$ would be a valid measurement in the absence of noise), and we decode assuming $s'$ was our syndrome. This procedure successfully corrects errors because classical tensor codes are \emph{locally testable} \cite{viderman_combination_2015}, meaning that the error $e'$ inducing syndrome $s'$ must be close to the true error on the code state, as long as $s'$ is close to the true error's syndrome. See \Cref{sec:errcorr} for more details.

\subsubsection{Distance Analysis}
\label{sec:distanceinf}
We described above how to perform fault-tolerant error-correction and transversal $\gCCX$ gates on subsystem products of Reed-Solomon codes. It remains to be shown that we can construct such codes of large (i.e.~almost-linear) distance. For factor codes $C^1,\dots,C^u$ of respective distances $d_1,\dots,d_u$, the quantum subsystem product $C$ in general may have distance $D$ between $\max_{i\in[u]}d_i$ and $\prod_{i\in[u]}d_i$. To obtain almost-linear distance $D\geq N^{1-o(1)}$ with respect to the block length $N$ of $C$, we want $D\approx\prod_{i\in[u]}d_i$ to roughly match the upper bound, assuming each $d_i$ grows linearly in the length of $C^i$. Note that classical tensor product codes, in contrast, always have distance $D=\prod_{i\in[u]}d_i$.

Closing the gap between these lower and upper distance bounds in the quantum case has proven to be challenging. Several works have studied this problem for various families of product codes \cite{bravyi_homological_2014,audoux_tensor_2018,zeng_minimal_2020,golowich_quantum_2025-2,golowich_near-asymptotically-good_2025}, though only the latter paper also required transversal non-Clifford gates. Specifically, \cite{golowich_near-asymptotically-good_2025} constructed $u=2$ carefully chosen Reed-Solomon codes, as well as $u=3$ randomly punctured product Reed-Solomon codes over exponentially large alphabets, for which the subsystem product has distance $D\geq N^{1-o(1)}$ and supports transversal non-Clifford gates. Their proof relied on a strong ``product-expansion'' property that is difficult to achieve for algebraic codes, and did not extend to $u>3$. Hence the parity-check weight was $\geq N^{1/3}$, which means that syndrome extraction requires a polynomial-depth circuit, and hence errors on a polynomially small fraction of qudits may lead to an entirely corrupted syndrome. \cite{golowich_near-asymptotically-good_2025} left it as an open question to extend these results to $u>3$.

We resolve this question by constructing a family of tuples of $u$ quantum Reed-Solomon codes for which we show the subsystem product has distance $D\geq N^{1-o(1)}$, where $u\rightarrow\infty$ as $N\rightarrow\infty$. We present our codes using the following notation for polynomial evaluations.

\begin{definition}[Restatement of \Cref{def:polyeval}]
  \label{def:polyevalinf}
  For $u\in\bN$ and $M\subseteq\bR^u$, we let $\bF_q[X_1,\dots,X_u]^M$ denote the space of $u$-variate polynomials over $\bF_q$ spanned by monomials of the form $X_1^{i_1}\cdots X_u^{i_u}$ for $(i_1,\dots,i_u)\in M\cap\bZ_{\geq 0}^u$. For $A\subseteq\bF_q^u$, we let $\bF_q[X_1,\dots,X_u]^M_A\subseteq\bF_q[X_1,\dots,X_u]^M$ denote the subspace of polynomials that have roots at all points in $A$. For a polynomial $f\in\bF_q[X_1,\dots,X_u]$ and a subset $E\subseteq\bF_q^u$, we let $\evl_E(f):=(f(x))_{x\in E}\in\bF_q^E$ denote the list of evaluations of $f$ at all points in $E$.
\end{definition}

For example, a \emph{Reed-Solomon code} is given by $\evl_E(\bF_q[X]^{[0,\ell)})\subseteq\bF_q^E$ for $\ell\in\bN$ and $E\subseteq\bF_q$.
We will prove large distance for the following family of subsystem products based on Reed-Solomon codes:

\begin{definition}[Informal statement of \Cref{def:pevprod}]
  For $u\in\bN$, integers $\ell_1,\dots,\ell_u\in\bN$, and subsets $A_1,\dots,A_u,E_1,\dots,E_u\subseteq\bF_q$ with each $A_i\cap E_i=\emptyset$ and $|A_i|\leq\ell_i\leq|E_i|$, we define a \emph{polynomial evaluation subsystem product code} $C=\bigotimes_{i\in[u]}C^i$ with each factor code $C^i=(C^i_X,C^i_Z)$ given by
  \begin{align*}
    C^i_X &= \evl_{E_i}(\bF_q[X]_{A_i}^{[0,\ell_i)}) \hspace{1em} \text{ and } \hspace{1em} C^i_Z = \evl_{E_i}(\bF_q[X]^{[0,\ell_i)}).
  \end{align*}
\end{definition}

As stated below, we show that if the field size $q$ is sufficiently large, then there exist explicit choices of sets $A_i,E_i$ for which the product code $C$ has large distance:

\begin{theorem}[Main distance bound; Informal statement of \Cref{thm:pevmain}]
  \label{thm:pevmaininf}
  For $i\in[u]$, fix positive integers $k_i<n_i$, and let $q=q_0^{\prod_{i\in[u]}n_i'}$ for a prime power $q_0\geq\max_{i\in[u]}n_i$ and integers $n_i'\geq n_i$. Then there exist explicit sets $A_i,E_i\subseteq\bF_q$ for $i\in[u]$ with $|A_i|=k_i$, $|E_i|=n_i$ such that for every $k_i\leq\ell_i\leq n_i$, the associated polynomial evaluation subsystem product code $C$ has $X$- and $Z$-distance
  \begin{align*}
    d_X &\geq \prod_{i\in[u]}(\ell_i-k_i+1) \hspace{1em} \text{ and } \hspace{1em} d_Z \geq \prod_{i\in[u]}(n_i-\ell_i+1).
  \end{align*}
\end{theorem}

In particular, we will set $u=\log n$, and then set every $n_i=n$ and $\ell_i,k_i=\Theta(n)$, so that $d_X,d_Z\geq\Theta(n)^{\log n}=n^{\log n}/\poly(n)$ is almost-linear in the product code's block length $n^{\log n}$.

We will describe our proof of the $X$-distance bound in \Cref{thm:pevmaininf}. The $Z$-distance bound then follows by the same argument, because the self-duality of Reed-Solomon codes implies that
\begin{equation}
  \label{eq:selfdualinf}
  (C^i_Z)^\perp = \beta_i * \evl_{E_i}(\bF_q[X]_{A_i}^{[0,n_i+k_i-\ell_i)}) \hspace{1em} \text{ and } \hspace{1em} (C^i_X)^\perp = \beta_i * \evl_{E_i}(\bF_q[X]^{[0,n_i+k_i-\ell_i)})
\end{equation}
for some fixed vectors of nonzero coefficients $\beta_i\in\bF_q^{E_i}$, which do not affect the distance analysis (see \Cref{lem:pevswapXZ}).

Let $A=A_1\times\cdots\times A_u$ and $E=E_1\times\cdots\times E_u$. Then the definition of the subsystem product code $C=(C_X,C_Z)$ implies that
\begin{align*}
  C_Z &= \evl_E(\bF_q[X_1,\dots,X_u])^{\prod_{i\in[u]}[0,\ell_i)}
\end{align*}
consists of evaluations of $u$-variate polynomials of individual degree $<\ell_i$, and
\begin{align}
  \label{eq:CZcapCXpinf}
  C_Z\cap C_X^\perp &= \evl_E(\bF_q[X_1,\dots,X_u])_A^{\prod_{i\in[u]}[0,\ell_i)}
\end{align}
is given by the subspace of those polynomials that vanish on all points in $A$.

\begin{proof}[Proof sketch of $X$-distance bound in \Cref{thm:pevmaininf}]
  It suffices to ensure that for every small support set $S \subseteq E$ of size $|S|<\prod_{i \in [u]}(\ell_i - k_i + 1)$, no codewords in $(C_X+C_Z^\perp)\setminus C_Z^\perp$ are supported inside~$S$. Any such codeword $c$ supported in $S$ would be orthogonal to all elements of $(C_X+C_Z^\perp)^\perp|_S=(C_Z\cap C_X^\perp)|_S$, but could not be orthogonal to all of $C_Z|_S$ (for otherwise we would have $c\in C_Z^\perp$). Hence it suffices to show that
  \begin{equation}
    \label{eq:distkeyinf}
    (C_Z\cap C_X^\perp)|_S = C_Z|_S.
  \end{equation}

  By assumption the space $\bF_q[X_1,\dots,X_u]^{\prod_{i\in[u]}[0,\ell_i-k_i]}$ of polynomials of individual degree $\leq\ell_i-k_i$ has dimension $>|S|$. Therefore there exists a nonzero polynomial $g_S$ in this space that vanishes on all of $S$. Now for every polynomial $f$ of individual degree $<\ell_i$, so that $\evl_S(f)\in C_Z|_S$, we want to choose a polynomial $h$ of individual degree $<k_i$ such that $\evl_A(h)=\evl_A(f)/\evl_A(g_S)$ (where `$/$' denotes pointwise division). Indeed, then $f-g_Sh$ has individual degree $<\ell_i$ and satisfies $\evl_A(f-g_Sh)=\evl_A(f)-\evl_A(f)=0$. Therefore $\evl_S(f-g_Sh)\in (C_Z\cap C_X^\perp)|_S$ by \Cref{eq:CZcapCXpinf}, and $\evl_S(f-g_Sh)=\evl_S(f)$ because $g_S$ vanishes on $S$. Thus we conclude that every $\evl_S(f)\in C_Z|_S$ also lies in $(C_Z\cap C_X^\perp)|_S$, so \Cref{eq:distkeyinf} holds.

  It only remains to be shown that we can ensure that each $g_S$ is nonzero on all of $A$, so that $\evl_A(h)=\evl_A(f)/\evl_A(g_S)$ is well-defined. For this purpose, we choose the evaluation sets $E_i$ to lie in a small base field $\bF_{q_0}$, so that the coefficients of $g_S$ also lie inside $\bF_{q_0}$. Then we choose the sets $A_i$ from successive extension fields, where each field extension has degree $n_i'$ exceeding the individual degree $\ell_i-k_i$ of $g_S$.
\end{proof}

Our use of successive extension fields above necessitates an exponentially large alphabet $q=q_0^{\prod_{i\in[u]}n_i'}$ with respect to the code's block length $\prod_{i\in[u]}n_i$. We leave it as an open question to reduce this alphabet size.

\subsubsection{Aside: Non-Subsystem Product Codes and Further Implications}
\label{sec:codeimpinf}
We believe that our codes are of independent interest, beyond the context of our fault-tolerance scheme. Indeed, the construction of quantum codes of almost-linear distance and dimension with low (e.g.~subpolynomial) check weight was itself a longstanding open question, which was resolved by the breakthrough line of works \cite{hastings_fiber_2021,panteleev_quantum_2022,breuckmann_balanced_2021,panteleev_asymptotically_2022} leading to asymptotically good quantum LDPC codes, which have linear distance and dimension and constant check weight, over constant-sized alphabets. All constructions to date of such codes \cite{breuckmann_balanced_2021,panteleev_asymptotically_2022,leverrier_quantum_2022-1,dinur_good_2023,lin_good_2022,hsieh_explicit_2025} consist of similar ingredients, namely the balanced product \cite{breuckmann_balanced_2021} (or the closely related lifted product \cite{panteleev_quantum_2022}) of classical codes, combined with a local robustness property called ``product-expansion'' \cite{panteleev_asymptotically_2022,kalachev_two-sided_2023}.

Even neglecting the transversal non-Clifford property (which was not obtained by the prior works above), our codes can be viewed as a contribution to this line of works with a fundamentally different construction and a simple, self-contained analysis. Just as the good qLDPC codes described above are non-subsystem codes, in \Cref{sec:homprod} we also give a non-subsystem version of our codes, which lose the transversal non-Clifford property, but still retain almost-linear distance and dimension with subpolynomial check weight over exponentially large alphabets. While a related construction was given in \cite{golowich_near-asymptotically-good_2025} that applied the high-dimensional product-expansion result of \cite{kalachev_maximally_2025} to random product codes, our construction is explicit, with a concise and self-contained analysis as described in \Cref{sec:distanceinf} above.

\subsection{Fault-Tolerance Scheme}
We now describe how we apply our polynomial evaluation subsystem product codes from \Cref{sec:codesinf} to construct our fault-tolerance scheme in \Cref{thm:maininf}. Our key idea here is to combine transversal gates on our subsystem product codes $C=\bigotimes_{i\in[u]}C^i$ with repeated code switching. In this code switching, for each $i\in[u]$, we un-encode the factor code $C^i$, and then re-encode into some $(C^i)'$ that may be different from $C^i$.

This code switching performs two purposes. First, it allows us to obtain a universal set of transversal gates, which according to the Eastin-Knill theorem must require some sort of code switching (see the discussion of transversal Hadamard below). Second, because we can apply transversal gates after un-encoding but before re-encoding, we are able to perform gates between any two qudits that lie in the same direction direction-$i$ column, for any $i$. Hence we are able to perform circuits with a $u$-dimensional hypercubic connectivity, which enables efficient logical qudit permutations, and hence targeted parallel gates. We provide more details below.

Related code switching techniques have been used in prior quantum fault-tolerance schemes, e.g.~\cite{bombin_dimensional_2016,jochym-oconnor_fault-tolerant_2019,tan_single-shot_2025,golowich_constant-overhead_2025,xu_batched_2025}. The classical fault-tolerance scheme of \cite{spielman_highly_1996} also uses similar ideas, and like our scheme is based on tensor products of Reed-Solomon codes. However, such schemes typically only used products of a constant number of factor codes, which leads to a smaller polynomial error tolerance. Our use of code switching is most similar to the classical adversarial fault-tolerance scheme of \cite{anshu2026classical}, which also used a growing number of factor codes.

\subsubsection{Transversal Gates}
Recall from \Cref{sec:subprodinf} that our codes support transversal $\gCCX$ gates. As (even subsystem) CSS codes, they naturally also support Pauli gates (\Cref{lem:paulis}), $\gCX$ gates (\Cref{lem:cx}), and Hadamard gates (\Cref{lem:hadamard}). In each case, the logical gate is induced by applying the physical gate across code states, which as a depth-1 circuit cannot propagate errors within a code state. For the Hadamard gate, the transversal application maps a code state of $C = (C_X, C_Z)$ to a code state of the dualized code $C^\perp = (C_Z, C_X)$. This dualized code is still a polynomial evaluation subsystem product code by the self-duality formula in \Cref{eq:selfdualinf} in \Cref{sec:distanceinf} above, so it still has large distance by \Cref{thm:pevmaininf}. We can then return to the original code by switching out the dualized factor codes for the original $C^i$ one at a time, as described below.

\subsubsection{Code Switching}
As described above, to perform code switching on a subsystem product code $C=\bigotimes_{i\in[u]}C^i$, we simply un-encode one factor $C^i$, and then re-encode, possibly into a different code $(C^i)'$. We subsequently perform error-correction, and then can repeat in the other directions, to sequentially switch all $u$ factor codes.

Recall that we take $u=\log n$ where $n$ is the length of each factor code $C^i$, so that $C$ has length $N=n^u=n^{\log n}$. Each switching operation simply applies a depth-$\poly(n)$ un-encoding and then re-encoding circuit within each direction-$i$ column. Therefore each error on a qudit can only propagate to the $n$ qudits in the same direction-$i$ column. Hence we can withstand $N/\poly(n)=N^{1-o(1)}$ errors per time step of this switching procedure without overloading the code's distance, which is also $N/\poly(n)$ (see \Cref{thm:pevmaininf}).

\subsubsection{Universal Fault-Tolerant Computation in Hypercubic Architecture}
We now describe how we combine the code switching and transversal gates described above to obtain our scheme for universal fault-tolerant computation.
Let each factor code $C^i$ have dimension $k$, so that $C$ has logical qudits labeled by $[k]^u$. After un-encoding $C^i$ but before re-encoding, our $k^u$ logical qudits are distributed across $k$ code states of $\bigotimes_{i'\in[u]\setminus\{i\}}C^{i'}$, each containing $k^{u-1}$ logical qudits. We can perform the transversal gates described above across these $k$ code states, before switching back up to the code $C$. 

We therefore want to compile our logical circuit into one consisting of a sequence of transversal gates, each acting along a single direction $i\in[u]$ of the $u$-dimensional hypercube $[k]^u$ of logical qudits. For this purpose, we apply an identical idea as in the classical case~\cite{anshu2026classical}, which also appeared in prior works such as \cite{spielman_highly_1996}: using bitonic sorting networks \cite{batcher_sorting_1968}, we can implement arbitrary permutations of qudits on the hypercube via short sequences of transversal gates. Hence we can route qudits into arbitrary positions, apply arbitrary desired transversal gates, and then route qudits back to their original positions. By routing qudits to appropriate positions with respect to the subsequent transversal gates, we can perform arbitrary targeted gates in parallel.
It follows from \cite{batcher_sorting_1968} that this compilation increases the circuit depth by a factor of at most $O(u^2 \log^2 k)$.

Note that the geometry aids fault-tolerance in multiple ways. 
First, even after the downwards switching step we are still protected by an almost linear distance when $u=\log n$, i.e.~we are always protected by the bulk.
Second, as gates are applied in a single direction, errors spread only within columns.

Every fault-tolerant gadget in our scheme follows a common pattern: 
we first run error-correction (\Cref{lem:errcorr}) to reduce the error weight, then apply the desired gate gadget, and finally run error-correction again to clean up any new errors introduced by the gate and any faults that occurred during its execution. We choose parameters to ensure that the number of errors always remains at most $N/\poly(n)=N^{1-o(1)}$.

\subsubsection{Fault-Tolerance Formalism}
\label{sec:faulttolinf}
The fault-tolerance notation and definitions we use are adapted from the composable gadget formalism of Nguyen and Pattison~\cite{nguyen_quantum_2025} and He, Nguyen, and Pattison~\cite{he_composable_2025}, and most closely follows the subsequent presentation in~\cite{golowich_constant-overhead_2025}.
We model a fault-tolerant gadget (\Cref{def:faulttol}) as a physical circuit~$\cQ$ that implements a logical operation $\bar{O}$ on states encoded into quantum error-correcting codes, provided that the fault, an arbitrary sequence of superoperators applied after each timestep, avoids certain ``bad sets'' $\cE_{\mathrm{run}}$ specifying the maximum tolerable error weight per timestep.
The input and output codes are each decorated with an encoding map $\Enc$, a set $\Gamma$ of allowed gauge qudit states, and a family $\cE$ of bad sets defining the maximum tolerable error on the encoded state.
A gadget is fault-tolerant if, whenever the input state is a low-weight deviation from a valid codeword and the fault avoids the bad sets, the output is again a low-weight deviation from the correctly encoded logical output.

Gadgets can be composed sequentially (\Cref{lem:seqcomp}) as long as each gadget's output error avoids all bad sets of the subsequent gadget's decorated input code. 
Parallel composition (\Cref{lem:parcomp}) is handled by including an \emph{external system} of $\ell$ auxiliary qudits into the definition that may be entangled with the encoded state, ensuring that fault-tolerance guarantees are preserved under tensor products with other components of a larger computation.
We discuss the more involved notion of simulative composition below, which we use to reduce the qudit dimension of our scheme.


While our fault-tolerance definitions involve Pauli errors and Pauli faults, they in fact capture protection against fully general quantum noise.
By the Pauli decomposition (\Cref{lem:paulidecomp}), every superoperator can be expressed as a linear combination of Pauli superoperators with supports contained in the support of the original superoperator.
This allows us to reduce the analysis of arbitrary quantum channels, including coherent errors, to Pauli errors (\Cref{lem:paulift}).
Furthermore, \Cref{lem:extft} shows that our formalism extends to an \emph{extended fault} model in which the adversary maintains a private register of ancilla qudits that can be entangled with the system throughout the computation, capturing the strongest possible noise model. Similar ideas were described in \cite{he_composable_2025}, which we adapt to our framework.

Our gate set $\cG = \{\gH^*, \gX^*, \gZ^*, \gCX^*, \gCCX^*, \gInit_*, \gCO_*\}$ includes, besides the familiar standard gates, \emph{classical function gates} $\gCO_*$ (\Cref{def:gates}), which perform a basis measurement followed by a noiseless classical computation and reinitialization.\footnote{While these gates allow noiseless queries to classical functions, the basis measurement ensures that the queries cannot be made in superposition, so they perform truly classical computations.} We use these gates to decode the noisy syndromes described in \Cref{sec:errcorrinf}, as well as to help handle Hadamard gates in the simulative composition described in \Cref{sec:alphredinf} below for alphabet reduction.

Because we prove fault-tolerance even against errors induced by non-physical (i.e.~non-channel) superoperators, an error following a classical function gate may induce a \emph{postselection} on our logical state. For instance, suppose we apply a classical function gate that measures a quantum code state in the standard basis, and outputs the measurement outcome of some logical qudit. A single-qudit error superoperator could then postselect on any given measurement outcome that has positive probability. As a result, our final fault-tolerance result (\Cref{thm:main}) only guarantees that the fault-tolerant circuit performs the desired logical computation under some postselected versions of the logical classical function gates.

Fortunately, logical circuits consisting of unitary gates and single-qudit mid-circuit measurements (i.e.~the ordinary quantum circuit model) never need to perform classical function gates, and hence avoid postselection entirely. Specifically, unitary gates never involve postselection. Meanwhile, a mid-circuit standard-basis measurement can be performed by applying $\gCX$ to couple the data qudit with an ancilla qudit initialized to $\ket{0}$, and then re-initializing (to effectively trace out) the ancilla qudit. Crucially, our initialization gadget (\Cref{lem:stateprep}) does \emph{not} induce any logical postselection, so we can fault-tolerantly perform such mid-circuit measurements without postselection. Thus, despite the presence of postselections in our formalism, we can fault-tolerantly compile non-unitary logical circuits without any logical postselection.


\subsubsection{Alphabet Reduction via Simulative Composition}
\label{sec:alphredinf}
The fault-tolerance scheme described above operates over qudits of dimension $q'$ that grows exponentially in the other parameters, due to the large alphabet required in our distance analysis in \Cref{thm:pevmaininf}.
To obtain a result over qudits of any \emph{fixed} prime power dimension $r$ we apply \emph{simulative composition} in \Cref{sec:alphred}.

We first show how a quantum circuit over a small field $\bF_r$ can be simulated by a circuit over a large extension field $\bF_{q'}$ via the natural inclusion map $\bF_r \hookrightarrow \bF_{q'}$, by embedding each $r$-dimensional qudit into a $q'$-dimensional qudit, see \Cref{sec:inclog}.
For most gates in our gate set, the simulation is straightforward:
Pauli gates $\gX^a$, $\gZ^a$, controlled gates $\gCX^a$, $\gCCX^a$, and initialization gates $\gInit_*$ all lift naturally from $\bF_r$ to $\bF_{q'}$.
The main subtlety is implementing the Hadamard gate~$\gH$ over~$\bF_r$ using operations over $\bF_{q'}$.
Our approach (\Cref{lem:inclog}) proceeds as follows:
we first apply $\gH$ over~$\bF_{q'}$, then use the Frobenius map $z\mapsto z^r$ to compute $z^r-z$, whose kernel is precisely $\bF_r\subseteq\bF_{q'}$.
By measuring $z^r-z$ and applying a correction derived from the finite field trace $\tr_{\bF_{q'}/\bF_r}$ (and computed using a classical function gate), we project the output back into $\bF_r$, effectively implementing~$\gH$ over the base field.
The resulting simulation increases the circuit depth by only an $O(\log r)$ factor.

We are then able to apply our large-alphabet fault-tolerance scheme described above (see \Cref{sec:ftschemelarge} for the details) to this compiled logical circuit over $\bF_{q'}$, as long as $q'$ is exponentially large in the number of logical qudits $\bar{N}$. Specifically, we take $q'=r^{2^{v'}}$ for $v'\approx\log\bar{N}$.

We subsequently show in \Cref{sec:alphabetreduce} how to reduce the physical qudit dimension from $q'$ back towards~$r$ by recursively composing our large-alphabet fault-tolerance scheme with smaller instances of itself.
The key technical tool is \Cref{cor:simcomp}, which shows that given a fault-tolerant gadget over $q'$-dimensional qudits with $q' = r^{2^{v'}}$, one can construct a fault-tolerant gadget over $q$-dimensional qudits with $q = r^{2^{v}}$ for $v = \lfloor(3/4+\epsilon)v'\rfloor$, by encoding each $q'$-dimensional qudit into a block of $q$-dimensional qudits using the codes from \Cref{sec:codes}. We fault-tolerantly simulate each gate on a $q'$-dimensional qudit using a smaller instance of our fault-tolerance scheme over $q$-dimensional qudits from \Cref{sec:ftschemelarge}. We then recursively repeat this procedure to obtain a sequence of fault-tolerant circuits over successively smaller alphabets $q=r^{2^v}$.

Crucially, each such composition step preserves the fault-tolerance properties up to controlled losses:
the code distance is multiplied by a factor of $2^{v - O(\sqrt{v})}$, the error threshold per time step is multiplied by $2^{v - O(\sqrt{v})}$, and the circuit depth increases by only a $2^{O(\sqrt{v})}$ factor.
We choose $\epsilon < 1/8$ so that each application strictly reduces the exponent $v$, and after $m = O(\log v')$ iterations, the exponent drops below a constant $v_0(\epsilon)$ that depends only on $\epsilon$.
At this point, the qudit dimension $\tilde{q} = r^{2^{v_0(\epsilon)}}$ is a constant (depending on $r$ and $\epsilon$). Our recursive simulative composition described above stops working at such small qudit dimension, so we instead apply a final, simpler alphabet reduction step (\Cref{lem:basiccomp}) that replaces each $\tilde{q}$-dimensional qudit with $2^{v_0(\epsilon)}$ $r$-dimensional qudits via the field isomorphism $\bF_{\tilde{q}} \cong \bF_r^{2^{v_0(\epsilon)}}$; this step incurs only a constant-factor overhead since $\tilde{q}$ is bounded.

The cumulative effect of all composition steps is controlled by a geometric series with ratio $3/4+\epsilon$ and starting value $v'\approx\log\bar{N}$. That is, let $v^{(1)},\dots,v^{(m)}$ be the sequence of values of $v$ across all steps. Then
the total block length overhead is roughly $2^{\sum_{i=1}^m v^{(i)}} \leq 2^{v'/(1/4-\epsilon)} = \bar{N}^{4 + O(\epsilon)}$, which is polynomial in $\bar{N}$, while the total depth overhead is $2^{O(\sum_{i=1}^m \sqrt{v^{(i)}})} = 2^{O(\sqrt{\log N})}$, which is subpolynomial in $\bar{N}$.
Similarly, the distance and error threshold degrade by at most subpolynomial factors of the form $2^{-O(\sqrt{\log N})}$ relative to the block length.

Observe that our final space overhead $\bar{N}^{4+O(\epsilon)}$ was determined by the ratio $3/4+\epsilon$ of the geometric series above. Ultimately, this ratio is bottlenecked by the alphabet requirement in \Cref{thm:pevmaininf}, along with a space-time tradeoff we use to efficiently apply linear transformations over the small field, which are needed to simulate $\gCX$, $\gCCX$, and $\gH$ gates over the large field. Specifically, our codes in \Cref{thm:pevmaininf} can have alphabet size roughly exponential in the number of logical qudits. Therefore we could hope to set each $v^{(i+1)}\approx v^{(i)}/2$, so that each $r^{2^{v^{(i)}}}$-dimensional qudit can be encoded in the $\approx 2^{v^{(i+1)}}$ logical qudits in an instance of our code over alphabet size $r^{2^{v^{(i+1)}}}$. However, we instead set $v^{(i+1)}\approx 3v^{(i)}/4$ to provide extra ancilla logical qudits at every level of composition. We use these ancilla qudits to apply linear transformations to our logical data qudits in logarithmic depth but with polynomial space overhead. This space-time tradeoff is important for maintaining low circuit depth at every level of concatenation, which ensures errors do not have time to accumulate during the recursive simulations. We leave it as an open question to reduce this space overhead (see \Cref{sec:future}).

\subsubsection{Proof of Main Result: Putting it all Together}
Combining the above ideas yields our main result \Cref{thm:maininf}.
To summarize, fix any prime power~$r$ and any $0 < \epsilon < 1/8$.
Given a logical circuit $\bar{\cQ}$ on $\bar{N}$ $r$-dimensional qudits of depth $\bar{T}$ using the universal gate set $\cG = \{\gH^*,\gX^*,\gZ^*,\gCX^*,\gCCX^*,\gInit_*,\gCO_*\}$, we construct a fault-tolerant physical circuit $\cQ$ over $r$-dimensional qudits as follows.
First, we embed the logical circuit into a circuit over a large extension field $\bF_{q'}$ via the inclusion $\bF_r \hookrightarrow \bF_{q'}$ using \Cref{lem:inclog}, increasing the depth by a factor of $O(\log r)$.
We then apply \Cref{thm:laft} to compile this large-alphabet circuit into a fault-tolerant circuit over $\bF_{q'}$-qudits using subsystem product codes, where the hypercubic architecture and direction-by-direction code switching provide mending fault-tolerant gadgets for every gate in the universal set.
Finally, the recursive alphabet reduction of \Cref{sec:alphabetreduce} brings the qudit dimension back down to~$r$ through $O(\log\log\bar{N})$ rounds of simulative composition (\Cref{cor:simcomp}), followed by a single application of direct alphabet reduction (\Cref{lem:basiccomp}).

The resulting physical circuit $\cQ$ uses $N \leq \bar{N}^{5+O(\epsilon)}$, and has depth $T \leq 2^{O(\sqrt{\log\bar{N}})} \cdot \bar{T}$.
The input and output logical states are encoded into CSS non-subsystem codes obtained by concatenating the polynomial evaluation subsystem product codes of \Cref{thm:pevmain} through the recursive composition, yielding code distance $d \geq 2^{-O(\sqrt{\log \bar{N}})} \cdot N$.
The fault-tolerant circuit guarantees correctness against any adversarial fault pattern affecting fewer than $2^{-O(\sqrt{\log \bar{N}})} \cdot N = N^{1-o(1)}$
physical qudits per time step, where the $o(1)$ term vanishes as $\bar{N} \to \infty$ for fixed $r$ and $\epsilon$.

By \Cref{lem:extft}, the adversary may apply arbitrary quantum channels, including coherent errors, at each time step, and may furthermore maintain a private register of ancilla qudits that can be entangled with the system throughout the computation.
This captures the strongest possible noise model:
the adversary has full knowledge of the circuit, chooses errors adaptively, and the errors need not be Markovian or local in any sense.
Despite this, the computation succeeds provided only that the total number of affected qudits per step is sublinear.

\subsection{Future Directions}
\label{sec:future}
Several directions for future work remain. First, it may be possible to improve the space overhead of our construction from polynomial to almost linear by encoding many small-alphabet qudits into each large-alphabet qudit during the simulative composition step, similarly to techniques in \cite{golowich_asymptotically_2025,nguyen_good_2025}. The main obstacles would be reducing the space overhead of the linear transformations used to implement the $\gCX$, $\gCCX$, and Hadamard gates in \Cref{lem:qimp}, as well as routing the small qudits within each large qudit. Alternatively, a larger goal would be to directly reduce the alphabet size of our codes in \Cref{thm:pevmaininf}.

Second, it is natural to ask whether the $N^{o(1)}$ time overhead can be reduced, for instance to $\poly\log N$, or whether the error threshold can be improved.
Such improvements could ultimately translate to quantum PCPs with larger soundness gaps.

Third, our construction does not address the question of decoder efficiency, as we assume the classical decoding problem is performed instantly by classical function gates. It remains unclear whether there exist efficient decoders for quantum product codes in which all factors are quantum codes or chain complexes of dimension $\geq 2$. \cite[Sections~5--6]{golowich_near-asymptotically-good_2025} provided some progress in this direction, by showing how to decode the product of a pair of quantum Reed-Solomon codes, though instantiated in a different parameter regime from ours.

Finally, our scheme may be naturally adaptable to run with nearest-neighbor interactions in $u$ spatial dimensions, but since we take~$u$ growing with $N$, this yields a tradeoff between spatial locality and error threshold for any fixed number of spatial dimensions. It would be interesting to understand whether there are ``dynamic BPT bounds'' that establish fundamental limitations on such tradeoffs, analogously to the bounds of \cite{bravyi_tradeoffs_2010} that limit (static) code parameters in a bounded number of spatial dimensions.

\subsection{Roadmap}
In \Cref{sec:preliminaries} we fix notation and provide the technical foundation and the construction of our fault-tolerance framework. 

In \Cref{sec:codes}, we introduce our primary coding-theoretic contribution: 
a family of quantum subsystem codes derived from tensor products of Reed--Solomon codes. 
In \Cref{sec:errcorr}, we show how the classical local testability of these products enables single-shot error correction, which is essential for maintaining stability against adversarial noise without repeated measurements.
We give the construction of a set of universal gates for these codes in the remainder of \Cref{sec:coregad}.

The architectural heart of our protocol is detailed in \Cref{sec:ftschemelarge}, where we describe the hypercubic arrangement of qudits and formalize our code-switching mechanism. 
This approach allows us to perform transversal operations and qudit routing along specific dimensions of the hypercube while maintaining the protection of the code's global distance.
The main result of this section is an adversarial fault-tolerance theorem where the local qudit dimension grows exponentially in the system size.

Finally, \Cref{sec:alphred} describes our alphabet reduction techniques. 
Specifically, we show how to simulate low-dimensional qudits using high-dimensional qudits in \Cref{sec:inclog}, and then we show how to recursively reduce the qudit dimension via simulative composition in \Cref{sec:alphabetreduce}. 
We combine these components in our main result, \Cref{thm:main}, which states the formal fault-tolerance theorem for quantum computation under adversarial noise.

\section{Preliminaries}
\label{sec:preliminaries}
\subsection{Notation}
\label{sec:notation}
For $n\in\bN$, we write $[n]=\{1,2,\dots,n\}$. For vectors $x,y\in\bF^n$ over an arbitrary field $\bF$, we write $x\propto y$ if there exists some $\alpha\in\bF$ such that $x=\alpha y$. For a prime power $q=p^e$, we let $\bF_q$ denote the finite field of order $q$, and we let $\bF_q^*=(\bF_q\setminus\{0\})$. 
For a subfield $\bF_{q'}\subseteq\bF_q$, we let $\tr_{\bF_q/\bF_{q'}}:\bF_q\rightarrow\bF_{q'}$ denote the $\bF_{q'}$-linear finite field trace map. 
For vectors $x,y\in\bF^n$, we let $x\cdot y=\sum_{i=1}^nx_iy_i$ denote the standard bilinear form, and we let $x*y=(x_iy_i)_{i\in[n]}\in\bF^n$ denote the component-wise product.

For a prime power $q=p^e$, we associate a $q$-dimensional qudit, with an element of $\bC^q\cong\bC^{\bF_q}=\spn\{\ket{x}:x\in\bF_q\}$. 
For a set $S\subseteq\bF_q^n$, we let $\ket{S}=\frac{1}{|S|}\sum_{x\in S}\ket{x}$ denote the uniform superposition over elements in $S$. We write $\ket{+}=\ket{\bF_q}=(1/\sqrt{q})\sum_{x\in\bF_q}\ket{x}$ to denote the single-qudit uniform superposition over all of $\bF_q$.

A \emph{density operator} on $n$ $q$-dimensional qudits is a self-adjoint positive semi-definite operator $\rho\in\bC^{\bF_q^n\times\bF_q^n}$ of trace~$1$. For a general operator $A\in\bC^{\bF_q^n\times\bF_q^n}$, the \emph{support} $S=\supp(A)$ is the minimal subset $S\subseteq[n]$ for which we can express $A=A_S\otimes I_{[n]\setminus S}$ for some operator $A_S\in\bC^{\bF_q^S\times\bF_q^S}$, where $I_{[n]\setminus S}\in\bC^{\bF_q^{([n]\setminus S)\times([n]\setminus S)}}$ denotes the identity operator.

A \emph{superoperator} is a linear map $O:\bC^{\bF_q^m\times\bF_q^m}\rightarrow\bC^{\bF_q^n\times\bF_q^n}$. A \emph{quantum channel} is a completely positive trace-preserving (CPTP) superoperator. If $m=n$, then the \emph{support} $S=\supp(O)$ of $O$ is the minimal subset $S\subseteq[m]$ for which we can express $O=O_S\otimes I_{[m]\setminus S}$ for some superoperator $O_S:\bC^{\bF_q^S\times\bF_q^S}\rightarrow\bC^{\bF_q^S\times\bF_q^S}$ acting on qudits in $S$, where $I_{[m]\setminus S}:\bC^{\bF_q^{([m]\setminus S)\times([m]\setminus S)}}\rightarrow\bC^{\bF_q^{([m]\setminus S)\times([m]\setminus S)}}$ denotes the identity superoperator acting on qudits in $[m]\setminus S$.

For an operator or superoperator $A$, we write $|A|=|\supp(A)|$.

\subsection{Error-Correcting Codes}
In this section we present definitions and basic results regarding the error correcting codes used in our fault-tolerance construction.
In particular, we make use of quantum codes constructed from tensor products of classical codes.
The tensor product structure is crucial for our fault-tolerance protocol.
One property in particular is that tensor products of classical codes are locally testable \cite{viderman_combination_2015}, which allows for the construction of efficient fault-tolerant error correction gadgets.

\begin{definition}
  \label{def:classcode}
  A \emph{classical (linear) code of length $n$ and dimension $k$ over $\bF_q$} is a $k$-dimensional linear subspace $C\subseteq\bF_q^n$. 
  The \emph{distance} of $C$ is $d:=\min_{c\in C\setminus\{0\}}|c|$. We summarize these parameters by saying that $C$ is a $[n,k,d]_q$ code.

  The \emph{dual} of $C$ is $C^\perp:=\{x\in\bF_q^n:x\cdot c=0\;\forall c\in C\}$.

  For codes $C\subseteq\bF_q^n$, $C'\subseteq\bF_q^{n'}$, the \emph{tensor product} is the code $C\otimes C'\subseteq\bF_q^{n\times n'}$ given by $C\otimes C'=\spn\{c\otimes c':c\in C,c'\in C'\}$. Equivalently, $C\otimes C'$ contains all $n\times n'$ matrices such that every column lies in $C$ and every row lies in $C'$.
\end{definition}

We often refer to classical linear codes as simply ``linear codes,'' as we will not consider nonlinear codes in this paper.

We will use the result of \cite{viderman_combination_2015} that high-dimensional tensor product codes are \emph{locally testable}, as stated below. Our specific presentation of this result of \cite{viderman_combination_2015} below mirrors that in \cite{anshu2026classical}, though we repeat the details for the reader's convenience.

\begin{definition}
  \label{def:pcpc}
  For $u\in\bN$, for $i\in[u]$ let $H^i\in\bF_q^{m_i\times n_i}$ be a matrix. Define sets
  \begin{equation*}
    N = [n_1]\times\cdots\times[n_u]
  \end{equation*}
   and $S=\bigsqcup_{i\in[u]}S^i$ with each
  \begin{equation*}
    S^i = [n_1]\times\cdots\times[n_{i-1}]\times[m_i]\times[n_{i+1}]\times\cdots\times[n_u].
  \end{equation*}
  The \emph{product-code parity-check matrix} $H\in\bF_q^{S\times N}$ associated to $H^1,\dots,H^u$ is given for $x\in\bF_q^N$ by
  \begin{equation*}
    Hx = ((I^{\otimes i-1}\otimes H^i\otimes I^{\otimes u-i})x)_{i\in[u]} \in \bigoplus_{i\in[u]}\bF_q^{S^i} = \bF_q^S.
  \end{equation*}
  This matrix is by definition a parity-check matrix for the tensor product of the codes $\ker(H^i)$ for $i\in[u]$, that is,
  \begin{equation*}
    \ker(H) = \bigotimes_{i\in[u]}\ker(H^i).
  \end{equation*}
\end{definition}

\begin{proposition}[Local testability of tensor product codes \cite{viderman_combination_2015}]
  \label{prop:loctest}
  For $u,n,d\in\bN$, for $i\in[u]$ let $C^i=\ker(H^i)$ be an $[n,\; k_i,\; d_i\geq d]_q$ code with full-rank parity-check matrix $H^i\in\bF_q^{(n-k_i)\times n}$. Let $H\in\bF_q^{S\times[N]}$ be the associated product-code parity-check matrix. Then for every $s\in\im(H)$, there exists some $x\in\bF_q^N$ with $Hx=s$ and $|x|\leq\mu(u,n,d)\cdot|s|$, where
  \begin{equation*}
    \mu(u,n,d) = \left(un\cdot\left(\frac{n}{d}\right)^{u}\right)^\eta
  \end{equation*}
  for an absolute constant $\eta>0$.
\end{proposition}


The parameter $\mu(u,n,d)$ is called the \emph{filling constant} of the tensor product code $\bigotimes_{i\in[u]}C^i$ and the inverse $1/\mu(u,n,d)$ is called the \emph{soundness}.


\begin{remark}
  \label{remark:viderman}
  We have stated \Cref{prop:loctest} slightly differently from \cite[Theorem 3.1]{viderman_combination_2015}, but our statement still follows directly from \cite{viderman_combination_2015}. Specifically:
  \begin{enumerate}
  \item \cite[Theorem 3.1]{viderman_combination_2015} is stated assuming all codes $C^i$ are equal, though as noted in \cite[Remark 4.5]{viderman_combination_2015}, the techniques extend flawlessly to different $C^i$.
  \item The tester in \cite[Theorem 3.1]{viderman_combination_2015} actually checks if any of the $n-k_i$ entries of $s$ in a given direction-$i$ column are nonzero, meaning that $|s|$ in \Cref{prop:loctest} would be replaced by the number of such direction-$i$ columns with a nonzero entry. Yet this quantity is always within a factor of $n$ of $|s|$, so we can absorb the factor of $n$ into the $n^\eta$ factor in $\mu(u,n,d)$.
  \item When $u$ is not a power of $3$, the local tester in \cite{viderman_combination_2015} samples different syndrome components with different probabilities, meaning that it may seem that we should add different weights for different components when computing the Hamming norm $|s|$ in \Cref{prop:loctest}. However, the result of \cite{viderman_combination_2015} naturally holds under all different permutations of the codes $C^1,\dots,C^u$, and all permutations of the $n$ columns and the $n-k_i$ rows in each $H^i$. Averaging over all of these permutations, we conclude that the result holds with uniform weights on the components of the syndrome $s$.
  \end{enumerate}
\end{remark}

The primary classical codes we consider are polynomial evaluation codes, defined below.
\begin{definition}
  \label{def:polyeval}
  For $u\in\bN$ and $M\subseteq\bR^u$, we let $\bF_q[X_1,\dots,X_u]^M$ denote the space of $u$-variate polynomials over $\bF_q$ spanned by monomials of the form $X_1^{i_1}\cdots X_u^{i_u}$ for $(i_1,\dots,i_u)\in M\cap\bZ_{\geq 0}^u$. For $A\subseteq\bF_q^u$, we let $\bF_q[X_1,\dots,X_u]^M_A\subseteq\bF_q[X_1,\dots,X_u]^M$ denote the subspace of polynomials that have roots at all points in $A$. For a polynomial $f\in\bF_q[X_1,\dots,X_u]$ and a subset $E\subseteq\bF_q^u$, we let $\evl_E(f):=(f(x))_{x\in E}\in\bF_q^E$ denote the list of evaluations of $f$ at all points in $E$.
\end{definition}

For $E\subseteq\bF_q$ and $k\in\bN$, the code $\evl_E(\bF_q[X]^{[0,k)})$ has parameters $[|E|,k,|E|-k+1]_q$, and is commonly called a \emph{Reed-Solomon code}.

The following two lemmas are well known, but for completeness we include proofs.
They show that the dual of (punctured) Reed-Solomon codes have a similar Reed-Solomon structure, which is useful for the construction of CSS codes.

\begin{lemma}
  \label{lem:RSdual}
  For every $E\subseteq\bF_q$, every nonnegative integer $\ell\leq n:=|E|$, and every vector of nonzero coefficients $\beta\in(\bF_q^*)^E$, there exists another vector $\beta'\in(\bF_q^*)^E$ such that
  \begin{equation*}
    (\beta*\evl_E(\bF_q[X]^{[0,\ell)}))^\perp = \beta' * \evl_E(\bF_q[X]^{[0,n-\ell)}).
  \end{equation*}
\end{lemma}
\begin{proof}
  By definition $\beta*\evl_E(\bF_q[X]^{[0,n-1)})$ is a length-$n$ code of dimension $n-1$, so there exists some nonzero vector $\beta'\in(\beta*\evl_E(\bF_q[X]^{[0,n-1)}))^\perp$. Assume for a contradiction that some $y\in E$ has $\beta'_y=0$. Fixing any $y'\in E$ with $\beta'_{y'}\neq 0$, then the polynomial $h(X)=\prod_{x\in E\setminus\{y,y'\}}(X-x)\in\bF_q[X]^{[0,n-1)}$ is nonzero at $y,y'$ and zero at all other points in $E$. Therefore $\beta'\cdot(\beta*\evl(h))=\beta'_{y'}\beta_{y'}h(y')\neq 0$, contradicting the assumption that $\beta'\in(\beta*\evl_E(\bF_q[X]^{[0,n-1)}))^\perp$. Therefore we must indeed have every $\beta'_y\neq 0$, that is, $\beta'\in(\bF_q^*)^n$.

  Now for every $f\in\bF_q[X]^{[0,\ell)}$ and $g\in\bF_q[X]^{[0,n-\ell)}$, then
  \begin{equation*}
    (\beta*\evl_E(f))\cdot(\beta'*\evl_E(g)) = \beta'\cdot(\beta*\evl_E(f)*\evl_E(g)) = \beta'\cdot(\beta*\evl_E(fg)) = 0,
  \end{equation*}
  where the final equality above holds because $\beta'\in(\beta*\evl_E(\bF_q[X]^{[0,n-1)}))^\perp$ and $fg\in\bF_q[X]^{[0,n-1)}$. Thus the code $(\beta*\evl_E(\bF_q[X]^{[0,\ell)}))^\perp$ contains $\beta'*\evl_E(\bF_q[X]^{[0,n-\ell)})$, and these two codes have the same dimension $n-\ell$, so they are equal.
\end{proof}

\begin{lemma}
  \label{lem:puncshort}
  For a code $C\subseteq\bF_q^n$ and a subset $A\subseteq[n]$, then
  \begin{equation*}
    (C|_{[n]\setminus A})^\perp = (C^\perp\cap(\{0\}^A\times\bF_q^{[n]\setminus A}))|_{[n]\setminus A}.
  \end{equation*}
\end{lemma}
\begin{proof}
  If $c\in(C|_{[n]\setminus A})^\perp$, then by definition $(c,0^A)\in\bF_q^n$ is orthogonal to every codeword in $C$, so $(c,0^A)\in C^\perp\cap(\{0\}^A\times\bF_q^{[n]\setminus A})$, and thus $c\in(C^\perp\cap(\{0\}^A\times\bF_q^{[n]\setminus A}))|_{[n]\setminus A}$.

  For the opposite inclusion, for every $(c,0^A)\in C^\perp\cap(\{0\}^A\times\bF_q^{[n]\setminus A})$, then by definition $c\in\bF_q^{[n]\setminus A}$ is orthogonal to every codeword in $C|_{[n]\setminus A}$, so $c\in(C|_{[n]\setminus A})^\perp$, as desired.
\end{proof}

The quantum codes we will consider are subsystem codes, specifically constructed via products, as defined below.

\begin{definition}
  \label{def:subsystemcode}
  A \emph{quantum (CSS)\footnote{All quantum codes we consider in this paper are CSS codes, so we often omit the ``CSS'' quantifier.} subsystem code of length $n$ over $\bF_q$} consists of a pair $C=(C_X,C_Z)$ of classical codes $C_X,C_Z\subseteq\bF_q^n$. The \emph{dimension} of $C$ is
  \begin{equation*}
    k := \dim(C_Z+C_X^\perp)-\dim(C_X^\perp) = \dim(C_Z)-\dim(C_Z\cap C_X^\perp),
  \end{equation*}
  and the \emph{distance $d$} of $C$ is the minimum $d=\min\{d_X,d_Z\}$ of the \emph{$X$-distance~$d_X$} and \emph{$Z$-distance~$d_Z$} defined by
  \begin{align*}
    d_X &:= \min_{c\in(C_X+C_Z^\perp)\setminus C_Z^\perp}|c| \\
    d_Z &:= \min_{c\in(C_Z+C_X^\perp)\setminus C_X^\perp}|c|.
  \end{align*}
  We summarize these parameters by saying that $C$ is a $[[n,k,d]]_q$ subsystem code.

  If $C_X^\perp\subseteq C_Z$, we say that $C$ is a \emph{quantum (CSS) non-subsystem code}.\footnote{Such codes are traditionally simply called CSS codes in the literature; in this paper we take more care to distinguish them from our less standard notion of CSS subsystem codes.}

  For $u\in\bN$ and CSS non-subsystem codes $(C^i=(C^i_X,C^i_Z))_{i\in[u]}$ over $\bF_q$, the \emph{subsystem product code} $C=(C_X,C_Z)=\bigotimes_{i\in[u]}C^i$ is given by
  \begin{align*}
    C_X &= \bigotimes_{i\in[u]}C^i_X \\
    C_Z &= \bigotimes_{i\in[u]}C^i_Z.
  \end{align*}
\end{definition}

The notion of subsystem product in \Cref{def:subsystemcode} was previously defined and studied in \cite{zeng_minimal_2020,golowich_near-asymptotically-good_2025}. In particular, \cite{golowich_near-asymptotically-good_2025} showed the following analogue of the K\"{u}nneth formula for subsystem products:

\begin{proposition}[\cite{golowich_near-asymptotically-good_2025}]
  \label{prop:spkunneth}
  For $u\in\bN$, let $(C^i=(C^i_X,C^i_Z))_{i\in[u]}$ be non-subsystem CSS codes over~$\bF_q$, and let $C=(C_X,C_Z)=\bigotimes_{i\in[u]}C^i$ be the subsystem product. 
  If each $C^i$ has dimension $k_i$, then~$C$ is a subsystem code of dimension $k=\prod_{i\in[u]}k_i$, and there is an isomorphism
  \begin{equation*}
    (C_Z+C_X^\perp)/C_X^\perp \cong \bigotimes_{i\in[u]}(C^i_Z/{C^i_X}^\perp)
  \end{equation*}
  Furthermore, for codewords $a_i\in C^i_Z$ for $i\in[u]$, the isomorphism above maps
  \begin{equation*}
    \bigotimes_{i\in[u]}a_i+C_X^\perp \mapsfrom \bigotimes_{i\in[u]}(a_i+{C^i_X}^\perp).
  \end{equation*}
\end{proposition}


While we use the CSS formalism in \Cref{def:subsystemcode}, subsystem codes are often described using the more general stabilizer formalism. 
In this language, the Pauli operators $X^c$ for $c\in C_X^\perp$ and~$Z^c$ for $c\in C_Z^\perp$ are called \emph{gauge operators}. 
Similarly, the operators $X^c$ for $c\in C_X^\perp\cap C_Z$ and $Z^c$ for $c\in C_Z^\perp\cap C_X$ are \emph{stabilizers}, and the operators $X^c$ for $c\in(C_Z+C_X^\perp)\setminus C_X^\perp$ and $Z^c$ for $c\in(C_X+C_Z^\perp)\setminus C_Z^\perp$ are \emph{logical operators}.

The following type of encoding map will be useful for relating a logical operation to its physical implementation in a fault-tolerance scheme.

\begin{definition}
  \label{def:CSSenc}
  For a $[[n,k]]_q$ CSS subsystem code $C=(C_X,C_Z)$ with
  \begin{align*}
    k &:= \dim(C_Z)-\dim(C_Z\cap C_X^\perp) \\
    k' &:= \dim(C_X^\perp)-\dim(C_Z\cap C_X^\perp),
  \end{align*}
  a \emph{CSS encoding map} is a $\bF_q$-linear isomorphism
  \begin{equation*}
    \Enc:\bF_q^k\oplus\bF_q^{k'}=\bF_q^{k+k'}\xrightarrow{\sim} (C_Z+C_X^\perp)/(C_Z\cap C_X^\perp) = C_Z/(C_Z\cap C_X^\perp)\oplus C_X^\perp/(C_Z\cap C_X^\perp)
  \end{equation*}
  that satisfies $\Enc(\bF_q^k,0)=C_Z/(C_Z\cap C_X^\perp)$ and $\Enc(0,\bF_q^{k'})=C_X^\perp/(C_Z\cap C_X^\perp)$.
  
  This map has a naturally associated isometry
  \begin{equation*}
    \Enc:\bC^{\bF_q^{k+k'}}\rightarrow\bC^{\bF_q^n}
  \end{equation*}
  given by $\Enc\ket{x} = \ket{\Enc(x)}$, which we also view as a channel
  \begin{equation*}
    \Enc:\bC^{\bF_q^{k+k'}\times\bF_q^{k+k'}}\rightarrow\bC^{\bF_q^n\times\bF_q^n}
  \end{equation*}
  by $\Enc(\rho) = \Enc\rho\Enc^\dagger$. Note that the overloading of $\Enc$ here will be resolved by context, specifically by the type of the argument (i.e.~$x$ vs $\ket{x}$ vs $\rho$).
\end{definition}

In \Cref{def:CSSenc}, the quantum code $C$ encodes $k$ logical qudits used to store a message, along with $k'$ additional ``gauge'' qudits that may not be protected from low-weight corruptions. Rather, the gauge qudits provide additional degrees of freedom that will be helpful for measuring syndromes to perform error correction.

We will use encoding maps for subsystem product codes arising from tensor products of encoding maps for the underlying factor codes, as defined below.

\begin{definition}
  \label{def:prodenc}
  For $u\in\bN$, for $i\in[u]$ let $C^i=(C^i_X,C^i_Z)$ be a CSS non-subsystem code with CSS encoding map $\Enc^i:\bF_q^{k_i}\rightarrow C^i_Z/{C^i_X}^\perp$. A \emph{product encoding map} is a CSS encoding map $\Enc:\bF_q^k\oplus\bF_q^{k'}\xrightarrow{\sim}(C_Z+C_X^\perp)/(C_Z\cap C_X^\perp)$ for the subsystem product code $C=\bigotimes_{i\in[u]}C^i$ such that for every $(x_i\in\bF_q^{k_i})_{i\in[u]}$, letting $x=\bigotimes_{i\in[u]}x_i\in\bigotimes_{i\in[u]}\bF_q^{k_i}=\bF_q^k$, then
  \begin{equation}
    \label{eq:prodenc}
    \Enc(x,0^{k'}) = \bigotimes_{i\in[u]}\Enc^i(x).
  \end{equation}
  Here we implicitly use the natural isomorphism
  \begin{equation*}
    \bigotimes_{i\in[u]}C^i_Z/{C^i_X}^\perp \cong C_Z/(C_Z\cap C_X^\perp).
  \end{equation*}
\end{definition}

Note that product encoding maps always exist, as \Cref{eq:prodenc} specifies a well-defined linear map on the domain $\bF_q^k$, which we may extend arbitrarily to a linear map on $\bF_q^k\oplus\bF_q^{k'}$.

\subsection{Quantum Circuits and Fault-Tolerance}
In this section, we provide basic definitions regarding quantum circuits and fault-tolerance. Our definitions are inspired by those in \cite{nguyen_quantum_2025,he_composable_2025}, and most closely resemble those in the subsequent work \cite{golowich_constant-overhead_2025}.
The main contribution of this section is \Cref{def:faulttol}, which formally defines what it means for a logical operation to be fault-tolerant.

\subsubsection{Gates}
In this section, we present the gates we will use in this paper.

\begin{definition}
  A \emph{gate} acting on $m$ $q$-dimensional qudits is a quantum channel $G:\bC^{\bF_q^m\times\bF_q^m}\rightarrow\bC^{\bF_q^m\times\bF_q^m}$. 
  A \emph{unitary gate} is a gate of the form $G(\rho)=U\rho U^\dagger$ for some unitary $U\in\bC^{\bF_q^m\times\bF_q^m}$.
\end{definition}

We now present the gate set that we will use. Here (and throughout the paper) we often refer to unitary gates $G(\rho)=U\rho U^\dagger$ by the unitary $U$, as opposed to the channel $G$.

\begin{definition}
  \label{def:gates}
  For a prime power $q=p^e$, we define the following unitary $q$-ary gates:
  \begin{itemize}
  \item The Hadamard gate $\gH\in\bC^{\bF_q\times\bF_q}$ and its inverse $\gH^{-1}=\gH^\dagger\in\bC^{\bF_q\times\bF_q}$, where
    \begin{align*}
      \gH\ket{x} &= \frac{1}{\sqrt{q}}\sum_{z\in\bF_q}e^{2\pi i\tr_{\bF_q/\bF_p}(xz)/p}\ket{z}.
    \end{align*}
  \item For $a\in\bF_q$, the Pauli gates $\gX^a,\gZ^a\in\bC^{\bF_q\times\bF_q}$ given by
    \begin{align*}
      \gX^a\ket{x} &= \ket{x+a} \\
      \gZ^a\ket{x} &= e^{2\pi i \tr_{\bF_q/\bF_p}(ax)/p}\ket{x}.
    \end{align*}
  \item For $a\in\bF_q$, the gate $\gCX^a\in\bC^{\bF_q^2\times\bF_q^2}$ (also sometimes called $\opn{CNOT}^a$) given by
    \begin{align*}
      \gCX^a\ket{x_1,\;x_2} &= \ket{x_1,\;ax_1+x_2}.
    \end{align*}
  \item For $a\in\bF_q$, the gate $\gCCX^a\in\bC^{\bF_q^3\times\bF_q^3}$ (also sometimes called $\opn{Toffoli}^a$) given by
    \begin{align*}
      \gCCX^a\ket{x_1,\;x_2,\;x_3} &= \ket{x_1,\;x_2,\;ax_1x_2+x_3}.
    \end{align*}
  \end{itemize}
  
  We also define the following non-unitary $q$-ary gates:
  \begin{itemize}
  \item The initialization gates $\gInit_X,\gInit_Z:\bC^{\bF_q\times\bF_q}\rightarrow\bC^{\bF_q\times\bF_q}$ given by
    \begin{align*}
      \gInit_X(\rho) &= \ket{+}\bra{+} = \gH\ket{0}\bra{0}\gH^\dagger \\
      \gInit_Z(\rho) &= \ket{0}\bra{0}
    \end{align*}
  \item For $m_1,m_2\in\bN$ with $m=\max\{m_1,m_2\}$ and a function $f:\bF_q^{m_1}\rightarrow\bF_q^{m_2}$, the \emph{classical function gates} $\gCO_{X,f},\gCO_{Z,f}:\bC^{\bF_q^m\times\bF_q^m}\rightarrow\bC^{\bF_q^m\times\bF_q^m}$ given as follows. For $\zeta\in\bC^{\bF_q^m}$, we define superoperators $\gCO_{X,f}^\zeta,\gCO_{Z,f}^\zeta:\bC^{\bF_q^m\times\bF_q^m}\rightarrow\bC^{\bF_q^m\times\bF_q^m}$, called \emph{$\zeta$-postselected classical function gates}, by
    \begin{align*}
      \gCO_{Z,f}^\zeta(\rho) &= \sum_{x\in\bF_q^m}\zeta_x\cdot\ket{f(x|_{[m_1]}),0^{m-m_2}}\bra{x}\rho\ket{x}\bra{f(x|_{[m_1]}),0^{m-m_2}} \\
      \gCO_{X,f}^\zeta(\rho) &= \gH^{\otimes m}\gCO_{Z,f}^\zeta\left((\gH^\dagger)^{\otimes m}\rho\gH^{\otimes m}\right)(\gH^\dagger)^{\otimes m}.
    \end{align*}
    We then define $\gCO_{X,f}=\gCO_{X,f}^{\vec{1}}$ and $\gCO_{Z,f}=\gCO_{Z,f}^{\vec{1}}$, where $\vec{1}\in\bF_q^m$ denotes the all-1s vector.
  \end{itemize}
  For the gates above, we sometimes replace the super/subscript with `$*$' to denote the set of all possible values, so that for instance $\{\gH^*,\gX^*,\gZ^*\}$ denotes the set containing $\gH,\gH^\dagger$ and all Pauli gates $\gX^a$ and $\gZ^a$ for $a\in\bF_q$. Note that $\gCO_*$ is an infinite set, as the number $m$ of qudits can be arbitrarily large.
\end{definition}


The following standard fact is a direct consequence of \Cref{def:gates}. Below, we let $\gCX^a_{i,j}$ denote the $\gCX^a$ gate with control qudit $i$ and target qudit $j$.

\begin{fact}[Well known]
  \label{fact:Hconj}
  For $a\in\bF_q$, we have:
  \begin{itemize}
  \item $\gH\gZ^a\gH^\dagger=\gX^{-a}$
  \item $\gH^\dagger\gZ^a\gH=X^a$
  \item $\gH^{\otimes 2}\gCX^a_{1,2}(\gH^\dagger)^{\otimes 2} = \gCX^{-a}_{2,1}$.
  \end{itemize}
\end{fact}

\begin{remark}
  We use the classical function gates $\gCO_{X,f}$ and $\gCO_{Z,f}$ in \Cref{def:gates} to capture the notion of noiseless classical side-computation within an otherwise noisy quantum circuit. 
  Specifically, they perform respectively either $X$- or $Z$-basis measurements on some number $m=\max\{m_1,m_2\}$ of qudits, and then noiselessly apply some classical function $f:\bF_q^{m_1}\rightarrow\bF_q^{m_2}$ to the first $m_1$ qudits of the $m$-qudit measurement outcome, and print the output in the first $m_2$ of the $m$ qudits. Setting $f$ to be the identity function on $\bF_q^m$ recovers ordinary $X$- or $Z$-basis measurements (with no additional classical processing).
  
  The $\zeta$-postselected versions $\gCO_{X,f}^\zeta,\gCO_{Z,f}^\zeta$ similarly perform measurements and then apply $f$, but they additionally apply postselection to the measurement outcome according to the amplitudes given by $\zeta$.

  We emphasize that because gates in $\gCO_*$ perform $X$- or $Z$-basis measurements, the classical function $f$ \emph{cannot} be queried in superposition. This definition is in contrast to oracles often studied in quantum complexity theory that can be queried in superposition.
\end{remark}


We will use the following standard notion of multi-qudit Pauli operators, which are simply defined as tensor products of single-qudit Pauli gates:

\begin{definition}
  An \emph{$n$-qudit Pauli operator $\bF_q$} is an operator $\gX^a\gZ^b:=\bigotimes_{i\in[n]}\gX^{a_i}\gZ^{b_i}\in\bC^{\bF_q^n\times\bF_q^n}$ for $a,b\in\bF_q^n$. An \emph{$n$-qudit Pauli superoperator} is a map of the form $\rho\mapsto P\rho P'$ for $n$-qudit Paulis~$P$ and $P'$.
\end{definition}

\Cref{lem:paulidecomp} below shows the well-known result that every operator and superoperator has a unique decomposition into a linear combination of Paulis. We include a proof for completeness, as this lemma is often only stated over $\bF_2$.

\begin{lemma}[Well known]
  \label{lem:paulidecomp}
  Every operator $A\in\bC^{\bF_q^n\times\bF_q^n}$ (resp.~superoperator $A:\bC^{\bF_q^n\times\bF_q^n}\rightarrow\bC^{\bF_q^n\times\bF_q^n}$) has a unique decomposition $A=\sum_P\alpha_PP$ as a linear combination of $n$-qudit Pauli operators $P$ (resp.~Pauli superoperators $P$), for some coefficients $\alpha_P\in\bC$. Furthermore, $\supp(A)=\bigcup_{P:\alpha_P\neq 0}\supp(P)$.
\end{lemma}
\begin{proof}
  There are $q^{2n}=\dim_{\bC}(\bC^{\bF_q^n\times\bF_q^n})$ Pauli operators, and they are all orthogonal with respect to the inner product $(A,B)\mapsto\tr(AB^\dagger)$. Thus Pauli operators form a basis for $\bC^{\bF_q^n\times\bF_q^n}$.

  Similarly, viewing superoperators as operators on $2n$ qudits via the isomorphism $\{A:\bC^{\bF_q^n\times\bF_q^n}\rightarrow\bC^{\bF_q^n\times\bF_q^n}\}\cong\bC^{\bF_q^{2n}\times\bF_q^{2n}}$, then we apply the same argument as above. That is, there are $q^{4n}=\dim_{\bC}(\bC^{\bF_q^{2n}\times\bF_q^{2n}})$ Pauli superoperators, and they are all orthogonal with respect to the inner product $(A,B)\mapsto\tr(AB^\dagger)$; here $AB^\dagger$ refers to the composition of $A$ and $B^\dagger$ as superoperators, or equivalently to their product as operators in $\bC^{\bF_q^{2n}\times\bF_q^{2n}}$. Thus Pauli superoperators form a basis for the space of all superoperators.

  For both operators and superoperators, the claim that $\supp(A)=\bigcup_{P:\alpha_P\neq 0}\supp(P)$ then follows from the fact that every $n$-qudit Pauli is a tensor product of $n$ single-qudit Paulis, and the identity is one of the single-qudit Paulis. Specifically, for a given qudit $i\in[n]$, if every $P$ with $\alpha_P\neq 0$ has the identity factor at position $I$, then we can factor out this identity to write $A=A_{[n]\setminus\{i\}}\otimes I_{\{i\}}$, so that $i\notin\supp(A)$. Conversely, if $i\notin\supp(A)$, then by definition we can write $A=A_{[n]\setminus\{i\}}\otimes I_{\{i\}}$, and hence every $P$ with $\alpha_P\neq 0$ has the identity factor at position $I$. Thus indeed $i\in\supp(A)$ iff $i\in\supp(P)$ for some $P$ with $\alpha_P\neq 0$.
\end{proof}

\subsubsection{Circuits}
\label{sec:circuits}
In this section, we define quantum circuits and faults.

\npb{Consider making quantum circuits and faults two definitions since this is pretty long and overwhelming. idk if this is referenced in other places, so wanted to check w you first.}
\begin{definition}
  \label{def:circuit}
  A \emph{quantum circuit} $\cQ=(Q_1,\dots,Q_T;N_{\mathrm{in}},N_{\mathrm{out}})$ acting on a set $N$ of $q$-dimensional qudits is a sequence $Q_1,\dots,Q_T$ of superoperators acting on qudits $N$, along with specified subsets $N_{\mathrm{in}},N_{\mathrm{out}}\subseteq N$. 
  We require that $Q_1$ applies $\gInit_X$ or $\gInit_Z$ to every qudit in $N\setminus N_{\mathrm{in}}$.\footnote{\label{footnote:initall}
    Unless explicitly stated otherwise. In particular, when describing a gadget, we will sometimes specify the initialization gate $\gInit_\alpha$ at a timestep $t>1$ for qudits in $N\setminus N_{\mathrm{in}}$ that are not used until time $t$. In such cases, it should be implicitly assumed that $\gInit_\alpha$ is also applied in timestep $1$.

    For circuits where we do explicitly state that $Q_1$ need not apply $\gInit_X$ or $\gInit_Z$ to every qudit in $N\setminus N_{\mathrm{in}}$, we will ensure that $\cQ$ is constructed so that the output $\cQ(\rho)$ still does not depend on the initial state $\rho'$ of the qudits in $N\setminus N_{\mathrm{in}}$. Specifically, we will compile certain logical circuits $\cQ$ to first perform swap gates to move qudits certain qudits in $N\setminus N_{\mathrm{in}}$ to a separate block of qudits before applying $\gInit_X$ or $\gInit_Z$.
  }
  We let $\cQ(\cdot):\bC^{\bF_q^{N_{\mathrm{in}}}\times\bF_q^{N_{\mathrm{in}}}}\rightarrow\bC^{\bF_q^{N_{\mathrm{out}}}\times\bF_q^{N_{\mathrm{out}}}}$ denote the superoperator
  \begin{equation*}
    \cQ(\rho) = \tr_{N\setminus N_{\mathrm{out}}}\left(Q_T\circ\cdots\circ Q_1\left(\rho\otimes\rho'\right)\right),
  \end{equation*}
  where $\rho'\in\bC^{\bF_q^{N\setminus N_{\mathrm{in}}}\times\bF_q^{N\setminus N_{\mathrm{in}}}}$ is a density operator, whose value does not affect $\cQ(\rho)$ by the definition of $Q_1$.
  We say $\cQ$ uses a gate set $\cG$ if each $Q_t$ can be decomposed as a tensor product of gates in~$\cG$ acting on disjoint sets of $q$-dimensional qudits.
  We also say $\cQ$ uses time $T$, and when $q$ is clear from context we say that~$\cQ$ uses space $N$ (or as a shorthand, space $|N|$).

  A \emph{fault} on the circuit $\cQ$ is a sequence $\cF=(F_1,\dots,F_T)$ of superoperators $F_t$ acting on qudits~$N$.
  We say~$\cF$ is a \emph{Pauli fault} if each $F_t$ is a Pauli superoperator, that is, $F_t(\rho)=F_t\rho F_t'$ for some Paulis~$F_t,F_t'$ acting on qudits $N$. The \emph{support} $\supp(\cF)\subseteq N\times[T]$ is defined by $\supp(\cF)=\supp(F_1)\sqcup\cdots\sqcup\supp(F_T)$. 
  
  The \emph{$\cF$-corrupted circuit} $\cQ[\cF]$ is defined by
  \begin{equation*}
    \cQ[\cF] = (Q_1,F_1,Q_2,F_2,\dots,Q_T,F_T;N_{\mathrm{in}},N_{\mathrm{out}}).
  \end{equation*}

  Let $n_{\gCO}$ denote the number of $\gCO_*$ gates in $\cQ$, and let the $i$th such gate $\gCO_{\alpha_i,f_i}$ act on $m_i$ qudits. Let $M=\bigsqcup_{i\in[n_{\gCO}]}\bF_q^{m_i}$. Then a \emph{postselection} on the circuit $\cQ$ is a vector $\zeta=(\zeta_1,\dots,\zeta_{n_{\gCO}})\in\bC^M$. The \emph{$\zeta$-postselected circuit} $\cQ[\zeta]$ is defined by replacing the $i$th $\gCO_*$ gate in $\cQ$ with its $\zeta_i$-postselected version $\gCO_{\alpha_i,f_i}^{\zeta_i}$ for every $i\in[n_{\gCO}]$, and otherwise leaving $\cQ$ unchanged.

  It will always be clear from context whether an argument to $\cQ[\cdot]$ is a fault or postselection. When both are present, we apply the postselection and then the fault, so that $\cQ[\cF,\zeta]=(\cQ[\zeta])[\cF]$. We also let $\cQ[*]=\{\cQ[\zeta]:\zeta\in\bC^M\}$ denote the set of all postselected versions of $\cQ$.
\end{definition}

We will typically consider quantum circuits using gate set $\cG=\{\gH^*,\gX^*,\gZ^*,\gCX^*,\gCCX^*,\gInit_*,\gCO_*\}$. Specifically, the following result from \cite{aharonov_fault-tolerant_1997} shows that the gates $\{\gH^*,\gX^*,\gZ^*,\gCX^*,\gCCX^*,\gInit_*\}$ are sufficient to achieve universal quantum computation over appropriate fields. We will use the gates $\gCO_*$ for decoding in our error-correction gadget (see \Cref{sec:errcorr}), as well as for simulating Hadamard gates over some $\bF_q$ using qudits over another $\bF_{q'}$ (see \Cref{sec:inclog}).

\begin{lemma}[\cite{aharonov_fault-tolerant_1997}]
  \label{lem:universal}
  For every prime $q\geq 5$, the gate set $\cG=\{\gH^*,\gX^*,\gZ^*,\gCX^*,\gCCX^*,\gInit_*\}$ over $q$-dimensional qudits is \emph{universal}. That is, for every unitary $U\in\bC^{\bF_q^n\times\bF_q^n}$ and every $\epsilon>0$, there exists a circuit $\cQ$ using gate set $\cG$ with $n$ input and output qudits such that for every density operator $\rho\in\bC^{\bF_q^n\times\bF_q^n}$, then
  \begin{equation}
    \label{eq:universal}
    \|\cQ(\rho)-U\rho U^\dagger\|\leq\epsilon.
  \end{equation}
  Furthermore, $\cQ$ only needs to call $\gInit_*$ in the first timestep.
\end{lemma}

\begin{remark}
  We intentionally do not specify the norm in \Cref{eq:universal}, as the result holds under all norms. Also note that \cite{aharonov_fault-tolerant_1997} prove universality for a slightly different gate set, which includes the swap gate $\ket{x_1}\ket{x_2}\mapsto\ket{x_2}\ket{x_1}$ and the scalar multiplication gates $\ket{x}\mapsto\ket{ax}$ for each $a\in\bF_q$. However, these gates can be implemented using a constant number of $\gCX^*$ gates along with a constant number of ancilla qudits initialized to $\ket{0}$.
\end{remark}

\subsubsection{Fault-Tolerant Gadgets}
Our goal in this paper is to construct \emph{fault-tolerant gadgets}, which we define in this section to be circuits that yield the correct output whenever the fault avoids certain bad sets. We begin by formally defining a notion of such bad sets below.

\npb{again, would be more readable if we break this definition up. this also allows for a more natural flow when talking about deconstructing general faults into paulis. but leaving it up to you as fixing references might be a pain}

\begin{definition}
  \label{def:Eavoid}
  For a set $N$ and a family of subsets $\cE\subseteq 2^N$ called \emph{bad sets}, we say that a set $S\subseteq N$ is \emph{$\cE$-avoiding} if no element of $\cE$ is contained in $S$, that is, $E\not\subseteq S$ for every $E\in\cE$.

  We extend this definition to Pauli faults in the natural way, so that a Pauli fault $\cF$ on a circuit using space $\subseteq N$ and time $\leq T$ is \emph{$\cE$-avoiding} for $\cE\subseteq 2^{N\times[T]}$ if $\supp(\cF)$ is $\cE$-avoiding.

  We then say that a fault $\cF=(F_1,\dots,F_T)$ is $\cE$-avoiding if each $F_t$ can be expressed as a linear combination $F_t=\sum_{i=1}^{m_t}\alpha_{t,i}F_{t,i}$ of Pauli superoperators $F_{t,1},\dots,F_{t,m_t}$ for coefficients $\alpha_{t,i}\in\bC$ such that the Pauli fault $(F_{1,i_1},\dots,F_{T,i_T})$ is $\cE$-avoiding for every tuple $(i_1,\dots,i_T)\in[m_1]\times\cdots\times[m_T]$.

  For families $\cE_1\subseteq 2^{N_1}$, $\cE_2\subseteq 2^{N_2}$, we let $\cE_1\sqcup\cE_2\subseteq 2^{N_1\sqcup N_2}$ denote the disjoint union of~$\cE_1$ and~$\cE_2$ on separate sets of underlying elements. Therefore for $\cE\subseteq 2^N$ and $T\in\bN$, we write $\cE^{\sqcup T}\subseteq 2^{N\times[T]}$ to denote $\cE\sqcup\cdots\sqcup\cE$ (with $T$ copies of $\cE$).
\end{definition}

\npb{isn't this remark redundant? everything is linear and we have a basis and that fact is already used in the definition.}\lzg{Hmm I think the point of the remark is that there's an equivalent formulation of the definition that does not rely on the basis, so the definition is actually basis-independent? (I've updated the wording to try to make this message more clear.)}
\begin{remark}
  \label{remark:paulibasis}
  At first glance, it may seem restrictive that the definition of a $\cE$-avoiding fault involves Paulis, as we ultimately want to protect against general noise and not simply Pauli noise. 
  However, recall from \Cref{lem:paulidecomp} that every superoperator can be expressed as a linear combination of Pauli superoperators, with supports inside that of the original superoperator. 
  Therefore we can equivalently say that $\cF=(F_1,\dots,F_T)$ is a $\cE$-avoiding fault if each $F_t$ can be expressed as a linear combination $F_t=\sum_{i=1}^{m_t}\alpha_{t,i}F_{t,i}$ of superoperators $F_{t,1},\dots,F_{t,m_t}$ for coefficients $\alpha_{t,i}\in\bC$ such that the support of the fault $(F_{1,i_1},\dots,F_{T,i_T})$ is $\cE$-avoiding for every tuple $(i_1,\dots,i_T)\in[m_1]\times\cdots\times[m_T]$.
  Therefore \Cref{def:Eavoid} is actually basis-independent, despite being stated in terms of the Pauli basis for superoperators.
\end{remark}


To specify a fault-tolerant gadget using a subsystem code $C$, we will decorate the code with an encoding map $\Enc$, a set $\Gamma$ of allowed states of the gauge qudits, and a family $\cE$ of bad sets of errors, as stated in \Cref{def:deccode} below.

\lzg{TODO: break up definition and add more commentary}

\begin{definition}
  \label{def:deccode}
  Let $C=(C_X,C_Z)$ be a $[[n,k]]_q$ CSS subsystem code with CSS encoding map $\Enc:\bF_q^{k+k'}\rightarrow(C_Z+C_X^\perp)/(C_Z\cap C_X^\perp)$. Let $\Gamma\subseteq\bC^{\bF_q^{k'}\times\bF_q^{k'}}$ be a set of operators, which we call \emph{allowed states} of the gauge qudits of $C$. Let $\cE\subseteq 2^{[n]}$ be a family of subsets of qudits, which we call \emph{bad sets} for $C$. We say the data $D=(C,\Enc,\Gamma,\cE)$ forms a \emph{decorated CSS subsystem code}. If $k'=0$, then $\Gamma$ is trivial, so we say $D=(C,\Enc,\cE)$ forms a \emph{decorated CSS non-subsystem code}. We let $\emptyset$ denote the trivial decorated code with $n=k=0$.

  For an operator $\rho_0\in\bC^{\bF_q^{k+k'}\times\bF_q^{k+k'}}$, we say an operator $\sigma\in\bC^{\bF_q^n\times\bF_q^n}$ is a \emph{Pauli $\cE$-deviation of $\Enc(\rho_0)$} if there exists a Pauli superoperator $F:\bC^{\bF_q^n\times\bF_q^n}\rightarrow\bC^{\bF_q^n\times\bF_q^n}$ such that $\sigma\propto F\circ\Enc(\rho_0)$, and such that $\supp(F)$ is $\cE$-avoiding.

  For a set $R\subseteq\bC^{\bF_q^{k+k'}\times\bF_q^{k+k'}}$, we say an operator $\sigma\in\bC^{\bF_q^n\times\bF_q^n}$ is a \emph{Pauli $\cE$-deviation of $\Enc(R)$} if there exists some $\rho_0\in R$ such that $\sigma$ is a Pauli $\cE$-deviation of $\Enc(\rho_0)$. In particular, for a set $\bar{\cO}$ of superoperators $\bar{O}:\bC^{\bF_q^k\times\bF_q^k}\rightarrow\bC^{\bF_q^k\times\bF_q^k}$, we say that $\sigma$ is a Pauli $\cE$-deviation of $\Enc(\bar{\cO}(\rho)\otimes\Gamma)$ if there exists $\bar{O}\in\bar{\cO}$ and $\gamma\in\Gamma$ such that $\sigma$ is a Pauli $\cE$-deviation of $\Enc(\bar{O}(\rho)\otimes\gamma)$.

  We say $\sigma$ is a \emph{$\cE$-deviation of $\Enc(\rho_0)$} (resp.~$\Enc(R)$) if $\sigma$ can be expressed as a linear combination $\sigma=\sum_{i=1}^m\sigma_i$ of Pauli $\cE$-deviations $\sigma_1,\dots,\sigma_m$ of $\Enc(\rho_0)$ (resp.~$\Enc(R)$).

\end{definition}

\begin{remark}
  \label{remark:paulibasis2}
  Similarly as described in \Cref{remark:paulibasis} regarding \Cref{def:Eavoid}, Pauli noise is not intrinsic to \Cref{def:deccode}. Specifically, by \Cref{lem:paulidecomp}, we can equivalently say that $\sigma$ is a $\cE$-deviation of $\Enc(\rho_0)$ if $\sigma$ can be expressed as a linear combination $\sigma=\sum_{i=1}^m\sigma_i$ of operators $\sigma_i$ satisfying $\tr_{[N]\setminus[n]}(\sigma_i)\propto F\circ\Enc(\rho_0)$ for some superoperator $F$ such that $\supp(F)$ is $\cE$-avoiding. An analogous Pauli-free formulation also holds for $\cE$-deviations of $\Enc(R)$.
\end{remark}

In \Cref{def:deccode}, it is helpful to think of $\Gamma$ to be a set of density operators, though the definition also allows for non-physical operators in $\Gamma$. In both \Cref{def:Eavoid} and \Cref{def:deccode}, we will often take $\cE$ to be the family of all subsets of $[n]$ of size above some threshold $\lambda$, which we denote $2^{[n]}|_{\geq\lambda}$ as defined below.

\begin{definition}
  For a set $N$, a family of subsets $\cE\subseteq 2^N$, and a real number $\lambda$, we let
  \begin{equation*}
    \cE|_{\geq\lambda} = \{E\in\cE:|E|\geq\lambda\}
  \end{equation*}
  denote the subfamily of those sets of size $\geq\lambda$.
\end{definition}

It will often be helpful to view multiple disjoint codes as a single larger code, for which we use the notation below.

\begin{definition}
  \label{def:codedisun}
  For $[[n_i,k_i]]_q$ decorated CSS subsystem codes $D_i=(C_i,\Enc_i,\Gamma_i,\cE_i)$ for $i\in[2]$, we define a $[[n_1+n_2,\;k_1+k_2]]_q$ \emph{disjoint union decorated CSS subsystem code}
  \begin{equation*}
    D_1\sqcup D_2 = (C_1\sqcup C_2,\;\Enc_1\sqcup\Enc_2,\;\Gamma_1\sqcup\Gamma_2,\;\cE_1\sqcup\cE_2)
  \end{equation*}
  by
  \begin{align*}
    (C_1\sqcup C_2)_\alpha &= C_{1,\alpha}\oplus C_{2,\alpha} \subseteq \bF_q^{n_1}\oplus\bF_q^{n_2} \hspace{1em}\text{ for }\alpha\in\{X,Z\} \\
    \Enc_1\sqcup\Enc_2 &= \Enc_1\oplus \Enc_2:\bF_q^{k_1+k_1'}\oplus\F_q^{k_2+k_2'}\rightarrow (C_{1,Z}+C_{1,X}^\perp)/(C_{1,Z}\cap C_{1,X}^\perp) \\
    &\hspace{16em}\oplus(C_{2,Z}+C_{2,X}^\perp)/(C_{2,Z}\cap C_{2,X}^\perp) \\
    \Gamma_1\sqcup\Gamma_2 &= \{\gamma_1\otimes\gamma_2:\gamma_1\in\Gamma_1,\gamma_2\in\Gamma_2\}.
  \end{align*}
\end{definition}


We are now ready to present our main definition of a fault-tolerant gadget.
Such a gadget implements a logical operation $\bar{O}$ where the input and output are encoded into codes $D_{\mathrm{in}}$ and $D_{\mathrm{out}}$ respectively.
\begin{definition}
  \label{def:faulttol}
  For $\alpha\in\{\mathrm{in},\mathrm{out}\}$, let $D_\alpha=(C_\alpha,\Enc_\alpha,\Gamma_\alpha,\cE_\alpha)$ be a decorated $[[n_\alpha,k_\alpha]]_q$ subsystem code. 
  Let $\bar{\cO}$ be a set of superoperators with $k_{\mathrm{in}}$ input qudits and $k_{\mathrm{out}}$ output qudits. 
  Let $\cQ=(Q_1,\dots,Q_T;N_{\mathrm{in}},N_{\mathrm{out}})$ be a quantum circuit on qudit set $N$, with associated isomorphisms $N_{\mathrm{in}}\cong[n_{\mathrm{in}}]$, $N_{\mathrm{out}}\cong[n_{\mathrm{out}}]$, and a family of \emph{bad sets} $\cE_{\mathrm{run}}\subseteq 2^{N\times[T]}$. 
  Here all qudits have a fixed prime power dimension $q$.
  
  We say the data $(\cQ,\cE_{\mathrm{run}},D_{\mathrm{in}},D_{\mathrm{out}})$ provides a \emph{fault-tolerant gadget} for $\bar{\cO}$ if for every $\ell\in\bN$, every operator $\rho\in\bC^{\bF_q^{k_{\mathrm{in}}+\ell}\times\bF_q^{k_{\mathrm{in}}+\ell}}$, every $\cE_{\mathrm{in}}\sqcup [\ell]$-deviation $\sigma\in\bC^{\bF_q^{n_{\mathrm{in}}+\ell}\times\bF_q^{n_{\mathrm{in}}+\ell}}$ of $\Enc_{\mathrm{in}}\otimes I_\ell(\rho\otimes\Gamma_{\mathrm{in}})$, and every $\cE_{\mathrm{run}}$-avoiding fault $\cF$ and postselection $\zeta$ for $\cQ$ we have that $\cQ[\cF,\zeta]\otimes I_\ell(\sigma)\in\bC^{\bF_q^{n_{\mathrm{out}}+\ell}\times\bF_q^{n_{\mathrm{out}}+\ell}}$ is a $\cE_{\mathrm{out}}\sqcup [\ell]$-deviation of $(\Enc_{\mathrm{out}}\circ\bar{\cO})\otimes I_\ell(\rho\otimes\Gamma_{\mathrm{out}})$.\footnote{As a point of notation, here $\rho$ consists of a $k_{\mathrm{in}}$-qudit register and an $\ell$-qudit register (the ``external system'').
  In the expression $\Enc_{\mathrm{in}}\otimes I_\ell(\rho\otimes\Gamma_{\mathrm{in}})$, the channel $\Enc_{\mathrm{in}}$ acts on the $k_{\mathrm{in}}$-qudit register of $\rho$ along with $\Gamma_{\mathrm{in}}$, while the channel $I_\ell$ acts on the $\ell$-qudit register of $\rho$. 
  This slight misalignment of register orderings will unfortunately persist in our notation throughout, but the meaning will be clear as $I_\ell$ will always act on the $\ell$-qudit external system.
  Also note that the $n_{\mathrm{in}}$-qudit register of $\sigma\in\bC^{\bF_q^{n_{\mathrm{in}}+\ell}\times\bF_q^{n_{\mathrm{in}}+\ell}}$ corresponds to qudits $N_{\mathrm{in}}\cong[n_{\mathrm{in}}]$.} 

  The gadget is furthermore said to be \emph{mending} if for every $\rho\in\bC^{\bF_q^{k_{\mathrm{in}}+\ell}\times\bF_q^{k_{\mathrm{in}}+\ell}}$, every $[\ell]$-deviation $\sigma\in\bC^{\bF_q^{n_{\mathrm{in}}+\ell}\times\bF_q^{n_{\mathrm{in}}+\ell}}$ of $\Enc_{\mathrm{in}}\otimes I_\ell(\rho\otimes\Gamma_{\mathrm{in}})$, and every $\cE_{\mathrm{run}}$-avoiding fault $\cF$ and postselection $\zeta$ for $\cQ$, then $\cQ[\cF,\zeta]\otimes I_\ell(\sigma)\in\bC^{\bF_q^{n_{\mathrm{out}}+\ell}\times\bF_q^{n_{\mathrm{out}}+\ell}}$ is of the form $\cQ[\cF]\otimes I_\ell(\sigma)=\sum_{i=1}^m\sigma_i$, where each $\sigma_i$ is a $\cE_{\mathrm{out}}\sqcup [\ell]$-deviation of $(\Enc_{\mathrm{out}}\circ\bar{\cO}\circ L_i)\otimes I_\ell(\rho\otimes\Gamma_{\mathrm{out}})$ for some superoperator $L_i:\bC^{\bF_q^{k_{\mathrm{in}}}\times\bF_q^{k_{\mathrm{in}}}}\rightarrow\bC^{\bF_q^{k_{\mathrm{in}}}\times\bF_q^{k_{\mathrm{in}}}}$.

  
\end{definition}

The additional $\ell$-qudit system in \Cref{def:faulttol}, which we call an \emph{external system}\footnote{\cite{nguyen_quantum_2025} instead used the term ``reference system.''}, is required to ensure that fault-tolerance is preserved under parallel composition and captures entanglement with other components in a larger computation, see \Cref{sec:parcomp}.


The \emph{mending} property\footnote{The ``mending property'' is referred to as the ``\emph{friendly} property'' in \cite{he_composable_2025,nguyen_quantum_2025}} is used to remedy leakage from the code space when recursively composing fault-tolerant gadgets. 
That is, a mending gadget will return to the code space under a low-weight fault, even if the input state was far from any code state.

\subsubsection{Basic Results on Fault-Tolerant Gadgets}
In this section, we present some basic results regarding fault-tolerant gadgets as defined in \Cref{def:faulttol}. 

The following lemma shows that for proving (mending) fault-tolerance, it suffices to consider Pauli deviations on the input state, as well as Pauli faults. 
Similar decompositions of general noise into Pauli noise are used throughout the fault-tolerance literature.

\begin{lemma}
  \label{lem:paulift}
  Define all variables as in \Cref{def:faulttol}. Then the data  $(\cQ,\cE_{\mathrm{run}},D_{\mathrm{in}},D_{\mathrm{out}})$ provides a fault-tolerant gadget for $\bar{\cO}$ if for every $\ell\in\bN$, every operator $\rho\in\bC^{\bF_q^{k_{\mathrm{in}}+\ell}\times\bF_q^{k_{\mathrm{in}}+\ell}}$, every Pauli $\cE_{\mathrm{in}}\sqcup [\ell]$-deviation $\sigma$ of $\Enc_{\mathrm{in}}\otimes I_\ell(\rho\otimes\Gamma_{\mathrm{in}})$, and every $\cE_{\mathrm{run}}$-avoiding Pauli fault $\cF$ and postselection~$\zeta$ for $\cQ$, then $\cQ[\cF,\zeta]\otimes I_\ell(\sigma)$ is a $\cE_{\mathrm{out}}\sqcup [\ell]$-deviation of $(\Enc_{\mathrm{out}}\circ\bar{\cO})\otimes I_\ell(\rho\otimes\Gamma_{\mathrm{out}})$.

  Furthermore, the gadget $(\cQ,\cE_{\mathrm{run}},D_{\mathrm{in}},D_{\mathrm{out}})$ is mending if for every $\ell\in\bN$, every operator $\rho\in\bC^{\bF_q^{k_{\mathrm{in}}+\ell}\times\bF_q^{k_{\mathrm{in}}+\ell}}$, every Pauli $[\ell]$-deviation $\sigma$ of $\Enc_{\mathrm{in}}\otimes I_\ell(\rho\otimes\Gamma_{\mathrm{in}})$, and every $\cE_{\mathrm{run}}$-avoiding Pauli fault $\cF$ and postselection $\zeta$ for $\cQ$, then $\cQ[\cF,\zeta]\otimes I_\ell(\sigma)$ is of the form $\cQ[\cF,\zeta]\otimes I_\ell(\sigma)=\sum_{i=1}^m\sigma_i$, where each $\sigma_i$ is a $\cE_{\mathrm{out}}\sqcup [\ell]$-deviation of $(\Enc_{\mathrm{out}}\circ\bar{\cO}\circ L_i)\otimes I_\ell(\rho\otimes\Gamma_{\mathrm{out}})$ for some superoperator $L_i:\bC^{\bF_q^{k_{\mathrm{in}}}\times\bF_q^{k_{\mathrm{in}}}}\rightarrow\bC^{\bF_q^{k_{\mathrm{in}}}\times\bF_q^{k_{\mathrm{in}}}}$.
\end{lemma}
\begin{proof}
  Consider an arbitrary $\cE_{\mathrm{in}}\sqcup [\ell]$-deviation $\sigma$ of $\Enc_{\mathrm{in}}\otimes I_\ell(\rho\otimes\Gamma_{\mathrm{in}})$, and an arbitrary $\cE_{\mathrm{run}}$-avoiding fault $\cF=(F_1,\dots,F_T)$ and postselection $\zeta$ for $\cQ$. By definition, we can write $\sigma=\sum_{i_0\in[m_0]}\sigma_{i_0}$ for Pauli $\cE_{\mathrm{in}}\sqcup [\ell]$-deviations $\sigma_{i_0}$ of $\Enc_{\mathrm{in}}\otimes I_\ell(\rho\otimes\Gamma_{\mathrm{in}})$, and we can write each $F_t=\sum_{i_t\in[m_t]}\alpha_{t,i_t}F_{t,i_t}$ for Pauli superoperators $F_{t,i_t}$ such that each $(F_{1,i_1},\dots,F_{T,i_T})$ is $\cE_{\mathrm{run}}$-avoiding. Thus
  \begin{equation*}
    \cQ[\cF,\zeta]\otimes I_\ell(\sigma) = \sum_{(i_0,\dots,i_T)\in[m_0]\times\cdots\times[m_T]}\alpha_{1,i_1}\cdots\alpha_{T,i_T}\cdot\cQ[(F_{1,i_1},\dots,F_{T,i_T}),\zeta]\otimes I_\ell(\sigma_{i_0})
  \end{equation*}
  Assuming the hypothesis in the lemma statement, then every term on the RHS above is a $\cE_{\mathrm{out}}\sqcup [\ell]$-deviation of $(\Enc_{\mathrm{out}}\circ\bar{\cO})\otimes I_\ell(\rho\otimes\Gamma_{\mathrm{out}})$, which by \Cref{remark:paulibasis2} implies that the LHS above is a $\cE_{\mathrm{out}}\sqcup [\ell]$-deviation of $(\Enc_{\mathrm{out}}\circ\bar{\cO})\otimes I_\ell(\rho\otimes\Gamma_{\mathrm{out}})$. Thus $(\cQ,\cE_{\mathrm{run}},D_{\mathrm{in}},D_{\mathrm{out}})$ is a fault-tolerant gadget for~$\bar{\cO}$, as desired.

  The proof of the mending condition is analogous, by again decomposing an arbitrary input error and fault into Paulis; we omit the details to avoid redundancy.
\end{proof}


In \Cref{lem:extft} below, we show that our notion of fault-tolerance in \Cref{def:faulttol} implies fault-tolerance against a stronger type of adversary, which is allowed access to its own private side-register. Similarly as in \Cref{lem:paulift}, the proof will use a standard Pauli decomposition, which we present for completeness.

\begin{definition}
  \label{def:extfault}
  Let $\cQ=(Q_1,\dots,Q_T;N_{\mathrm{in}},N_{\mathrm{out}})$ be a quantum circuit acting on a set $N$ of $q$-dimensional qudits. An \emph{extended fault} on $\cQ$ is a sequence $\cF=(F_1,\dots,F_T)$ of superoperators $F_t$ acting on qudits $N\sqcup A$, where $A$ is a set of ancilla $q$-dimensional qudits. Similarly as with ordinary faults, we say $\cF$ is an \emph{extended Pauli fault} if each $F_t$ is a Pauli superoperator. The \emph{support} $\supp(\cF)\subseteq N\times[T]$ is defined by
  \begin{equation*}
    \supp(\cF)=(\supp(F_1)\cap N)\sqcup\cdots\sqcup(\supp(F_T)\cap N).
  \end{equation*}
  For $\cE\subseteq 2^{N\times[T]}$, we say $\cF$ is \emph{$\cE$-avoiding} if $\supp(\cF)$ is $\cE$-avoiding.

  The \emph{$\cF$-corrupted circuit $\cQ[\cF]$} is defined by
  \begin{align*}
    \cQ[\cF] &= (Q_1\otimes\gInit_Z^{\otimes A},\; F_1\; ,Q_2\otimes I_A,\; F_2,\; Q_3\otimes I_A,\; F_3,\; \dots,\; Q_T\otimes I_A,\; F_T;\; N_{\mathrm{in}},\; N_{\mathrm{out}}).
  \end{align*}
\end{definition}

\begin{lemma}
  \label{lem:extft}
  For $\alpha\in\{\mathrm{in},\mathrm{out}\}$, let $D_\alpha=(C_\alpha,\Enc_\alpha,\Gamma_\alpha,\cE_\alpha)$ be a decorated $[[n_\alpha,k_\alpha]]_q$ subsystem code. Let $(\cQ,\cE_{\mathrm{run}},D_{\mathrm{in}},D_{\mathrm{out}})$ be a fault-tolerant gadget for a set of superoperators $\bar{\cO}$.

  Then for every $\ell\in\bN$, every operator $\rho\in\bC^{\bF_q^{k_{\mathrm{in}}+\ell}\times\bF_q^{k_{\mathrm{in}}+\ell}}$, every $\cE_{\mathrm{in}}\sqcup [\ell]$-deviation $\sigma$ of $\Enc_{\mathrm{in}}\otimes I_\ell(\rho\otimes\Gamma_{\mathrm{in}})$, and every $\cE_{\mathrm{run}}$-avoiding extended fault $\cF$ and postselection $\zeta$ for $\cQ$, we have that $\cQ[\cF,\zeta]\otimes I_\ell(\sigma)$ is a $\cE_{\mathrm{out}}\sqcup [\ell]$-deviation of $(\Enc_{\mathrm{out}}\circ\bar{\cO})\otimes I_\ell(\rho\otimes\Gamma_{\mathrm{out}})$.

  Similarly, if $(\cQ,\cE_{\mathrm{run}},D_{\mathrm{in}},D_{\mathrm{out}})$ is furthermore mending, then for every $\rho\in\bC^{\bF_q^{k_{\mathrm{in}}+\ell}\times\bF_q^{k_{\mathrm{in}}+\ell}}$, every $[\ell]$-deviation $\sigma$ of $\Enc_{\mathrm{in}}\otimes I_\ell(\rho\otimes\Gamma_{\mathrm{in}})$, and every $\cE_{\mathrm{run}}$-avoiding extended fault $\cF$ and postselection $\zeta$ for $\cQ$, then $\cQ[\cF,\zeta]\otimes I_\ell(\sigma)$ is of the form $\cQ[\cF,\zeta]\otimes I_\ell(\sigma)=\sum_{i=1}^m\sigma_i$, where each $\sigma_i$ is a $\cE_{\mathrm{out}}\sqcup [\ell]$-deviation of $(\Enc_{\mathrm{out}}\circ\bar{\cO}\circ L_i)\otimes I_\ell(\rho\otimes\Gamma_{\mathrm{out}})$ for some superoperator $L_i:\bC^{\bF_q^{k_{\mathrm{in}}}\times\bF_q^{k_{\mathrm{in}}}}\rightarrow\bC^{\bF_q^{k_{\mathrm{in}}}\times\bF_q^{k_{\mathrm{in}}}}$.
\end{lemma}
\begin{proof}
  We begin with the non-mending statement. The proof is similar to that of \Cref{lem:paulift}. Specifically, we can write $\sigma=\sum_{i_0\in[m_0]}\sigma_{i_0}$ for Pauli $\cE_{\mathrm{in}}\sqcup [\ell]$-deviations $\sigma_{i_0}$ of $\Enc_{\mathrm{in}}\otimes I_\ell(\rho\otimes\Gamma_{\mathrm{in}})$, and we can write each $F_t=\sum_{i_t\in[m_t]}\alpha_{t,i_t}F_{t,i_t}$ for Pauli superoperators $F_{t,i_t}$ such that each $(F_{1,i_1},\dots,F_{T,i_T})$ is a $\cE_{\mathrm{run}}$-avoiding extended Pauli fault. Thus
  \begin{equation}
    \label{eq:extdecomp}
    \cQ[\cF,\zeta]\otimes I_\ell(\sigma) = \sum_{(i_0,\dots,i_T)\in[m_0]\times\cdots\times[m_T]}\alpha_{1,i_1}\cdots\alpha_{T,i_T}\cdot\cQ[(F_{1,i_1},\dots,F_{T,i_T}),\zeta]\otimes I_\ell(\sigma_{i_0}).
  \end{equation}
  Let $\cQ$ act on qudit set $N$, and let $A$ denote the additional ancilla qudits acted on by the extended fault $\cF$. Then we can write each $F_{t,i_t}=F_{t,i_t,N}\otimes F_{t,i_t,A}$, where $F_{t,i_t,N},F_{t,i_t,A}$ are Pauli superoperators acting inqudits $N,A$ respectively. It follows by definition
  \begin{align*}
    \hspace{1em}&\hspace{-1em}\cQ[(F_{1,i_1},\dots,F_{T,i_T}),\zeta]\otimes I_\ell(\sigma_{i_0}) \\
    &= \tr_A\left(\cQ[(F_{1,i_1,N},\dots,F_{T,i_T,N}),\zeta]\otimes I_\ell(\sigma_{i_0}) \otimes F_{T,i_T,A}\circ\cdots\circ F_{1,i_1,A}\left(\ket{0^A}\bra{0^A}\right)\right) \\
    &\propto \cQ[(F_{1,i_1,N},\dots,F_{T,i_T,N}),\zeta]\otimes I_\ell(\sigma_{i_0}).
  \end{align*}
  By the fault-tolerance of $(Q,\cE_{\mathrm{run}},D_{\mathrm{in}},D_{\mathrm{out}})$, the RHS above is a $\cE_{\mathrm{out}}\sqcup [\ell]$-deviation of $(\Enc_{\mathrm{out}}\circ\bar{\cO})\otimes I_\ell(\rho\otimes\Gamma_{\mathrm{out}})$. As the RHS of \Cref{eq:extdecomp} is a linear combination of such states, \Cref{remark:paulibasis2} implies that the LHS of \Cref{eq:extdecomp} is a $\cE_{\mathrm{out}}\sqcup [\ell]$-deviation of $(\Enc_{\mathrm{out}}\circ\bar{\cO})\otimes I_\ell(\rho\otimes\Gamma_{\mathrm{out}})$, as desired.

  The proof of the mending condition is analogous, by again decomposing an arbitrary input error and fault into Paulis; we omit the details to avoid redundancy.
\end{proof}


\subsection{Gadget Composition}
In this section, we present results showing how fault-tolerance properties are preserved under sequential and parallel composition of gadgets, as defined below. 
Similar results are shown in \cite{he_composable_2025,nguyen_quantum_2025}; we simply adapt these results to our setting and notation, and provide proofs for completeness. A more involved form of composition, namely simulative composition, will be addressed in \Cref{sec:alphred}.

\subsubsection{Sequential Composition}
\Cref{lem:seqcomp} below considers sequential composition, in which gadgets are run sequentially on the same set of qudits.

\begin{lemma}[Similar to Proposition 4.9 of \cite{he_composable_2025}]
  \label{lem:seqcomp}
  Let $N$ be a set of qudits. For $m\in\bN$ and $i\in[m]$ let $(\cQ^i,\cE_{\mathrm{run}}^i,D_{\mathrm{in}}^i,D_{\mathrm{out}}^i)$ be a fault-tolerant gadget for a set of superoperators $\bar{\cO}^i$ using space $N_i\subseteq N$ and time $T_i$. Writing $\cQ^i=(Q_1^i,\dots,Q_{T_i}^i;N_{\mathrm{in}}^i,N_{\mathrm{out}}^i)$ and $D_\alpha^i=(C_\alpha^i,\Enc_\alpha^i,\Gamma_\alpha^i,\cE_\alpha^i)$, for $i\in[m-1]$ assume that $N_{\mathrm{out}}^i=N_{\mathrm{in}}^{i+1}$, $(C_{\mathrm{out}}^i,\Enc_{\mathrm{out}}^i)=(C_{\mathrm{in}}^{i+1},\Enc_{\mathrm{in}}^{i+1})$, $\Gamma_{\mathrm{out}}^i\subseteq\Gamma_{\mathrm{in}}^{i+1}$, and $\cE_{\mathrm{out}}^i\supseteq\cE_{\mathrm{in}}^{i+1}$. We view each $\cQ^i$ as a circuit on qudits $N\supseteq N_i$ by letting $Q_1^i$ apply $\gInit_Z$ gates to all qudits in $N\setminus N_i$, and letting $Q_2^i,\dots,Q_{T_i}^i$ act as the identity on qudits in $N\setminus N_i$. Then letting
  \begin{align*}
    \cQ^m\circ\cdots\circ\cQ^1
    &= (Q^1_1,\dots,Q^1_{T_1},Q^2_1,\dots,Q^2_{T_2},\dots,Q^m_1,\dots,Q^m_{T_m};N_{\mathrm{in}}^1,N_{\mathrm{out}}^m),
  \end{align*}
  it follows that
  \begin{align*}
    \left(\cQ=\cQ^m\circ\cdots\circ\cQ^1,\; \cE_{\mathrm{run}}=\bigsqcup_{i\in[m]}\cE_{\mathrm{run}}^i,\; D_{\mathrm{in}}^1,\; D_{\mathrm{out}}^m\right)
  \end{align*}
  is a fault-tolerant gadget for $\bar{\cO}=\bar{\cO}^m\circ\cdots\circ\bar{\cO}^1$ using space $N$ (so $|N|\geq\max_{i\in[m]}|N_i|$) and time $T=\sum_{i\in[m]}T_i$. Furthermore, if $(\cQ^1,\cE_{\mathrm{run}}^1,D_{\mathrm{in}}^1,D_{\mathrm{out}}^1)$ is mending, then $(\cQ,\cE_{\mathrm{run}},D_{\mathrm{in}}^1,D_{\mathrm{out}}^m)$ is mending.
\end{lemma}
\begin{proof}
  By definition $(\cQ,\cE_{\mathrm{run}},D_{\mathrm{in}}^1,D_{\mathrm{out}}^m)$ uses space $N$ and time $T=\sum_{i\in[m]}T_i$, so it suffices to show that it forms a (mending) fault-tolerant gadget for $\bar{\cO}$. For this purpose, for $\alpha\in\{\mathrm{in},\mathrm{out}\}$, let each $D_\alpha^i$ be a $[[n_\alpha^i,k_\alpha^i]]_q$ code.

  To show (non-mending) fault-tolerance, it suffices to show that for every $\ell\in\bN$, every operator $\rho^0\in\bC^{\bF_q^{k_{\mathrm{in}}^1+\ell}\times\bF_q^{k_{\mathrm{in}}^1+\ell}}$, every $\cE_{\mathrm{in}}^1\sqcup [\ell]$-deviation $\sigma^0$ of $\Enc_{\mathrm{in}}^1\otimes I(\rho^0\otimes\Gamma_{\mathrm{in}}^1)$, and every $\cE_{\mathrm{run}}$-avoiding fault~$\cF$ and postselection $\zeta$ for $\cQ$, then $\cQ[\cF]\otimes I_\ell(\sigma^0)$ is a $\cE_{\mathrm{out}}^m\sqcup [\ell]$-deviation of $(\Enc_{\mathrm{out}}^m\circ\bar{O})\otimes I_\ell(\rho^0\otimes\Gamma_{\mathrm{out}}^m)$. 
  By definition, we can write $\cF=(\cF^1,\dots,\cF^m)$ as the sequential composition (i.e.~concatenation) of $\cE_{\mathrm{run}}^i$-avoiding faults $\cF^i$ for $\cQ^i$. Similarly, we can write $\zeta=(\zeta^1,\dots,\zeta^m)$ as the sequential composition of postselections for the $\cQ^i$. Inductively defining $R^0=\{\rho^0\}$, $R^i=\bar{\cO}^i\otimes I_\ell(R^{i-1})$ and $\sigma^i=\cQ^i[\cF^i,\zeta^i]\otimes I_\ell(\sigma^{i-1})$, then the fault-tolerance of $(\cQ^i,\cE_{\mathrm{run}}^i,D_{\mathrm{in}}^i,D_{\mathrm{out}}^i)$ implies that $\sigma^i$ is a $\cE_{\mathrm{out}}^i\sqcup [\ell]$-deviation of $\Enc_{\mathrm{out}}^i\otimes I(R^i\otimes\Gamma_{\mathrm{out}}^i)$, and hence also a $\cE_{\mathrm{in}}^{i+1}\sqcup [\ell]$-deviation of $\Enc_{\mathrm{in}}^{i+1}\otimes I(R^i\otimes\Gamma_{\mathrm{in}}^{i+1})$.
  Here we use the fact that the first timestep of each $\cQ^{i+1}$ by definition applies a $\gInit_*$ gate to every qudit in $N\setminus N_{\mathrm{in}}^{i+1}$ (see \Cref{def:circuit}), which has the effect of tracing out qudits $N\setminus N_{\mathrm{out}}^i=N\setminus N_{\mathrm{in}}^{i+1}$ following the final timestep of $\cQ^i$. Hence the state passed as input to $\cQ^{i+1}$ within the execution of $\cQ$ is indeed $\sigma^i=\cQ^i[\cF^i,\zeta^i]\otimes I_\ell(\sigma^{i-1})$.
  Thus the output $\sigma^m=\cQ[\cF]\otimes I_\ell(\sigma^0)$ is a $\cE_{\mathrm{out}}^m\sqcup [\ell]$-deviation of $(\Enc_{\mathrm{out}}^m\circ\bar{\cO})\otimes I(\rho^0\otimes\Gamma^i_{\mathrm{out}})$, as desired.

  For the proof of the mending property, we now assume that $\sigma^0$ is a $[\ell]$-deviation of $\Enc_{\mathrm{in}}^1\otimes I_\ell(\rho^0\otimes\Gamma_{\mathrm{in}})$. 
  The mending property of $(\cQ^1,\cE_{\mathrm{run}}^1,D_{\mathrm{in}}^1,D_{\mathrm{out}}^i)$ implies that $\sigma^1=\cQ^1[\cF^1,\zeta^1]\otimes I_\ell(\sigma^0)$ is a linear combination of $\cE_{\mathrm{out}}^1\sqcup [\ell]$-deviations of $(\Enc_{\mathrm{out}}^1\circ\bar{\cO}^1\circ L)\otimes I_\ell(\rho^0\otimes\Gamma_{\mathrm{out}}^1)$ for superoperators $L:\bC^{\bF_q^{k_{\mathrm{in}}^1}\times\bF_q^{k_{\mathrm{in}}^1}}\rightarrow\bC^{\bF_q^{k_{\mathrm{in}}^1}\times\bF_q^{k_{\mathrm{in}}^1}}$.
  Again inducting over $i\in[m]$, we conclude that each $\sigma^i=\cQ^i[\cF^i]\otimes I_\ell(\sigma^{i-1})$ is a linear combination of $\cE_{\mathrm{out}}^i\sqcup [\ell]$-deviations of $(\Enc_{\mathrm{out}}^i\circ\bar{\cO}^i\circ\cdots\circ\bar{\cO}^1\circ L)\otimes I_\ell(\rho^0\otimes\Gamma_{\mathrm{out}}^i)$. 
  The $i=m$ case then implies the desired mending property.
\end{proof}

\subsubsection{Parallel Composition}
\label{sec:parcomp}
\Cref{lem:parcomp} below considers parallel composition, in which gadgets are run at the same time on disjoint sets of qudits. For simplicity in \Cref{lem:parcomp}, we assume that all input gadgets have the same running time. In general we could instead pad gadgets with idling timesteps (i.e.\ identity gates) to ensure this condition is satisfied, though at the cost of potentially increasing the output error.

\begin{lemma}[Similar to Proposition 4.9 of \cite{he_composable_2025}]
  \label{lem:parcomp}
  For $m\in\bN$, for $i\in[m]$ let $(\cQ^i,\cE_{\mathrm{run}}^i,D_{\mathrm{in}}^i,D_{\mathrm{out}}^i)$ be a fault-tolerant gadget for a set of superoperators $\bar{\cO}^i$ using space $N_i$ and time $T$. Then
  \begin{align*}
    \left(\cQ=\bigsqcup_{i\in[m]}\cQ^i,\; \cE_{\mathrm{run}}=\bigsqcup_{i\in[m]}\cE_{\mathrm{run}}^i,\; D_{\mathrm{in}}=\bigsqcup_{i\in[m]}D_{\mathrm{in}}^i,\; D_{\mathrm{out}}=\bigsqcup_{i\in[m]}D_{\mathrm{out}}^i\right)
  \end{align*}
  is a fault-tolerant gadget for $\bar{\cO}=\bigotimes_{i\in[m]}\bar{\cO}^i$ using space $N=\bigsqcup_{i\in[m]}N_i$ and time $T$, where $\cQ=\bigsqcup_{i\in[m]}\cQ^i$ denotes the circuit that runs $\cQ^1,\dots,\cQ^m$ in parallel on disjoint sets of qudits. Furthermore, if $(\cQ^i,\cE_{\mathrm{run}}^i,D_{\mathrm{in}}^i,D_{\mathrm{out}}^i)$ is mending for every $i\in[m]$, then $(\cQ,\cE_{\mathrm{run}},D_{\mathrm{in}},D_{\mathrm{out}})$ is mending.
\end{lemma}
\begin{proof}
  By definition $(\cQ,\cE_{\mathrm{run}},D_{\mathrm{in}},D_{\mathrm{out}})$ uses space $N$ and time $T$, so it suffices to show that it forms a (mending) fault-tolerant gadget for $\bar{\cO}$. For this purpose, for $\alpha\in\{\mathrm{in},\mathrm{out}\}$, let each $D_\alpha^i$ be a $[[n_\alpha^i,k_\alpha^i]]_q$ codes, and let $n_\alpha=\sum_{i\in[m]}n_\alpha^i$ and $k_\alpha=\sum_{i\in[m]}k_\alpha^i$.
  
  By \Cref{lem:paulift}, to show (non-mending) fault-tolerance, it suffices to show that for every $\ell\in\bN$, every operator $\rho\in\bC^{\bF_q^{k_{\mathrm{in}}+\ell}\times\bF_q^{k_{\mathrm{in}}+\ell}}$, every Pauli $\cE_{\mathrm{in}}\sqcup [\ell]$-deviation $\sigma$ of $\Enc_{\mathrm{in}}\otimes I_\ell(\rho\otimes\Gamma_{\mathrm{in}})$, and every $\cE_{\mathrm{run}}$-avoiding Pauli fault $\cF$ and postselection $\zeta$ for $\cQ$, then $\cQ[\cF,\zeta]\otimes I_\ell(\sigma)$ is a $\cE_{\mathrm{out}}\sqcup [\ell]$-deviation of $(\Enc_{\mathrm{out}}\circ\bar{O})\otimes I_\ell(\rho\otimes\Gamma_{\mathrm{out}})$. Because $\cF=(F_1,\dots,F_T)$ is a Pauli fault, we can write each $F_t=\bigotimes_{i\in[m]}F^i_t$, where $F^i_t$ is a Pauli superoperator acting on qudits $N_i$. Writing $\cF^i=(F^i_1,\dots,F^i_T)$, then by definition $\cF^i$ is $\cE_{\mathrm{run}}^i$-avoiding. We can also write $\zeta=(\zeta^1,\dots,\zeta^m)$, where each $\zeta^i$ is the restriction of $\zeta$ to components associated to $\gCO_*$ gates in $\cQ^i$. Then
  \begin{align*}
    \cQ[\cF,\zeta]
    &= \bigotimes_{i\in[m]}\cQ^i[\cF^i,\zeta^i] = \prod_{i\in[m]}\left(I^{\otimes i-1}\otimes\cQ^i[\cF^i,\zeta^i]\otimes I^{\otimes m-i}\right).
  \end{align*}
  If we apply the RHS above to $\sigma$ one factor at a time, then after applying the $j$th factor, the resulting state $\sigma_j$ equals a linear combination of states of the form
  \begin{align*}
    &\bigotimes_{i=1}^j(E^i\circ\Enc_{\mathrm{out}}\circ\bar{O}^i)\otimes\bigotimes_{i=j+1}^m(E^i\circ\Enc_{\mathrm{in}})\otimes I_\ell\left(\rho\otimes\bigotimes_{i\in[m]}\gamma^i\right),
  \end{align*}
  such that for $i\leq j$ then $E^i$ is some $\cE_{\mathrm{out}}$-avoiding Pauli and $\gamma^i\in\Gamma_{\mathrm{out}}^i$ and $\bar{O}^i\in\bar{\cO}^i$, and for $i>j$ then $E^i$ is some $\cE_{\mathrm{in}}$-avoiding Pauli and $\gamma^i\in\Gamma_{\mathrm{in}}^i$. Specifically, this conclusion holds by applying the fault-tolerance of $(\cQ^i,\cE_{\mathrm{run}}^i,D_{\mathrm{in}}^i,D_{\mathrm{out}}^i)$ for each $i=1,\dots,j$ with external system $\bigsqcup_{i'=1}^{i-1}[k_{\mathrm{out}}^i]\sqcup\bigsqcup_{i'=i+1}^m[k_{\mathrm{in}}^i]\sqcup[\ell]$. Thus the final state $\sigma_m=\cQ[\cF,\zeta](\sigma)$ is a linear combination of Pauli $\cE_{\mathrm{out}}$-deviations of $(\Enc_{\mathrm{out}}\circ\bar{\cO})(\rho\otimes\Gamma_{\mathrm{out}})$, as desired.

  The proof of the mending property is analogous as above, though we now assume $\sigma$ is a Pauli $[\ell]$-deviation of $\Enc_{\mathrm{in}}\otimes I_\ell(\rho\otimes\Gamma_{\mathrm{in}})$, and we replace the superoperator $E^i\circ\Enc_{\mathrm{out}}\circ\bar{O}^i$ with $E^i\circ\Enc_{\mathrm{out}}\circ\bar{O}^i\circ L^i$ for a superoperator $L^i$; we omit the details to avoid redundancy.
\end{proof}

\section{Code Construction and Properties}
\label{sec:codes}
In this section, we present the subsystem codes underlying our fault-tolerance scheme, and analyze their properties. We construct these codes from tensor products of Reed-Solomon codes, or equivalently, multivariate polynomial evaluation codes, over a large alphabet.

\subsection{Construction and Result Statement}
\label{sec:construct}
The family of subsystem product codes (see \Cref{def:subsystemcode}) we consider are defined below.

\begin{definition}
  \label{def:pevprod}
  For a field $\bF_q$, positive integers $u$ and $\ell_1,\dots,\ell_u$, subsets $A_1,\dots,A_u,\;E_1,\dots,E_u\subseteq\bF_q$ with each $A_i\cap E_i=\emptyset$ and $|A_i|\leq\ell_i\leq|E_i|$, and vectors $\beta_i\in(\bF_q^*)^{E_i}$ let $C^i=(C^i_X,C^i_Z)$ be the $[[n_i=|E_i|,\;k_i=|A_i|]]_q$ CSS non-subsystem code given by
  \begin{align*}
    {C^i_X}^\perp &= \beta_i*\evl_{E_i}(\bF_q[X]^{[0,\ell_i)}_{A_i}) \\
    C^i_Z &= \beta_i*\evl_{E_i}(\bF_q[X]^{[0,\ell_i)}).
  \end{align*}
  We define an associated \emph{polynomial evaluation subsystem product code} $C=C(q,u,(\ell_i,A_i,E_i,\beta_i)_{i\in[u]})$ by
  \begin{equation*}
    C = (C_X,C_Z) := \bigotimes_{i\in[u]}C^i.
  \end{equation*}
  For $I\subseteq[u]$, we write $C^I(q,u,(\ell_i,A_i,E_i,\beta_i)_{i\in[u]}):=C(q,|I|,(\ell_i,A_i,E_i,\beta_i)_{i\in I})$. We also write $A=A_1\times\cdots\times A_u$ and $E=E_1\times\cdots\times E_u$.
\end{definition}

The polynomial evaluation subsystem product code $C^I$ in \Cref{def:pevprod} has dimension $k^I=\prod_{i\in I}k_i$ by \Cref{prop:spkunneth}. Our main result constructing such codes of large distance is stated below.

\begin{theorem}
  \label{thm:pevmain}
  Consider positive integers $u$, as well as $k_i<n_i$ for $i\in[u]$, and $q=q_0^{\prod_{i\in[u]}n_i'}$ with $q_0\geq\max_{i\in[u]}n_i$ a prime power and each $n_i'\geq n_i$ an integer. Then there exist explicit sets $A_i,E_i\subseteq\bF_q$ with $|A_i|=k_i$, $|E_i|=n_i$ for $i\in[u]$ such that for every integer $k_i\leq\ell_i\leq n_i$, every $\beta_i\in(\bF_q^*)^{E_i}$, and every $I\subseteq[u]$, we have a well-defined polynomial evaluation subsystem product code $C^I=C^I(q,u,(\ell_i,A_i,E_i,\beta_i)_{i\in[u]})$ with $X$-distance
  \begin{equation}
    \label{eq:pevmaindX}
    d^I_X \geq \prod_{i\in I}(\ell_i-k_i+1)
  \end{equation}
  and $Z$-distance
  \begin{equation}
    \label{eq:pevmaindZ}
    d^I_Z \geq \prod_{i\in I}(n_i-\ell_i+1).
  \end{equation}
\end{theorem}

We prove \Cref{thm:pevmain} in \Cref{sec:codeanal} below. In \Cref{sec:homprod}, we show that an appropriate gauge-fixing of the subsystem codes in \Cref{thm:pevmain} yields non-subsystem codes with low-weight stabilizers.

\subsection{Analysis}
\label{sec:codeanal}
In this section, we analyze the polynomial evaluation subsystem codes in \Cref{def:pevprod}, and specifically prove \Cref{thm:pevmain}.

We begin with the following basic lemma, which shows that the family of codes $C=(C_X,C_Z)$ in \Cref{def:pevprod} is closed under swapping $C_X$ and $C_Z$.

\begin{lemma}
  \label{lem:pevswapXZ}
  For a polynomial evaluation subsystem product code
  \begin{equation*}
    (C_X,C_Z)=C(q,u,(\ell_i,A_i,E_i,\beta_i)_{i\in[u]}),
  \end{equation*}
  then there exist $\beta_i'\in(\bF_q^*)^{E_i}$ for $i\in[u]$ such that
  \begin{equation*}
    (C_Z,C_X)=C(q,u,(n_i+k_i-\ell_i,A_i,E_i,\beta_i')_{i\in[u]}).
  \end{equation*}
\end{lemma}
\begin{proof}
  For each $i\in[u]$, let $\bar{\beta}_i\in(\bF_q^*)^{A_i\cup E_i}$ be some vector with $\bar{\beta}_i|_{E_i}=\beta_i$. By \Cref{lem:RSdual}, there exists $\bar{\beta}_i'\in(\bF_q^*)^{A_i\cup E_i}$ such that
  \begin{equation*}
    (\bar{\beta}_i*\evl_{A_i\cup E_i}(\bF_q[X]^{[0,\ell_i)}))^\perp = \bar{\beta}_i'*\evl_{A_i\cup E_i}(\bF_q[X]^{[0,n_i+k_i-\ell_i)}).
  \end{equation*}
  Then by \Cref{lem:puncshort},
  \begin{align*}
    C^i_X
    &= (\bar{\beta}_i*\evl_{A_i\cup E_i}(\bF_q[X]^{[0,\ell_i)}))^\perp|_{E_i} \\
    &= \bar{\beta}_i'|_{E_i} * \evl_{E_i}(\bF_q[X]^{[0,n_i+k_i-\ell_i)})\\
    {C^i_Z}^\perp
    &= ((\bar{\beta}_i*\evl_{A_i\cup E_i}(\bF_q[X]^{[0,\ell_i)}))^\perp\cap(\{0\}^{A_i}\times\bF_q^{E_i}))|_{E_i} \\
    &= \bar{\beta}_i'|_{E_i} * \evl_{E_i}(\bF_q[X]^{[0,n_i+k_i-\ell_i)}_{A_i}).
  \end{align*}
  Thus we obtain the desired result with $\beta_i'=\bar{\beta}_i'|_{E_i}$.
\end{proof}

The following lemma gives convenient expressions for $C_Z$ and $C_X^\perp$. 

\begin{lemma}
  \label{lem:QZcQXp}
  Define all variables as in \Cref{def:pevprod}, and let $\beta=\bigotimes_{i\in[u]}\beta_i$. Then
  \begin{align}
    \label{eq:QZcQXp}
    \begin{split}
      C_Z &= \beta*\evl_E(\bF_q[X_1,\dots,X_u]^{\prod_{i\in[u]}[0,\ell_i)}) \\
      C_Z\cap C_X^\perp &= \beta*\evl_E(\bF_q[X_1,\dots,X_u]^{\prod_{i\in[u]}[0,\ell_i)}_A).
    \end{split}
  \end{align}
\end{lemma}
\begin{proof}
  By dividing all codes in \Cref{eq:QZcQXp} component-wise by $\beta$, we may assume without loss of generality that all $\beta_i$ (and hence also $\beta$) are all-$1$s vectors.
  
  Now the expression in \Cref{eq:QZcQXp} for $C_Z$ follows directly by the definition of $C_Z=\bigotimes_{i\in[u]}C^i_Z$, so it only remains to prove the expression for $C_Z\cap C_X^\perp$. For this purpose, for $i\in[u]$ let $L_i'=\prod_{j=1}^{i-1}[0,n_j)\times[0,\ell_i)\times\prod_{j=i+1}^u[0,n_j)$ and $A_i'=\bF_q^{i-1}\times A_i\times\bF_q^{n-i}$. Then by definition,
  \begin{align*}
    C_X^\perp
    &= \sum_{i\in[u]}\bF_q^{E_1\times\cdots\times E_{i-1}} \otimes {C^i_X}^\perp \otimes \bF_q^{E_{i+1}\times\cdots\times E_u} \\
    &= \sum_{i\in[u]}\evl_E(\bF_q[X_1,\dots,X_u]^{L_i'}_{A_i'}) \\
    &\subseteq \evl_E(\bF_q[X_1,\dots,X_u]^{L_1'\cup\cdots\cup L_u'}_{A_1'\cap\cdots\cap A_u'}) \\
    &\subseteq \evl_E(\bF_q[X_1,\dots,X_u]^{\prod_{i\in[u]}[0,n_i)}_A).
  \end{align*}
  Thus
  \begin{align*}
    C_Z\cap C_X^\perp
    &\subseteq \evl_E(\bF_q[X_1,\dots,X_u]^{\prod_{i\in[u]}[0,\ell_i)}) \cap \evl_E(\bF_q[X_1,\dots,X_u]^{\prod_{i\in[u]}[0,n_i)}_A) \\
    &= \evl_E(\bF_q[X_1,\dots,X_u]^{\prod_{i\in[u]}[0,\ell_i)}_A).
  \end{align*}

  For the opposite inclusion, consider an arbitrary $f\in\bF_q[X_1,\dots,X_u]^{\prod_{i\in[u]}[0,\ell_i)}_A$. Our goal is to show that $\evl_E(f)\in C_Z\cap C_X^\perp$. If $f\in\bF_q[X_1,\dots,X_u]^{\prod_{i\in[u]}[0,k_i)}$, then because each $|A_i|=k$ and $f$ vanishes inside $A$, we must have $f=0$. Otherwise, if $f\notin\bF_q[X_1,\dots,X_u]^{\prod_{i\in[u]}[0,k_i)}$, then there exists some $j=(j_1,\dots,j_u)\in\prod_{i\in[u]}[0,\ell_i)\setminus\prod_{i\in[u]}[0,k_i)$ such that the associated monomial $X_1^{j_1}\cdots X_u^{j_u}$ has a nonzero coefficient $f_j\neq 0$ in $f$. Assume without loss of generality that $j$ is the lexicographically largest such tuple. As $j\notin\prod_{i\in[u]}[0,k_i)$, some $\bar{i}\in[u]$ has $j_{\bar{i}}\geq k_{\bar{i}}$. Then
  \begin{equation*}
    f'(X_1,\dots,X_u) : =f(X_1,\dots,X_u)-f_j\cdot X_{\bar{i}}^{j_{\bar{i}}-k_{\bar{i}}}\cdot\prod_{a_{\bar{i}}\in A_{\bar{i}}}(X_{\bar{i}}-a_{\bar{i}})\cdot\prod_{i\in[u]\setminus\{\bar{i}\}}X_i^{j_i}
  \end{equation*}
  also lies inside $\bF_q[X_1,\dots,X_u]^{\prod_{i\in[u]}[0,\ell_i)}_A$. Note that the second term on the RHS above lies in $\bF_q[X_1,\dots,X_u]^{\prod_{i\in[u]}[0,\ell_i)}_{A_i'}$. By construction, all nonzero monomials $X_1^{j_1'}\cdots X_u^{j_u'}$ of $f'$ for $j'=(j_1',\dots,j_u')\in\prod_{i\in[u]}[0,\ell_i)\setminus\prod_{i\in[u]}[0,k_i)$ have $j'$ lexicographically smaller than $j$. We may repeat the above procedure until we arrive at some $f''\in\bF_q[X_1,\dots,X_u]^{\prod_{i\in[u]}[0,k_i)}_A=\{0\}$ so that $f''=0$. Summing up the terms added in each step, we conclude that
  \begin{equation*}
    f \in \sum_{i\in[u]}\bF_q[X_1,\dots,X_u]^{\prod_{i\in[u]}[0,\ell_i)}_{A_i'}.
  \end{equation*}
  Thus $\evl_E(f)\in\evl_E(\bF_q[X_1,\dots,X_u]^{\prod_{i\in[u]}[0,\ell_i)})=C_Z$ and $\evl_E(f)\in\sum_{i\in[u]}\evl_E(\bF_q[X_1,\dots,X_u]^{L_i'}_{A_i'})=C_X^\perp$, so $\evl_E(f)\in C_Z\cap C_X^\perp$, as desired.
\end{proof}

Our key lemma for bounding the distance of polynomial evaluation subsystem product codes is given below.

\npb{discuss tidying these to lemmas up}
\begin{lemma}
  \label{lem:dXkey}
  Define all variables as in \Cref{def:pevprod}. For every subset $S\subseteq E$ of size $|S|<\prod_{i\in[u]}(\ell_i-k_i+1)$, let $g_S(X_1,\dots,X_u)\in\bF_q[X_1,\dots,X_u]^{\prod_{i\in[u]}[0,\ell_i-k_i]}_S$ be a nonzero polynomial with each individual $X_i$-degree $\leq\ell_i-k_i$ that vanishes at every point in $S$; such $g_S$ always exists, as the valid space of coefficients of $g_S$ is specified by a system of $|S|<\prod_{i\in[u]}(\ell_i-k_i+1)$ homogeneous linear equations in $\prod_{i\in[u]}(\ell_i-k_i+1)$ variables.

  If it holds for every $S\subseteq E$ of size $|S|<\prod_{i\in[u]}(\ell_i-k_i+1)$ and every point $a\in A$ that $g_S(a)\neq 0$, then $C$ has $X$-distance
  \begin{equation*}
    d_X \geq \prod_{i\in[u]}(\ell_i-k_i+1)
  \end{equation*}
\end{lemma}
\begin{proof}
  Assume for a contradiction that $d_X<\prod_{i\in[u]}(\ell_i-k_i+1)$. Then there exists some $c\in(C_X+C_Z^\perp)\setminus C_Z^\perp$ of weight $|c|<\prod_{i\in[u]}(\ell_i-k_i+1)$, meaning that the support $S=\supp(c)$ has size $|S|<\prod_{i\in[u]}(\ell_i-k_i+1)$. 
  Because $(C_X+C_Z^\perp)^\perp=C_Z\cap C_X^\perp$, it follows that $c|_S\in((C_Z\cap C_X^\perp)|_S)^\perp$, but $c|_S\notin(C_Z|_S)^\perp$. 
  Therefore $((C_Z\cap C_X^\perp)|_S)^\perp \supsetneq (C_Z|_S)^\perp$, or equivalently,
  \begin{equation}
    \label{eq:sepcont}
    (C_Z\cap C_X^\perp)|_S \subsetneq C_Z|_S.
  \end{equation}

  We will now obtain a contradiction by showing that every element of $C_Z|_S$ in fact lies in $(C_Z\cap C_X^\perp)|_S$. 
  By \Cref{lem:QZcQXp}, every element of $C_Z|_S$ can be expressed in the form $\evl_S(f)$ for some $f\in\bF_q[X_1,\dots,X_u]^{\prod_{i\in[u]}[0,\ell_i)}$.
  Because each $|A_i|=k_i$, there exists some polynomial $h\in\bF_q[X_1,\dots,X_u]^{\prod_{i\in[u]}[0,k_i)}$ such that $\evl_A(h)=\evl_A(f/g_S)$, where here we may extend the notation $\evl_A$ to the rational function $f/g_S$ because by assumption $g_S$ is non-vanishing at every point in $A$. 
  Then by construction the polynomial $f-g_Sh$ lies in $\bF_q[X_1,\dots,X_u]^{\prod_{i\in[u]}[0,\ell_i)}$, and satisfies
  \begin{align*}
    \evl_A(f-g_Sh) &= \evl_A(f)-\evl_A(f) = 0 \\
    \evl_S(f-g_Sh) &= \evl_S(f)-\evl_S(g_Sh) = \evl_S(f),
  \end{align*}
  where the final equality above holds because by definition $\evl_S(g_S)=0$. 
  Thus
  \begin{equation*}
    f-g_Sh\in\bF_q[X_1,\dots,X_u]^{\prod_{i\in[u]}[0,\ell_i)}_A
  \end{equation*}
  and so by \Cref{lem:QZcQXp} we have that $\evl_E(f-g_Sh)$ lies in $C_Z\cap C_X^\perp$. 
  Therefore $\evl_S(f)=\evl_S(f-g_Sh)\in(C_Z\cap C_X^\perp)|_S$. 
  Thus we have shown that every element of $C_Z|_S$ lies inside $(C_Z\cap C_X^\perp)|_S$, contradicting \Cref{eq:sepcont}, so the assumption that $d_X<\prod_{i\in[u]}(\ell_i-k_i+1)$ was false, as desired.
\end{proof}

The following lemma shows how we can construct sets $A_i$ and $E_i$ satisfying the conditions in \Cref{lem:dXkey}.

\begin{lemma}
  \label{lem:extfieldsdis}
  Define all variables as in \Cref{def:pevprod}. Assume that there exists a sequence of nested fields $\bF_{q_0}\subseteq\bF_{q_1}\subseteq\cdots\subseteq\bF_{q_u}=\bF_q$ such that for $i\in[u]$, $E_i$ is a subset of $\bF_{q_0}$, and $A_i$ is a subset of those elements of $\bF_{q_i}$ with degree (of the minimal polynomial) $\geq\ell_i-k_i+1$ over $\bF_{q_{i-1}}$. Then $C$ has $X$-distance
  \begin{equation*}
    d_X \geq \prod_{i\in[u]}(\ell_i-k_i+1).
  \end{equation*}
\end{lemma}
\begin{proof}
  Because $E\subseteq\bF_{q_0}^u\subseteq\bF_q^u$, we may choose the polynomials $g_S\in\bF_q[X_1,\dots,X_u]^{\prod_{i\in[u]}[0,\ell_i-k_i]}_S$ in \Cref{lem:dXkey} to in fact lie in $g_S\in\bF_{q_0}[X_1,\dots,X_u]^{\prod_{i\in[u]}[0,\ell_i-k_i]}_S$. Then by \Cref{lem:dXkey}, to prove the desired $X$-distance bound, it suffices to show that $g_S(a)\neq 0$ for every $S\subseteq E$ with $|S|<\prod_{i\in[u]}(\ell_i-k_i+1)$ and every $a\in A$.
  
  For this purpose, we will show inductively that for every $0\leq i\leq u$, the polynomial
  \begin{equation*}
    g_S(a_1,\dots,a_i,X_{i+1},\dots,X_u)\in\bF_{q_i}[X_{i+1},\dots,X_u]^{\prod_{j\in[u]}[0,\ell_j-k_j]}
  \end{equation*}
  is nonzero. For the base case, when $i=0$ then by assumption $g_S(X_1,\dots,X_u)\neq 0$. For the inductive step, if $g_S(a_1,\dots,a_{i-1},X_i,\dots,X_u)\neq 0$, then viewing this polynomial as an element of $\bF_{q_{i-1}}[X_i][X_{i+1},\dots,X_u]$, it must have some nonzero coefficient $f(X_i)\in\bF_{q_{i-1}}[X_i]^{[0,\ell-k]}$. By assumption $a_i\in A_i$ cannot be a root of such a polynomial $f(X_i)$, so $g_S(a_1,\dots,a_i,X_{i+1},\dots,X_u)\neq 0$, completing the inductive step. Thus $g_S(a_1,\dots,a_u)\neq 0$, as desired.
\end{proof}

\begin{remark}
  In \Cref{lem:extfieldsdis}, we must have $q_0\geq n$ in order to have $E_i\subseteq\bF_{q_0}$, and for $i\in[u]$ we must have $q_i=q_{i-1}^{r_i}$ for some $r_i\geq\ell-k+1$ in order for $\bF_{q_i}$ to contain elements of degree $\geq\ell-k+1$ over $\bF_{q_{i-1}}$.
\end{remark}

We are now ready to complete the proof of \Cref{thm:pevmain}.

\begin{proof}[Proof of \Cref{thm:pevmain}]
  For $i\in[u]$, inductively define $q_i=q_{i-1}^{n_i'}$, so that $q=q_u$. Let $E_i\subseteq\bF_{q_0}$ be some subset of size $n_i$. Let $\alpha_i\in\bF_{q_i}$ be an element with an irreducible degree-$n_i'\geq n_i$ minimal polynomial $p_i(X)\in\bF_{q_{i-1}}[X]$, so that $\bF_{q_i}=\bF_{q_{i-1}}[\alpha_i]=\bF_{q_{i-1}}[X]/(p_i(X))$. Let $A_i$ be some size-$k_i$ subset of the set $\alpha_i+\bF_{q_{i-1}}\subseteq\bF_{q_i}$.

  Our goal is to show that for every $k_i\leq\ell_i\leq n_i$, every $\beta_i\in(\bF_q^*)^{E_i}$, and every $I\subseteq[u]$, the polynomial evaluation subsystem product code $C^I=(C^I_X,C^I_Z)=C^I(q,u,(\ell_i,A_i,E_i,\beta_i)_{i\in[u]})$ satisfies the $X$- and $Z$-distance bounds in \Cref{eq:pevmaindX,eq:pevmaindZ}.

  We first show the $X$-distance bound in \Cref{eq:pevmaindX}. By definition each $n_i'\geq n_i\geq\ell_i-k_i+1$. Letting $I=\{i_1<\cdots<i_{|I|}\}$, then each $a_{i_j}\in A_{i_j}$ has the form $a_{i_j}=\alpha_{i_j}+\alpha'$ for some $\alpha'\in\F_{q_{i_j-1}}$. Because $\alpha_{i_j}$ is not the root of any polynomial in $\bF_{q_{i_j-1}}[X]^{[0,n_{i_j}')}$, then $a_{i_j}$ is also not the root of any polynomial in $\bF_{q_{i_j-1}}[X]^{[0,n_{i_j}')}$, and in particular, in $\bF_{q_{i_{j-1}}}[X]^{[0,n_{i_j}')}$. Thus \Cref{eq:pevmaindX} follows by \Cref{lem:extfieldsdis}.

  By \Cref{lem:pevswapXZ}, there exist $\beta_i'\in(\bF_q^*)^{E_i}$ for $i\in[u]$ such that $(C^I_Z,C^I_X)=C^I(q,u,(n_i+k_i-\ell_i,A_i,E_i,\beta'_i)_{i\in[u]})$. The $Z$-distance bound \Cref{eq:pevmaindZ} then follows by applying \Cref{lem:extfieldsdis} to this code analogously as for the $X$-distance bound described above.
\end{proof}

\section{Core Gadgets over Large Alphabets}
\label{sec:coregad}
In this section, we present our core fault-tolerant gadgets over large alphabets, implementing error correction (\Cref{sec:errcorr}), initialization (\Cref{sec:initialization}), code switching (\Cref{sec:codeswitching}), logical gates Hadamard (\Cref{sec:hadamard}), Paulis (\Cref{sec:Paulis}), CNOTs (\Cref{sec:cnot}), Toffolis (\Cref{sec:toffoli}), as well as classical function gates (\Cref{sec:classicalfunctiongates}). 
Throughout this section we assume that all qudits have the same prime power dimension $q$, which may be growing in other parameters such as the block lengths of the relevant codes.
We will later show how to recursively compose these gadgets to reduce the qudit dimension.

\begin{remark}
  \label{remark:lacirind}
  The fault-tolerant gadgets in this section are provided in \Cref{lem:errcorr,lem:stateprep,lem:codeswitch,lem:hadamard,lem:paulis,lem:cx,lem:ccx,lem:co}. In each of these lemmas, we fix a parameter $\lambda_{\mathrm{run}}$ specifying the maximum weight per timestep of a fault under which the gadget returns an appropriate deviation of the desired output state. However, the physical circuit in each of these gadgets does not depend on the choice of $\lambda_{\mathrm{run}}$.
\end{remark}

\subsection{Error Correction}
\label{sec:errcorr}
In this section, we present our error-correction gadget for subsystem product codes.
We construct a quantum circuit $\cQ$ that performs error correction in the $X$ and $Z$ basis sequentially.
The construction crucially makes use of the local testability of tensor products of classical codes (see~\Cref{prop:loctest}) to deal with measurement errors.

\begin{lemma}
  \label{lem:errcorr}
  For $u,n,d\in\bN$, for $i\in[u]$ let $C^i=(C^i_X,C^i_Z)$ be a $[[n_i=n,\; k_i]]_q$ CSS non-subsystem code with specified CSS encoding map $\Enc^i:\bF_q^{k_i}\rightarrow C^i_Z/{C^i_X}^\perp$, such that the classical codes $C^i_X,C^i_Z$ have distance $\geq d$. 
  Define $\mu=\mu(u,n,d)$ to be the filling constant defined in \Cref{prop:loctest}.
  Let $N=\prod_{i\in[u]}[n_i]$, $K=\prod_{i\in[u]}[k_i]$, and let $C=\bigotimes_{i\in[u]}C^i$ be the $[[|N|,|K|,D]]_q$ subsystem product code, with product encoding map $\Enc:\bF_q^K\oplus\bF_q^{K'}\rightarrow(C_Z+C_X^\perp)/(C_Z\cap C_X^\perp)$. 
  For $\lambda_{\mathrm{in}},\lambda_{\mathrm{run}}\geq 0$ satisfying
  \begin{align}
    \label{eq:ecparams}
    \begin{split}
      2\lambda_{\mathrm{in}} + 4n(1+\mu u)(1+un^2)\cdot\lambda_{\mathrm{run}} &< D,
    \end{split}
  \end{align}
  let
  \begin{align}
    \label{eq:eclamout}
    \lambda_{\mathrm{out}} &= 10n(1+2\mu u)(1+un^2)\cdot\lambda_{\mathrm{run}},
  \end{align}
  and define decorated $[[|N|,|K|]]_q$ subsystem codes
  \begin{align*}
    D_{\mathrm{in}} &= \left(C,\; \Enc,\; \Gamma_{\mathrm{in}}=\bC^{\bF_q^{K'}\times\bF_q^{K'}},\; \cE_{\mathrm{in}}=2^{N}|_{\geq \lambda_{\mathrm{in}}}\right) \\
    D_{\mathrm{out}} &= \left(C,\; \Enc,\; \Gamma_{\mathrm{out}}=\left\{\ket{0^{K'}}\bra{0^{K'}}\right\},\; \cE_{\mathrm{out}}=2^{N}|_{\geq \lambda_{\mathrm{out}}}\right)
  \end{align*}
  Then there exists a quantum circuit $\cQ$ using gate set $\{\gCX^*,\gInit_*,\gCO_*\}$, space $|N'|\leq (u+1)n^u$, and time $T\leq 2(3+un^2)$ such that $(\cQ,2^{N'}|_{\geq \lambda_{\mathrm{run}}}^{\sqcup T},D_{\mathrm{in}},D_{\mathrm{out}})$ forms a mending fault-tolerant gadget for the $|K|$-qudit identity channel $\bar{O}=I_K:\bC^{\bF_q^K\times\bF_q^K}\rightarrow\bC^{\bF_q^K\times\bF_q^K}$.
\end{lemma}

\begin{algorithm}
  \caption{\label{alg:syndext} Syndrome extraction circuit for subsystem product codes. We refer to the qudits labeled by $N=\prod_{i\in[u]}[n_i]$ comprising the input $\sigma$ as ``data qudits.'' We call the other qudits ``ancilla qudits,'' and label them by a set $S=\bigsqcup_{i\in[u]}S^i$ defined as in \Cref{def:pcpc} by
    \begin{equation*}
      S^i=[n_1]\times\cdots\times[n_{i-1}]\times[m_i]\times[n_{i+1}]\times\cdots\times[n_u],
    \end{equation*}
    Furthermore, for a tuple $j=(j_1,\dots,j_u)$, we let $j_{-i}=(j_{i'})_{i'\in[u]\setminus\{i\}}$, and we write $j=(j_i,j_{-i})$. Within a given step of the outer loop in \Cref{li:seseq}, each step of the inner loop in \Cref{li:separr} acts on disjoint qudits, and hence this entire inner loop can be performed using parallel $\gCX$ gates in a single timestep.}
  \SetKwInOut{Input}{Input}
  \SetKwInOut{Output}{Output}

  \SetKwFunction{FnSyndExt}{SyndExt}
  \SetKwProg{Fn}{Function}{:}{}

  \Input{State $\sigma\in\bC^{\bF_q^N\times\bF_q^N}$, basis $\alpha\in\{X,Z\}$, parity-check matrices $H^i\in\bF_q^{m_i\times n_i}$ for $i\in[u]$}
  \Output{Outcome of applying the isometry $\ket{x}\mapsto\ket{x}\otimes\ket{Hx}$ to $\sigma$, where $H\in\bF_q^{S\times N}$ is the product-code parity-check matrix associated to $H^1,\dots,H^u$ (see \Cref{def:pcpc}).}

  \Fn{\FnSyndExt{$\sigma;\alpha,(H^i)_{i\in[u]}$}}{
    Initialize $|S|$ ancilla qudits, labeled by the set $S$, to $\ket{0}$ (if $\alpha=Z$) or $\ket{+}$ (if $\alpha=X$) \\
    \ForEach{$i\in[u],\; j_i\in[n_i],\; j_i'\in[m_i]$}{ \label{li:seseq}
      \ForEach{$j_{-i}\in\prod_{i'\in[u]\setminus\{i\}}[n_{i'}]$}{ \label{li:separr}
        \If{$\alpha=X$}{
          Apply $\gCX^{-H^i_{j_i',j_i}}$ to ancilla qudit $(j_i',j_{-i})\in S^i$ and data qudit $(j_i,j_{-i})$
        }
        \If{$\alpha=Z$}{
          Apply $\gCX^{H^i_{j_i',j_i}}$ to data qudit $(j_i,j_{-i})$ and ancilla qudit $(j_i',j_{-i})\in S^i$
        }
      }
    }
    \Return{All $|N|$ data qudits and $|S|$ ancilla qudits}
  }
\end{algorithm}

\begin{algorithm}
  \caption{\label{alg:errcorr} Error correction algorithm. Below we define $S=\bigsqcup_{i\in[u]}S^i$ as in \Cref{alg:syndext}. The specific classical decoding function $\Dec$ we will use is described in \Cref{lem:errcorr}. Each step of the loop in \Cref{li:ecfor} acts on disjoint qudits, and hence this entire loop can be performed using parallel $\gCX$ gates in a single timestep.}
  \SetKwInOut{Input}{Input}
  \SetKwInOut{Output}{Output}

  \SetKwFunction{FnErrCorr}{ErrCorr}
  \SetKwProg{Fn}{Function}{:}{}

  \Input{State $\sigma$ on ``data qudits'' $N=\prod_{i\in[u]}[n_i]$, basis $\alpha\in\{X,Z\}$, parity-check matrices $H^i\in\bF_q^{m_i\times n_i}$ for $i\in[u]$, classical decoding function $\Dec:\bF_q^S\rightarrow\bF_q^N$}
  \Output{State $\sigma'$ obtained by running $\alpha$-basis error correction on $\sigma$}

  \Fn{\FnErrCorr{$\sigma;\alpha,(H^i)_{i\in[u]},\Dec$}}{
    Let $\sigma_1\gets\FnSyndExt{$\sigma;\alpha,(H^i)_{i\in[u]}$}$, so that $\sigma_1\in\bC^{\bF_q^N\times\bF_q^N}\otimes\bC^{\bF_q^S\times\bF_q^S}$ \\ \label{li:ecse}
    Let $\sigma_2\in\bC^{\bF_q^N\times\bF_q^N}$ be the output from running $\gCO_{\alpha,\Dec}$ on the $S$-register of $\sigma_1$, along with $|N|-|S|$ ancilla qudits if $|N|>|S|$ \\ \label{li:ecco}
    \ForEach{$j\in N$}{ \label{li:ecfor}
      \If{$\alpha=X$}{
        Apply $\gCX^{+1}$ to the $j$th qudit in the $N$-register of $\sigma_1$ and the $j$th qudit in $\sigma_2$
      }
      \If{$\alpha=Z$}{
        Apply $\gCX^{-1}$ to the $j$th qudit in $\sigma_2$ and the $j$th qudit in the $N$-register of $\sigma_1$
      }
    }
    \Return{$\sigma'=\tr_S(\sigma_1)$ (i.e.~the $N$-register of $\sigma_1$)}
  }
\end{algorithm}

\begin{proof}
  We first describe the desired circuit $\cQ$, and then we will prove that it exhibits the desired fault-tolerance properties. For each basis $\alpha\in\{X,Z\}$, fix a full-rank parity-check matrix $H^i_\alpha\in\bF_q^{m_i\times n_i}$ for $C^i_\alpha$, so that $C^i_\alpha=\ker(H^i_\alpha)$. 
  Let $H_\alpha\in\bF_q^{S_\alpha\times N}$ be the associated product-code parity-check matrix with $S_\alpha=\bigsqcup_{i\in[u]}S^i_\alpha$ defined as in \Cref{def:pcpc}. 
  That is, $H_\alpha$ is a parity-check matrix for $C_\alpha=\bigotimes_{i\in[u]}C^i_\alpha$.
  Furthermore, \Cref{alg:syndext} provides a circuit using space $|N|+|S_\alpha|$ and time $\leq 1+un^2$ that on a pure input state $\ket{x}$ for $x\in\bF_q^N$ will (in the absence of a fault) output $\ket{x}\otimes\ket{H_\alpha x}$.

  We now define a classical function $\Dec_\alpha:\bF_q^{S_\alpha}\rightarrow\bF_q^N$ as follows. For $s\in\bF_q^{S_\alpha}$, then $\Dec_\alpha(s)$ first computes the vector $s'\in\im(H_\alpha)$ that is closest to $s$ in Hamming distance. Let $\alpha'$ be the opposite basis to $\alpha$, so that $\{\alpha,\alpha'\}=\{X,Z\}$. $\Dec_\alpha(s)$ subsequently computes and outputs the vector $c\in\bF_q^N$ satisfying $H_\alpha c=s'$ that is closest to $C_{\alpha'}^\perp$ in Hamming distance.

  On input $\sigma\in\bC^{\bF_q^N\times\bF_q^N}$, letting $\cQ_\alpha$ denote the circuit \FnErrCorr{$\sigma;\alpha,(H_\alpha^i)_{i\in[u]},\Dec_\alpha$} (see \Cref{alg:errcorr}), then we define our error-correction circuit $\cQ$ to invoke $\cQ_X$ and followed by $\cQ_Z$. In words, for each basis $\alpha\in\{X,Z\}$, we have defined $\cQ$ to do the following:
  \begin{enumerate}
  \item Perform syndrome extraction on the parity-check matrix $H_\alpha$
  \item Compute the nearest valid syndrome $s'$ to the measured syndrome $s$
  \item Compute the vector $c$ inducing syndrome $s'=H_\alpha(c)$ that is as close as possible to $C_{\alpha'}^\perp$
  \item Apply a Pauli correction ${\alpha'}^{-c}$ to the code state
  \end{enumerate}
  As we can think of elements of $C_{\alpha'}^\perp$ as gauge operators, the correction $c$ can be viewed as the sum of a gauge operator with the lowest possible weight error.

  We now analyze the space and time usage of $\cQ$. 
  Recall here that every $n_i=n$.
  As mentioned above, the call to \FnErrCorr{$\sigma;\alpha,(H^i)_{i\in[u]}$} in \Cref{li:ecse} of \Cref{alg:errcorr} uses space $|N|+|S_\alpha|$ and time $\leq 1+un^2$. \Cref{li:ecco} of \Cref{alg:errcorr} uses time $1$ while requiring an additional $|N|-|S_\alpha|$ qudits if $|N|>|S_\alpha|$, bringing the total space usage of \Cref{alg:errcorr} to $|N|+\max\{|N|,|S_\alpha|\}$. 
  The loop in \Cref{li:ecfor} uses time $1$ and requires no additional qudits. Thus \Cref{alg:errcorr} has overall space usage $|N|+\max\{|N|,|S_\alpha|\}\leq(u+1)n^u$ and time $\leq 3+un^2$, so $\cQ$ uses space $|N'|\leq (u+1)n^u$ and time $T\leq 2(3+un^2)$. 
  Note that here $N'$ denotes the set of all qudits used by $\cQ$.
  
  It remains to be shown that $(\cQ,2^{N'}|_{\geq \lambda_{\mathrm{run}}}^{\sqcup T},D_{\mathrm{in}},D_{\mathrm{out}})$ forms a mending fault-tolerant gadget for the $K$-qudit identity channel $\bar{O}=I_K$. For this purpose, we will use the following key claim which characterizes the output state after running the $Z$-basis error correction $\cQ_Z$.

  \begin{claim}
    \label{claim:eckey}
    For some $\ell\in\bN$, let $\rho\in\bC^{\bF_q^{|K|+\ell}\times\bF_q^{|K|+\ell}}$. Let $g_X,g_X'\in C_X^\perp$, and let $e_X,e_X',e_Z,e_Z'\in\bF_q^N$. Define $\sigma\in\bC^{\bF_q^{|N|+\ell}\times\bF_q^{|N|+\ell}}$ by
    \begin{equation*}
      \sigma = X^{g_X+e_X}Z^{e_Z}\Enc\left(\rho\otimes\ket{0^{K'}}\bra{0^{K'}}\right)\Enc^\dagger(Z^\dagger)^{e_Z'}(X^\dagger)^{g_X'+e_X'},
    \end{equation*}
    where above the encoding isometry $\Enc$ (see \Cref{def:CSSenc}) acts on the $K$-register of $\rho$ along with $\ket{0^{K'}}\bra{0^{K'}}$. Let $\cF$ be a $2^{N'}|_{\geq \lambda_{\mathrm{run}}}^{\sqcup T/2}$-avoiding Pauli fault for $\cQ_Z$. 
    Let $\zeta$ be a postselection for $\cQ_Z$.
    Then the output $\tilde{\sigma}=\cQ_Z[\cF,\zeta]\otimes I_\ell(\sigma)$
    is of the form
    \begin{equation*}
      \tilde{\sigma} \propto X^{\tilde{c}_X+\tilde{e}_X}Z^{\tilde{e}_Z}\Enc\left(\rho\otimes\ket{0^{K'}}\bra{0^{K'}}\right)\Enc^\dagger(Z^\dagger)^{\tilde{e}_Z'}(X^\dagger)^{\tilde{c}_X'+\tilde{e}_X'},
    \end{equation*}
    for some $\tilde{c}_X,\tilde{c}_X'\in C_Z$ and some $\tilde{e}_X,\tilde{e}_X',\tilde{e}_Z,\tilde{e}_Z'\in\bF_q^N$ satisfying
    \begin{align*}
      |\tilde{c}_X-C_X^\perp| &\leq 2|e_X|+ 4\mu un(1+un^2)\cdot\lambda_{\mathrm{run}} \\
      |\tilde{c}_X'-C_X^\perp| &\leq 2|e_X'|+ 4\mu un(1+un^2)\cdot\lambda_{\mathrm{run}} \\
      |\tilde{e}_X|,|\tilde{e}_X'| &\leq 4\mu un(1+un^2)\cdot\lambda_{\mathrm{run}} \\
      |\tilde{e}_Z-e_Z|,|\tilde{e}_Z'-e_Z'| &\leq 2n(1+un^2)\cdot\lambda_{\mathrm{run}},
    \end{align*}
    where we denote $|a-B|=\min_{b\in B}|a-b|$.
  \end{claim}

  We will first complete the proof of mending fault-tolerance assuming \Cref{claim:eckey}, and then we will prove the claim. 
  For this purpose, for some $\ell\in\bN$, let $\rho\in\bC^{\bF_q^{|K|+\ell}\times\bF_q^{|K|+\ell}}$, let $\gamma\in\Gamma_{\mathrm{in}}$, let $\sigma_0\in\bC^{\bF_q^{|N|+\ell}\times\bF_q^{|N|+\ell}}$ be a Pauli $[\ell]$-deviation of $\Enc\otimes I_\ell(\rho\otimes\gamma)$, and let $\cF$ be a $2^{N'}|_{\geq \lambda_{\mathrm{run}}}^{\sqcup T}$-avoiding Pauli fault and $\zeta$ be a postselection for $\cQ$. 
  Let $\zeta_X$ denote the component of $\zeta$ corresponding to~$\gCO_*$ gates in $\cQ_X$ and, vice versa, $\zeta_Z$ for the component of $\zeta$ corresponding to~$\gCO_*$ gates in $\cQ_Z$.
  To show mending, by \Cref{lem:paulift} it suffices to show that the resulting output $\cQ[\cF,\zeta]\otimes I_\ell(\sigma_0)$ is a linear combination of $\cE_{\mathrm{out}}\sqcup [\ell]$-deviations of $\Enc\otimes I_\ell(L(\rho)\otimes\ket{0^{K'}}\bra{0^{K'}})$ for superoperators $L:\bC^{\bF_q^K\times\bF_q^K}\rightarrow\bC^{\bF_q^K\times\bF_q^K}$. To furthermore show fault-tolerance, by \Cref{lem:paulift} it suffices to show that if in fact
  \begin{equation}
    \label{eq:ecft}
    \sigma_0 \text{ is a Pauli }\cE_{\mathrm{in}}\sqcup [\ell]\text{-deviation of }\Enc\otimes I_\ell(\rho\otimes\gamma),
  \end{equation}
  then the output $\cQ[\cF,\zeta]\otimes I_\ell(\sigma_0)$ is a $\cE_{\mathrm{out}}\sqcup [\ell]$-deviation of $\Enc\otimes I_\ell(\rho\otimes\ket{0^{K'}}\bra{0^{K'}})$.

  An arbitrary operator $\gamma\in\Gamma_{\mathrm{in}}=\bC^{\bF_q^{K'}\times\bF_q^{K'}}$ can be written as a linear combination of operators of the form $Z^{f_Z}\ket{+^{K'}}\bra{+^{K'}}(Z^\dagger)^{f_Z'}$ for $f_Z,f_Z'\in\bF_q^{K'}$. 
  Furthermore, by the definition of the code~$C$ and associated encoding map~$\Enc$, for every such $f_Z,f_Z'$, there exist $g_Z,g_Z'\in C_Z^\perp$ for which
  \begin{align*}
Z^{g_Z}\Enc\left(\rho\otimes\ket{+^{K'}}\bra{+^{K'}}\right)\Enc^\dagger(Z^\dagger)^{g_Z}
    &= \Enc\left(\rho\otimes Z^{f_Z}\ket{+^{K'}}\bra{+^{K'}}(Z^\dagger)^{f_Z'}\right)\Enc^\dagger.
  \end{align*}
  Here $\Enc$ refers to the encoding isometry; in the language of stabilizer codes we are simply translating unencoded gauge operators $Z^{f_Z},(Z^\dagger)^{f_Z'}$ to their respective encoded representatives $Z^{g_Z},(Z^\dagger)^{g_Z'}$.

  Thus we may decompose $\sigma_0$ into a linear combination of $\sigma$ of the form
  \begin{align*}
    \sigma
    &= X^{e_X}Z^{g_Z+e_Z}\Enc\left(\rho\otimes\ket{+^{K'}}\bra{+^{K'}}\right)\Enc^\dagger(Z^\dagger)^{g_Z'+e_Z'}X^{e_X'},
  \end{align*}
  where each $e_X,e_X',e_Z,e_Z'\in\bF_q^N$, and each $g_Z,g_Z'\in C_Z^\perp$. In the case where \Cref{eq:ecft} holds, then we furthermore have that $|e_X,|e_X'|,|e_Z|,|e_Z'|\leq\lambda_{\mathrm{in}}$. Now as \Cref{claim:eckey} is completely symmetric with respect to the $X$ and $Z$ bases, the claim also holds when swapping the bases. Applying this basis-swapped version of \Cref{claim:eckey}, we conclude that when $\cQ_X[(F_1,\dots,F_{T/2}),\zeta_X]$ is applied to $\sigma$, the result $\tilde{\sigma}$ is of the form
  \begin{align*}
    \tilde{\sigma}
    &\propto X^{\tilde{e}_X}Z^{\tilde{c}_Z+\tilde{e}_Z}\Enc\left(\rho\otimes\ket{+^{K'}}\bra{+^{K'}}\right)\Enc^\dagger(Z^\dagger)^{\tilde{c}_Z'+\tilde{e}_Z'}(X^\dagger)^{\tilde{e}_X'}
  \end{align*}
  for some $\tilde{c}_Z,\tilde{c}_Z'\in C_X$ and some $\tilde{e}_X,\tilde{e}_X',\tilde{e}_Z,\tilde{e}_Z'\in\bF_q^N$ satisfying
  \begin{align}
    \label{eq:echalfway}
    \begin{split}
      |\tilde{c}_Z-C_Z^\perp| &\leq 2|e_Z|+ 4\mu un(1+un^2)\cdot\lambda_{\mathrm{run}} \\
      |\tilde{c}_Z'-C_Z^\perp| &\leq 2|e_Z'|+ 4\mu un(1+un^2)\cdot\lambda_{\mathrm{run}} \\
      |\tilde{e}_Z|,|\tilde{e}_Z'| &\leq 4\mu un(1+un^2)\cdot\lambda_{\mathrm{run}} \\
      |\tilde{e}_X| &\leq |e_X|+2n(1+un^2)\cdot\lambda_{\mathrm{run}} \\
      |\tilde{e}_X'| &\leq |e_X'|+2n(1+un^2)\cdot\lambda_{\mathrm{run}}.
    \end{split}
  \end{align}
  By the definition of a subsystem code $C$ with associated encoding isometry $\Enc$, given $\tilde{c}_Z,\tilde{c}_Z'\in C_X$, there must exist $\tilde{b}_Z,\tilde{b}_Z'\in\bF_q^K$ for which
  \begin{equation*}
    Z^{\tilde{c}_Z}\Enc = \Enc Z^{\tilde{b}_Z} \hspace{1em} \text{and} \hspace{1em} Z^{\tilde{c}_Z'}\Enc = \Enc Z^{\tilde{b}_Z'}.
  \end{equation*}
  In the language of stabilizer codes were are simply translating between unencoded logical operators $Z^{\tilde{b}_Z},Z^{\tilde{b}_Z'}$ and their respective encoded representatives $Z^{\tilde{c}_Z},Z^{\tilde{c}_Z'}$.
  Furthermore, by \Cref{eq:ecparams}, we have $2\lambda_{\mathrm{in}}+ 4\mu un(1+un^2)\cdot\lambda_{\mathrm{run}} < D$. 
  Hence by the definition of the subsystem code distance $D$, if \Cref{eq:ecft} holds so that $|e_Z|,|e_Z'|\leq\lambda_{\mathrm{in}}$, then $\tilde{c}_Z,\tilde{c}_Z'\in C_Z^\perp$. 
  But $\tilde{c}_Z,\tilde{c}_Z'$ also lie in $C_X$, so they lie in $C_X\cap C_Z^\perp$. Hence $Z^{\tilde{c}_Z},Z^{\tilde{c}_Z'}$ are stabilizers of $C$, that is, they act trivially on the codespace $\im(\Enc)$, or equivalently, $\tilde{b}_Z,\tilde{b}'_Z=0$. 
  Thus $\tilde{\sigma}$ has the form
  \begin{align*}
    \tilde{\sigma}
    &\propto X^{\tilde{e}_X}Z^{\tilde{e}_Z}\Enc\left(Z^{\tilde{b}_Z}\rho(Z^\dagger)^{\tilde{b}_Z'}\otimes\ket{+^{K'}}\bra{+^{K'}}\right)\Enc^\dagger(Z^\dagger)^{\tilde{e}_Z'}(X^\dagger)^{\tilde{e}_X'},
  \end{align*}
  with $\tilde{b}_Z,\tilde{b}'_Z=0$ if \Cref{eq:ecft} holds.

  We now analyze the action of $\cQ_Z[(F_{T/2+1},\dots,F_T),\zeta_Z]$ on $\tilde{\sigma}$ using a similar argument as above, but with the $X$ and $Z$ bases swapped. Specifically, we decompose $\ket{+^{K'}}\bra{+^{K'}}$ into a linear combination of operators $X^{f_X}\ket{0^{K'}}\bra{0^{K'}}(X^\dagger)^{f_X'}$ for $f_X,f_X'\in\bF_q^{K'}$. For every such $f_X,f_X'$, we choose appropriate $g_X,g_X'\in C_X^\perp$ such that we obtain $\tilde{\sigma}$ as a linear combination of operators of the form
  \begin{equation*}
    X^{g_X+\tilde{e}_X}Z^{\tilde{e}_Z}\Enc\left(Z^{\tilde{b}_Z}\rho(Z^\dagger)^{\tilde{b}_Z'}\otimes\ket{0^{K'}}\bra{0^{K'}}\right)\Enc^\dagger(Z^\dagger)^{\tilde{e}_Z'}(X^\dagger)^{g_X'+\tilde{e}_X'}.
  \end{equation*}
  (In fact here we may choose $g_X\in\Enc(0,f_X)$, $g_X'\in\Enc(0,f_X')$ to be coset representatives given by the $\bF_q$-linear encoding map $\Enc$.) By \Cref{claim:eckey}, applying $\cQ_Z[(F_{T/2+1},\dots,F_T),\zeta_Z]$ to the operator above yields a result $\titi{\sigma}$ of the form
  \begin{align*}
    \titi{\sigma}
    &\propto X^{\tilde{c}_X+\titi{e}_X}Z^{\titi{e}_Z}\Enc\left(Z^{\tilde{b}_Z}\rho(Z^\dagger)^{\tilde{b}_Z'}\otimes\ket{0^{K'}}\bra{0^{K'}}\right)\Enc^\dagger(Z^\dagger)^{\titi{e}_Z'}(X^\dagger)^{\tilde{c}_X'+\titi{e}_X'}
  \end{align*}
  for some $\tilde{c}_X,\tilde{c}_X'\in C_Z$ and some $\titi{e}_X,\titi{e}_X',\titi{e}_Z,\titi{e}_Z'\in\bF_q^N$ satisfying 
  \begin{align*}
    |\tilde{c}_X-C_X^\perp| &\leq 2|\tilde{e}_X|+ 4\mu un(1+un^2)\cdot\lambda_{\mathrm{run}} \\
    |\tilde{c}_X'-C_X^\perp| &\leq 2|\tilde{e}_X'|+ 4\mu un(1+un^2)\cdot\lambda_{\mathrm{run}} \\
    |\titi{e}_X|,|\titi{e}_X'| &\leq 4\mu un(1+un^2)\cdot\lambda_{\mathrm{run}} \\
    |\titi{e}_Z| &\leq |\tilde{e}_Z|+2n(1+un^2)\cdot\lambda_{\mathrm{run}} \\
    |\titi{e}_Z'| &\leq |\tilde{e}_X'|+2n(1+un^2)\cdot\lambda_{\mathrm{run}}.
  \end{align*}
  By the definition of $\Enc$, given $\tilde{c}_X,\tilde{c}_X'\in C_Z$, there must exist $\tilde{b}_X,\tilde{b}_X'\in\bF_q^K$ for which
  \begin{equation}
    \label{eq:ecloginout}
    X^{\tilde{c}_X}\Enc = \Enc X^{\tilde{b}_X} \hspace{1em} \text{and} \hspace{1em} X^{\tilde{c}_X'}\Enc = \Enc X^{\tilde{b}_X'}.
  \end{equation}
  Furthermore, in the case where \Cref{eq:ecft} holds so that $|e_X|,|e_X'|\leq\lambda_{\mathrm{in}}$, then by \Cref{eq:echalfway} and \Cref{eq:ecparams}, we have
  \begin{align*}
    |\tilde{c}_X-C_X^\perp|,|\tilde{c}'_X-C_X^\perp|
    &\leq 2\max\{|\tilde{e}_X|,|\tilde{e}_X'|\} + 4\mu un(1+un^2)\cdot\lambda_{\mathrm{run}} \\
    &\leq 2\max\{|e_X|,|e_X'|\} + 4n(1+un^2)\cdot\lambda_{\mathrm{run}} + 4\mu un(1+un^2)\cdot\lambda_{\mathrm{run}} \\
    &\leq 2\lambda_{\mathrm{in}} + 4n(1+\mu u)(1+un^2)\cdot\lambda_{\mathrm{run}} \\
    &< D,
  \end{align*}
  and hence $\tilde{c}_X,\tilde{c}_X'\in C_Z\cap C_X^\perp$ so that $\tilde{b}_X,\tilde{b}_X'=0$. Thus we conclude that when we apply $\cQ[\cF,\zeta]$ to input $\sigma$, the final output is a linear combination of operators $\titi{\sigma}$ of the form
  \begin{align*}
    \titi{\sigma}
    &\propto X^{\titi{e}_X}Z^{\titi{e}_Z}\Enc\left(X^{\tilde{b}_X}Z^{\tilde{b}_Z}\rho(Z^\dagger)^{\tilde{b}_Z'}(X^\dagger)^{\tilde{b}_X'}\otimes\ket{0^{K'}}\bra{0^{K'}}\right)\Enc^\dagger(Z^\dagger)^{\titi{e}_Z'}(X^\dagger)^{\titi{e}_X'}
  \end{align*}
  with
  \begin{align*}
    |\titi{e}_X|,|\titi{e}_X'| &\leq 4\mu un(1+un^2)\cdot\lambda_{\mathrm{run}} \\
    |\titi{e}_Z|,|\titi{e}_Z'| &\leq 2n(1+2\mu u)(1+un^2)\cdot\lambda_{\mathrm{run}},
  \end{align*}
  and with $\tilde{b}_X,\tilde{b}_X',\tilde{b}_Z,\tilde{b}_Z'=0$ if \Cref{eq:ecft} holds. The Pauli superoperator $\tau\mapsto X^{\titi{e}_X}Z^{\titi{e}_Z}\tau(Z^\dagger)^{\titi{e}_Z'}(X^\dagger)^{\titi{e}_X'}$ therefore has weight $\leq 8n(1+2\mu u)(1+un^2)\cdot\lambda_{\mathrm{run}}<\lambda_{\mathrm{out}}$, so by \Cref{lem:paulift}, $(\cQ,2^{N'}|_{\geq \lambda_{\mathrm{run}}}^{\sqcup T},D_{\mathrm{in}},D_{\mathrm{out}})$ forms a mending fault-tolerant gadget for the $K$-qudit identity channel $\bar{O}=I_K$, as desired.

  It only remains to prove \Cref{claim:eckey}, which we do below.
  
  \begin{proof}[Proof of \Cref{claim:eckey}]
    The first timestep of $\cQ_Z$ by definition initializes ancilla qudits to $\ket{0}\bra{0}$, while leaving the input qudits labeled $N\sqcup[\ell]$ in the state $\sigma$. The remainder of $\cQ_Z$ simply consists of $\gCX$ gates, along with a single classical function gate. By definition, conjugating a $X$ or $Z$ Pauli operator by a $\gCX$ gate yields another $X$ or $Z$ Pauli operator, respectively. Furthermore, such conjugation can only expand the support of a Pauli $X$ (resp.~Pauli $Z$) if the original operator's support contained the control (resp.~target) qudit.

    It follows that just prior to executing the classical function gate in \Cref{li:ecco} of \Cref{alg:errcorr}, the algorithm's state is proportional to
    \begin{align*}
      &X^{g_X+e_X+e_{X,1}}Z^{e_Z+e_{Z,1}}\Enc\left(\rho\otimes\ket{0^{K'}}\bra{0^{K'}}\right)\Enc^\dagger(Z^\dagger)^{e_Z'+e_{Z,1}'}(X^\dagger)^{g_X'+e_X'+e_{X,1}'} \\
      &\hspace{1em}\otimes \ket{H_Z(g_X+e_X)+f_1}\bra{H_Z(g_X'+e_X')+f_1'}
    \end{align*}
    for some $e_{X,1},e_{X,1}',e_{Z,1},e_{Z,1}'\in\bF_q^N$ satisfying
    \begin{align*}
      |e_{X,1}|,|e_{X,1}'| &\leq \lambda_{\mathrm{run}}\cdot(1+un^2) \\
      |e_{Z,1}|,|e_{Z,1}'| &\leq \lambda_{\mathrm{run}}\cdot(1+un^2)\cdot n
    \end{align*}
    and some $f_1,f_1'\in\bF_q^{S_Z}$ satisfying
    \begin{align*}
      |f_1|,|f_1'| &\leq \lambda_{\mathrm{run}}\cdot(1+un^2)\cdot un.
    \end{align*}
    Specifically, the $\lambda_{\mathrm{run}}\cdot(1+un^2)$ factor in each of the RHSs above arises because the syndrome extraction circuit runs for $\leq 1+un^2$ timesteps, each of which could introduce up to $\lambda_{\mathrm{run}}$ new Pauli errors from the fault $\cF$. The additional factor of $n$ in the bound on $|e_{Z,1}|,|e_{Z,1}'|$ arises because each syndrome qudit in $S_Z$ is the target of $\leq n$ $\gCX$ gates, each of which could propagate a $Z$ error on that qudit to the control. Similarly, the factor of $un$ on the bound on $|f_1|,|f_1'|$ arises because each code qudit in $N$ is the control of $\leq un$ $\gCX$ gates, each of which could propagate a $X$ error on that qudit to the target.

    Now if $H_Z(g_X+e_X)+f_1\neq H_Z(g_X'+e_X')+f_1'$, then the $Z$-basis measurements in the classical function gate in \Cref{li:ecco} of \Cref{alg:errcorr} collapse the state down to $0$, and the desired claim is trivially satisfied. Therefore assume instead that some $s\in\bF_q^{S_Z}$ has
    \begin{equation*}
      H_Z(g_X+e_X)+f_1 = s = H_Z(g_X'+e_X')+f_1'.
    \end{equation*}
    The $\zeta$-postselected classical function gate $\gCO_{Z,\Dec_Z}^\zeta$ in \Cref{li:ecco} of \Cref{alg:errcorr} then computes the $s'\in\im(H_Z)$ that is closest to $s$ in Hamming distance, before computing and outputting the vector $g_2+e_2\in\bF_q^N$ with $g_2\in C_X^\perp$ and $|e_2|$ as small as possible while satisfying $H_Z(g_2+e_2)=s'$. Note that because we are applying $\gCO_{Z,\Dec_Z}^\zeta$ to a state that is already in the computational basis, the postselection $\zeta$ simply introduces a global phase.

    Thus the final output of $\cQ_Z$ is proportional to
    \begin{align}
      \label{eq:ecoutraw}
      &X^{(g_X+e_X)-(g_2+e_2)+e_{X,3}}Z^{\tilde{e}_Z}\Enc\left(\rho\otimes\ket{0^{K'}}\bra{0^{K'}}\right)\Enc^\dagger(Z^\dagger)^{\tilde{e}_Z'}(X^\dagger)^{(g_X'+e_X')-(g_2+e_2)+e_{X,3}'},
    \end{align}
    where $e_{X,3},e_{X,3}',\tilde{e}_Z,\tilde{e}_Z'\in\bF_q^N$ satisfy
    \begin{align*}
      |e_{X,3}-e_{X,1}|,|e_{X,3}'-e_{X,1}'| &\leq \lambda_{\mathrm{run}}\cdot 2 \\
      |\tilde{e}_Z-(e_Z+e_{Z,1})|,|\tilde{e}_Z'-(e_Z'+e_{Z,1}')| &\leq \lambda_{\mathrm{run}}\cdot 2.
    \end{align*}

    Now
    \begin{equation}
      \label{eq:HZdiff}
      H_Z((g_X+e_X)-(g_2+e_2)) = s-s'-f_1.
    \end{equation}
    By definition $H_Z(g_X+e_X)$ was a valid choice for $s'$ in the execution of $\Dec_Z$ (though perhaps not the closest to $s=H_Z(g_X+e_X)+f_1$), meaning that $|s-s'|\leq|f_1|$. Therefore $s-s'-f_1\in\im(H_Z)$ has $|f_1+s'-s|\leq 2|f_1|$. Recalling that $\mu=\mu(u,n,d)$ as in \Cref{prop:loctest}, then \Cref{prop:loctest} implies that there is some $e_4\in\bF_q^N$ of weight $|e_4|\leq\mu\cdot 2|f_1|$ such that $H_Ze_4=s-s'-f_1$. Thus by \Cref{eq:HZdiff},
    \begin{equation*}
      (g_X+e_X)-(g_2+e_2) = \tilde{c}_X+e_4
    \end{equation*}
    for some $\tilde{c}_X\in C_Z$. Furthermore, we have
    \begin{align*}
      |\tilde{c}_X-C_X^\perp|
      &\leq |\tilde{c}_X-(g_X-g_2)| \\
      &= |e_X-e_2-e_4|.
    \end{align*}
    Now $g_X+(e_X-e_4)$ was a valid choice for $g_2+e_2$ in the execution of $\Dec_Z$ (though perhaps not the choice with the smallest $|e_2|$), meaning that $|e_2|\leq|e_X-e_4|$. Thus
    \begin{align*}
      |\tilde{c}_X-C_X^\perp|
      &\leq 2|e_X-e_4| \\
      &\leq 2|e_X|+4\mu|f_1| \\
      &\leq 2|e_X|+ 4\mu un(1+un^2)\cdot\lambda_{\mathrm{run}}.
    \end{align*}

    Applying the exact same reasoning as above but with $g_X'+e_X'$ in place of $g_X+e_X$ and $f_1'$ in place of $f_1$, we conclude that there exists some $e_4'\in\bF_q^N$ of weight $|e_4'|\leq\mu\cdot 2|f_1'|$ such that
    \begin{equation*}
      (g_X'+e_X')-(g_2+e_2) = \tilde{c}_X'+e_4'
    \end{equation*}
    for some $\tilde{c}_X'\in C_Z$ satisfying
    \begin{align*}
      |\tilde{c}_X'-C_X^\perp|
      &\leq 2|e_X'|+ 4\mu un(1+un^2)\cdot\lambda_{\mathrm{run}}.
    \end{align*}

    Thus letting
    \begin{align*}
      \tilde{e}_X &= e_4+e_{X,3} \\
      \tilde{e}_X' &= e_4'+e_{X,3}',
    \end{align*}
    then the output state in \Cref{eq:ecoutraw} takes on the desired form
    \begin{equation*}
      X^{\tilde{c}_X+\tilde{e}_X}Z^{\tilde{e}_Z}\Enc\left(\rho\otimes\ket{0^{K'}}\bra{0^{K'}}\right)\Enc^\dagger(Z^\dagger)^{\tilde{e}_Z'}(X^\dagger)^{\tilde{c}_X'+\tilde{e}_X'},
    \end{equation*}
    with
    \begin{align*}
      |\tilde{e}_X|,|\tilde{e}_X'|
      &\leq 2\mu|f_1|+2\lambda_{\mathrm{run}}+(1+un^2)\lambda_{\mathrm{run}} \\
      &\leq 4\mu un(1+un^2)\cdot\lambda_{\mathrm{run}}
    \end{align*}
    and
    \begin{align*}
      |\tilde{e}_Z-e_Z|,|\tilde{e}_Z'-e_Z'|
      &\leq 2\lambda_{\mathrm{run}}+n(1+un^2)\lambda_{\mathrm{run}} \\
      &\leq 2n(1+un^2)\lambda_{\mathrm{run}}.
    \end{align*}
  \end{proof}
  This establishes \Cref{claim:eckey} and thus we have finished the proof of \Cref{lem:errcorr}.
\end{proof}

\subsection{Initialization}
\label{sec:initialization}
In this section, we present our gadget for initializing logical $\ket{0}$ and $\ket{+}$ states in subsystem product codes. The core subroutine of this gadget is the same as was used in \Cref{sec:errcorr} for error correction, namely \Cref{alg:errcorr}.

\begin{lemma}
  \label{lem:stateprep}
  For $u,n,d\in\bN$, for $i\in[u]$ let $C^i=(C^i_X,C^i_Z)$ be a $[[n_i=n,\; k_i]]_q$ CSS non-subsystem code with specified CSS encoding map $\Enc^i:\bF_q^{k_i}\rightarrow C^i_Z/{C^i_X}^\perp$, such that the classical codes $C^i_X,C^i_Z$ have distance $\geq d$. Define $\mu=\mu(u,n,d)$ to be the expression in \Cref{prop:loctest}.
  Let $N=\prod_{i\in[u]}[n_i]$, $K=\prod_{i\in[u]}[k_i]$, and let $C=\bigotimes_{i\in[u]}C^i$ be the $[[|N|,|K|,D]]_q$ subsystem product code, with product encoding map $\Enc:\bF_q^K\oplus\bF_q^{K'}\rightarrow(C_Z+C_X^\perp)/(C_Z\cap C_X^\perp)$. For $\lambda_{\mathrm{in}},\lambda_{\mathrm{run}}\geq 0$, let
  \begin{align}
    \label{eq:splamout}
    \lambda_{\mathrm{out}} &= 16\mu un(1+un^2)\cdot\lambda_{\mathrm{run}}.
  \end{align}
  Let $\Gamma_{\mathrm{out}}^X=\left\{\ket{0^{K'}}\bra{0^{K'}}\right\}$ and $\Gamma_{\mathrm{out}}^Z=\left\{\ket{+^{K'}}\bra{+^{K'}}\right\}$. Then the following hold:
  \begin{enumerate}
  \item\label{it:spnew} For $\alpha\in\{X,Z\}$, define decorated subsystem codes
    \begin{align*}
      D_{\mathrm{in}} &= \emptyset \\
      D_{\mathrm{out}}^\alpha &= \left(C,\; \Enc,\; \Gamma_{\mathrm{out}}^\alpha,\; \cE_{\mathrm{out}}=2^{N}|_{\geq \lambda_{\mathrm{out}}}\right).
    \end{align*}
    Then there exists a quantum circuit $\cQ_\alpha$ using gate set $\{\gCX^*,\gInit_*,\gCO_*\}$, space $|N'|\leq (u+1)n^u$, and time $T\leq 4+un^2$ such that $(\cQ_\alpha,2^{N'}|_{\geq \lambda_{\mathrm{run}}}^{\sqcup T},D_{\mathrm{in}},D_{\mathrm{out}}^\alpha)$ forms a fault-tolerant gadget for the channel $\bar{O}_\alpha:\bC\rightarrow\bC^{\bF_q^K\times\bF_q^K}$ given by $\bar{O}_X(1)=\ket{+^K}\bra{+^K}$ and $\bar{O}_Z(1)=\ket{0^K}\bra{0^K}$.
  \item\label{it:spreplace} For $\alpha\in\{X,Z\}$, letting $\lambda_{\mathrm{in}}=D$, then define decorated $[[|N|,|K|,D]]_q$ subsystem codes
    \begin{align*}
      D_{\mathrm{in}} &= \left(C,\; \Enc,\; \Gamma_{\mathrm{in}}=\bC^{\bF_q^{K'}\times\bF_q^{K'}},\; \cE_{\mathrm{in}}=2^{N}|_{\geq \lambda_{\mathrm{in}}}\right) \\
      D_{\mathrm{out}}^\alpha &= \left(C,\; \Enc,\; \Gamma_{\mathrm{out}}^\alpha,\; \cE_{\mathrm{out}}=2^{N}|_{\geq \lambda_{\mathrm{out}}}\right).
    \end{align*}
    Then there exists a quantum circuit $\cQ_\alpha$ using gate set $\{\gCX^*,\gInit_*,\gCO_*\}$, space $|N'|\leq (u+1)n^u$, and time $T\leq 4+un^2$ such that $(\cQ_\alpha,2^{N'}|_{\geq \lambda_{\mathrm{run}}}^{\sqcup T},D_{\mathrm{in}},D_{\mathrm{out}}^\alpha)$ forms a fault-tolerant gadget for the channel $\bar{O}_\alpha=\gInit_\alpha^{\otimes K}:\bC\rightarrow\bC^{\bF_q^K\times\bF_q^K}$.
  \end{enumerate}
\end{lemma}

\begin{remark}
  \label{remark:init}
  The gadgets in \Cref{it:spnew,it:spreplace} are given by identical physical circuits. \Cref{it:spnew} has trivial input code $D_{\mathrm{in}}$, i.e.~no input qudits. 
  In contrast, \Cref{it:spreplace} applies $\gInit_*$ gates to an input code state and performing a ``reset'' operation, thereby tracing out the logical input qudits. 
  Because we allow errors that are non-physical (i.e.~non-channel) superoperators, \Cref{it:spreplace} requires a more careful analysis to ensure that logical errors are not introduced to any ``external system'' that is entangled with the input code state.
\end{remark}

\begin{proof}[Proof of \Cref{lem:stateprep}]
  We will prove the result for the $\alpha=X$ case; the proof for the $\alpha=Z$ case is exactly analogous by symmetry. We consider the two statements in the lemma separately:
  \begin{enumerate}
  \item Define the parity-check matrices $(H^i)_{i\in[u]}$, the set $S_Z=\bigsqcup_{i\in[u]}S^i_Z$, and the classical function $\Dec_Z$ as in the proof of \Cref{lem:errcorr}. Our desired state preparation circuit $\cQ_X$ uses a set of qudits labeled by $N'=N\sqcup S_Z$. $\cQ_X$ applies $\gInit_X$ to all qudits in $N'$ to obtain an initial state $\sigma\in\bC^{\bF_q^{N'}\times\bF_q^{N'}}$, and then runs \FnErrCorr{$\tr_{S_Z}(\sigma);Z,(H^i)_{i\in[u]},\Dec_Z$} from \Cref{alg:errcorr} and outputs the returned state in $\bC^{\bF_q^N\times\bF_q^N}$.

    As described in the proof of \Cref{lem:errcorr}, \FnErrCorr{$\tr_{S_Z}(\sigma);Z,(H^i)_{i\in[u]},\Dec_Z$} uses space $|N'|\leq(u+1)n^u$ and time $\leq 3+un^2$, so our state preparation gadget $\cQ_X$ here has the same space usage $N'$, and time usage $T\leq 4+un^2$.
    
    It only remains to be shown that $(\cQ_X,2^{N'}|_{\geq \lambda_{\mathrm{run}}}^{\sqcup T},D_{\mathrm{in}},D_{\mathrm{out}})$ is a fault-tolerant gadget for $\bar{O}_X$. For this purpose, fix a $2^{N'}|_{\geq \lambda_{\mathrm{run}}}^{\sqcup T}$-avoiding Pauli fault $\cF=(F_1,\dots,F_T)$, and fix a postselection~$\zeta$ for $\cQ_X$. By \Cref{lem:paulift}, it suffices to show that for every $\ell\in\bN$ and every $\rho\in\bC^{\bF_q^\ell\times\bF_q^\ell}$, the output $\cQ_X[\cF,\zeta]\otimes I_\ell(\rho)$ is a $\cE_{\mathrm{out}}\sqcup [\ell]$-deviation of $\Enc(\ket{+^K}\bra{+^K}\otimes\ket{0^{K'}}\bra{0^{K'}})\otimes\rho$.
    By definition the $\ell$-qudit external system remains in the state $\rho$ throughout the execution, so we may ignore this external system and just focus on the qudits in $N'=N\sqcup S_Z$.
    After the first timestep, the state of these qudits is $F_1(\ket{+^{N'}}\bra{+^{N'}})$. Because $F_1$ is a Pauli superoperator and
    \begin{align*}
      \ket{+^N}\bra{+^N}
      &= \sum_{e_X,e_X'\in\bF_q^N}X^{e_X}\Enc\left(\ket{+^K}\bra{+^K}\otimes\ket{0^{K'}}\bra{0^{K'}}\right)\Enc^\dagger(X^\dagger)^{e_X'},
    \end{align*}
    we can therefore express the state $\tr_{S_Z}(\sigma)=\tr_{S_Z}\left(F_1\left(\ket{+^{N'}}\bra{+^{N'}}\right)\right)$ on qudits in $N$ that is passed as input to \FnErrCorr{$\tr_{S_Z}(\sigma);Z,(H^i)_{i\in[u]},\Dec_Z$} as a linear combination of states of the form
    \begin{align*}
      &X^{e_X}Z^{e_Z}\Enc\left(\ket{+^K}\bra{+^K}\otimes\ket{0^{K'}}\bra{0^{K'}}\right)\Enc^\dagger(Z^\dagger)^{e_Z'}(X^\dagger)^{e_X'}
    \end{align*}
    for $e_X,e_X',e_Z,e_Z'\in\bF_q^N$ with $|e_Z|,|e_Z'|\leq|\supp(F_1)|\leq\lambda_{\mathrm{run}}$. By \Cref{claim:eckey}, it follows that under fault $\cF$ and postselection $\zeta$, \FnErrCorr{$\tr_{S_Z}(\sigma);Z,(H^i)_{i\in[u]},\Dec_Z$} outputs a linear combination of states of the form
    \begin{align}
      \label{eq:sppostec}
      X^{\tilde{c}_X+\tilde{e}_X}Z^{\tilde{e_Z}}\Enc\left(\ket{+^K}\bra{+^K}\otimes\ket{0^{K'}}\bra{0^{K'}}\right)\Enc^\dagger(Z^\dagger)^{\tilde{e}_Z'}(X^\dagger)^{\tilde{c}_X'+\tilde{e}_X'}
    \end{align}
    for $\tilde{c}_X,\tilde{c}_X'\in C_Z$ and $\tilde{e}_X,\tilde{e}_X',\tilde{e}_Z,\tilde{e}_Z'\in\bF_q^N$ with
    \begin{align}
      \label{eq:spebounds}
      \begin{split}
        |\tilde{e}_X|,|\tilde{e}_X'| &\leq 4\mu un(1+un^2)\cdot\lambda_{\mathrm{run}} \\
        |\tilde{e}_Z|,|\tilde{e}_Z'| &\leq \lambda_{\mathrm{run}} + 2n(1+un^2)\cdot\lambda_{\mathrm{run}}.
      \end{split}
    \end{align}
    By \Cref{eq:ecloginout}, the $X^{\tilde{c}_X}$ and $X^{\tilde{c}_X'}$ operators above can equivalently be replaced by Pauli~$X$ operators acting on the $\ket{+^K}\bra{+^K}$ factor inside the parentheses in \Cref{eq:sppostec}, and hence these operators do not affect the state; that is, the state in \Cref{eq:sppostec} is equal to
    \begin{equation*}
      X^{\tilde{e}_X}Z^{\tilde{e_Z}}\Enc\left(\ket{+^K}\bra{+^K}\otimes\ket{0^{K'}}\bra{0^{K'}}\right)\Enc^\dagger(Z^\dagger)^{\tilde{e}_Z'}(X^\dagger)^{\tilde{e}_X'}
    \end{equation*}
    which is indeed a $\cE_{\mathrm{out}}$-deviation of $\Enc\left(\ket{+^K}\bra{+^K}\otimes\ket{0^{K'}}\bra{0^{K'}}\right)\Enc^\dagger$ by \Cref{eq:spebounds} and \Cref{eq:splamout}. Thus $(\cQ_X,2^{N'}|_{\geq \lambda_{\mathrm{run}}}^{\sqcup T},D_{\mathrm{in}},D_{\mathrm{out}}^\alpha)$ is a fault-tolerant gadget for $\bar{O}_X$, as desired.
  \item In the case where we wish to reset a given set of qudits, our desired state preparation circuit $\cQ_X$ is exactly the same as in the proof of \Cref{it:spnew} above. 
  However, now $\gInit_X$ is applied to the input code state with qudits labeled by $N$, instead of to ancilla qudits. 
  Fix $\ell\in\bN$, a Pauli $\cE_{\mathrm{in}}\sqcup [\ell]$-deviation $\sigma_0\in\bC^{\bF_q^{|N|+\ell}\times\bF_q^{|N|+\ell}}$ of $\Enc\otimes I_\ell(\rho\otimes\Gamma_{\mathrm{in}})$, and a $2^{N'}|_{\geq \lambda_{\mathrm{run}}}^{\sqcup T}$-avoiding Pauli fault $\cF=(F_1,\dots,F_T)$ and a postselection $\zeta$ for $\cQ_X$. 
  By \Cref{lem:paulift}, it suffices to show that the output $\cQ_X[\cF,\zeta]\otimes I_\ell(\sigma_0)$ is a $\cE_{\mathrm{out}}\sqcup [\ell]$-deviation of
    \begin{align}
      \label{eq:sprend}
      (\Enc\circ\bar{\cO})\otimes I_\ell\left(\rho\otimes\ket{0^{K'}}\bra{0^{K'}}\right)
      &= \Enc\left(\ket{+^K}\bra{+^K}\otimes\ket{0^{K'}}\bra{0^{K'}}\right)\otimes\tr_N(\rho).
    \end{align}
    Recall that the first timestep of $\cQ_X$ applies $\gInit_X^{\otimes N}$ to the input code state $\sigma_0$ to obtain
    \begin{align}
      \label{eq:sprstart}
      \gInit_Z^{\otimes N}\otimes I_\ell(\sigma_0)
      &= \ket{+^N}\bra{+^N}\otimes\tr_N(\sigma_0).
    \end{align}
    If
    \begin{align}
      \label{eq:sprtr}
      \tr_N(\sigma_0) &= \tr_N(\rho),
    \end{align}
    then the RHS of \Cref{eq:sprstart} is precisely the same state as in the proof of \Cref{it:spnew} above after the first timestep if the external system were initialized to $\tr_N(\rho)$. Hence assuming \Cref{eq:sprtr} holds, then by the proof of \Cref{it:spnew} above, the output $\cQ_X[\cF,\zeta]\otimes I_\ell(\sigma_0)$ is indeed a $\cE_{\mathrm{out}}\sqcup [\ell]$-deviation of the state in \Cref{eq:sprend}, and \Cref{it:spreplace} in the lemma statement holds.

    Thus it remains to prove \Cref{eq:sprtr}. For this purpose, we can express
    \begin{align*}
      \rho
      &= \sum_{x,x'\in\bF_q^K}\ket{x}\bra{x'}\otimes\rho_{x,x'}
    \end{align*}
    for some $\rho_{x,x'}\in\bC^{\bF_q^\ell\times\bF_q^\ell}$.
    Then by definition we can express $\sigma_0$ as a linear combination of states of the form
    \begin{align*}
      \sigma
      &= \sum_{x,x'\in\bF_q^K}Z^{e_Z}\ket{\Enc(x,0)+g_X+e_X}\bra{\Enc(x',0)+g_X'+e_X'}(Z^\dagger)^{e_Z'}\otimes\rho_{x,x'}
    \end{align*}
    for some $g_X,g_X'\in C_X^\perp$ and some $e_X,e_X',e_Z,e_Z'\in\bF_q^N$ with
    \begin{equation*}
      |\supp(e_X)\cup\supp(e_X')\cup\supp(e_Z)\cup\supp(e_Z')| < \lambda_{\mathrm{in}} = D.
    \end{equation*}
    Specifically, we arrive at this decomposition by expressing $\gamma=\sum_{z,z'\in\bF_q^{K'}}\ket{z}\bra{z'}\cdot\gamma_{z,z'}$ for $\gamma_{z,z'}\in\bC$, and then considering each term in the sum over $z,z'$ separately, while choosing $g_X\in\Enc(0,z)$ and $g_X'\in\Enc(0,z')$. Then
    \begin{align}
      \label{eq:trsigma}
      \tr_N(\sigma)
      &= \sum_{x,x'\in\bF_q^K,y\in\bF_q^N}\bra{y}Z^{e_Z}\ket{\Enc(x,0)+g_X+e_X}\bra{\Enc(x',0)+g_X'+e_X'}(Z^\dagger)^{e_Z'}\ket{y}\otimes\rho_{x,x'}
    \end{align}
    
    \begin{claim}
      \label{claim:geCZ}
      If $(g_X-g_X')+(e_X-e_X')\in C_Z$, then $(g_X-g_X')+(e_X-e_X')\in C_Z\cap C_X^\perp$.
    \end{claim}
    \begin{proof}
      By definition $g_X-g_X'\in C_X^\perp$, so if $(g_X-g_X')+(e_X-e_X')\in C_Z$, then $e_X-e_X'\in C_Z+C_X^\perp$. However, because $|e_X-e_X'|<D$, then $e_X-e_X'\notin(C_Z+C_X^\perp)\setminus C_X^\perp$ by the definition of subsystem code distance. Thus $e_X-e_X'\in C_X^\perp$, so $(g_X-g_X')+(e_X-e_X')$ lies in $C_X^\perp$ (and in $C_Z$), and thus in $C_Z\cap C_X^\perp$.
    \end{proof}
    
    If $(g_X-g_X')+(e_X-e_X')\notin C_Z\cap C_X^\perp$, then by \Cref{claim:geCZ} $(g_X-g_X')+(e_X-e_X')\notin C_Z$, and hence as $\Enc(x,0),\Enc(x',0)\subseteq C_Z$, the expression inside the sum in \Cref{eq:trsigma} vanishes, and $\tr_N(\sigma)=0$. Therefore assume that $(g_X-g_X')+(e_X-e_X')\in C_Z\cap C_X^\perp$. Then the expression inside the sum in \Cref{eq:trsigma} is nonvanishing only when $\Enc(x,0)=\Enc(x',0)$, or equivalently when $x=x'$. Therefore letting $p$ denote the characteristic of $\bF_q$ and $\omega_p=e^{2\pi i/p}$, then
    \begin{align}
      \label{eq:trsigfinal}
      \begin{split}
        \tr_N(\sigma)
        &\propto \sum_{x\in\bF_q^K} \rho_{x,x} \cdot \sum_{y\in\Enc(x,0)+g_X+e_X} \omega_p^{\tr_{\bF_q/\bF_p}((e_Z-e_Z')\cdot y)} \\
        &\propto \sum_{x\in\bF_q^K} \rho_{x,x} \cdot \sum_{y\in\Enc(x,0)} \omega_p^{\tr_{\bF_q/\bF_p}((e_Z-e_Z')\cdot y)}.
      \end{split}
    \end{align}
    Now by the definition of subsystem code distance, because $|e_Z-e_Z'|<D$ we have $e_Z-e_Z'\notin(C_X+C_Z^\perp)\setminus C_Z^\perp$. If $e_Z-e_Z'\in C_Z^\perp$, then for every $y\in\Enc(x,0)\subseteq C_Z$ we have $(e_Z-e_Z')\cdot y=0$, and hence the RHS of \Cref{eq:trsigfinal} is proportional to $\sum_{x\in\bF_q^K}\rho_{x,x}=\tr_N(\rho)$. Otherwise, we must have $e_Z-e_Z'\notin(C_X+C_Z^\perp)=(C_Z\cap C_X^\perp)^\perp$, so as $\Enc(x,0)$ is a coset in $C_Z/(C_Z\cap C_X^\perp)$, the phases in the sum over $y\in\Enc(x,0)$ on the RHS of \Cref{eq:trsigfinal} vanish for each fixed $x\in\bF_q^K$, and we have $\tr_N(\sigma)=0$. Thus in either case we have $\tr_N(\sigma)\propto\tr_N(\rho)$, so \Cref{eq:sprtr} holds, as desired.
  \end{enumerate}
\end{proof}

\subsection{Code Switching}
\label{sec:codeswitching}
In this section, we present our gadget for switching between different product codes by encoding or unencoding in a single direction at a time.

\begin{lemma}
  \label{lem:codeswitch}
  For $u\in\bN$, for $i\in[u]$ let $C^i=(C^i_X,C^i_Z)$ be a $[[n_i,k_i]]_q$ CSS non-subsystem code with specified CSS encoding map $\Enc^i:\bF_q^{k_i}\rightarrow C^i_Z/{C^i_X}^\perp$. For $I\subseteq[u]$, let $N^I=\prod_{i\in I}[n_i]$, $K^I=\prod_{i\in I}[k_i]$, and let $C^I=\bigotimes_{i\in I}C^i$ be the $[[|N^I|,|K^I|,D^I]]_q$ subsystem product code, with product encoding map $\Enc^I:\bF_q^{K^I}\oplus\bF_q^{{K^I}'}\rightarrow(C^I_Z+{C^I_X}^\perp)/(C^I_Z\cap{C^I_X}^\perp)$. Let $\Gamma^I=\left\{\ket{0^{{K^I}'}}\bra{0^{{K^I}'}}\right\}$. Fix an arbitrary $\bar{i}\in[u]$. Then the following hold:
  \begin{enumerate}
  \item\label{it:switchdown} (Downwards switching) For $\lambda_{\mathrm{in}},\lambda_{\mathrm{run}}\geq 0$, let
    \begin{align*}
      \lambda_{\mathrm{out}} &= n_{\bar{i}}(\lambda_{\mathrm{in}}+(2n_{\bar{i}}^2+4)\lambda_{\mathrm{run}}),
    \end{align*}
    and define decorated subsystem codes
    \begin{align*}
      D_{\mathrm{in}} &= \left(C^{[u]},\; \Enc^{[u]},\; \Gamma_{\mathrm{in}}=\Gamma^{[u]},\; \cE_{\mathrm{in}}=2^{N^{[u]}}|_{\geq\lambda_{\mathrm{in}}}\right) \\
      D_{\mathrm{out}} &= \left(C^{[u]\setminus\{\bar{i}\}},\; \Enc^{[u]\setminus\{\bar{i}\}},\; \Gamma_{\mathrm{out}}=\Gamma^{[u]\setminus\{\bar{i}\}},\; \cE_{\mathrm{out}}=2^{N^{[u]\setminus\{\bar{i}\}}}|_{\geq\lambda_{\mathrm{out}}}\right).
    \end{align*}
    Then there exists a quantum circuit $\cQ$ using gate set $\{CX^*,\gInit_*\}$, space $|N'|=2|N^{[u]}|$, and time $T\leq 2n_{\bar{i}}^2+4$ such that $(\cQ,\cE_{\mathrm{run}}=2^{N'}|_{\geq\lambda_{\mathrm{run}}}^{\sqcup T},D_{\mathrm{in}},D_{\mathrm{out}}^{\sqcup k_{\bar{i}}})$ forms a fault-tolerant gadget for the $K^{[u]}$-qudit identity channel $\bar{O}=I_{K^{[u]}}:\bC^{\bF_q^{K^{[u]}}\times\bF_q^{K^{[u]}}}\rightarrow\bC^{\bF_q^{K^{[u]}}\times\bF_q^{K^{[u]}}}$.
  \item\label{it:switchup} (Upwards switching) For $n,d\in\bN$, assume that every $C^i$ has length $n_i=n$, and that $C^i_X,C^i_Z$ have distance $\geq d$. Define $\mu(u-1,n,d)$ to be the filling constant in \Cref{prop:loctest}. 
  For $\lambda_{\mathrm{in}},\lambda_{\mathrm{run}}\geq 0$, let
    \begin{align*}
      \lambda_{\mathrm{out}} &= n\cdot\lambda_{\mathrm{in}} + 50\mu(u-1,n,d)u^2n^5\cdot\lambda_{\mathrm{run}},
    \end{align*}
    and define decorated subsystem codes
    \begin{align*}
      D_{\mathrm{in}} &= \left(C^{[u]\setminus\{\bar{i}\}},\; \Enc^{[u]\setminus\{\bar{i}\}},\; \Gamma_{\mathrm{out}}=\Gamma^{[u]\setminus\{\bar{i}\}},\; \cE_{\mathrm{out}}=2^{N^{[u]\setminus\{\bar{i}\}}}|_{\geq\lambda_{\mathrm{in}}}\right) \\
      D_{\mathrm{out}} &= \left(C^{[u]},\; \Enc^{[u]},\; \Gamma_{\mathrm{in}}=\Gamma^{[u]},\; \cE_{\mathrm{in}}=2^{N^{[u]}}|_{\geq\lambda_{\mathrm{out}}}\right).
    \end{align*}
    Then there exists a quantum circuit $\cQ$ using gate set $\{\gCX^*,\gInit_*,\gCO_*\}$, space $|N'|\leq(u+2)|N^{[u]}|$, and time $T\leq(u+2)n^2+7$ such that $(\cQ,\cE_{\mathrm{run}}=2^{N'}|_{\geq\lambda_{\mathrm{run}}}^{\sqcup T},D_{\mathrm{in}}^{\sqcup k_{\bar{i}}},D_{\mathrm{out}})$ forms a fault-tolerant gadget for the $K^{[u]}$-qudit identity channel $\bar{O}=I_{K^{[u]}}:\bC^{\bF_q^{K^{[u]}}\times\bF_q^{K^{[u]}}}\rightarrow\bC^{\bF_q^{K^{[u]}}\times\bF_q^{K^{[u]}}}$.
  \end{enumerate}
\end{lemma}

In defining $\bar{O}=I_{K^{[u]}}$ in \Cref{lem:codeswitch}, we use the fact that $K^{[u]}$ is by definition equal to $k_{\bar{i}}$ copies of the set $K^{[u]\setminus\{\bar{i}\}}$.

\begin{remark}
  \Cref{lem:codeswitch} should hold with $\Gamma_{\mathrm{in}}$ containing more general states, rather than simply the all-0s state as stated. However, for simplicity we only consider the all-0s state, which will be sufficient because we can always run the error-correction gadget in \Cref{lem:errcorr} prior to the code-switching gadget in \Cref{lem:codeswitch}.
\end{remark}

\begin{proof}[Proof of \Cref{lem:codeswitch}]
  To begin, for every $I\subseteq[u]$, let $M^I_X=(\prod_{i\in I}[\dim(C^i_Z)])\setminus K^I$ and $M^I_Z=N^I\setminus(\prod_{i\in I}[\dim(C^I_Z)])$, so that $|M^I_X|=\dim(C^I_Z\cap{C^I_X}^\perp)$, $|M^I_Z|=|N^I|-\dim(C^I_Z)$, and $K^I\sqcup M^I_X\sqcup M^I_Z=N^I$. Then we define an invertible linear map
  \begin{equation*}
    E^I : \bF_q^{K^I}\oplus\bF_q^{M^I_X}\oplus\bF_q^{M^I_Z}=\bF_q^{N^I} \rightarrow \bF_q^{N^I}
  \end{equation*}
  such that for every $x\in\bF_q^{K^I}$,
  \begin{equation}
    \label{eq:csbe}
    E^I(x,\bF_q^{M^I_X},0^{M^I_Z}) = \Enc^I(x,0^{{K^I}'}) \in C^I_Z/(C^I_Z\cap{C^I_X}^\perp).
  \end{equation}
  Specifically, for singleton $I=\{i\}$ (so that\footnote{Although every $n_i=n$ here, for clarity we will often write $[n_i]$ to refer to the set of labels for the code components of $C^i$.} $N^i=[n_i]$ and $K^i=[k_i]$), by definition we may specify an appropriate injective map for the restriction of $E^i$ to $\bF_q^{K^i}\oplus\bF_q^{M^i_X}\oplus 0^{M^i_Z}$ such that \Cref{eq:csbe} is satisfied. We then extend this map to one on all of $\bF_q^{N^i}$ in some way that maintains injectivity. Then for general subsets $I\subseteq[u]$, we define
  \begin{equation}
    \label{eq:cspe}
    E^I=\bigotimes_{i\in I}E^i.
  \end{equation}
  \Cref{def:prodenc} implies that \Cref{eq:csbe} is satisfied for every $I\subseteq[u]$. Specifically, for every $y=\bigotimes_{i\in[u]}y_i\in\bF_q^{K^I\sqcup M^I_X}$ with each $y_i\in\bF_q^{[k_i]\sqcup M^i_X}$, then \Cref{eq:cspe} and \Cref{def:prodenc} together imply that $E^I(y,0^{M^I_Z})\in\Enc^I(y|_{K^I},0^{{K^I}'})$. By linearity we conclude that every $x\in\bF_q^{K^I}$ has $E^I(x,\bF_q^{M^I_X},0^{M^I_Z})\subseteq\Enc^I(x,0^{{K^I}'})$. Then because $E^I$ is invertible and $|M^I_X|=\dim(C^I_Z\cap{C^I_X}^\perp)$, this inclusion must be an equality, that is, \Cref{eq:csbe} holds.

  We now apply the above definitions to prove the two items in the lemma statement separately:
  \begin{enumerate}
  \item\label{it:csdown} (Downwards switching) The desired circuit $\cQ$ will simply unencode the $\bar{i}$th factor of the product code $C^{[u]}$.
    Specifically, $\cQ$ acts on a set of qudits labeled by $N'=N^{[u]}\sqcup N^{[u]}$, consisting of an input block $N^{[u]}$ and an ancilla block $N^{[u]}$, the latter of which $\cQ$ initializes to $\ket{0^{N^{[u]}}}\bra{0^{N^{[u]}}}$ by running $\gInit_Z^{\otimes N^{[u]}}$. We may view $N^{[u]}\cong[n_{\bar{i}}]\times N^{[u]\setminus\{\bar{i}\}}$ as a collection of $|N^{[u]\setminus\{\bar{i}\}}|$ direction-$\bar{i}$ columns labeled by $N^{[u]\setminus\{\bar{i}\}}$, each of length $n_{\bar{i}}$. For each element of $N^{[u]\setminus\{\bar{i}\}}$, we therefore have a direction-$\bar{i}$ column in each of the input and ancilla blocks, across which $\cQ$ then applies the sequence of unitaries
    \begin{align}
      \label{eq:csdownmaps}
      \begin{split}
        \ket{y}_{\mathrm{inp}}\ket{z}_{\mathrm{anc}} &\mapsto \ket{y}_{\mathrm{inp}}\ket{z+(E^{\bar{i}})^{-1}y}_{\mathrm{anc}} \\
        \ket{y'}_{\mathrm{inp}}\ket{z'}_{\mathrm{anc}} &\mapsto \ket{y'-E^{\bar{i}}z'}_{\mathrm{inp}}\ket{z'}_{\mathrm{anc}} \\
        \ket{y''}_{\mathrm{inp}}\ket{z''}_{\mathrm{anc}} &\mapsto \ket{z''}_{\mathrm{inp}}\ket{y''}_{\mathrm{anc}}
      \end{split}
    \end{align}
    for $y,y',y'',z,z',z''\in\bF_q^{n_{\bar{i}}}$. After initialization of the ancilla block, within each direction-$\bar{i}$ column these unitaries have the combined effect of mapping
    \begin{equation*}
      \ket{y}_{\mathrm{inp}}\ket{0}_{\mathrm{anc}} \mapsto \ket{y}_{\mathrm{inp}}\ket{(E^{\bar{i}})^{-1}y}_{\mathrm{anc}} \mapsto \ket{0}_{\mathrm{inp}}\ket{(E^{\bar{i}})^{-1}y}_{\mathrm{anc}} \mapsto \ket{(E^{\bar{i}})^{-1}y}_{\mathrm{inp}}\ket{0}_{\mathrm{anc}}.
    \end{equation*}
    The circuit $\cQ$ then returns the first $k_{\bar{i}}$ qudits of every direction-$\bar{i}$ column in the resulting state of the input block.

    The first two unitaries in \Cref{eq:csdownmaps} are each of the form $\ket{y}\ket{z}\mapsto\ket{y}\ket{z+Ay}$ for a matrix $A\in\bF_q^{n_{\bar{i}}\times n_{\bar{i}}}$, which by definition can be implemented by applying the gate $\gCX^{A_{i,j}}$ to qudits $\ket{y_j}$ and $\ket{z_i}$ for every $(i,j)\in [n_{\bar{i}}]^2$. The third unitary in \Cref{eq:csdownmaps} simply swaps $n_{\bar{i}}$ pairs of qudits; each such swap can be implemented using three $\gCX$ gates. Also adding in an initial timestep for ancilla initialization, $\cQ$ has overall time usage $T\leq 1+2n_{\bar{i}}^2+3=2n_{\bar{i}}^2+4$ and space usage $|N'|=2|N^{[u]}|$. 

    We first analyze the behavior of $\cQ$ in the absence of errors. By definition $\cQ$ implements the unitary $\ket{y}\mapsto\ket{((E^{\bar{i}})^{-1}\otimes I^{\otimes[u]\setminus\{\bar{i}\}})y}$, where here we let $(E^{\bar{i}})^{-1}$ act on the $\bar{i}$th factor of $\bigotimes_{i\in[u]}\bF_q^{n_i}=\bF_q^{N^{[u]}}$. For $\ell\in\bN$ and $\rho\in\bC^{\bF_q^{|K^{[u]}|+\ell}\times\bF_q^{|K^{[u]}|+\ell}}$ given by
    \begin{align}
      \label{eq:csrho}
      \rho
      &= \sum_{x,x'\in\bF_q^{K^{[u]}}}\ket{x}\bra{x'}\otimes\rho_{x,x'}
    \end{align}
    with each $\rho_{x,x'}\in\bC^{\bF_q^\ell\times\bF_q^\ell}$, define $\sigma\in\bC^{\bF_q^{|N^{[u]}|+\ell}\times\bF_q^{|N^{[u]}|+\ell}}$ by
    \begin{align*}
      \sigma
      &= \Enc^{[u]}\otimes I_\ell\left(\rho\otimes\ket{0^{{K^{[u]}}'}}\bra{0^{{K^{[u]}}'}}\right) \\
      &= \sum_{x,x'\in\bF_q^{K^{[u]}}}\ket{\Enc^{[u]}(x,0^{{K^{[u]}}'})}\bra{\Enc^{[u]}(x',0^{{K^{[u]}}'})}\otimes\rho_{x,x'}.
    \end{align*}
    Now for $x\in\bF_q^{K^{[u]}}$, by \Cref{eq:csbe} and \Cref{eq:cspe},
    \begin{align*}
      ((E^{\bar{i}})^{-1}\otimes I^{\otimes[u]\setminus\{\bar{i}\}})\Enc^{[u]}(x,0^{{K^{[u]}}'})
      &= ((E^{\bar{i}})^{-1}\otimes I^{\otimes[u]\setminus\{\bar{i}\}})(E^{\bar{i}}\otimes E^{[u]\setminus\{\bar{i}\}})(x,\bF_q^{M^{[u]}_X},0^{M^{[u]}_Z}) \\
      &= (I_{n_{\bar{i}}}\otimes E^{[u]\setminus\{\bar{i}\}})(x,\bF_q^{M^{[u]}_X},0^{M^{[u]}_Z}).
    \end{align*}
    The restriction of the RHS above to entries in $[k_{\bar{i}}]\times N^{[u]\setminus\{\bar{i}\}}$ by definition equals
    \begin{equation*}
      (I_{k_{\bar{i}}}\otimes E^{[u]\setminus\{\bar{i}\}})(x,\bF_q^{M^{[u]\setminus\{\bar{i}\}}_X},0^{M^{[u]\setminus\{\bar{i}\}}_Z}) = \Enc^{[u]\setminus\{\bar{i}\}}(x,0^{{K^{[u]\setminus\{\bar{i}\}}}'}),
    \end{equation*}
    while the restriction to entries in $([n_{\bar{i}}]\setminus[k_{\bar{i}}])\times N^{[u]\setminus\{\bar{i}\}}$ does not depend on $x$.
    Thus under a noiseless execution, the returned state $\tr_{(N^{[u]}\setminus([k_{\bar{i}}]\times N^{[u]\setminus\{\bar{i}\}}))\sqcup N^{[u]}}(\cQ\otimes I_\ell(\sigma))$ is proportional to
    \begin{align*}
      &\sum_{x,x'\in\bF_q^{K^{[u]}}}\ket{(\Enc^{[u]\setminus\{\bar{i}\}})^{\sqcup k_{\bar{i}}}(x,0^{{K^{[u]\setminus\{\bar{i}\}}}'})}\bra{(\Enc^{[u]\setminus\{\bar{i}\}})^{\sqcup k_{\bar{i}}}(x',0^{{K^{[u]\setminus\{\bar{i}\}}}'})} \otimes \rho_{x,x'} \\
      &\hspace{1em}= (\Enc^{[u]\setminus\{\bar{i}\}})^{\sqcup k_{\bar{i}}}\otimes I_\ell\left(\rho\otimes\left(\ket{0^{{K^{[u]\setminus\{\bar{i}\}}}'}}\bra{0^{{K^{[u]\setminus\{\bar{i}\}}}'}}\right)^{\otimes k_{\bar{i}}}\right).
    \end{align*}
    Thus we have shown that in the absence of a fault, when $\cQ$ is given as input the state $\rho$ encoded in $D_{\mathrm{in}}$, it uses the desired space $N'$ and time $T$, and outputs $\rho$ encoded in $D_{\mathrm{out}}$.

    It only remains to analyze the behavior of $\cQ$ under input errors and faults; there are no $\gCO_*$ gates and therefore no postselections to consider. By \Cref{lem:paulift}, it suffices to show for every Pauli $\cE_{\mathrm{in}}\sqcup [\ell]$-deviation $\sigma$ of $\Enc_{\mathrm{in}}\otimes I_\ell(\rho\otimes\Gamma_{\mathrm{in}})$ and every $\cE_{\mathrm{run}}$-avoiding Pauli fault $\cF$ for $\cQ$, then $\cQ[\cF]\otimes I_\ell(\sigma)$ is a $\cE_{\mathrm{out}}\sqcup [\ell]$-deviation of $\Enc_{\mathrm{out}}\otimes I_\ell(\rho\otimes\Gamma_{\mathrm{out}})$. After the $\gInit_Z$ gates in the first timestep, all gates in $\cQ$ are $\gCX$, which are Clifford, meaning that conjugating a Pauli by a $\gCX$ gate yields another Pauli. Thus the output $\cQ[\cF]\otimes I_\ell(\sigma)$ must be a Pauli deviation of the output from the noiseless execution, which we showed above is $\Enc_{\mathrm{out}}\otimes I_\ell(\rho\otimes\Gamma_{\mathrm{out}})$. Furthermore, as all $\gCX$ gates are performed within a direction-$\bar{i}$ column, such conjugation can only propagate Pauli errors within direction-$\bar{i}$ columns. As there are $<\lambda_{\mathrm{in}}$ qudits in the support of the input error and $<T\lambda_{\mathrm{run}}$ qudits in the support of the fault, and every direction-$\bar{i}$ column has $n_{\bar{i}}$ qudits, the weight of the output error is less than
    \begin{align*}
      n_{\bar{i}}(\lambda_{\mathrm{in}}+T\lambda_{\mathrm{run}})
      &\leq \lambda_{\mathrm{out}},
    \end{align*}
    as desired.
  \item (Upwards switching) The desired circuit $\cQ$ will simply encode the $\bar{i}$th factor of the product code $C^{[u]}$. Specifically, $\cQ$ takes as input a block of qudits labeled by $B_1=[k_{\bar{i}}]\times N^{[u]\setminus\{\bar{i}\}}$, which (in an error-free execution) contains $k_{\bar{i}}$ code states of $C^{[u]\setminus\{\bar{i}\}}$. First, $\cQ$ executes the $X$-basis (i.e.~$\ket{+}$-state) initialization gadget in \Cref{it:spnew} of \Cref{lem:stateprep} for the code $C^{[u]\setminus\{\bar{i}\}}$ with encoding map $\Enc^{[u]\setminus\{\bar{i}\}}$ a total of $|M^{\bar{i}}_X|$ times, to create a block of qudits labeled by $B_2=M^{\bar{i}}_X\times N^{[u]\setminus\{\bar{i}\}}$ in the state
    \begin{equation*}
      \left(\Enc^{[u]\setminus\{\bar{i}\}}\left(\ket{+^{K^{[u]\setminus\{\bar{i}\}}}}\bra{+^{K^{[u]\setminus\{\bar{i}\}}}}\otimes \ket{0^{{K^{[u]\setminus\{\bar{i}\}}}'}}\bra{0^{{K^{[u]\setminus\{\bar{i}\}}}'}}\right)\right)^{\otimes M^{\bar{i}}_X}
    \end{equation*}
    $\cQ$ also calls initializes a block of qudits labeled by $B_3=M^{\bar{i}}_Z\times N^{[u]\setminus\{\bar{i}\}}$ to the all-0s state by calling $\gInit_Z^{\otimes B_3}$. These three blocks combine to form a single larger block of qudits labeled by $N^{[u]}=B_1\sqcup B_2\sqcup B_3$, on which $\cQ$ applies the unitary $\ket{y}\mapsto\ket{(E^{\bar{i}}\otimes I^{\otimes[u]\setminus\{\bar{i}\}})y}$ similarly as in \Cref{it:csdown} above (but with $E^{\bar{i}}$ here replacing $(E^{\bar{i}})^{-1}$). That is, $\cQ$ calls $\gInit_Z^{\otimes N^{[u]}}$ to initialize an ancilla block of qudits labeled by $N^{[u]}$. Then within every pair of direction-$\bar{i}$ columns across these two blocks (``inp'' and ``anc'') sharing the same label in $N^{[u]\setminus\{\bar{i}\}}$, $\cQ$ executes the sequence of unitaries
    \begin{align}
      \label{eq:csupmaps}
      \begin{split}
        \ket{y}_{\mathrm{inp}}\ket{z}_{\mathrm{anc}} &\mapsto \ket{y}_{\mathrm{inp}}\ket{z+E^{\bar{i}}y}_{\mathrm{anc}} \\
        \ket{y'}_{\mathrm{inp}}\ket{z'}_{\mathrm{anc}} &\mapsto \ket{y'-(E^{\bar{i}})^{-1}z'}_{\mathrm{inp}}\ket{z'}_{\mathrm{anc}} \\
        \ket{y''}_{\mathrm{inp}}\ket{z''}_{\mathrm{anc}} &\mapsto \ket{z''}_{\mathrm{inp}}\ket{y''}_{\mathrm{anc}}.
      \end{split}
    \end{align}
    Finally, $\cQ$ outputs the resulting state in the $N^{[u]}$-block ``inp.'' Similarly as in \Cref{it:csdown}, the unitaries in \Cref{eq:csupmaps} can collectively be implemented using $\gCX$ gates in time $\leq 2n^2+3$ and space $2|N^{[u]}|=2n^u$, as every $n_i=n$. Meanwhile, the calls to the initialization gadget in \Cref{lem:stateprep} collectively use time $\leq un^2+4$ and space $\leq|M^{\bar{i}}_X|\cdot un^{u-1}\leq u|N^{[u]}|$. Thus the overall circuit $\cQ$ uses time $T\leq(u+2)n^2+7$ and space $|N'|\leq(u+2)|N^{[u]}|$.

    We first analyze the behavior of $\cQ$ in the absence of errors. Similarly as in \Cref{it:csdown}, fix $\ell\in\bN$ and $\rho\in\bC^{\bF_q^{|K^{[u]}|+\ell}\times\bF_q^{|K^{[u]}|+\ell}}$ given by \Cref{eq:csrho}, and assume that $\cQ$ is given as input
    \begin{align*}
      \sigma
      &= (\Enc^{[u]\setminus\{\bar{i}\}})^{\sqcup k_{\bar{i}}}\otimes I_\ell\left(\rho\otimes\ket{0^{[k_{\bar{i}}]\times{K^{[u]\setminus\{\bar{i}\}}}'}}\bra{0^{[k_{\bar{i}}]\times{K^{[u]\setminus\{\bar{i}\}}}'}}\right).
    \end{align*}
    in block $B_1$.
    By definition, after the calls to the initialization gadget in \Cref{lem:stateprep}, the collective state of qudits in $B_1\sqcup B_2\sqcup B_3=N^{[u]}$ is 
    \begin{align}
      \label{eq:csupinitstate}
      \begin{split}
        &(\Enc^{[u]\setminus\{\bar{i}\}})^{\sqcup k_{\bar{i}}}\otimes I_\ell\left(\rho\otimes\ket{0^{[k_{\bar{i}}]\times{K^{[u]\setminus\{\bar{i}\}}}'}}\bra{0^{[k_{\bar{i}}]\times{K^{[u]\setminus\{\bar{i}\}}}'}}\right) \\
        &\hspace{1em} \otimes (\Enc^{[u]\setminus\{\bar{i}\}})^{\sqcup k_{\bar{i}}}\left(\ket{+^{M^{\bar{i}}_X\times{K^{[u]\setminus\{\bar{i}\}}}}}\bra{+^{M^{\bar{i}}_X\times{K^{[u]\setminus\{\bar{i}\}}}}}\otimes\ket{0^{[k_{\bar{i}}]\times{K^{[u]\setminus\{\bar{i}\}}}'}}\bra{0^{[k_{\bar{i}}]\times{K^{[u]\setminus\{\bar{i}\}}}'}}\right) \\
        &\hspace{1em} \otimes \ket{0^{M^{\bar{i}}_Z\times N^{[u]\setminus\{\bar{i}\}}}}\bra{0^{M^{\bar{i}}_Z\times N^{[u]\setminus\{\bar{i}\}}}} \\
        &\propto \sum_{x,x'\in\bF_q^{K^{[u]}}} \ket{(\Enc^{[u]\setminus\{\bar{i}\}})^{\sqcup k_{\bar{i}}}(x,0)}\bra{(\Enc^{[u]\setminus\{\bar{i}\}})^{\sqcup k_{\bar{i}}}(x,0)} \\
        &\hspace{1em} \otimes \ket{(\Enc^{[u]\setminus\{\bar{i}\}})^{\sqcup M^{\bar{i}}_X}(\bF_q^{M^{\bar{i}}_X\times K^{[u]\setminus\{\bar{i}\}}},0)}\bra{(\Enc^{[u]\setminus\{\bar{i}\}})^{\sqcup M^{\bar{i}}_X}(\bF_q^{M^{\bar{i}}_X\times K^{[u]\setminus\{\bar{i}\}}},0)} \\
        &\hspace{1em} \otimes \ket{0^{M^{\bar{i}}_Z\times N^{[u]\setminus\{\bar{i}\}}}}\bra{0^{M^{\bar{i}}_Z\times N^{[u]\setminus\{\bar{i}\}}}} \otimes \rho_{x,x'}.
      \end{split}
    \end{align}
    For each term on the RHS above with a given $x\in\bF_q^{K^{[u]}}$, collecting the kets yields
    \begin{align*}
      &\ket{(\Enc^{[u]\setminus\{\bar{i}\}})^{\sqcup k_{\bar{i}}}(x,0)} \otimes \ket{(\Enc^{[u]\setminus\{\bar{i}\}})^{\sqcup M^{\bar{i}}_X}(\bF_q^{M^{\bar{i}}_X\times K^{[u]\setminus\{\bar{i}\}}},0)} \otimes \ket{0^{M^{\bar{i}}_Z\times N^{[u]\setminus\{\bar{i}\}}}} \\
      &\hspace{1em}= \ket{(I_{k_{\bar{i}}}\otimes E^{[u]\setminus\{\bar{i}\}})(x,\bF_q^{[k_{\bar{i}}]\times M^{[u]\setminus\{\bar{i}\}}_X},0)} \otimes \ket{(I_{M^{\bar{i}}_X}\otimes E^{[u]\setminus\{\bar{i}\}})(\bF_q^{M^{\bar{i}}_X\times K^{[u]\setminus\{\bar{i}\}}},\bF_q^{M^{\bar{i}}_X\times M^{[u]\setminus\{\bar{i}\}}_X},0)} \\
      &\hspace{17.5em} \otimes \ket{0^{M^{\bar{i}}_Z\times N^{[u]\setminus\{\bar{i}\}}}} \\
      &\hspace{1em}= \ket{(I_{n_{\bar{i}}}\otimes E^{[u]\setminus\{\bar{i}\}})(x,\bF_q^{M^{[u]}_X},0)},
    \end{align*}
    where here we use the definition of $E^{[u]\setminus\{\bar{i}\}}$, and the fact that by definition
    \begin{align*}
      ([k_{\bar{i}}]\times M^{[u]\setminus\{\bar{i}\}}_X) \sqcup (M^{\bar{i}}_X\times K^{[u]\setminus\{\bar{i}\}}) \sqcup (M^{\bar{i}}_X\times M^{[u]\setminus\{\bar{i}\}}_X) &= M^{[u]}_X.
    \end{align*}
    Applying similar reasoning to the bras on the RHS of \Cref{eq:csupinitstate}, we conclude that the RHS of \Cref{eq:csupinitstate} is equal to
    \begin{align*}
      \sum_{x,x'\in\bF_q^{K^{[u]}}} \ket{(I_{n_{\bar{i}}}\otimes E^{[u]\setminus\{\bar{i}\}})(x,\bF_q^{M^{[u]}_X},0)}\bra{(I_{n_{\bar{i}}}\otimes E^{[u]\setminus\{\bar{i}\}})(x',\bF_q^{M^{[u]}_X},0)} \otimes \rho_{x,x'}.
    \end{align*}
    As $\cQ$ simply applies the unitary $\ket{y}\mapsto\ket{(E^{\bar{i}}\otimes I^{\otimes[u]\setminus\{\bar{i}\}})y}$ to this state and outputs the result, the output of $\cQ\otimes I_\ell(\sigma)$ is
    \begin{align}
      \label{eq:csupfinalstate}
      \begin{split}
        &\sum_{x,x'\in\bF_q^{K^{[u]}}} \ket{(E^{\bar{i}}\otimes E^{[u]\setminus\{\bar{i}\}})(x,\bF_q^{M^{[u]}_X},0)}\bra{(E^{\bar{i}}\otimes E^{[u]\setminus\{\bar{i}\}})(x',\bF_q^{M^{[u]}_X},0)} \otimes \rho_{x,x'} \\
        &\hspace{1em}= \Enc^{[u]}\otimes I_\ell\left(\rho\otimes\ket{0^{{K^{[u]}}'}}\bra{0^{{K^{[u]}}'}}\right),
      \end{split}
    \end{align}
    where above we use the fact that by definition $E^{\bar{i}}\otimes E^{[u]\setminus\{\bar{i}\}}=E^{[u]}$.
    Thus we have shown that in the absence of a fault, when $\cQ$ is given as input the state $\rho$ encoded in $D_{\mathrm{in}}$, it uses the desired space $N'$ and time $T$, and outputs $\rho$ encoded in $D_{\mathrm{out}}$.

    It only remains to analyze the behavior of $\cQ$ under input errors, faults, and postselections. By \Cref{lem:paulift}, it suffices to show for every Pauli $\cE_{\mathrm{in}}\sqcup [\ell]$-deviation $\sigma$ of $\Enc_{\mathrm{in}}\otimes I_\ell(\rho\otimes\Gamma_{\mathrm{in}})$, and every $\cE_{\mathrm{run}}$-avoiding Pauli fault $\cF$ and postselection $\zeta$ for $\cQ$, then $\cQ[\cF,\zeta]\otimes I_\ell(\sigma)$ is a $\cE_{\mathrm{out}}\sqcup [\ell]$-deviation of $\Enc_{\mathrm{out}}\otimes I_\ell(\rho\otimes\Gamma_{\mathrm{out}})$. By \Cref{lem:stateprep} and \Cref{lem:parcomp}, following the calls to the initialization gadget in \Cref{lem:stateprep}, the state of the qudits in $B_1\sqcup B_2\sqcup B_3=N^{[u]}$ can be expressed as a linear combination of states given by applying Pauli superoperators $P_1,P_2,P_3$ to qudits $B_1,B_2,B_3$ respectively of the state in \Cref{eq:csupinitstate}, with
    \begin{align*}
      |P_2| &\leq |M^{\bar{i}}_X|\cdot 16\mu(u-1,n,d)un(1+un^2)\cdot\lambda_{\mathrm{run}} \\
      |P_1|+|P_3| &\leq \lambda_{\mathrm{in}}+(un^2+4)\lambda_{\mathrm{run}}.
    \end{align*}
    The remaining $\leq 2n^2+3$ timesteps of $\cQ$ simply consist of $\gCX$ gates that only act (and hence propagate errors) within direction-$\bar{i}$ columns in $N^{[u]}\sqcup N^{[u]}$. Similarly as described for \Cref{it:csdown} above, these $\gCX$ gates can therefore only propagate a Pauli error to other Pauli errors within the same direction-$\bar{i}$ column. Thus the final state $\cQ[\cF,\zeta]\otimes I_\ell(\sigma)$ is a linear combination of states given by applying a Pauli superoperator $P$ to qudits $N^{[u]}$ of the state in \Cref{eq:csupfinalstate}, with
    \begin{align*}
      |P|
      &\leq n_{\bar{i}}\cdot\left(|M^{\bar{i}}_X|\cdot 16\mu(u-1,n,d)un(1+un^2)\cdot\lambda_{\mathrm{run}} + \lambda_{\mathrm{in}}+(un^2+4)\lambda_{\mathrm{run}} + (2n^2+3)\lambda_{\mathrm{run}}\right) \\
      &< n\cdot\lambda_{\mathrm{in}} + 50\mu(u-1,n,d)u^2n^5\cdot\lambda_{\mathrm{run}} \\
      &= \lambda_{\mathrm{out}}
    \end{align*}
    Note that the second inequality above holds because $|M^{\bar{i}}_X|\leq n_{\bar{i}}=n$. Therefore we have shown that $\cQ[\cF,\zeta]\otimes I_\ell(\sigma)$ is a $\cE_{\mathrm{out}}\sqcup [\ell]$-deviation of $\Enc_{\mathrm{out}}\otimes I_\ell(\rho\otimes\Gamma_{\mathrm{out}})$, as desired.
  \end{enumerate}
\end{proof}

\subsection{Hadamard}
\label{sec:hadamard}
In this section, we present a basic gadget for a logical Hadamard gate by applying physical Hadamard gates to all qudits in a code state. Note that this gadget outputs a state encoded in a dualized version $C^\perp=(C_Z,C_X)$ of the input code $C=(C_X,C_Z)$. We will later show how to return to the original code by repeated applications of the code switching and error correction gadgets in \Cref{lem:codeswitch} and \Cref{lem:errcorr}, respectively.

We will first need the notion of dual encoding maps for dualized codes in \Cref{def:dualenc} below. Below, for subpaces $A,B\subseteq\bF_q^N$ with $A^\perp\subseteq B$, and for cosets $a+B^\perp\in A/B^\perp$, $b+A^\perp\in B/A^\perp$, we write $(a+B^\perp)\cdot(b+A^\perp)=a\cdot b\in\bF_q$. This notation is well-defined because the dot product of every $a+b'\in a+B^\perp$ with every $b+a'\in b+A^\perp$ equals $(a+b')\cdot(b+a')=a\cdot b$.

\begin{definition}
  \label{def:dualenc}
  Let $C=(C_X,C_Z)$ be a CSS subsystem code with CSS encoding map
  \begin{equation*}
    \Enc:\bF_q^k\oplus\bF_q^{k'}\xrightarrow{\sim}(C_Z+C_X^\perp)/(C_Z\cap C_X^\perp).
  \end{equation*}
  We call $C^\perp:=(C_Z,C_X)$ the \emph{dualization}\footnote{We reserve the term ``dual code'' to refer to duals of classical codes as defined in \Cref{def:classcode}.} of $C$.
  Let
  \begin{equation*}
    \Enc':\bF_q^k\oplus\bF_q^{k'}\xrightarrow{\sim}(C_X+C_Z^\perp)/(C_X\cap C_Z^\perp)
  \end{equation*}
  be an encoding map for $C^\perp$. We say the maps $\Enc,\Enc'$ are \emph{dual} to each other if it holds for every $x,x'\in\bF_q^{k+k'}$ that
  \begin{equation}
    \label{eq:dualenc}
    \Enc(x) \cdot \Enc'(x') = x\cdot x'.
  \end{equation}
\end{definition}

\Cref{lem:dualenc} below shows that every encoding map $\Enc$ has a unique dual encoding map $\Enc^\perp$, and that such duals respect subsystem product structure.

\begin{lemma}
  \label{lem:dualenc}
  Let $C$ be a CSS subsystem code with CSS encoding map $\Enc$. Then there exists a unique encoding map $\Enc^\perp$ for $C^\perp$ that is dual to $\Enc$.

  Furthermore, if $C=\bigotimes_{i\in[u]}C^i$ is the subsystem product of codes $C^i$ with encoding maps $\Enc^i$, and $\Enc$ is a product encoding map (see \Cref{def:subsystemcode,def:prodenc}), then $C^\perp=\bigotimes_{i\in[u]}{C^i}^\perp$ is the subsystem product of dualized codes ${C^i}^\perp$ with dual encoding maps ${\Enc^i}^\perp$, and $\Enc^\perp$ is a product encoding map.
\end{lemma}
\begin{proof}
  Fix a basis $\{x_1,\dots,x_{k+k'}\}$ for $\bF_q^{k+k'}$. For each $i\in[k+k']$, plugging these basis vectors into $x$ in \Cref{eq:dualenc} while letting $x'=x_i$ yields a system of $k+k'$ linear equations in the $k+k'$ degrees of freedom of $\Enc'(x_i)\in (C_X+C_Z^\perp)/(C_X\cap C_Z^\perp)$. Because $\Enc:\bF_q^k\oplus\bF_q^{k'}\xrightarrow{\sim}(C_Z+C_X^\perp)/(C_Z\cap C_X^\perp)$ is by definition an isomorphism, this system has a unique solution for $\Enc'(x_i)$. These solutions across all $i\in[k+k']$ then uniquely define the encoding map $\Enc^\perp=\Enc'$ dual to $\Enc$.

  Now assume that $C=\bigotimes_{i\in[u]}C^i$ is the subsystem product of $[[n_i,k_i]]$ codes $C^i$ with product encoding map $\Enc$. For each $i\in[u]$, fix a basis $\{x^i_1,\dots,x^i_{k_i}\}$ for $\bF_q^{k_i}$, and recall that $\bF_q^k\cong\bigotimes_{i\in[u]}\bF_q^{k_i}$. Consider $x'=\bigotimes_{i\in[u]}x^i_{j_i'}\in\bF_q^k\oplus 0^{k'}$ with $j_i'\in[k_i]$ for $i\in[u]$. If
  \begin{equation}
    \label{eq:dualprodenc}
    \Enc'(x') = \bigotimes_{i\in[u]}{\Enc^i}^\perp(x^i_{j_i'})\in C_X/(C_X\cap C_Z^\perp),
  \end{equation}
  then \Cref{eq:dualenc} by definition holds for every $x=\bigotimes_{i\in[u]}x^i_{j_i}\in\bF_q^k\oplus 0^{k'}$ with $j_i\in[k_i]$ for $i\in[u]$. Furthermore, for every $x\in 0^k\oplus\bF_q^{k'}$, then by definition $\Enc(x)\in C_X^\perp/(C_Z\cap C_X^\perp)$, so \Cref{eq:dualenc} holds as both sides vanish. Thus we have shown that the value of $\Enc'(x')$ in \Cref{eq:dualprodenc} satisfies the system of $k+k'$ linear equations given by letting $x$ in \Cref{eq:dualenc} range across a basis of $\bF_q^k\oplus\bF_q^{k'}$. But as $\Enc^\perp(x')$ is the unique solution to this system, we must have $\Enc^\perp(x')=\Enc'(x')$ given by \Cref{eq:dualprodenc} for every $x'=\bigotimes_{i\in[u]}x^i_{j_i'}$. Thus $\Enc^\perp$ is a product encoding map (see \Cref{def:prodenc}), as desired.
\end{proof}

We are now ready to present our Hadamard gadget in \Cref{lem:hadamard} below. While the gadget applies to arbitrary subsystem codes, we will specifically apply it to subsystem product codes, while using the characterization of dual encoding maps in \Cref{lem:dualenc}.

\begin{lemma}
  \label{lem:hadamard}
  Let $C=(C_X,C_Z)$ be a $[[n,k]]_q$ CSS subsystem code with specified CSS encoding map $\Enc:\bF_q^{k+k'}\rightarrow(C_Z+C_X^\perp)/(C_Z\cap C_X^\perp)$, and define $C^\perp,\Enc^\perp$ as in \Cref{def:dualenc,lem:dualenc}. For $\lambda_{\mathrm{in}},\lambda_{\mathrm{run}}\geq 0$, let $\lambda_{\mathrm{out}}=\lambda_{\mathrm{in}}+\lambda_{\mathrm{run}}$. For $\Gamma_{\mathrm{in}}\subseteq\bC^{\bF_q^{k'}\times\bF_q^{k'}}$, let $\Gamma_{\mathrm{out}}=\{\gH^{\otimes k'}\gamma(\gH^\dagger)^{\otimes k'}:\gamma\in\Gamma_{\mathrm{in}}\}$. Define decorated subsystem codes
  \begin{align*}
    D_{\mathrm{in}} &= \left(C,\; \Enc,\; \Gamma_{\mathrm{in}},\; \cE_{\mathrm{in}}=2^{[n]}|_{\geq\lambda_{\mathrm{in}}}\right) \\
    D_{\mathrm{out}} &= \left(C^\perp,\; \Enc^\perp,\; \Gamma_{\mathrm{out}},\; \cE_{\mathrm{in}}=2^{[n]}|_{\geq\lambda_{\mathrm{out}}}\right).
  \end{align*}
  Then there exists a quantum circuit $\cQ$ using gate set $\{\gH\}$, space $|N|=n$, and time $T=1$ such that $(\cQ,\cE_{\mathrm{run}}=2^{[n]}|_{\geq\lambda_{\mathrm{run}}},D_{\mathrm{in}},D_{\mathrm{out}})$ forms a fault-tolerant gadget for the channel $\bar{O}:\bC^{\bF_q^k\times\bF_q^k}\rightarrow\bC^{\bF_q^k\times\bF_q^k}$ that applies the unitary $\gH^{\otimes k}$.
\end{lemma}
\begin{proof}
  The desired circuit $\cQ$ simply applies $\gH^{\otimes n}$ in its single timestep. We first analyze the behavior of $\cQ$ in the absence of errors. Given $\rho\in\bC^{\bF_q^{k+\ell}\times\bF_q^{k+\ell}}$ and $\gamma\in\bC^{\bF_q^{k'}\times\bF_q^{k'}}$, we may express
  \begin{equation*}
    \rho\otimes\gamma = \sum_{x,x'\in\bF_q^{k+k'}}\ket{x}\bra{x'} \otimes \nu_{x,x'},
  \end{equation*}
  for some $\nu_{x,x'}\in\bC^{\bF_q^\ell\times\bF_q^\ell}$. Letting $p$ denote the characteristic of $\bF_q$ and writing $\omega_p=e^{2\pi i/p}$, then by definition,
  \begin{align}
    \label{eq:hadconj}
    \begin{split}
      \hspace{1em}&\hspace{-1em}\gH^{\otimes n}\left(\Enc\otimes I_\ell(\rho\otimes\gamma)\right)(\gH^\dagger)^{\otimes n} \\
                  &= \sum_{x,x'\in\bF_q^{k+k'}}\gH^{\otimes n}\ket{\Enc(x)}\bra{\Enc(x')}(\gH^\dagger)^{\otimes n} \otimes \nu_{x,x'} \\
                  &\propto \sum_{x,x'\in\bF_q^{k+k'},\;y,y'\in C_X+C_Z^\perp} \omega_p^{\tr_{\bF_q/\bF_p}(\Enc(x)\cdot y)-\tr_{\bF_q/\bF_p}(\Enc(x')\cdot y')}\ket{y}\bra{y'} \otimes \nu_{x,x'} \\
                  &\propto \sum_{x,x',z,z'\in\bF_q^{k+k'}} \omega_p^{\tr_{\bF_q/\bF_p}(\Enc(x)\cdot\Enc^\perp(z))-\tr_{\bF_q/\bF_p}(\Enc(x')\cdot\Enc^\perp(z'))}\ket{\Enc^\perp(z)}\bra{\Enc^\perp(z')} \otimes \nu_{x,x'} \\
                  &= \sum_{x,x',z,z'\in\bF_q^{k+k'}} \omega_p^{\tr_{\bF_q/\bF_p}(x\cdot z)-\tr_{\bF_q/\bF_p}(x'\cdot z')}\ket{\Enc^\perp(z)}\bra{\Enc^{\perp}(z')} \otimes \nu_{x,x'} \\
                  &= \Enc^\perp(\gH^{\otimes k+k'}\rho\otimes\gamma(\gH^\dagger)^{\otimes k+k'}).
    \end{split}
  \end{align}
  The first $\propto$ above holds because while the definition of the Hadamard gate requires a sum over all $y,y'\in\bF_q^n$, the terms associated to any $y$ or $y'$ lying outside of $C_X+C_Z^\perp=(C_Z\cap C_X)^\perp$ must vanish, as $\ket{\Enc(x)}$ and $\bra{\Enc(x')}$ are uniform superpositions over $C_Z\cap C_X^\perp$-cosets (i.e.~they are cosets in $(C_Z+C_X^\perp)/(C_Z\cap C_X^\perp)$). The second $\propto$ above holds because $\Enc^\perp$ is an encoding map for $C^\perp=(C_X,C_Z)$. The second equality above then holds by the definition of the dual encoding map $\Enc^\perp$. The final equality holds by the definition of $\gH$.

  Thus we have shown that in the absence of errors, given as input $\Enc\otimes I_\ell(\rho\otimes\gamma)$, then $\cQ$ outputs $\Enc^\perp(\gH^{\otimes k+k'}\rho\otimes\gamma(\gH^\dagger)^{\otimes k+k'})$. It then follows by \Cref{lem:paulift} that $(\cQ,\cE_{\mathrm{run}},D_{\mathrm{in}},D_{\mathrm{out}})$ is a fault-tolerant gadget for $\bar{O}$, as $\cQ$ applies single-qudit unitary gates for a single timestep, and hence an output qudit can receive an error only if it was one of the $<\lambda_{\mathrm{in}}$ qudits in the support of the input (Pauli) error, or one of the $<\lambda_{\mathrm{run}}$ qudits in the support of the (Pauli) fault.
\end{proof}

\begin{remark}
  \label{remark:hadamardinv}
  Swapping the role of $\gH$ and $\gH^\dagger$ in \Cref{lem:hadamard} yields a fault-tolerant gadget for $\bar{O}=(\gH^\dagger)^{\otimes k}$ using gate set $\{\gH^\dagger\}$; the parameters and the proof are exactly analogous.
\end{remark}


\subsection{Paulis}
\label{sec:Paulis}
In this section, we present a basic gadget that applies a logical Pauli operator to a code state. This gadget is entirely standard, and here we simply present it using our fault-tolerance notation, along with a brief proof for completeness.

\begin{lemma}
  \label{lem:paulis}
  Let $C=(C_X,C_Z)$ be a $[[n,k]]_q$ CSS subsystem code with CSS encoding map $\Enc$. For $\lambda_{\mathrm{in}},\lambda_{\mathrm{run}}\geq 0$, let $\lambda_{\mathrm{out}}=\lambda_{\mathrm{in}}+\lambda_{\mathrm{run}}$. For $\Gamma\subseteq\bC^{\bF_q^{k'}\times\bF_q^{k'}}$, define decorated subsystem codes
  \begin{align*}
    D_{\mathrm{in}} &= \left(C,\; \Enc,\; \Gamma,\; \cE_{\mathrm{in}}=2^{[n]}|_{\geq\lambda_{\mathrm{in}}}\right) \\
    D_{\mathrm{out}} &= \left(C,\; \Enc,\; \Gamma,\; \cE_{\mathrm{out}}=2^{[n]}|_{\geq\lambda_{\mathrm{out}}}\right).
  \end{align*}
  Then for every $\alpha\in\{\gX,\gZ\}$ and every $a\in\bF_q^k$, there exists a quantum circuit $\cQ$ using gate set $\{\alpha^*\}$, space $|N|=n$, and time $T=1$ such that $(\cQ,\cE_{\mathrm{run}}=2^{[n]}|_{\geq\lambda_{\mathrm{run}}},D_{\mathrm{in}},D_{\mathrm{out}})$ forms a fault-tolerant gadget for the channel $\bar{O}:\bC^{\bF_q^k\times\bF_q^k}\rightarrow\bC^{\bF_q^k\times\bF_q^k}$ that applies the unitary Pauli $\alpha^a$.
\end{lemma}
\begin{proof}
  First assume that $\alpha=\gX$. Then the desired circuit $\cQ$ simply applies $\gX^y$ for some $y\in\Enc(a,0^{k'})$. We first analyze the behavior of $\cQ$ in the absence of errors. Given $\rho\in\bC^{\bF_q^{k+\ell}\times\bF_q^{k+\ell}}$ and $\gamma\in\bC^{\bF_q^{k'}\times\bF_q^{k'}}$, we may express
  \begin{equation*}
    \rho\otimes\gamma = \sum_{x,x'\in\bF_q^{k+k'}}\ket{x}\bra{x'} \otimes \nu_{x,x'},
  \end{equation*}
  for some $\nu_{x,x'}\in\bC^{\bF_q^\ell\times\bF_q^\ell}$. Then
  \begin{align*}
    \gX^a(\Enc\otimes I_\ell(\rho\otimes\gamma))(\gX^a)^\dagger
    &= \sum_{x,x'\in\bF_q^{k+k'}}\ket{\Enc(x+(a,0))}\bra{\Enc(x'+(a,0))} \otimes \nu_{x,x'} \\
    &= \Enc\otimes I_\ell(\gX^a\rho(\gX^a)^\dagger\otimes\gamma),
  \end{align*}
  so $\gX$ applies the desired unitary $\gX^a$. The fact then $(\cQ,\cE_{\mathrm{run}},D_{\mathrm{in}},D_{\mathrm{out}})$ is a fault-tolerant gadget for $\bar{O}$ then follows from the fact that $\cQ$ consists of a single timestep of single-qudit gates, similarly as in the proof of \Cref{lem:hadamard}.

  If $\alpha=\gZ$, then $\cQ$ instead applies $\gZ^y$ for some $y\in\Enc^\perp(a,0)$, where $\Enc^\perp$ is the dual encoding map of $\Enc$ (see \Cref{lem:dualenc}). We may then apply the same argument as above for the $\alpha=\gX$ case, but with all states conjugated by Hadamard gates, which has the effect of replacing $\gX$ with $\gZ$ and $\Enc$ with $\Enc^\perp$ (see \Cref{fact:Hconj,eq:hadconj}).
\end{proof}

\subsection{CX}
\label{sec:cnot}
In this section, we present a basic gadget that applies $\gCX^a$ to all qudits in a code state. This gadget is well known (often under the name ``transversal $\gCX$''); we provide a proof for completeness.

\begin{lemma}
  \label{lem:cx}
  Let $C=(C_X,C_Z)$ be a $[[n,k]]_q$ CSS subsystem code with CSS encoding map $\Enc$. For $\lambda_{\mathrm{in}},\lambda_{\mathrm{run}}\geq 0$, let $\lambda_{\mathrm{out}}=2\lambda_{\mathrm{in}}+\lambda_{\mathrm{run}}$. Define decorated subsystem codes
  \begin{align*}
    D_{\mathrm{in}} &= \left(C,\; \Enc,\; \Gamma_{\mathrm{in}}=\left\{\ket{0^{k'}}\bra{0^{k'}}\right\},\; \cE_{\mathrm{in}}=2^{[n]}|_{\geq\lambda_{\mathrm{in}}}\right) \\
    D_{\mathrm{out}} &= \left(C,\; \Enc,\; \Gamma_{\mathrm{out}}=\left\{\ket{0^{k'}}\bra{0^{k'}}\right\},\; \cE_{\mathrm{out}}=2^{[n]}|_{\geq\lambda_{\mathrm{out}}}\right).
  \end{align*}
  Then for every $a\in\bF_q$, there exists a quantum circuit $\cQ$ using gate set $\{\gCX^a\}$, space $|N|=2n$, and time $T=1$ such that $(\cQ,\cE_{\mathrm{run}}=2^{[n]^{\sqcup 2}}|_{\geq\lambda_{\mathrm{run}}},D_{\mathrm{in}}^{\sqcup 2},D_{\mathrm{out}}^{\sqcup 2})$ forms a fault-tolerant gadget for the channel $\bar{O}:\bC^{\bF_q^{[k]^{\sqcup 2}}\times\bF_q^{[k]^{\sqcup 2}}}\rightarrow\bC^{\bF_q^{[k]^{\sqcup 2}}\times\bF_q^{[k]^{\sqcup 2}}}$ that applies the unitary $(\gCX^a)^{\otimes k}$.
\end{lemma}

\begin{remark}
  \Cref{lem:cx} can be easily extended to allow for more general $\Gamma_{\mathrm{in}},\Gamma_{\mathrm{out}}$. We omit such a generalization for simplicity; it will be sufficient to assume $\Gamma_{\mathrm{in}}=\left\{\ket{0^{k'}}\bra{0^{k'}}\right\}$ because we can always run the error-correction gadget in \Cref{lem:errcorr} prior to applying \Cref{lem:cx}.
\end{remark}

\begin{proof}[Proof of \Cref{lem:cx}]
  The desired circuit $\cQ$ simply applies $(\gCX^a)^{\otimes n}$ in its single timestep. We first analyze the behavior of $\cQ$ in the absence of errors.
  Given $\rho\in\bC^{\bF_q^{2k+\ell}\times\bF_q^{2k+\ell}}$, we may express
  \begin{equation*}
    \rho = \sum_{x_1,x_2,x_1',x_2'\in\bF_q^k}\ket{x_1}\ket{x_2}\bra{x_1'}\bra{x_2'} \otimes \nu_{x_1,x_2,x_1',x_2'},
  \end{equation*}
  for some $\nu_{x_1,x_2,x_1',x_2'}\in\bC^{\bF_q^\ell\times\bF_q^\ell}$. Then
  \begin{align*}
    \hspace{1em}&\hspace{-1em}(\gCX^a)^{\otimes n}\left(\Enc^{\sqcup 2}\otimes I_\ell\left(\rho\otimes\ket{0^{2k'}}\bra{0^{2k'}}\right)\right)({\gCX^a}^\dagger)^{\otimes n} \\
    &= \sum_{x_1,x_2,x_1',x_2'\in\bF_q^k}(\gCX^a)^{\otimes n}\ket{\Enc(x_1,0)}\ket{\Enc(x_2,0)}\bra{\Enc(x_1',0)}\bra{\Enc(x_2',0)}({\gCX^a}^\dagger)^{\otimes n} \otimes \rho_{x_1,x_2,x_1',x_2'} \\
    &= \sum_{x_1,x_2,x_1',x_2'\in\bF_q^{k}}\ket{\Enc(x_1,0)}\ket{\Enc(ax_1+x_2,0)}\bra{\Enc(x_1',0)}\bra{\Enc(ax_1'+x_2',0)} \otimes \rho_{x_1,x_2,x_1',x_2'} \\
    &= \Enc^{\sqcup 2}\otimes I_\ell\left((\gCX^a)^{\otimes k}\rho({\gCX^a}^\dagger)^{\otimes k}\otimes\ket{0^{2k'}}\bra{0^{2k'}}\right),
  \end{align*}
  where the second equality above holds by the linearity of $\Enc$.

  Thus we have shown that $\cQ$ has the desired behavior of applying $\bar{O}$ in the absence of errors. It then follows by \Cref{lem:paulift} that $(\cQ,\cE_{\mathrm{run}},D_{\mathrm{in}},D_{\mathrm{out}})$ is a fault-tolerant gadget for $\bar{O}$, as $\cQ$ applies 2-qudit unitary gates for a single timestep. Specifically, an output qudit labeled $i\in[n]$ in either $n$-qudit block can receive an error only if it or the qudit labeled $i$ in the other $n$-qudit block was one of the $<\lambda_{\mathrm{in}}$ qudits in the support of the input (Pauli) error, or if it was one of the $<\lambda_{\mathrm{run}}$ qudits in the support of the (Pauli) fault. Hence the output error has weight $<2\lambda_{\mathrm{in}}+\lambda_{\mathrm{run}}=\lambda_{\mathrm{out}}$.
\end{proof}

\subsection{CCX}
\label{sec:toffoli}
In this section, we present a gadget that applies $\gCCX^a$ to all qudits across three code states, assuming that the three (possibly different) codes satisfy an appropriate multiplicative structure. Below, recall from \Cref{sec:notation} that for $x,y\in\bF_q^n$, we let $x*y=(x_iy_i)_{i\in[n]}\in\bF_q^n$ denote the component-wise product. We also extend this notation to sets $X,Y\subseteq\bF_q^n$, so that $X*Y=\{x*y:x\in X,y\in Y\}$.

\begin{lemma}
  \label{lem:ccx}
  For $i\in[3]$, let $C^i=(C^i_X,C^i_Z)$ be a $[[n,k]]_q$ CSS subsystem code with CSS encoding map $\Enc^i:\bF_q^{k+k_i'}\rightarrow(C^i_Z+{C^i_X}^\perp)/(C^i_Z\cap{C^i_X}^\perp)$. Assume that for every $x_1,x_2\in\bF_q^k$,
  \begin{equation}
    \label{eq:multprop}
    \Enc^1(x_1,0^{k_1'})*\Enc^2(x_2,0^{k_2'}) \subseteq \Enc^3(x_1*x_2,0^{k_3'}).
  \end{equation}
  For $\lambda_{\mathrm{in}},\lambda_{\mathrm{run}}\geq 0$, let $\lambda_{\mathrm{out}}=3\lambda_{\mathrm{in}}+\lambda_{\mathrm{run}}$. For $i\in[3]$, define decorated subsystem codes
  \begin{align*}
    D^i_{\mathrm{in}} &= \left(C^i,\; \Enc^i,\; \Gamma^i_{\mathrm{in}}=\left\{\ket{0^{k_i'}}\bra{0^{k_i'}}\right\},\; \cE_{\mathrm{in}}=2^{[n]}|_{\geq\lambda_{\mathrm{in}}}\right) \\
    D^i_{\mathrm{out}} &= \left(C^i,\; \Enc^i,\; \Gamma^i_{\mathrm{out}}=\left\{\ket{0^{k_i'}}\bra{0^{k_i'}}\right\},\; \cE_{\mathrm{out}}=2^{[n]}|_{\geq\lambda_{\mathrm{out}}}\right).
  \end{align*}
  Then for every $a\in\bF_q$, there exists a quantum circuit $\cQ$ using gate set $\{\gCCX^a\}$, space $|N|=3n$, and time $T=1$ such that
  \begin{equation*}
    \left(\cQ,\; \cE_{\mathrm{run}}=2^{[n]^{\sqcup 3}}|_{\geq\lambda_{\mathrm{run}}},\; \bigsqcup_{i\in[3]}D^i_{\mathrm{in}},\; \bigsqcup_{i\in[3]}D^i_{\mathrm{out}}\right)
  \end{equation*}
  forms a fault-tolerant gadget for the channel $\bar{O}:\bC^{\bF_q^{[k]^{\sqcup 3}}\times\bF_q^{[k]^{\sqcup 3}}}\rightarrow\bC^{\bF_q^{[k]^{\sqcup 3}}\times\bF_q^{[k]^{\sqcup 3}}}$ that applies the unitary $(\gCCX^a)^{\otimes k}$.
\end{lemma}
\begin{proof}
  The proof is similar to that of \Cref{lem:cx}
  The desired circuit $\cQ$ simply applies $(\gCCX^a)^{\otimes n}$ in its single timestep. We first analyze the behavior of $\cQ$ in the absence of errors.
  Given $\rho\in\bC^{\bF_q^{3k+\ell}\times\bF_q^{3k+\ell}}$, we may express
  \begin{equation*}
    \rho = \sum_{x=(x_1,x_2,x_3),x'\in(x_1',x_2',x_3')\in(\bF_q^k)^3}\ket{x_1}\ket{x_2}\ket{x_3}\bra{x_1'}\bra{x_2'}\bra{x_3'} \otimes \nu_{x,x'},
  \end{equation*}
  for some $\nu_{x,x'}\in\bC^{\bF_q^\ell\times\bF_q^\ell}$. Letting $\Enc=\bigsqcup_{i\in[3]}\Enc^i$, then
  \begin{align*}
    \hspace{1em}&\hspace{-1em}(\gCCX^a)^{\otimes n}\left(\Enc\otimes I_\ell\left(\rho\otimes\ket{0^{k_1'+k_2'+k_3'}}\bra{0^{k_1'+k_2'+k_3'}}\right)\right)({\gCCX^a}^\dagger)^{\otimes n} \\
                &= \sum_{x,x'\in(\bF_q^k)^3}(\gCCX^a)^{\otimes n}\ket{\Enc^1(x_1,0)}\ket{\Enc^2(x_2,0)}\ket{\Enc^3(x_3,0)} \\
                &\hspace{10em}\bra{\Enc^1(x_1',0)}\bra{\Enc^2(x_2',0)}\bra{\Enc^3(x_3',0)}({\gCCX^a}^\dagger)^{\otimes n} \otimes \rho_{x,x'} \\
                &= \sum_{x,x'\in(\bF_q^k)^3}\ket{\Enc^1(x_1,0)}\ket{\Enc^2(x_2,0)}\ket{\Enc^3(ax_1*x_2+x_3,0)} \\
                &\hspace{10em}\bra{\Enc^1(x_1',0)}\bra{\Enc^2(x_2',0)}\bra{\Enc^3(ax_1'*x_2'+x_3',0)} \otimes \rho_{x,x'} \\
                &= \Enc\otimes I_\ell\left((\gCCX^a)^{\otimes k}\rho({\gCCX^a}^\dagger)^{\otimes k}\otimes\ket{0^{k_1'+k_2'+k_3'}}\bra{0^{k_1'+k_2'+k_3'}}\right),
  \end{align*}
  where the second equality above holds by \Cref{eq:multprop} along with the linearity of $\Enc^3$.

  Thus we have shown that $\cQ$ has the desired behavior of applying $\bar{O}$ in the absence of errors. It then follows by \Cref{lem:paulift} that $(\cQ,\cE_{\mathrm{run}},D_{\mathrm{in}},D_{\mathrm{out}})$ is a fault-tolerant gadget for $\bar{O}$, as $\cQ$ applies 3-qudit unitary gates for a single timestep. Specifically, an output qudit labeled $i\in[n]$ in any of the three $n$-qudit blocks can receive an error only if that qudit was one of the $<\lambda_{\mathrm{run}}$ qudits in the support of the (Pauli) fault, or if any of the qudits labeled $i$ across the three $n$-qudit blocks was one of the $<\lambda_{\mathrm{in}}$ qudits in the support of the input (Pauli) error. Hence the output error has weight $<3\lambda_{\mathrm{in}}+\lambda_{\mathrm{run}}=\lambda_{\mathrm{out}}$. Note that as $\gCCX$ is non-Clifford, even under Pauli input and fault errors, the output error may be given by non-Pauli unitaries, though by \Cref{lem:paulidecomp} it can be decomposed into a sum of Pauli errors with the same (or smaller) supports.
\end{proof}

The following lemma shows that a triple of subsystem product codes satisfies the condition in \Cref{eq:multprop} if the respective triples of factor codes satisfy this condition.

\begin{lemma}
  \label{lem:hadfactor}
  For $i\in[3]$ and $j\in[u]$, let $C^{i,j}$ be a $[[n_j,k_j]]_q$ CSS non-subsystem code with CSS encoding map $\Enc^{i,j}:\bF_q^{k_j}\rightarrow C^{i,j}_Z/{C^{i,j}_X}^\perp$. For $i\in[3]$, let $C^i=\bigotimes_{j\in[u]}C^{i,j}$ be the $[[|N|,|K|]]_q$ subsystem product code with product encoding map $\Enc^i:\bF_q^K\oplus\bF_q^{K_i'}\rightarrow(C^i_Z+{C^i_X}^\perp)/(C^i_Z\cap{C^i_X}^\perp)$, where $N=\prod_{j\in[u]}[n_j]$ and $K=\prod_{j\in[u]}[k_j]$. Assume that for every $j\in[u]$ and every $x_1,x_2\in\bF_q^{k_j}$,
  \begin{equation}
    \label{eq:hadfactor}
    \Enc^{1,j}(x_1)*\Enc^{2,j}(x_2) \subseteq \Enc^{3,j}(x_1*x_2).
  \end{equation}
  Then for every $x_1,x_2\in\bF_q^K$,
  \begin{equation*}
    \Enc^1(x_1,0^{K_1'})*\Enc^2(x_2,0^{K_2'}) \subseteq \Enc^3(x_1*x_2,0^{K_3'}).
  \end{equation*}
\end{lemma}
\begin{proof}
  By definition for $i\in[1,2]$, we can decompose $x_i\in\bF_q^K$ into a sum of rank-$1$ tensors $x_i=\sum_{m_i}\bigotimes_{j\in[u]}x_{i,j,m_i}$ for $x_{i,j,m_i}\in\bF_q^{k_j}$. Then
  \begin{align*}
    \hspace{1em}&\hspace{-1em}\Enc^1(x_1,0^{K_1'})*\Enc^2(x_2,0^{K_2'})+(C^3_Z\cap{C^3_X}^\perp) \\
                &= \sum_{m_1,m_2}\Enc^1\left(\bigotimes_{j\in[u]}x_{1,j,m_1},\; 0^{K_1'}\right)*\Enc^2\left(\bigotimes_{j\in[u]}x_{2,j,m_2},\; 0^{K_2'}\right)+(C^3_Z\cap{C^3_X}^\perp) \\
                &= \sum_{m_1,m_2}\bigotimes_{j\in[u]}(\Enc^{1,j}(x_{1,j,m_1})*\Enc^{2,j}(x_{2,j,m_2}))+(C^3_Z\cap{C^3_X}^\perp) \\
                &\subseteq \sum_{m_1,m_2}\bigotimes_{j\in[u]}\Enc^{3,j}(x_{1,j,m_1}*x_{2,j,m_2})+(C^3_Z\cap{C^3_X}^\perp) \\
                &= \sum_{m_1,m_2}\Enc^{3,j}\left(\left(\bigotimes_{j\in[u]}x_{1,j,m_1}\right)*\left(\bigotimes_{j\in[u]}x_{2,j,m_2}\right),\; 0^{K_3'}\right) \\
                &= \Enc^{3,j}\left(x_1*x_2,\; 0^{K_3'}\right),
  \end{align*}
  as desired, where the inclusion of the LHS in the RHS above implies equality, as the RHS is by definition a coset in $C^3_Z/(C^3_Z+{C^3_X}^\perp)$. Note that the second and third equalities above use the fact that the $\Enc^i$ are product encoding maps (see \Cref{def:prodenc}), while the inclusion above applies \Cref{eq:hadfactor}.
\end{proof}


\subsection{Classical Function Gates}
\label{sec:classicalfunctiongates}
In this section, we present a gadget for performing a logical classical function gate on an entire code state. We will later show how to perform logical permutations that allow us to apply such gates on desired subsets of the logical qudits.

\begin{lemma}
  \label{lem:co}
  For $\bar{b},u,n,d\in\bN$, for $b\in[\bar{b}]$ and $i\in[u]$ let $C_b^i=((C_b^i)_X,(C_b^i)_Z)$ be a $[[n_i=n,\; k_i]]_q$ CSS non-subsystem code with specified CSS encoding map $\Enc_b^i:\bF_q^{k_i}\rightarrow (C_b^i)_Z/{(C_b^i)_X}^\perp$, such that the classical codes $(C_b^i)_X,(C_b^i)_Z$ have distance $\geq d$. Define $\mu=\mu(u,n,d)$ to be the expression in \Cref{prop:loctest}.
  Let $C_b=\bigotimes_{i\in[u]}C_b^i$ be the $[[|N_b|,\;|K_b|,\;\geq D]]_q$ subsystem product code with product encoding map $\Enc_b$, where every $N_b=\prod_{i\in[u]}[n_i]$ and $K_b=\prod_{i\in[u]}[k_i]$, and. Then let
  \begin{equation*}
    (C,\Enc) = \bigsqcup_{b\in[\bar{b}]}(C_b,\Enc_b).
  \end{equation*}
  be the $[[|N|,\;|K|,\;\geq D]]_q$ disjoint union code, where $N=\bigsqcup_{b\in[\bar{b}]}N_b$ and $K=\bigsqcup_{b\in[\bar{b}]}K_b$, so that $\Enc:\bF_q^K\oplus\bF_q^{K'}\rightarrow(C_Z+C_X)/(C_Z\cap C_X^\perp)$ is a disjoint union of product encoding maps.
  Let $\lambda_{\mathrm{in}}=D/2$, and for $\lambda_{\mathrm{run}}\geq 0$ let
  \begin{align*}
    \lambda_{\mathrm{out}} &= 50\bar{b}\mu u^2n^3\cdot\lambda_{\mathrm{run}}.
  \end{align*}
  Let $\Gamma_{\mathrm{out}}^X=\left\{\ket{0^{K'}}\bra{0^{K'}}\right\}$ and $\Gamma_{\mathrm{out}}^Z=\left\{\ket{+^{K'}}\bra{+^{K'}}\right\}$. For $\alpha\in\{X,Z\}$, let $\Gamma_{\mathrm{in}}^\alpha=\bC^{\bF_q^{K'}\times\bF_q^{K'}}$, and define decorated $[[|N|,|K|,D]]_q$ subsystem codes
  \begin{align*}
    D_{\mathrm{in}}^\alpha &= \left(C,\; \Enc,\; \Gamma_{\mathrm{in}}^\alpha,\; \cE_{\mathrm{in}}=2^{N}|_{\geq \lambda_{\mathrm{in}}}\right). \\
    D_{\mathrm{out}}^\alpha &= \left(C,\; \Enc,\; \Gamma_{\mathrm{out}}^\alpha,\; \cE_{\mathrm{out}}=2^{N}|_{\geq \lambda_{\mathrm{out}}}\right).
  \end{align*}
  Then for every function $f:\bF_q^K\rightarrow\bF_q^K$, there exists a quantum circuit $\cQ_{\alpha,f}$ using gate set $\{\gCX^*,\gInit_*,\gCO_*\}$, space $|N'|\leq\bar{b}\cdot(u+2)n^u$, and time $T\leq 5+un^2$ such that $(\cQ_{\alpha,f},\; \cE_{\mathrm{run}}=2^{N'}|_{\geq \lambda_{\mathrm{run}}}^{\sqcup T},\; D_{\mathrm{in}},\; D_{\mathrm{out}})$ forms a fault-tolerant gadget for the set of superoperators $\bar{\cO}_{\alpha,f}=\{\gCO_{\alpha,f}^{\bar{\zeta}}:\bar{\zeta}\in\bC^{\bF_q^K}\}$.
\end{lemma}
\begin{proof}
  Let $\Enc_Z=\Enc$ and $\Enc_X=\Enc^\perp$ (see \Cref{lem:dualenc}). For $\alpha\in\{X,Z\}$, fix a linear map $\Enc^0_\alpha:\bF_q^K\rightarrow C_\alpha$ such that $\Enc^0_\alpha(x)\in\Enc_\alpha(x,0^{K'})$ for every $x\in\bF_q^K$; such a map $\Enc^0_\alpha$ exists by the definition of $\Enc_\alpha$. For each $y\in\bF_q^N$, define $x^{\min}_\alpha(y)\in\bF_q^K$ by\footnote{\label{footnote:setham} Recall here that for a set $S\subseteq\bF_q^N$, we write $|y-S|=\min_{y'\in S}|y-y'|$.}
  \begin{align*}
    x^{\min}_\alpha(y) = \argmin_{x\in\bF_q^K}|y-\Enc_\alpha(x,\bF_q^{K'})|.
  \end{align*}
  Then define a function $F_\alpha:\bF_q^N\rightarrow\bF_q^N$ by
  \begin{equation*}
    F_\alpha(y) = \Enc^0_\alpha(f(x^{\min}_\alpha(y))).
  \end{equation*}

  We now define the circuit $\cQ_{\alpha,f}$ as follows. In the first timestep, $\cQ_{\alpha,f}$ applies $\gCO_{\alpha,F}$ to the block of input qudits labeled by $N$. On separate ancilla qudits, $\cQ_{\alpha,f}$ runs the gadget in \Cref{it:spnew} of \Cref{lem:stateprep} $\bar{b}$ times in parallel, once for $(C_b,\Enc_b)$ for each $b\in[\bar{b}]$, to initialize a block of ancilla qudits labeled by $N$ to the state $\Enc(\ket{+^K}\bra{+^K}\otimes\ket{0^{K'}}\bra{0^{K'}})$ if $\alpha=X$, or to the state $\Enc(\ket{0^K}\bra{0^K}\otimes\ket{+^{K'}}\bra{+^{K'}})$ if $\alpha=Z$. Once this initialization completes, $\cQ_{X,f}$ then for every $i\in N$ applies a $\gCX^{-1}$ gate with control qudit $i$ in the ancilla block and target qudit $i$ in the input block; $\cQ_{Z,f}$ instead applies $\gCX^{+1}$ with control qudit $i$ in the input block and target qudit $i$ in the ancilla block. Finally, $\cQ_{\alpha,f}$ returns the ancilla block.

  By \Cref{lem:stateprep}, $\cQ_{\alpha,f}$ uses space $|N'|\leq\bar{b}(n^u+(u+1)n^u)=\bar{b}\cdot(u+2)n^u$ and time $T\leq(4+un^2)+1=5+un^2$, as desired. It remains to be shown that $(\cQ_{\alpha,f},\cE_{\mathrm{run}},D_{\mathrm{in}},D_{\mathrm{out}})$ forms a fault-tolerant gadget for $\bar{\cO}_{\alpha,f}$. Assume that $\alpha=Z$; the $\alpha=X$ case will be analogous. For $\rho\in\bC^{\bF_q^{|K|+\ell}\times\bF_q^{|K|+\ell}}$ and $\gamma\in\Gamma_{\mathrm{in}}^Z$, let $\sigma_0$ be a Pauli $\cE_{\mathrm{in}}\sqcup [\ell]$-deviation of $\Enc\otimes I_\ell(\rho\otimes\gamma)$, and let $\cF$ be a $\cE_{\mathrm{run}}$-avoiding Pauli fault and $\zeta$ be a postselection for $\cQ_{Z,f}$. By \Cref{lem:paulift}, it suffices to show that the resulting output $\cQ_{Z,f}[\cF,\zeta]\otimes I_\ell(\sigma_0)$ is a $\cE_{\mathrm{out}}\sqcup [\ell]$-deviation of $(\Enc\circ\bar{\cO}_{Z,f})\otimes I_\ell(\rho\otimes\ket{+^{K'}}\bra{+^{K'}})$. 

  We can express
  \begin{align*}
    \rho
    &= \sum_{x,x'\in\bF_q^K}\ket{x}\bra{x'}\otimes\rho_{x,x'}
  \end{align*}
  for some $\rho_{x,x'}\in\bC^{\bF_q^\ell\times\bF_q^\ell}$.
  Then by definition we can express $\sigma_0$ as a linear combination of states of the form
  \begin{align*}
    \sigma
    &= \sum_{x,x'\in\bF_q^K}Z^{e_Z}\ket{\Enc(x,0)+g_X+e_X}\bra{\Enc(x',0)+g_X'+e_X'}(Z^\dagger)^{e_Z'}\otimes\rho_{x,x'}
  \end{align*}
  for some $g_X,g_X'\in C_X^\perp$ and some $e_X,e_X',e_Z,e_Z'\in\bF_q^N$ with
  \begin{equation*}
    |\supp(e_X)\cup\supp(e_X')\cup\supp(e_Z)\cup\supp(e_Z')| < \lambda_{\mathrm{in}} = D/2.
  \end{equation*}
  Specifically, we arrive at this decomposition by expressing $\gamma=\sum_{z,z'\in\bF_q^{K'}}\ket{z}\bra{z'}\cdot\gamma_{z,z'}$ for $\gamma_{z,z'}\in\bC$, and then considering each term in the sum over $z,z'$ separately, while choosing $g_X\in\Enc(0,z)$ and $g_X'\in\Enc(0,z')$.

  We now analyze the behavior of $\cQ_{Z,f}[\cF,\zeta]$ on input $\sigma$. First, $\cQ_{Z,f}$ applies $\gCO_{Z,F}^\zeta$ to $\sigma$, yielding
  \begin{align}
    \label{eq:cosigma}
    \begin{split}
      \gCO_{Z,F}^\zeta(\sigma)
    &= \sum_{x,x'\in\bF_q^K,y\in\bF_q^N}\zeta_y\cdot\ket{F_Z(y)}\bra{F_Z(y)}\otimes\rho_{x,x'} \\
        &\hspace{7em} \cdot \bra{y}Z^{e_Z}\ket{\Enc(x,0)+g_X+e_X}\bra{\Enc(x',0)+g_X'+e_X'}(Z^\dagger)^{e_Z'}\ket{y}.
    \end{split}
  \end{align}

  We will now apply \Cref{claim:geCZ} from the proof of \Cref{lem:stateprep} above; we restate the claim below for convenience, though the proof is identical.
  
  \begin{claim}
    \label{claim:geCZrestate}
    If $(g_X-g_X')+(e_X-e_X')\in C_Z$, then $(g_X-g_X')+(e_X-e_X')\in C_Z\cap C_X^\perp$.
  \end{claim}
  
  If $(g_X-g_X')+(e_X-e_X')\notin C_Z\cap C_X^\perp$, then by \Cref{claim:geCZrestate} $(g_X-g_X')+(e_X-e_X')\notin C_Z$, and hence as $\Enc(x,0),\Enc(x',0)\subseteq C_Z$, the expression on the second line of \Cref{eq:cosigma} vanishes, and $\gCO_{Z,F}^\zeta(\sigma)=0$. Therefore assume that $(g_X-g_X')+(e_X-e_X')\in C_Z\cap C_X^\perp$. Then the expression on the second line of \Cref{eq:cosigma} is nonvanishing only when $\Enc(x,0)=\Enc(x',0)$, or equivalently when $x=x'$. Therefore letting $p$ denote the characteristic of $\bF_q$ and $\omega_p=e^{2\pi i/p}$, then
  \begin{align*}
    \gCO_{Z,F}^\zeta(\sigma)
    &\propto \sum_{x\in\bF_q^K}\sum_{y\in\Enc(x,0)+g_X+e_X}\zeta_y\cdot\ket{F_Z(y)}\bra{F_Z(y)}\otimes\rho_{x,x}\cdot\omega_p^{\tr_{\bF_q/\bF_p}((e_Z-e_Z')\cdot y)}.
  \end{align*}
  Now recall that $\Enc(x,0)+g_X+e_X=\Enc(x,z)+e_X$ for some $z\in\bF_q^{K'}$, with $|e_X|<D/2$. By the definition of subsystem code distance, for every $y\in\Enc(x,z)+e_X$, then every $x'\in\bF_q^K\setminus\{x\}$ has $|y-\Enc(x',\bF_q^{K'})|>D/2$, and hence $x^{\min}_Z(y)=x$ so that $F_Z(y)=\Enc^0_Z(f(x))$. Thus
  \begin{align}
    \label{eq:cosigma2}
    \gCO_{Z,F}^\zeta(\sigma)
    &\propto \sum_{x\in\bF_q^K}\ket{\Enc^0_Z(f(x))}\bra{\Enc^0_Z(f(x))}\otimes\rho_{x,x}\cdot\sum_{y\in\Enc(x,0)+g_X+e_X}\zeta_y\cdot\omega_p^{\tr_{\bF_q/\bF_p}((e_Z-e_Z')\cdot y)}.
  \end{align}
  Then defining $\bar{\zeta}\in\bF_q^K$ by
  \begin{equation*}
    \bar{\zeta}_x = \sum_{y\in\Enc(x,0)+g_X+e_X}\zeta_y\cdot\omega_p^{\tr_{\bF_q/\bF_p}((e_Z-e_Z')\cdot y)},
  \end{equation*}
  then
  \begin{align*}
    \gCO_{Z,F}^\zeta(\sigma) &\propto \sum_{x\in\bF_q^K}\bar{\zeta}_x\cdot\ket{\Enc^0_Z(f(x))}\bra{\Enc^0_Z(f(x))}\otimes\rho_{x,x}.
  \end{align*}

  Now in the absence of a fault, $\cQ_{\alpha,f}\otimes I_\ell(\sigma)$ then prepares an ancilla code state
  \begin{equation*}
    \Enc\left(\ket{0^K}\bra{0^K}\otimes\ket{+^{K'}}\bra{+^{K'}}\right) = \ket{\Enc(0^K,\bF_q^{K'})}\bra{\Enc(0^K,\bF_q^{K'})},
  \end{equation*}
  applies $\gCX^{\otimes N}$ to $\gCO_{Z,F}(\sigma)$ and this ancilla state to obtain (using the definition of $\Enc^0_Z$)
  \begin{align*}
    \sum_{x\in\bF_q^K}\bar{\zeta}_x\cdot\ket{\Enc^0_Z(f(x))}\bra{\Enc^0_Z(f(x))}\otimes\ket{\Enc(f(x),\bF_q^{K'})}\bra{\Enc(f(x),\bF_q^{K'})}\otimes\rho_{x,x},
  \end{align*}
  and then returns the resulting ancilla state (i.e.~traces out over the original code block), yielding output
  \begin{align}
    \label{eq:cooutput}
    \sum_{x\in\bF_q^K}\bar{\zeta}_x\cdot\ket{\Enc(f(x),\bF_q^{K'})}\bra{\Enc(f(x),\bF_q^{K'})}\otimes\rho_{x,x}
    &= (\Enc\circ\gCO_{Z,f}^{\bar{\zeta}})\otimes I_\ell\left(\rho\otimes\ket{+^{K'}}\bra{+^{K'}}\right).
  \end{align}
  Thus $\cQ_{Z,f}$ has the desired output on input $\sigma$ in the absence of a fault. Now allowing for a $\cE_{\mathrm{run}}$-avoiding Pauli fault, then just prior to the final timestep (that runs $\gCX^{\otimes n}$), the input block differs from $\gCO_{Z,F}^\zeta(\sigma)$ by a Pauli error of weight $\leq(T-1)\cdot\lambda_{\mathrm{run}}$, while by \Cref{lem:stateprep}, the ancilla block differs from its desired state $\Enc\left(\ket{0^K}\bra{0^K}\otimes\ket{+^{K'}}\bra{+^{K'}}\right)$ by a Pauli error of weight $<\bar{b}\cdot 16\mu un(1+un^2)\cdot\lambda_{\mathrm{run}}$. Thus following the final timestep, the resulting state is of the form
  \begin{align}
    \label{eq:cooutputnoisy}
    \begin{split}
      \sum_{x\in\bF_q^K} &\bar{\zeta}_x \cdot X^{e_{1,X}}Z^{e_{1,Z}}\ket{\Enc^0_Z(f(x))}\bra{\Enc^0_Z(f(x))}(Z^\dagger)^{e_{1,Z}'}(X^\dagger)^{e_{1,X}'} \\
                         &\otimes X^{e_{2,X}}Z^{e_{2,Z}}\ket{\Enc(f(x),\bF_q^{K'})}\bra{\Enc(f(x),\bF_q^{K'})}(Z^\dagger)^{e_{2,Z}'}(X^\dagger)^{e_{2,X}'} \otimes \rho_{x,x}
    \end{split}
  \end{align}
  for some $e_{1,X},e_{1,X}',e_{1,Z},e_{1,Z}',e_{2,X},e_{2,X}',e_{2,Z},e_{2,Z}'\in\bF_q^N$ with
  \begin{align}
    \label{eq:coouterr}
    \begin{split}
      |\supp(e_{2,X})\cup\supp(e_{2,X}')\cup\supp(e_{2,Z})\cup\supp(e_{2,Z}')|
      &\leq T\cdot\lambda_{\mathrm{run}} + \bar{b}\cdot 16\mu un(1+un^2)\cdot\lambda_{\mathrm{run}} \\
      &< 50\bar{b}\mu u^2n^3\cdot\lambda_{\mathrm{run}} \\
      &= \lambda_{\mathrm{out}}.
    \end{split}
  \end{align}
  The final noisy output $\cQ_{Z,f}[\cF,\zeta]\otimes I_\ell(\sigma)$ equals the state in \Cref{eq:cooutputnoisy} with the first register (on the first line of \Cref{eq:cooutputnoisy}) traced out. If $e_{1,X}\neq e_{1,X'}$, then tracing out this first register gives $0$, and we are done. Otherwise, if $e_{1,X}=e_{1,X'}$, then defining $\bar{\zeta}'\in\bF_q^K$ by
  \begin{equation*}
    \bar{\zeta}_x' = \bar{\zeta}_x\cdot\omega_p^{\tr_{\bF_q/\bF_p}((e_{1,Z}-e_{1,Z}')\cdot\Enc^0_Z(f(x)))},
  \end{equation*}
  tracing out the first register of \Cref{eq:cooutputnoisy} gives
  \begin{align*}
      \sum_{x\in\bF_q^K} &\bar{\zeta}_x' \cdot X^{e_{2,X}}Z^{e_{2,Z}}\ket{\Enc(f(x),\bF_q^{K'})}\bra{\Enc(f(x),\bF_q^{K'})}(Z^\dagger)^{e_{2,Z}'}(X^\dagger)^{e_{2,X}'} \otimes \rho_{x,x},
  \end{align*}
  which by \Cref{eq:coouterr} differs from the desired output $(\Enc\circ\gCO_{Z,f}^{\bar{\zeta}'})\otimes I_\ell\left(\rho\otimes\ket{+^{K'}}\bra{+^{K'}}\right)$ by a Pauli error of weight $<\lambda_{\mathrm{out}}$, as desired. Thus indeed $(\cQ_{Z,f},\cE_{\mathrm{run}},D_{\mathrm{in}},D_{\mathrm{out}})$ forms a fault-tolerant gadget for $\bar{\cO}_{Z,f}$.

  For the $\alpha=X$ case, we can conjugate all states by Hadamard gates, in order to switch from the $Z$-basis to the $X$-basis, and from $C,\Enc$ to their dualized versions $C^\perp,\Enc^\perp$; then we can simply apply the same proof as above for the $\alpha=Z$ case. We omit the details to avoid redundancy.
\end{proof}

\section{Fault-Tolerance Scheme over Large Alphabets}
\label{sec:ftschemelarge}
In this section, we combine our core gadgets in \Cref{sec:coregad} to obtain a complete fault-tolerance scheme over sufficiently large (i.e.~exponential-sized) alphabets. 
Specifically, we prove the following.


\begin{theorem}
  \label{thm:laft}
  For every $u,n,k\in\bN$ with $u\geq 4$, $n\geq 8$, $k\leq n/8$, every $q=q_0^{\prod_{i\in[u]}n_i'}$ with $q_0\geq n$ a prime power and each $n_i'\geq n$ an integer, and every subset $K\subseteq[k]^u$, there is a CSS non-subsystem code $C(q,u,n,k,K)$ with associated encoding map $\Enc_{C(q,u,n,k,K)}$ which satisfy the following properties. Below, for $i\in\bN$ we let $d_i=d_i(n)=(n/8)^i$.
  \begin{enumerate}
  \item\label{it:laparam} $C(q,u,n,k,K)$ is a $[[n^u,\; |K|,\; \geq d_u]]_q$ code with logical qudits labeled by the set $K$.
  \item\label{it:lascheme} Let 
    \begin{align}
      \label{eq:lamrun}
      \bar{\lambda}_{\mathrm{run}} = \bar{\lambda}_{\mathrm{run}}(u,n) &= \frac{d_{u-2}}{2^{12+4u}\cdot u^2n^9\cdot\mu(u,n,d_1)},
    \end{align}
    where $\mu(u,n,d_1)$ is the filling constant in \Cref{prop:loctest}.
    For every $\lambda_{\mathrm{run}}\in[0,\bar{\lambda}_{\mathrm{run}}]$,
    let
    \begin{align}
      \label{eq:lamminmax}
      \begin{split}
        \lambda_{\min} = \lambda_{\min}(u,n,\lambda_{\mathrm{run}}) &= \eta_{\min}\cdot\lambda_{\mathrm{run}} \hspace{1em} \text{ for } \hspace{1em} \eta_{\min} = \eta_{\min}(u,n) = 2^{8+2u}\cdot u^2n^5\cdot\mu(u,n,d_1) \\
        \lambda_{\max} = \lambda_{\max}(u,n) &= \frac{d_{u-2}}{4}.
      \end{split}
    \end{align}
    Then for every quantum circuit $\bar{\cQ}=(\bar{Q}_1,\dots,\bar{Q}_{\bar{T}};K_{\mathrm{in}},K_{\mathrm{out}})$ acting on a set of $q$-dimensional qudits labeled by $K_{[3]}:=K_1\sqcup K_2\sqcup K_3$ with each $K_b=[k]^u$, using time $\bar{T}$ and gate set $\cG=\{\gH^*,\gX^*,\gZ^*,\gCX^*,\gCCX^*,\gInit_*,\gCO_*\}$, for $\alpha\in\{\mathrm{in},\mathrm{out}\}$ and $b\in[3]$ let $K_{\alpha,b}=K_\alpha\cap K_b$, and define associated decorated codes $D_{\alpha,b}$ by
    \begin{align*}
      D_{\mathrm{in},b} &= (C(q,u,n,k,K_{\mathrm{in},b}),\; \Enc_{C(q,u,n,k,K_{\mathrm{in},b})},\; 2^{[n]^u}|_{\geq\lambda_{\max}}) \\
      D_{\mathrm{out},b} &= (C(q,u,n,k,K_{\mathrm{out},b}),\; \Enc_{C(q,u,n,k,K_{\mathrm{out},b})},\; 2^{[n]^u}|_{\geq\lambda_{\min}})
    \end{align*}
    if $K_{\alpha,b}\neq\emptyset$ and $D_{\alpha,b}=\emptyset$ if $K_{\alpha,b}=\emptyset$.
    Then there exists a mending fault-tolerant gadget
    \begin{equation*}
      \left(\cQ,\; \cE_{\mathrm{run}}=2^{N'}|_{\geq\lambda_{\mathrm{run}}}^{\sqcup T},\; D_{\mathrm{in}}=D_{\mathrm{in},1}\sqcup D_{\mathrm{in},2}\sqcup D_{\mathrm{in},3},\; D_{\mathrm{out}}=D_{\mathrm{out},1}\sqcup D_{\mathrm{out},2}\sqcup D_{\mathrm{out},3}\right)
    \end{equation*}
    for $\bar{\cQ}[*](\cdot)$, where $\cQ$ is a quantum circuit using space $|N'|\leq O(n)^u$, time $T\leq O(\bar{T}\cdot u^4n^2\log^2n)$, and gate set~$\cG$.\footnote{Recall that for a quantum circuit $\bar{\cQ}$, we let $\bar{\cQ}[*]$ denote the set containing the postselected circuits $\bar{\cQ}[\bar{\zeta}]$ for every postselection $\bar{\zeta}$ for $\bar{\cQ}$.}
    Furthermore, $\cQ$ does not depend on the choice of $\lambda_{\mathrm{run}}$.
  \end{enumerate}
\end{theorem}

\begin{remark}
  In \Cref{it:lascheme} in \Cref{thm:laft}, it would be sufficient to consider $D_{\mathrm{in}},D_{\mathrm{out}}$ each given by disjoint unions of $2$ (instead of $3$) codes, as we can sequentially compose gadgets with $|K_{\mathrm{in},1}|+|K_{\mathrm{in},2}|=|K_{\mathrm{out},1}|$ to accumulate the logical qudits across any number $m$ of code states into a single code state. Similarly, we can compose gadgets with $|K_{\mathrm{in},1}|=|K_{\mathrm{out},1}|+|K_{\mathrm{out},2}|$ to distribute the logical qudits from one code state across $m$ code states. We choose to consider disjoint unions of $3$ codes in \Cref{thm:laft} for notational convenience with the $\gCCX$ gate, which acts on $3$ qudits.
\end{remark}

\begin{remark}
  \Cref{it:lascheme} in \Cref{thm:laft} can be generalized to allow $D_{\mathrm{in},b}$ and $D_{\mathrm{out},b}$ to be given by codes $C(q,u,n,k,K_{\mathrm{in},b})$ and $C(q,u',n',k',K_{\mathrm{out},b})$ respectively with different values of $(u,n,k)\neq(u',n',k')$. However, we restrict to the case where $(u,n,k)=(u',n',k')$ for simplicity.
\end{remark}

It will sometimes be helpful to refer to the family of decorated codes constructed from $C(q,u,n,k,K)$ in \Cref{thm:laft}, which we call $\cD(q,u,n,k,\lambda_{\mathrm{run}})$ below. We emphasize that for a fixed set of logical qudits, then the decorated codes we use do not depend on the choice of logical circuit $\bar{\cQ}$.

\begin{definition}
  \label{def:ladec}
  Define $q,u,n,k,\lambda_{\mathrm{run}},\;\lambda_{\min}=\lambda_{\min}(u,n,\lambda_{\mathrm{run}}),\;\lambda_{\max}=\lambda_{\max}(u,n)$ as in \Cref{thm:laft}. We define the family $\cD(q,u,n,k,\lambda_{\mathrm{run}})$ of decorated CSS non-subsystem codes by
  \begin{equation*}
    \cD(q,u,n,k,\lambda_{\mathrm{run}}) = \left\{\left(C(q,u,n,k,K),\; \Enc_{C(q,u,n,k,K)},\; 2^{[n]^u}|_{\geq\lambda}\right):K\subseteq[k]^u,\;\lambda\in\{\lambda_{\min},\lambda_{\max}\}\right\}.
  \end{equation*}
  For $D=(C(q,u,n,k,K),\; \Enc_{C(q,u,n,k,K)},\; 2^{[n]^u}|_{\geq\lambda})\in\cD(q,u,n,k,\lambda_{\mathrm{run}})$, we let $K(D)=K$.
\end{definition}

The remainder of this section is dedicated to proving \Cref{thm:laft}. As in \Cref{sec:coregad}, in this section we assume that all qudits have the same prime power dimension $q$, which may be growing in other parameters such as the block lengths of the relevant codes.

\subsection{Code Family}
\label{sec:ftlacodes}
In this section, we define the codes $C(q,u,n,k,K)$ and associated encoding maps $\Enc_{C(q,u,n,k,K)}$ in \Cref{thm:laft}. 
Specifically, in \Cref{def:ftsuitable} we define families of subsystem product codes that are \emph{suitable for fault-tolerance}, meaning they fulfill a set of criteria that are sufficient for our fault-tolerance protocol. 
We then give an explicit construction of such codes in \Cref{lem:ftsuitable} using polynomial evaluation subsystem product codes as defined in \Cref{def:pevprod}. 
Finally, in \Cref{def:lacodes} we define $C(q,u,n,k,K)$ to be such a suitable code with certain logical and gauge qudits fixed to $\ket{0}$.

\begin{definition}
  \label{def:ftsuitable}
  For a prime power $q$, for $u,n,k\in\bN$, and for $i\in[u]$, let $\cC^i$ be a family of pairs $(C^i,\Enc_{C^i})$ consisting of an $[[n,k]]_q$ CSS non-subsystem code $C^i$ with associated CSS encoding map $\Enc_{C^i}:\bF_q^k\rightarrow C^i_Z/{C^i_X}^\perp$. For $I\subseteq[u]$, let
  \begin{equation*}
    \cC^I = \left\{(C,\Enc_C):C=\bigotimes_{i\in I}C^i\text{ for }(C^i,\Enc_{C^i})\in\cC^i\text{ with product encoding map }\Enc_C\right\}.
  \end{equation*}
  For $d=(d_1,\dots,d_u)\in\bN^u$, we say that the family $\cC^*=\bigsqcup_{I\subseteq[u]}\cC^I$ is \emph{suitable for fault-tolerance with parameters $(q,u,n,k,d=(d_1,\dots,d_u))$} if the following hold:
  \begin{enumerate}
  \item\label{it:ftsdisc} For every $i\in[u]$ and every $C^i\in\cC^i$, the classical codes $C^i_X,C^i_Z$ have distance $\geq d_1$.
  \item\label{it:ftsdisq} For every $I\subseteq[u]$ and every $C\in\cC^I$, the CSS subsystem code $C$ has distance $\geq d_{|I|}$.
  \item\label{it:ftshad} For every $i\in[u]$, there exists $(C^i,\Enc_{C^i})\in\cC^i$ such that $({C^i}^\perp,\Enc_{C^i}^\perp)\in\cC^i$, where ${C^i}^\perp,\Enc_{C^i}^\perp$ are defined as in \Cref{def:dualenc,lem:dualenc}.
  \item\label{it:ftsccx} For every $i\in[u]$, there exist $(C^i_j,\Enc_{C^i_j})\in\cC^i$ for $j\in[3]$ such that for every $x_1,x_2\in\bF_q^k$, then $\Enc_{C^i_1}(x_1)*\Enc_{C^i_2}(x_2)\subseteq\Enc_{C^i_3}(x_1*x_2)$.
  \end{enumerate}
\end{definition}

\begin{remark}
  In \Cref{def:ftsuitable}, for a single code $C$ there may be multiple encoding maps $\Enc_C$ for which $(C,\Enc_C)\in\cC^I$.

  Also note that we will only ever use the distance bounds in \Cref{it:ftsdisc,it:ftsdisq} in \Cref{def:ftsuitable} given by $d_1,d_{u-2},d_{u-1},d_u$.
\end{remark}

\begin{lemma}
  \label{lem:ftsuitable}
  For every $u,n,k\in\bN$ with $n\geq 8$, $k\leq n/8$ and every $q=q_0^{\prod_{i\in[u]}n_i'}$ with $q_0\geq n$ a prime power and each $n_i'\geq n$ an integer, there exists a family $\cC^*$ that is suitable for fault-tolerance with parameters $(q,u,n,k,d=(d_1,\dots,d_u))$ for $d_i=(n/8)^i$.
\end{lemma}

We prove \Cref{lem:ftsuitable} using polynomial evaluation subsystem product codes as defined in \Cref{def:pevprod}. For this purpose, we need the following definition of encoding maps for such codes.

\begin{definition}
  \label{def:pevenc}
  For $I\subseteq[u]$, let $C=\bigotimes_{i\in I}C^i=C^I(q,u,(\ell_i,A_i,E_i,\beta_i)_{i\in[u]})$ be a polynomial evaluation subsystem product code as defined in \Cref{def:pevprod}, with associated sets $A_I=\prod_{i\in I}A_i$, $E_I=\prod_{i\in I}E_i$. For every $\bar{\beta}_i\in(\bF_q^*)^{A_i\cup E_i}$ with $\bar{\beta_i}|_{E_i}=\beta_i$ for $i\in[u]$, letting $\bar{\beta}_I=\bigotimes_{i\in I}\bar{\beta}_i$, we define an \emph{associated CSS encoding map} $\Enc_{C,\bar{\beta}_I}:\bF_q^{A_I}\oplus\bF_q^{K'_C}\rightarrow (C_Z+C_X^\perp)/(C_Z\cap C_X^\perp)$ by\footnote{We allow $\Enc_{C,\bar{\beta}_I}(0,g)$ for $g\in\bF_q^{K'_C}$ to be defined arbitrarily. 
  }
  \begin{align*}
    \Enc_{C,\bar{\beta}_I}(x,0^{K'_C})
    &= \left\{\bar{\beta}_I|_{E_I}*\evl_{E_I}(f):f\in\bF_q[(X_i)_{i\in I}]^{\prod_{i\in I}[0,\ell_i)},\;\bar{\beta}_I|_{A_I}*\evl_{A_I}(f)=x\right\}.
  \end{align*}
  By \Cref{lem:QZcQXp}, $\Enc_{C,\bar{\beta}_I}$ is a product encoding map (see \Cref{def:prodenc}), where for each $C^i$, we define an associated CSS encoding map $\Enc_{C^i,\bar{\beta}_i}:\bF_q^{A_i}\rightarrow C_Z/C_X^\perp$ by
  \begin{align*}
    \Enc_{C^i,\bar{\beta}_i}(x)
    &= \{\bar{\beta}_i|_{E_i}*\evl_{E_i}(f):f\in\bF_q[X]^{[0,\ell_i)},\;\bar{\beta}_i|_{A_i}*\evl_{A_i}(f)=x\}.
  \end{align*}
\end{definition}

\begin{proof}[Proof of \Cref{lem:ftsuitable}]
  Define sets $A=\prod_{i\in[u]}A_i$, $E=\prod_{i\in[u]}E_i$ for $A_i,E_i\subseteq\bF_q$ with $|A_i|=k$, $|E_i|=n$ as given in \Cref{thm:pevmain}. For $I\subseteq[u]$, we define
  \begin{align*}
    \cC^I &= \left\{(C,\Enc_{C,\bar{\beta}_I}):C=C^I(q,u,(\ell_i,A_i,E_i,\bar{\beta}_i|_{E_i})_{i\in[u]})\text{ for }\ell_i\in\bZ\cap[n/4,\;3n/4+k],\;\bar{\beta}_i\in(\bF_q^*)^{A_i\cup E_i}\right\},
  \end{align*}
  where $C^I(\cdot),\Enc_{C,\bar{\beta}_I}$ are as defined in \Cref{def:pevprod,def:pevenc}. \Cref{it:ftsdisc} in \Cref{def:ftsuitable} then follows from the fact that each $C^i_X$ and $C^i_Z$ is by definition a Reed-Solomon code of length $n$ and dimension $\leq 7n/8$. \Cref{it:ftsdisq} in \Cref{def:ftsuitable} follows by \Cref{thm:pevmain}. \Cref{it:ftsccx} in \Cref{def:ftsuitable} follows by setting $\bar{\beta}_i=(1,\dots,1)$ to be the all-$1$s vector, and setting $\ell_{i,1}=\ell_{i,2}=\lceil n/4\rceil$ and $\ell_{i,3}=\lceil n/2\rceil$, so that $C^i_j=(\evl_{E_i}(\bF_q[X]^{[0,\ell_{i,j})}_{A_i})^\perp,\; \evl_{E_i}(\bF_q[X]^{[0,\ell_{i,j})}))$ for $j\in[3]$.

  It remains to prove \Cref{it:ftshad} in \Cref{def:ftsuitable}. For this purpose, fix some $(C^i,\Enc_{C^i,\bar{\beta}_i})\in\cC^i$.
  By \Cref{lem:RSdual} and \Cref{lem:puncshort} (see the proof of \Cref{lem:pevswapXZ}), there exists $\bar{\beta}_i'\in(\bF_q^*)^{A_i\cup E_i}$ such that for some $\ell_i\in\bZ\cap[n/4,\;3n/4+k]$,
  \begin{align}
    \label{eq:ftscodes}
    \begin{split}
      C^i_Z &= \bar{\beta}_i|_{E_i}*\evl_{E_i}(\bF_q[X]^{[0,\ell_i)}) \\
      {C^i_X}^\perp &= \bar{\beta}_i|_{E_i}*\evl_{E_i}(\bF_q[X]^{[0,\ell_i)}_{A_i}) \\
      ({C^i}^\perp)_Z = C^i_X &= \bar{\beta}_i'|_{E_i}*\evl_{E_i}(\bF_q[X]^{[0,n+k-\ell_i)}) \\
      ({C^i}^\perp)_X^\perp = {C^i_Z}^\perp &= \bar{\beta}_i'|_{E_i}*\evl_{E_i}(\bF_q[X]^{[0,n+k-\ell_i)}_{A_i})
    \end{split}
  \end{align}
  and
  \begin{equation}
    \label{eq:ftsdual}
    (\bar{\beta}_i*\evl_{A_i\cup E_i}(\bF_q[X]^{[0,\ell_i)}))^\perp = \bar{\beta}_i'*\evl_{A_i\cup E_i}(\bF_q[X]^{[0,n_i+k_i-\ell_i)}).
  \end{equation}
  Define $\bar{\beta}''\in\bF_q^{A_i\cup E_i}$ by $\bar{\beta}''|_{A_i}=\bar{\beta}'|_{A_i}$ and $\bar{\beta}''|_{E_i}=-\bar{\beta}'|_{E_i}$. Then \Cref{eq:ftscodes} still holds with $\bar{\beta}''$ replacing $\bar{\beta}'$. By definition $({C^i}^\perp,\Enc_{{C^i}^\perp,\bar{\beta}_i''})\in\cC^i$. Furthermore, for every $x,x'\in\bF_q^{A_i}$ and every
  \begin{align*}
    \bar{\beta}_i|_{E_i}*\evl_{E_i}(f) &\in \Enc_{C^i,\bar{\beta}_i}(x) \\
    \bar{\beta}_i''|_{E_i}*\evl_{E_i}(f') &\in \Enc_{{C^i}^\perp,\bar{\beta}_i''}(x),
  \end{align*}
  so that $f\in\bF_q[X]^{[0,\ell_i)}$, $f'\in\bF_q[X]^{[0,n+k-\ell_i)}$ with $x=\bar{\beta}_i|_{A_i}*\evl_{E_i}(f)$, $x'=\bar{\beta}_i''|_{A_i}*\evl_{E_i}(f')$, then
  \begin{align*}
    \hspace{1em}&\hspace{-1em} x\cdot x' - (\bar{\beta}_i|_{E_i}*\evl_{E_i}(f))\cdot(\bar{\beta}_i''|_{E_i}*\evl_{E_i}(f')) \\
                &= (\bar{\beta}_i|_{A_i}*\evl_{A_i}(f))\cdot(\bar{\beta}_i'|_{A_i}*\evl_{A_i}(f')) + (\bar{\beta}_i|_{E_i}*\evl_{E_i}(f))\cdot(\bar{\beta}_i'|_{E_i}*\evl_{E_i}(f')) = 0,
  \end{align*}
  where the final equality above follows by \Cref{eq:ftsdual}. Thus for every $x,x'\in\bF_q^{A_i}$ we have
  \begin{align*}
    \Enc_{C^i,\bar{\beta}_i}(x)\cdot\Enc_{{C^i}^\perp,\bar{\beta}_i''}(x) &= 0,
  \end{align*}
  so $\Enc_{C^i,\bar{\beta}_i}^\perp=\Enc_{{C^i}^\perp,\bar{\beta}_i''}$. Therefore \Cref{it:ftshad} in \Cref{def:ftsuitable} in fact holds for every $(C^i,\Enc_{C^i,\bar{\beta}_i})\in\cC^i$, as we have shown that $({C^i}^\perp,\;\Enc_{C^i,\bar{\beta}_i}^\perp=\Enc_{{C^i}^\perp,\bar{\beta}_i''})\in\cC^i$.
\end{proof}

We are now ready to define the code $C(q,u,n,k,K)$ with encoding map $\Enc_{C(q,u,n,k,K)}$ in \Cref{thm:laft}.

\begin{definition}
  \label{def:lacodes}
  Define $q,u,n,k$ as in \Cref{thm:laft}. Let $\cC^*(q,u,n,k)$ be the code family that is suitable for fault-tolerance with parameters $(q,u,n,k,d=(d_1,\dots,d_u))$ given by \Cref{lem:ftsuitable}.
  We now fix some $(C,\Enc_C)\in\cC^{[u]}(q,u,n,k)$ and for $K\subseteq[k]^u$, define\footnote{While $\Enc_C(\bF_q^K,0)$ is strictly speaking a set of cosets, in a slight abuse of notation here we write $\Enc_C(\bF_q^K,0)$ to denote the union of these cosets, which is a linear subspace of $\bF_q^E$.}
  \begin{align*}
    C(q,u,n,k,K) &= C|_K := (C_X+C_Z^\perp,\;\Enc_C(\bF_q^K,0)) \\
    \Enc_{C(q,u,n,k,K)} &= \Enc_{C|_K} := \Enc_C|_K.
  \end{align*}
\end{definition}

\begin{proof}[Proof of \Cref{it:laparam} in \Cref{thm:laft}]
  \Cref{def:lacodes} implies that the logical qudits of $C(q,u,n,k,K)$ are labeled by $K$.
  Meanwhile, because $\Enc_C(\bF_q^K,0)\subseteq\Enc_C(\bF_q^{[k]^u},0)=C_Z$, the non-subsystem code $C(q,u,n,k,K)=(C_X+C_Z^\perp,\;\Enc_C(\bF_q^K,0))$ has distance at least that of the subsystem code $C$. Thus the distance lower bound of $d_u$ follows by \Cref{thm:pevmain}.
\end{proof}

It only remains to prove \Cref{it:lascheme} in \Cref{thm:laft}, which we do in the sections below.

\subsection{Logical Circuit Compilation}
In this section, we show how to compile an arbitrary logical circuit $\bar{\cQ}$ into one whose structure is amenable to our fault-tolerant gadgets in \Cref{sec:coregad}, with only small blowups in the space and time usage. This compilation is similar to the one used in the classical fault-tolerance scheme of \cite{anshu2026classical}.

We first need the following notation.

\begin{definition}
  For a set $K=\prod_{i\in[u]}[k_i]$, for every $i\in[u]$ and $j\in[k_i]$ we define a subset $K|_{i\rightarrow j}\subseteq K$ by $K|_{i\rightarrow j}=\{\kappa\in K:\kappa_i=j\}$.

  Given an $m$-qudit gate $G$ and disjoint sets of qudits $S_1\cong\cdots\cong S_m$ with fixed isomorphisms, the \emph{transversal application} $G^{S_1,\dots,S_m}$ of $G$ to $S_1,\dots,S_m$ is the $m|S_b|$-qudit gate given by applying $G$ to every $m$-tuple of qudits $s_1\in S_1,\dots,s_m\in S_m$ that are associated under the fixed isomorphisms.

  In particular, the transversal application of $G$ to $K|_{i\rightarrow j_1},\dots,K|_{i\rightarrow j_m}$ applies $G$ to all $m$-tuples of qudits in $K|_{i\rightarrow j_1}\times\cdots\times K|_{i\rightarrow j_m}$ whose labels agree on all coordinates in $[u]\setminus\{i\}$.
\end{definition}

We are now ready to state our logical circuit compilation result. We emphasize that we do not consider any fault-tolerance properties in this section.

\begin{lemma}
  \label{lem:compile}
  For $u,k\in\bN$ with $u\geq 4$, let $\bar{\cQ}=(\bar{Q}_1,\dots,\bar{Q}_{\bar{T}};K_{\mathrm{in}},K_{\mathrm{out}})$ be a quantum circuit acting on $3k^u$ qudits labeled by the set $K_{[3]}=K_1\sqcup K_2\sqcup K_3$ with each $K_b=[k]^u$, using time $\bar{T}$ and gate set $\cG=\{\gH^*,\gX^*,\gZ^*,\gCX^*,\gCCX^*,\gInit_*,\gCO_*\}$. Letting $\bar{b}=2\cdot 4^u$, then there exists a quantum circuit\footnote{We do not require that $\bar{\cQ}'$ initialize all qudits outside of $K_{\mathrm{in}}$ in the first timestep; see \Cref{footnote:initall}.} $\bar{\cQ}'=(\bar{Q}_1',\dots,\bar{Q}_{\bar{T}'}';K_{\mathrm{in}},K_{\mathrm{out}})$ acting on $\bar{b}\cdot k^u$ qudits labeled by $K_{[\bar{b}]}=K_1\sqcup\cdots\sqcup K_{\bar{b}}$ with each $K_b=[k]^u$, using time $\bar{T'}\leq O(\bar{T}\cdot(u\log k)^2)$ and gate set $\cG$, with the same associated spaces of postselected superoperators $\spn_{\bC}\{\bar{\cQ}'[*](\cdot)\}=\spn_{\bC}\{\bar{\cQ}[*](\cdot)\}$ mapping $\bC^{\bF_q^{K_{\mathrm{in}}}\times\bF_q^{K_{\mathrm{in}}}}\rightarrow\bC^{\bF_q^{K_{\mathrm{out}}}\times\bF_q^{K_{\mathrm{out}}}}$; in particular the unpostselected channels $\bar{\cQ}'(\cdot)=\bar{\cQ}(\cdot)$ are equal. Furthermore, every $\bar{Q}_t'$ for $t\in[\bar{T}']$ either consists entirely of $\gX^*$ gates, entirely of $\gZ^*$ gates, or else entirely of calls to a single gate $G\in\cG$ such that:
  \begin{enumerate}
  \item If $G\in\{\gH^*,\gCX^*,\gCCX^*,\gInit_*\}$, then for some $i\in[u]$, all calls to $G$ in $\bar{Q}_t'$ come as transversal applications to sets $K_b|_{i\rightarrow j}$ for arbitrary values of $b\in[\bar{b}]$ and $j\in[k]$.
  \item If $G\in\gCO_*$, then for some subset $B\subseteq[\bar{b}]$, letting $K_B=\bigsqcup_{b\in B}K_b$, we have $G=\gCO_{\alpha,f}$ for some $\alpha\in\{X,Z\}$ and some $f:\bF_q^{K_B}\rightarrow\bF_q^{K_B}$, and $G$ is applied to the qudits in $K_B$, that is, $\bar{Q}_t'=G\otimes I_{K_{[\bar{b}]}\setminus K_B}$.
  \end{enumerate}
\end{lemma}

To prove \Cref{lem:compile}, we will use the following known result on efficiently implementing permutations in circuits with a hypercubic connectivity structure. Below, we define the unitary gate $\gSwap\in\bC^{\bF_q^2\times\bF_q^2}$ in the ordinary way $\gSwap\ket{x_1,x_2}=\ket{x_2,x_1}$. Recall that $\gSwap$ can for instance be implemented with $6$ $\gCX^*$ gates, using an ancilla qudit initialized to $\ket{0}$.\footnote{If $\bF_q$ has characteristic $2$, there is a simpler implementation with $3$ $\gCX^*$ gates and no ancillas.}

\begin{lemma}[\cite{batcher_sorting_1968}]
  \label{lem:route}
  For $s\in\bN$, let $S=[2]^s$, and let $\pi:S\rightarrow S$ be a permutation. Then there exists a circuit $\cQ$ acting on qudits labeled by $S$ using time $T\leq O(s^2)$  
  and gate set $\{\gSwap\}$, such that $\cQ(\cdot)$ implements $\pi$, that is, $\cQ(\ket{x}\bra{x'})=\ket{\pi(x)}\bra{\pi(x')}$. Furthermore, for every time step $t\in[T]$, there exists some $i(t)\in[s]$ such that every $\gSwap$ gate in $Q_t$ acts on a pair of qudits whose labels differ only in coordinate $i(t)$.
\end{lemma}

\Cref{lem:route} allows us to implement arbitrary logical permutations via transversal controlled-swap gates, which in turn can be implemented using transversal $\gCCX$ gates.
Therefore to prove \Cref{lem:compile}, we will use \Cref{lem:route} to move qudits into appropriate positions, such that the desired gates can then be implemented transversally. We provide the details below; we will also use the following basic fact.

\begin{fact}
  \label{fact:CCXnocoeff}
  For every $a\in\bF_q$, the gate $\gCX^a$ can be implemented using one ancilla qudit initialized to $\ket{0}$ along with one $\gCCX^{+1}$ gate and two $\gX^*$ gates.

  Furthermore, the gate $\gCCX^a$ can be implemented using two ancilla qudits initialized to $\ket{0}$ along with two $\gCCX^{+1}$ gates, one $\gCCX^{-1}$ gate, and two $\gX^*$ gates.
\end{fact}
\begin{proof}
  We implement $\gCX^a$ by performing
  \begin{align*}
    \ket{x_1,x_2,0}
    &\xrightarrow{\gX^a} \ket{x_1,x_2,a} \xrightarrow{\gCCX^{+1}} \ket{x_1,x_2+ax_1,a} \xrightarrow{\gX^{-a}} \ket{x_1,x_2,0}.
  \end{align*}
  We implement $\gCCX^a$ by performing
  \begin{align*}
    \ket{x_1,x_2,x_3,0,0}
    &\xrightarrow{\gX^a} \ket{x_1,x_2,x_3,0,a} \xrightarrow{\gCCX^{+1}} \ket{x_1,x_2,x_3,x_1x_2,a} \xrightarrow{\gCCX^{+1}} \ket{x_1,x_2,x_3+ax_1x_2,x_1x_2,a} \\
    &\xrightarrow{\gCCX^{-1}} \ket{x_1,x_2,x_3+ax_1x_2,0,a} \xrightarrow{\gX^{-a}} \ket{x_1,x_2,x_3+ax_1x_2,0,0}.
  \end{align*}
\end{proof}

\begin{proof}[Proof of \Cref{lem:compile}]
  For $B\subseteq[\bar{b}]$, we let $K_B=\bigsqcup_{b\in B}K_b$. Also let $\mathrm{data}=\{1,\dots,4^u\}$ and $\mathrm{anc}=\{4^u+1,\dots,2\cdot 4^u\}$. We assume the blocks $K_1,\dots,K_{4^u}$ are arranged in a $u$-dimensional hypercube of side length $4$, so that we have a fixed isomorphism $[4k]^u\cong([4]\times[k])^u\cong K_{\mathrm{data}}$. Under this isomorphism, we assume that $K_{[16]}\subseteq[2k]^u$ (which is possible as $u\geq 4$). Let $s=\lceil\log_2 k\rceil+1$, and fix an isomorphism $[2^s]\cong[2]^s$, so that $[2^s]^u\cong[2]^{su}$. Note that $2k\leq 2^s\leq 4k$.

  We begin by applying \Cref{lem:route} to show \Cref{claim:comperm} below.
  In the remainder of this proof below, we say a circuit acting on dits $K_{[\bar{b}]}=K_{\mathrm{data}}\sqcup K_{\mathrm{anc}}$ (or on a subset thereof) has \emph{transversal structure} if each timestep of the circuit has the structure required of $\bar{Q}_t'$ in the statement of \Cref{lem:compile}.

  \begin{claim}
    \label{claim:comperm}
    For every permutation $\pi:[2^s]^u\times[2^s]^u$, there exists a circuit using gate set $\{\gInit_Z,\gX^*,\gCCX^*\}\subseteq\cG$ with transversal structure that permutes the qudits in $[2^s]^u$ by $\pi$ in time $O(su)^2$. Qudits in $K_{[\bar{b}]}\setminus[2^s]^u$ may be acted upon arbitrarily.
  \end{claim}
  \begin{proof}
    \Cref{lem:route} gives a circuit on qudits $[2^s]^u\cong[2]^{su}$ that implements the permutation $\pi$ using time $O(su)^2$ and gate set $\{\gSwap\}$, where each timestep $t$ applies a subset of the $\gSwap$ gates performed in transversal applications to sets $K_b|_{i\rightarrow j}$ for some $i=i(t)\in[u]$. That is, for various disjoint pairs of sets $K_{b_1}|_{i\rightarrow j_1},K_{b_2}|_{i\rightarrow j_2}\subseteq K_{\mathrm{data}}$, this circuit from \Cref{lem:route} applies $\gSwap$ to some (but maybe not all) pairs of qudits $\kappa_1\in K_{b_1}|_{i\rightarrow j_1},\; \kappa_2\in K_{b_2}|_{i\rightarrow j_2}$ whose labels in $[k]^u$ agree on coordinates $[u]\setminus\{i\}$. For each such $(b_1,j_1),(b_2,j_2)$, we may instead initialize a block $K_{b_3}|_{i\rightarrow j_3}\subseteq K_{\mathrm{anc}}$ to contain the indicator vector for the $\gSwap$ gates we want to perform on $K_{b_1}|_{i\rightarrow j_1},K_{b_2}|_{i\rightarrow j_2}$. This initialization can be done with one layer of $\gInit_Z$ gates followed by one layer of $\gX^*$ gates. Then the desired $\gSwap$ gates can be implemented by performing transversal controlled-$\gSwap$ across the three blocks $K_{b_3}|_{i\rightarrow j_3},K_{b_1}|_{i\rightarrow j_1},K_{b_2}|_{i\rightarrow j_2}$. Just as $\gSwap$ can be implemented using an ancilla qudit initialized to $\ket{0}$ along with $6$ $\gCX^*$ gates, controlled-$\gSwap$ can be implemented using an ancilla qudit initialized to $\ket{0}$ along with $6$ $\gCCX^*$ gates. Therefore we initialize an additional block $K_{b_4}|_{i\rightarrow j_4}\subseteq K_{\mathrm{anc}}$ for these additional ancilla qudits, and then we implement one layer of the desired $\gSwap$ gates using gates $\{\gInit_Z,\gX^*,\gCCX^*\}\subseteq\cG$ in a circuit with transversal structure. Furthermore, by construction this circuit acts on dits $K_{[\bar{b}]}$ (as the number of ancilla blocks in $K_{\mathrm{anc}}$ we use equals the number of data blocks in $K_{\mathrm{data}}$) and runs in time $O(1)$. It follows that the entire permutation circuit, which consists of $O(su)^2$ layers of $\gSwap$ gates, can be implemented in a circuit with transversal structure using gate set $\{\gInit_Z,\gX^*,\gCCX^*\}\subseteq\cG$ in time $O(su)^2$, as desired.
  \end{proof}



  We thus compile $\bar{\cQ}$ into $\bar{\cQ}'$ as follows. First, increasing the running time by at most a factor of~$5$, we can assume that every timestep in $\bar{\cQ}$ consists entirely of gates from one of the sets $\gX^*$, $\gZ^*$, $\gH^*$, $\gCX^*$, $\gCCX^*$, $\gInit_X$, $\gInit_Z$, $\gCO_{X,*}$, or $\gCO_{Z,*}$. We now compile each such timestep into a sub-circuit of $\bar{\cQ}'$ of the desired transversal structure, using time $O(su)^2$.
  Specifically, we consider timesteps of $\bar{\cQ}$ with each type of gates separately, where below $\alpha\in\{X,Z\}$:
  \begin{enumerate}
  \item $\gX^*$ (resp.~$\gZ^*$): Simply apply the desired $\gX^*$ (resp.~$\gZ^*$) gates to qudits in $K_{[3]}$.
  \item $\gH^*$ (resp.~$\gInit_\alpha$) on qudits $A\subseteq K_{1,2,3}$: Use \Cref{claim:comperm} to swap some $|A|$ qudits in $K_{4,5,6}\subseteq K_{\mathrm{data}}$ with the qudits $A\subseteq K_{1,2,3}$, while preserving the positions of the other qudits in $K_{[6]}$. Then run $\gH^*$ (resp.~$\gInit_\alpha$) on $K_{4,5,6}$, and subsequently apply \Cref{claim:comperm} again to permute the qudits in $K_{[6]}$ back to their original positions.
  \item $\gCX^*$ on various pairs of qudits in $K_{1,2,3}$: First, apply \Cref{claim:comperm} to permute the qudits in $K_{[6]}$ such that every $\gCX^*$ gate now has control qudit in $K_{1,2,3}$, and target qudit in the respective position in $K_{4,5,6}$. For qudits acted on the by identity, we simply assign their associated gate to be $\gCX^0=I_2$. We then implement the desired $\gCX^*$ gates using transversal gates in $\cG$ on $K_1,\dots,K_9$. Specifically, we implement each gate $\gCX^a$ using the procedure in \Cref{fact:CCXnocoeff}, where $K_{7,8,9}$ provides the ancilla qudits, so that each timestep either performs $\gX^*$ gates, or else performs transversal $\gCCX^{+1}$ or $\gInit_Z$ gates across different blocks $K_b$. Finally, we apply \Cref{claim:comperm} again to permute the qudits in $K_{[6]}$ back to their original positions.
  \item $\gCCX^*$ on various triples of qudits in $K_{1,2,3}$: First, apply \Cref{claim:comperm} to permute the qudits in $K_{[9]}$ such that every $\gCX^*$ gate now has first control qudit in $K_{1,2,3}$, second control qudit in the respective position in $K_{4,5,6}$, and target qudit in the respective position in $K_{7,8,9}$. For qudits acted on the by identity, we simply assign their associated gate to be $\gCCX^0=I_3$. We then implement the desired $\gCCX^*$ gates using transversal gates in $\cG$ on $K_1,\dots,K_{15}$. Specifically, we implement each gate $\gCCX^a$ using the procedure in \Cref{fact:CCXnocoeff}, where $K_{10,\dots,16}$ provides the ancilla qudits, so that each timestep either performs $\gX^*$ gates, or else performs transversal $\gCCX^{+1}$, $\gCCX^{-1}$, or $\gInit$ gates across different blocks $K_b$. Finally, we apply \Cref{claim:comperm} again to permute the qudits in $K_{[9]}$ back to their original positions.
  \item $\gCO_{\alpha,*}$ on qudits $A\subseteq K_{1,2,3}$: First, run $\gInit_\alpha$ to every qudit in $K_{4,5,6}$. Then we use \Cref{claim:comperm} to swap some $|A|$ qudits in $K_{4,5,6}$ with the qudits $A\subseteq K_{1,2,3}$, while preserving the positions of the other qudits in $K_{[6]}$. By definition, every tensor product of gates in $\gCO_{\alpha,*}$ is equivalent to a single gate in $\gCO_{\alpha,*}$ acting on the union $A$ of the qudits involved. Thus we can perform in $\bar{\cQ}'$ a single such gate in $\gCO_{\alpha,*}$ on $K_{4,5,6}$, in order to induce the same action as the associated $\gCO_{\alpha,*}$ gates in $\bar{\cQ}$. Finally, we apply \Cref{claim:comperm} again to permute the qudits in $K_{[6]}$ back to their original positions.
    Let $\zeta$ be a postselection on our $\gCO_{\alpha,*}$ gate in $\bar{\cQ}'$ that decomposes into a tensor product $\zeta=\bigotimes_{j\in K_{4,5,6}}\zeta_j$ of postselections $\zeta_j$ on individual qudits. Because we initialized qudits in $K_{4,5,6}$ to $\alpha$-basis states as described above, the effect of the postselection $\zeta$ on our $\gCO_{\alpha,*}$ gate in $\bar{\cQ}'$ is equivalent to that of the restriction of $\zeta$ to qudits $A$ on the respective $\gCO_{\alpha,*}$ gates in $\bar{\cQ}$, where this restriction implicitly applies to the permuted locations of the qudits in $A$.
  \end{enumerate}

  By construction, in all cases above, the resulting circuit acting on qudits $K_{[\bar{b}]}$ has the desired transversal structure, and by \Cref{claim:comperm} uses time $O(su)^2$. For each timestep of the original circuit $\bar{\cQ}$, the above construction for $\bar{\cQ}'$ by definition induces the same superoperator, even when including postselections that decompose into postselections on tensor products of individual qudits. Therefore $\bar{\cQ}'(\cdot)=\bar{\cQ}(\cdot)$, and because such pure tensor postselections span the entire vector space of all postselections, we also have $\spn_{\bC}\{\bar{\cQ}'[*](\cdot)\}=\spn_{\bC}\{\bar{\cQ}[*](\cdot)\}$, as desired.
\end{proof}

\subsection{Fault-Tolerant Circuits}
In this section, we apply our gadgets in \Cref{sec:coregad} to fault-tolerantly implement circuits given by the compiler in \Cref{lem:compile}, thereby completing the proof of \Cref{thm:laft}. Throughout this section, we define $q,u,n,k,\lambda_{\mathrm{run}},\bar{\lambda}_{\mathrm{run}},\lambda_{\min},\lambda_{\max},d=(d_1,\dots,d_u)$ as in \Cref{thm:laft}, we define $\cC^*(q,u,n,k),C(q,u,n,k,K)$ as in \Cref{def:lacodes}, we define $\cD(q,u,n,k,\lambda_{\mathrm{run}})$ as in \Cref{def:ladec}, and we define $\bar{b}=2\cdot 4^u$ as in \Cref{lem:compile}.

We begin by defining a family $\cD'(q,u,n,k,\lambda_{\mathrm{run}})$ of subsystem codes in \Cref{def:lacodes2} below, which we will use to perform the fault-tolerant computations. These codes are closely related to the codes $C(q,u,n,k,K)$ in \Cref{def:lacodes}; indeed, \Cref{lem:switchDDp} below shows how to fault-tolerantly switch between the two families of codes.

\begin{definition}
  \label{def:lacodes2}
  For $I\subseteq[u]$, we define the family
  \begin{align*}
    \cB^I(q,u,n,k) &= \left\{\bigsqcup_{j\in[\bar{b}]\times[k]^{[u]\setminus I}}(C_j,\Enc_{C_j}) : (C_j,\Enc_{C_j})\in\cC^I(q,u,n,k)\;\forall j\in[\bar{b}]\times[k]^{[u]\setminus I}\right\},
  \end{align*}
  so that each $(C,\Enc)\in\cB^I(q,u,n,k)$ has $\bar{b}\cdot k^{u-|I|}n^{|I|}$ physical qudits labeled by the set $N_C=[\bar{b}]\times[k]^{[u]\setminus I}\times[n]^I$, and $\bar{b}\cdot k^u$ logical qudits labeled by the set $[\bar{b}]\times[k]^u$.
  
  For $i\in[u]\setminus I$, we also define $\cB^{I;i}(q,u,n,k)\subseteq\cB^I(q,u,n,k)$ to contain those codes $\bigsqcup_{j\in[\bar{b}]\times[k]^{[u]\setminus I}}(C_j,\Enc_{C_j})\in\cB^I(q,u,n,k)$ such that for every $j'\in[\bar{b}]\times[k]^{[u]\setminus(I\cup\{i\})}$, the $k$ codes $(C_j,\Enc_{C_j})$ indexed by $j=(j',j'')$ for $j''\in[k]^{\{i\}}\cong[k]$ are equal.
  
  Then we let
  \begin{align*}
    \cB'(q,u,n,k) &= \bigsqcup_{I\subseteq[u]:|I|\geq u-1}\cB^I(q,u,n,k) \sqcup \bigsqcup_{I\subseteq[u]:|I|=u-2,\; i\in[u]\setminus I}\cB^{I;i}(q,u,n,k)
  \end{align*}
  and
  \begin{align*}
    \cD'(q,u,n,k,\lambda_{\mathrm{run}}) &= \left\{\left(C,\; \Enc,\; \left\{\ket{0^{K'_C}}\bra{0^{K'_C}}\right\},\; 2^{N_C}|_{\geq\lambda}\right) : (C,\Enc)\in\cB'(q,u,n,k),\; \lambda\in[\lambda_{\min},\lambda_{\max}]\right\}.
  \end{align*}
\end{definition}

\Cref{lem:ftall} below shows how to perform fault-tolerant logical gates on the codes in \Cref{def:lacodes2}.

\begin{lemma}
  \label{lem:ftall}
  Fix some $D_{\mathrm{in}}=(C_{\mathrm{in}},\Enc_{\mathrm{in}},\Gamma_{\mathrm{in}},\cE_{\mathrm{in}})\in\cD'(q,u,n,k,\lambda_{\mathrm{run}})$, so that for some $I\subseteq\{u\}$ with $|I|\geq u-2$, we have $C_{\mathrm{in}},\Enc_{\mathrm{in}}\in\cB^I(q,u,n,k)$ and hence
  \begin{align*}
    (C_{\mathrm{in}},\Enc_{\mathrm{in}})
    &= \bigsqcup_{j\in[\bar{b}]\times[k]^{[u]\setminus I}}(C_j,\Enc_{C_j})
  \end{align*}
  for some $(C_j,\Enc_{C_j})\in\cC^I(q,u,n,k)$.
  Let $\cQ$ be a quantum circuit that performs the following:
  \begin{enumerate}
  \item\label{it:ftallec1} Run the error-correction gadget in \Cref{lem:errcorr} on $(C_j,\Enc_{C_j})\in\cC^{I}(q,u,n,k)$ for every $j\in[\bar{b}]\times[k]^{[u]\setminus I}$.
  \item\label{it:ftallgad} Perform one of the following:
    \begin{enumerate}
    \item\label{it:ftainit} For every $j\in[\bar{b}]\times[k]^{[u]\setminus I}$, apply to $(C_j,\Enc_{C_j})$ either the $X$- or $Z$-basis initialization gadget in \Cref{it:spreplace} of \Cref{lem:stateprep}, or idle (i.e.~apply identity gates for the number of timesteps used in the initialization gadget). 
    \item\label{it:ftaswitchdown} If $|I|\geq u-1$: for every $j\in[\bar{b}]\times[k]^{[u]\setminus I}$, apply the downwards code-switching gadget in \Cref{it:switchdown} in \Cref{lem:codeswitch} to $(C_j,\Enc_{C_j})$.
    \item\label{it:ftaswitchup} If $(C_{\mathrm{in}},\Enc_{\mathrm{in}})\in\cB^{I;i}(q,u,n,k)$ for some $i\in[u]\setminus I$: for every $j'\in[\bar{b}]\times[k]^{[u]\setminus(I\cup\{i\})}$, apply the upwards code-switching gadget in \Cref{it:switchup} in \Cref{lem:codeswitch}, taking as input the $k$ codes $(C_j,\Enc_{C_j})$ for $j=(j',j'')$ with $j''$ ranging over all values in $[k]^{\{i\}}\cong[k]$. The execution of the gadget at different indices $j'$ may switch to different codes $(C^i,\Enc_{C^i})\in\cC^i(q,u,n,k)$ in the (new) $i$th direction.
    \item\label{it:ftahad} For every $j\in[\bar{b}]\times[k]^{[u]\setminus I}$ for which $(C_j,\Enc_{C_j})=\bigotimes_{i\in I}(C_j^i,\Enc_{C_j^i})$ has $({C_j^i}^\perp,\Enc_{C_j^i}^\perp)\in\cC^i(q,u,n,k)$ for $i\in I$, and hence $(C_j^\perp,\Enc_{C_j}^\perp)\in\cC^I(q,u,n,k)$ by \Cref{lem:dualenc}, apply either the $\gH$ or $\gH^\dagger$ gadget in \Cref{lem:hadamard} (see \Cref{remark:hadamardinv}), or idle (i.e.~apply identity gates).
    \item\label{it:ftapauli} For every $j\in[\bar{b}]\times[k]^{[u]\setminus I}$ apply to $(C_j,\Enc_{C_j})$ the gadget in \Cref{lem:paulis} that induces the logical Pauli $\gX^{a_j}$ or $\gZ^{a_j}$ for some $a_j\in\bF_q^{[k]^I}$.
    \item\label{it:ftacx} For some set of disjoint pairs $(j_1,j_2)\in([\bar{b}]\times[k]^{[u]\setminus I})^2$ (meaning that each $j\in[\bar{b}]\times[k]^{[u]\setminus I}$ appears in at most one pair) satisfying $(C_{j_1},\Enc_{C_{j_1}})=(C_{j_2},\Enc_{C_{j_2}})$, and for some $a_{j_1,j_2}\in\bF_q$, apply the transversal $\gCX^{a_{j_1,j_2}}$ gadget in \Cref{lem:cx} to $(C_{j_1},\Enc_{C_{j_1}})$, $(C_{j_2},\Enc_{C_{j_2}})$.
    \item\label{it:ftaccx} For some set of disjoint triples $(j_1,j_2,j_3)\in([\bar{b}]\times[k]^{[u]\setminus I})^3$ such that the triple $(C_{j_1},\Enc_{C_{j_1}})$, $(C_{j_2},\Enc_{C_{j_2}})$, $(C_{j_3},\Enc_{C_{j_3}})$ satisfies the condition in \Cref{eq:multprop} (for which it is sufficient that the triples of factor codes in the subsystem products satisfy \Cref{eq:multprop} by \Cref{lem:hadfactor}), and some $a_{j_1,j_2,j_3}\in\bF_q$, apply the transversal $\gCCX^{a_{j_1,j_2,j_3}}$ gadget in \Cref{lem:ccx} to $(C_{j_1},\Enc_{C_{j_1}}),(C_{j_2},\Enc_{C_{j_2}}),(C_{j_3},\Enc_{C_{j_3}})$.
    \item\label{it:ftaco} For disjoint subsets $J_X,J_Z\subseteq[\bar{b}]\times[k]^{[u]\setminus I}$ and functions $f_\alpha:\bF_q^{J_\alpha\times[k]^I}\rightarrow\bF_q^{J_\alpha\times[k]^I}$ for $\alpha\in\{X,Z\}$, apply the gadget in \Cref{lem:co} to $\bigsqcup_{j\in j_\alpha}(C_j,\Enc_{C_j})$ to induce the logical gate $\gCO_{\alpha,f_\alpha}$.
    \end{enumerate}
    Define $(C_{\mathrm{out}},\Enc_{\mathrm{out}})\in\cB^{I'}(q,u,n,k)$ to be the code resulting from applying one of the above choices, so that
    \begin{align*}
      (C_{\mathrm{out}},\Enc_{\mathrm{out}})
      &= \bigsqcup_{j\in[\bar{b}]\times[k]^{[u]\setminus I'}}(C_j,\Enc_{C_j}).
    \end{align*}
  \item\label{it:ftallec2} Run the error-correction gadget in \Cref{lem:errcorr} on $(C_j,\Enc_{C_j})\in\cC^{I'}(q,u,n,k)$ for every $j\in[\bar{b}]\times[k]^{[u]\setminus I'}$.
  \end{enumerate}
  Let $N'$ be the set of qudits acted on by $\cQ$. Then for every $\lambda_{\mathrm{out}}\in[\lambda_{\min},\lambda_{\max}]$, letting $D_{\mathrm{out}}=(C_{\mathrm{out}},\Enc_{\mathrm{out}},\Gamma_{\mathrm{out}},\cE_{\mathrm{out}})$ for $\Gamma_{\mathrm{out}}=\ket{0^{K_C'}}\bra{0^{K_C'}}$ and $\cE_{\mathrm{out}}=2^{N_{C_{\mathrm{out}}}}|_{\geq\lambda_{\mathrm{out}}}$, then $D_{\mathrm{out}}\in\cD'(q,u,n,k,\lambda_{\mathrm{run}})$. Furthermore, $(\cQ,2^{N'}|_{\geq\lambda_{\mathrm{run}}}^{\sqcup T},D_{\mathrm{in}},D_{\mathrm{out}})$ is a mending fault-tolerant gadget for a set $\bar{\cO}$ of superoperators $\bar{O}:\bC^{\bF_q^{[\bar{b}]\times[k]^u}\times\bF_q^{[\bar{b}]\times[k]^u}}\rightarrow\bC^{\bF_q^{[\bar{b}]\times[k]^u}\times\bF_q^{[\bar{b}]\times[k]^u}}$ determined by the choice of gadget in \Cref{it:ftallgad} above. In particular, $\bar{\cO}$ consists of a single channel $\bar{O}$ for all choices except \Cref{it:ftaco}, in which case $\bar{\cO}$ consists of a set of postselected $\gCO_*$ gates. This gadget uses $|N'|\leq 2(u+2)(4n)^u$, time $T\leq 30un^2$, and gate set $\cG\subseteq\{\gH^*,\gX^*,\gZ^*,\gCX^*,\gCCX^*,\gInit_*,\gCO_*\}$, where the precise subset of gates used is determined by the choice in \Cref{it:ftallgad} above.
\end{lemma}

\begin{remark}
  \label{remark:ftacirind}
  By \Cref{remark:lacirind}, the circuit $\cQ$ in \Cref{lem:ftall} does not depend on the choice of $\lambda_{\mathrm{run}}$ or on the families of bad sets $\cE_{\mathrm{in}},\cE_{\mathrm{out}}$.
\end{remark}

\begin{proof}[Proof of \Cref{lem:ftall}]
  The gadgets in \Cref{lem:errcorr,lem:stateprep,lem:codeswitch,lem:hadamard,lem:paulis,lem:cx,lem:ccx,lem:co} each use space $\leq\bar{b}\cdot(u+2)n^u$ and time $\leq 10un^2$ when applied to the $\leq\bar{b}\cdot k^{u-|I|}$ codes $(C_j,\Enc_{C_j})$ in parallel, so the overall circuit $\cQ$ uses space $|N'|\leq\bar{b}\cdot(u+2)n^u=2(u+2)(4n)^u$ and time $T\leq 30un^2$.

  Recall the definitions of $\bar{\lambda}_{\mathrm{run}}$ in \Cref{eq:lamrun} and $\lambda_{\min}=\lambda_{\min}(u,n,\lambda_{run})$, $\lambda_{\max}=\lambda_{\max}(u,n)$ in \Cref{eq:lamminmax}, for some $\lambda_{\mathrm{run}}\in[0,\bar{\lambda}_{\mathrm{run}}]$. By \Cref{lem:errcorr} and \Cref{lem:parcomp}, the error-correction gadgets in \Cref{it:ftallec1} form a mending fault-tolerant gadget for the identity channel with input code $D_{\mathrm{in}}$ and output code
  \begin{equation*}
    D_1=(C_{\mathrm{in}},\Enc_{\mathrm{in}},\Gamma_{\mathrm{in}},2^{N'}|_{\geq\lambda_{\mathrm{\min}}}),
  \end{equation*}
assuming the bad sets for the fault at every time step are given by $2^{N'}|_{\geq\lambda_{\mathrm{run}}}$. Specifically, this claim holds because $\cE_{\mathrm{in}}=2^{N_{C_{\mathrm{in}}}}|_{\geq\lambda_{\mathrm{in}}}$ for some $\lambda_{\mathrm{in}}\leq\lambda_{\max}$, so that \Cref{eq:ecparams} holds by the definition of $\lambda_{\max}$ and $\lambda_{\mathrm{run}}\leq\bar{\lambda}_{\mathrm{run}}$, as $C_{\mathrm{in}}$ has distance $\geq d_{u-2}$. Therefore the ($\leq\bar{b}n^2$-fold) parallel composition of the error-correction gadgets in \Cref{it:ftallec1} have output error weight at most $\bar{b}n^2\cdot\lambda_{\mathrm{out}}$ for $\lambda_{\mathrm{out}}$ given by \Cref{eq:eclamout}, and by definition $\bar{b}n^2\cdot\lambda_{\mathrm{out}}\leq\lambda_{\min}$.

  Now by \Cref{lem:stateprep,lem:codeswitch,lem:hadamard,lem:paulis,lem:cx,lem:ccx,lem:co} and \Cref{lem:parcomp}, the gadgets in \Cref{it:ftallgad} then form a fault-tolerant gadget with input code $D_1$ and output code
  \begin{equation*}
    D_2 = (C_{\mathrm{out}},\Enc_{\mathrm{out}},\bC^{\bF_q^{K'_{C_{\mathrm{out}}}}\times\bF_q^{K'_{C_{\mathrm{out}}}}},2^{N_{C_{\mathrm{out}}}}|_{\geq\lambda_{\max}}),
  \end{equation*}
  again assuming the bad sets for the fault are given by $2^{N'}|_{\geq\lambda_{\mathrm{run}}}$; the logical action $\bar{\cO}$ of this gadget is determined by the choice in \Cref{it:ftallgad}. Specifically, this claim holds because when any of the gadgets in \Cref{lem:stateprep,lem:codeswitch,lem:hadamard,lem:paulis,lem:cx,lem:ccx,lem:co} is applied to some subset of the $[\bar{b}]\times[k]^{[u]\setminus I}$ codes $(C_j,\Enc_{C_j})$, when the input error weight is $\leq\lambda_{\mathrm{in}}\leq\lambda_{\min}$ and the fault weight per timestep is $\leq\lambda_{\mathrm{run}}\leq\bar{\lambda}_{\mathrm{run}}$, then the output error is at most
  \begin{align*}
    \bar{b}n^2\cdot(n\cdot\lambda_{\mathrm{in}}+50u^2n^5\mu(u,n,d_1)\cdot\lambda_{\mathrm{run}}) &\leq \lambda_{\max}.
  \end{align*}
  Note that the LHS above is simply $\bar{b}n^2$ (i.e.~the maximum number of gadgets composed in parallel) times the maximum output error weight among the gadgets in \Cref{lem:stateprep,lem:codeswitch,lem:hadamard,lem:paulis,lem:cx,lem:ccx,lem:co}, and the inequality above then holds by the definition of $\lambda_{\min}$ and $\bar{\lambda}_{\mathrm{run}}$. These lemmas also imply that following the application of the gadgets in \Cref{it:ftallgad}, the resulting code $(C_{\mathrm{out}},\Enc_{\mathrm{out}})\in\cB'(q,u,n,k)$.

  Finally, by the same reasoning as used above for \Cref{it:ftallec1}, by \Cref{lem:errcorr} and \Cref{lem:parcomp} the error-correction gadgets in \Cref{it:ftallec2} form a fault-tolerant gadget for the identity channel with input code $D_2$ and output code $D_{\mathrm{out}}$, again assuming the bad sets for the fault are given by $2^{N'}|_{\geq\lambda_{\mathrm{run}}}$. Also by definition $D_{\mathrm{out}}\in\cD'(q,u,n,k,\lambda_{\mathrm{run}})$.

  Thus by \Cref{lem:seqcomp,lem:parcomp}, $(\cQ,2^{N'}|_{\geq\lambda_{\mathrm{run}}}^{\sqcup T},D_{\mathrm{in}},D_{\mathrm{out}})$ is a mending fault-tolerant gadget for a set of superoperators $\bar{\cO}$ given by the choice in \Cref{it:ftallgad}, as desired.
\end{proof}

\Cref{lem:switchdec} below applies \Cref{lem:ftall} to construct a gadget for switching between any two codes in $\cD'(q,u,n,k,\lambda_{\mathrm{run}})$.

\begin{lemma}
  \label{lem:switchdec}
  For every $D_{\mathrm{in}},D_{\mathrm{out}}\in\cD'(q,u,n,k,\lambda_{\mathrm{run}})$, there exists a quantum circuit $\cQ$ using space $|N'|\leq O(n)^u$, time $T\leq O(u^2n^2)$ and gate set $\{\gCX^*,\gInit_*,\gCO_*\}$ such that $(\cQ,\;\cE_{\mathrm{run}}=2^{N'}|_{\geq\lambda_{\mathrm{run}}}^{\sqcup T},\; D_{\mathrm{in}},\; D_{\mathrm{out}})$ forms a mending fault-tolerant gadget for the $\bar{b}\cdot k^u$-qudit identity channel $\bar{O}=I_{[\bar{b}]\times[k]^u}$. Furthermore, $\cQ$ does not depend on $\lambda_{\mathrm{run}}$ or on the families of bad sets for $D_{\mathrm{in}},D_{\mathrm{out}}$.
\end{lemma}
\begin{proof}
  The desired circuit $\cQ$ will simply be a serial composition of gadgets in \Cref{lem:ftall} using the choice in \Cref{it:ftaswitchdown} or \Cref{it:ftaswitchup}. It will therefore follow by \Cref{remark:ftacirind} that $\cQ$ does not depend on the choice of $\lambda_{\mathrm{run}}$ or on the families of bad sets.
  
  Specifically, for $\alpha\in\{\mathrm{in},\mathrm{out}\}$, let $D_\alpha=(C_\alpha,\Enc_\alpha,\Gamma_\alpha,\cE_\alpha)$, so that $(C_\alpha,\Enc_\alpha)\in\cB^{I_\alpha}(q,u,n,k)$ for some $I_\alpha\subseteq[u]$ of size $|I_\alpha|\geq u-2$, with $(C_\alpha,\Enc_\alpha)\in\cB^{I_\alpha;i_\alpha}(q,u,n,k)$ for some $i_\alpha\in[u]\setminus I$ if $|I_\alpha|=u-2$. If $|I_{\mathrm{in}}|=u$ then $\cQ$ first applies the gadget in \Cref{lem:ftall} using \Cref{it:ftaswitchdown}, while if $|I_{\mathrm{in}}|=u-2$ then $\cQ$ first applies the gadget in \Cref{lem:ftall} using \Cref{it:ftaswitchup}; in either case, $\cQ$ has switched from $D_{\mathrm{in}}$ to some code $D_0=(C_0,\Enc_0,\Gamma_0,\cE_0)\in\cD'(q,u,n,k,\lambda_{\mathrm{run}})$ with $(C_0,\Enc_0)\in\cB^I(q,u,n,k)$ for some $I\subseteq[u]$ of size $|I|=u-1$.

  Now we can switch from $D_0$ to any other code $D_1=(C_1,\Enc_1,\Gamma_1,\cE_1)\in\cD'(q,u,n,k,\lambda_{\mathrm{run}})$ with $(C_1,\Enc_1)\in\cB^I(q,u,n,k)$. To do so, we loop through $i\in I$, and for every $i$ we apply \Cref{lem:ftall} twice, using \Cref{it:ftaswitchdown} followed by \Cref{it:ftaswitchup}, with both switches applied to the $i$th factor of the product codes. Specifically, for $\alpha\in\bN\cup\{\mathrm{in},\mathrm{out}\}$ writing
  \begin{align*}
    (C_\alpha,\Enc_\alpha) &= \bigsqcup_{j\in[\bar{b}]\times[k]^{[u]\setminus I}}(C_{\alpha,j},\Enc_{\alpha,j})
  \end{align*}
  with each
  \begin{align*}
    (C_{\alpha,j},\Enc_{\alpha,j}) &= \bigotimes_{i'\in I}(C_{\alpha,j}^{i'},\Enc_{\alpha,j}^{i'}),
  \end{align*}
  then for each $j\in[\bar{b}]\times[k]^{[u]\setminus I}$, the downwards switch from \Cref{it:ftaswitchdown} unencodes the $i$th factor out of $(C_{0,j}^i,\Enc_{0,j}^i)$, and we choose the upwards switch from \Cref{it:ftaswitchup} to encode the $i$th factor into $(C_{1,j}^i,\Enc_{1,j}^i)$.

  Thus if $I_{\mathrm{out}}=I$, we can choose $D_1=D_{\mathrm{out}}$, and we have finished specifying $\cQ$. If $I_{\mathrm{out}}=[u]$, we can apply the above procedure to switch to some $D_1$ with all factors in $[u]\setminus I$ of $(C_{1,j},\Enc_{1,j})$ matching the respective factors of the product code at the same position $j'$ in $D_{\mathrm{out}}$. In this case we must have $D_1\in\cB^{I;[u]\setminus I}(q,u,n,k)$. We then apply \Cref{lem:ftall} using \Cref{it:ftaswitchup} one more time to encode the unique factor in $[u]\setminus I$ into its respective value in each product code making up $D_{\mathrm{out}}$.

  If $|I_{\mathrm{out}}|=u-1$ but $I_{\mathrm{out}}\neq I$, we can choose some $D_1\in\cB^{I;[u]\setminus I}(q,u,n,k)$. Then we can apply \Cref{lem:ftall} twice, using \Cref{it:ftaswitchup} followed by \Cref{it:ftaswitchdown}, with the switch up applied to the unique factor in $[u]\setminus I$, and the switch down applied to the unique factor in $[u]\setminus I_{\mathrm{out}}$, to arrive at some $D_2\in\cB^{I_{\mathrm{out}}}(q,u,n,k)$. Then using an analogous procedure as described above for switching from $D_0$ to $D_1$, we can switch from $D_2$ to $D_{\mathrm{out}}$ by looping through $i\in I_{\mathrm{out}}$, for every $i$ applying \Cref{lem:ftall} twice, using \Cref{it:ftaswitchdown} followed by \Cref{it:ftaswitchup}, with both switches applied to the $i$th factor.

  The only remaining case is when $|I_{\mathrm{out}}|=u-2$, so that $(C_{\mathrm{out}},\Enc_{\mathrm{out}})\in\cB^{I_{\mathrm{out}};i_{\mathrm{out}}}(q,u,n,k)$. In this case, writing $I'=I_{\mathrm{out}}\cup\{i_{\mathrm{out}}\}$, so that $|I'|=u-1$, we can first use an analogous procedure as described above for the $|I_{\mathrm{out}}|=u-1$, $I_{\mathrm{out}}\neq I$ case to switch to some $D_2=(C_2,\Enc_2,\Gamma_2,\cE_2)\in\cB^{I'}(q,u,n,k)$. Furthermore, we can choose $D_2$ such that for every $i\in I_{\mathrm{out}}$ and every $j\in[\bar{b}]\times[k]^{[u]\setminus I'}$, the $i$th factor $(C_{2,j}^i,\Enc_{2,j}^i)$ in each product code $(C_{2,j},\Enc_{2,j})$ matches the $i$th factor $(C_{\mathrm{out},j'}^i,\Enc_{\mathrm{out},j'}^i)$ of the $k$ respective product codes $(C_{\mathrm{out},j'},\Enc_{\mathrm{out},j'})$ corresponding to the $k$ values of $j'\in[\bar{b}]\times[k]^{[u]\setminus I_{\mathrm{out}}}$ that agree with $j$ outside of the component labeled $i_{\mathrm{out}}$. Such a choice of $D_2$ is possible because these $k$ product codes $(C_{\mathrm{out},j'},\Enc_{\mathrm{out},j'})$ are equal by the assumption that $(C_{\mathrm{out}},\Enc_{\mathrm{out}})\in\cB^{I_{\mathrm{out}};i_{\mathrm{out}}}(q,u,n,k)$.

Thus we have shown that for every $D_{\mathrm{in}},D_{\mathrm{out}}\in\cD'(q,u,n,k,\lambda_{\mathrm{run}})$, there exists a circuit $\cQ$ consisting of the sequential composition of $O(u)$ applications of \Cref{lem:ftall} using \Cref{it:ftaswitchdown,it:ftaswitchup} that switches from the code $D_{\mathrm{in}}$ to $D_{\mathrm{out}}$. \Cref{it:ftaswitchdown,it:ftaswitchup} in \Cref{lem:ftall} simply consists of parallel compositions of the code switching gadget from \Cref{lem:codeswitch}, which induces the logical identity channel $\bar{O}=I$. Thus letting $N'$ denote the set of all qudits acted on by $\cQ$, then \Cref{lem:ftall,lem:seqcomp} imply that $(\cQ,2^{N'}|_{\geq\lambda_{\mathrm{run}}},D_{\mathrm{in}},D_{\mathrm{out}})$ is a mending fault-tolerant gadget for the identity channel $\bar{O}=I_{[\bar{b}]\times[k]^u}$ using space $|N'|\leq 2(u+2)(4n)^u\leq u\cdot O(n)^u=O(n)^u$ and time $T\leq O(u^2n^2)$, as desired.
\end{proof}

\Cref{lem:switchDDp} below applies \Cref{lem:ftall,lem:switchdec} to provide fault-tolerant gadgets for switching between decorated codes in $\cD(q,u,n,k,\lambda_{\mathrm{run}})$ (see \Cref{def:ladec}) and $\cD'(q,u,n,k,\lambda_{\mathrm{run}})$.

\begin{lemma}
  \label{lem:switchDDp}
  Let $D=D_1\sqcup D_2\sqcup D_3$ for some $D_b\in\cD(q,u,n,k,\lambda_{\mathrm{run}})\cup\{\emptyset\}$ for $b\in[3]$, and let $K(D)=K(D_1)\sqcup K(D_2)\sqcup K(D_3)\subseteq[3]\times[k]^u$ (see \Cref{def:ladec}). Let $D'\in\cD'(q,u,n,k,\lambda_{\mathrm{run}})$. Then there exist quantum circuits $\cQ_{D\rightarrow D'},\cQ_{D'\rightarrow D}$, which do not depend on $\lambda_{\mathrm{run}}$, respectively using space $|N'_{D\rightarrow D'}|,|N'_{D'\rightarrow D}|\leq O(n)^u$, time $T_{D\rightarrow D'},T_{D'\rightarrow D}\leq O(u^4n^2\log^2k)$, and gate set $\{\gX^*,\gZ^*,\gCX^*,\gCCX^*,\gInit_*,\gCO_*\}$ such that:
  \begin{enumerate}
  \item\label{it:DDP1} $(\cQ_{D\rightarrow D'},\;\cE_{\mathrm{run}}=2^{N'_{D\rightarrow D'}}|_{\geq\lambda_{\mathrm{run}}}^{\sqcup T_{D\rightarrow D'}},\; D,\; D')$ forms a mending fault-tolerant gadget for the channel $\bar{O}_{D\rightarrow D'}:\bC^{\bF_q^{K(D)}\times\bF_q^{K(D)}}\rightarrow\bC^{\bF_q^{[\bar{b}]\times[k]^u}\times\bF_q^{[\bar{b}]\times[k]^u}}$ defined by $\bar{O}_{D\rightarrow D'}(\rho)=\rho\otimes(\ket{0}\bra{0})^{\otimes[\bar{b}]\times[k]^u\setminus K(D)}$. That is, $\bar{O}_{D\rightarrow D'}$ applies identity gates to all input qudits in $K(D)$, and initializes all qudits in $[\bar{b}]\times[k]^u\setminus K(D)$ to $\ket{0}\bra{0}$.
  \item\label{it:DDP2} $(\cQ_{D'\rightarrow D},\;\cE_{\mathrm{run}}=2^{N'_{D'\rightarrow D}}|_{\geq\lambda_{\mathrm{run}}}^{\sqcup T_{D'\rightarrow D}},\; D',\; D)$ forms a mending fault-tolerant gadget for the channel $\bar{O}_{D'\rightarrow D}:\bC^{\bF_q^{[\bar{b}]\times[k]^u}\times\bF_q^{[\bar{b}]\times[k]^u}}\rightarrow\bC^{\bF_q^{K(D)}\times\bF_q^{K(D)}}$ defined by $\bar{O}_{D'\rightarrow D}(\rho)=\tr_{[\bar{b}]\times[k]^u\setminus K(D)}(\rho)$. That is, $\bar{O}_{D'\rightarrow D}$ outputs those input qudits that lie in $K(D)\subseteq[3]\times[k]^u$, and throws out (i.e.~traces out) the rest.
  \end{enumerate}
\end{lemma}
\begin{proof}
  For $b\in[3]$, write $D_b=(C_b,\Enc_{C_b},\Gamma_b,\cE_b)$. For every $b\in[\bar{b}]\setminus B_{\mathrm{in}}$, choose an arbitrary code $(C_b,\Enc_{C_b})\in\cC^{[u]}(q,u,n,k)$. Define $D''\in\cD'(q,u,n,k,\lambda_{\mathrm{run}})$ by
  \begin{align*}
    D'' &= \left((C'',\Enc'')=\bigsqcup_{b\in[\bar{b}]}(C_b,\Enc_{C_b}),\; \left\{\ket{0^{K'_{C''}}}\bra{0^{K'_{C''}}}\right\},\; \cE''=2^N|_{\geq\lambda_{\min}}\right).
  \end{align*}
  By definition, $D''$ has physical qudits labeled by $N=\bigsqcup_{b\in[\bar{b}]}N_b$ with each $N_b=\{b\}\times[n]^u$ associated to the code $(C_b,\Enc_{C_b})\in\cC^{[u]}(q,u,n,k)$, and has logical qudits labeled by $K=\bigsqcup_{b\in[\bar{b}]}K_b$ with each $K_b=\{b\}\times[k]^u$.
  
  We now prove the two items in the lemma statement separately:
  \begin{description}
  \item[\Cref{it:DDP1}] Define a circuit
    \begin{equation*}
      \bar{\cQ}_{D\rightarrow D'}=(\bar{Q}_{D\rightarrow D',1};K(D),[3]\times[k]^u)
    \end{equation*}
    using space $[3]\times[k]^u$ and time $1$ that simply applies $\gInit_Z$ to all qudits in $[3]\times[k]^u\setminus K(D)$. Let
    \begin{equation*}
      \bar{\cQ}_{D\rightarrow D'}' = (\bar{Q}_{D\rightarrow D',1},\dots,\bar{Q}_{D\rightarrow D',\bar{T}_{D\rightarrow D'}'};K(D),[3]\times[k]^u)
    \end{equation*}
    be the compiled circuit using space $K$ given by applying \Cref{lem:compile} to $\bar{\cQ}_{D\rightarrow D'}$.

    Let $B_{\mathrm{in}}=\{b\in[3]:D_b\neq\emptyset\}$. We then define $\cQ_{D\rightarrow D'}$ to perform the following:
    \begin{enumerate}
    \item\label{it:DDP1start} Run the error-correction gadget in \Cref{lem:errcorr} on each $D_b$ with $b\in B_{\mathrm{in}}$. Meanwhile, apply $\gInit_Z$ to a set of qudits labeled by $\bigsqcup_{b\in[\bar{b}]\setminus B_{\mathrm{in}}}N_b$. Therefore we now have a set of physical qudits labeled by $N=\bigsqcup_{b\in[\bar{b}]}N_b$ such that each $N_b=\{b\}\times[n]^u$ has some associated $(C_b,\Enc_{C_b})\in\cC^{[u]}(q,u,n,k)$. (We also have some ancilla qudits used by the error-correction gadgets).
      We may view the state in qudits $N$ as a corrupted encoding into $C''$ of a state on logical qudits $K$, where the corruption on $N_b$ may be arbitrarily high-weight if $b\in[\bar{b}]\setminus B_{\mathrm{in}}$.
    \item\label{it:DDP1loginit1} Apply the gadget in \Cref{lem:ftall} with choice \Cref{it:ftainit} to induce logical $\gInit_Z$ on logical qudits $\bigsqcup_{b\in[\bar{b}]\setminus B_{\mathrm{in}}}K_b$ within the code $C''$.
    \item\label{it:DDP1logcirc} For every $t\in[\bar{T}_{D\rightarrow D'}']$, perform the appropriate option below:
      \begin{itemize}
      \item If $\bar{Q}_{D\rightarrow D',t}'$ consists of applications of a gate in $\gX^*$, $\gZ^*$, or $\gCO_*$, (which act on entire blocks $K_b$ for $\gCO_*$; see \Cref{lem:compile}) then apply \Cref{lem:ftall} with the appropriate respective choice in \Cref{it:ftapauli,it:ftacx,it:ftainit,it:ftaco} to induce $\bar{Q}_{D\rightarrow D',t}'$ on the logical qudits $K$.
      \item If for some $i\in[u]$, $\bar{Q}_{D\rightarrow D',t}'$ consists of transversal applications of a gate in $\gInit_*$ or $\gCX^*$ to sets $K_b|_{i\rightarrow j}$ (see \Cref{lem:compile}), let $D'''\in\cD'(q,u,n,k,\lambda_{\mathrm{run}})$ be a disjoint union of copies of an arbitrary code in $\cC^{[u]\setminus\{i\}}(q,u,n,k)$. Then apply \Cref{lem:switchdec} to switch from $D''$ to $D'''$, apply \Cref{lem:ftall} using \Cref{it:ftainit} or \Cref{it:ftacx} to induce $\bar{Q}_{D\rightarrow D',t}'$ on the logical qudits $K$, and apply \Cref{lem:switchdec} to switch back to $D''$.
      \item If for some $i\in[u]$, $\bar{Q}_{D\rightarrow D',t}'$ consists of transversal applications of a gate in $\gH^*$ to sets $K_b|_{i\rightarrow j}$, let $D'''\in\cD'(q,u,n,k,\lambda_{\mathrm{run}})$ be a disjoint union of codes in $\cC^{[u]\setminus\{i\}}(q,u,n,k)$ given by \Cref{it:ftshad} in \Cref{def:ftsuitable}. Then apply \Cref{lem:switchdec} to switch from $D''$ to $D'''$, apply \Cref{lem:ftall} using \Cref{it:ftahad} to induce $\bar{Q}_{D\rightarrow D',t}'$ on the logical qudits $K$, and apply \Cref{lem:switchdec} to switch back to $D''$.
      \item If for some $i\in[u]$, $\bar{Q}_{D\rightarrow D',t}'$ consists of transversal applications of a gate in $\gCCX^*$ to sets $K_b|_{i\rightarrow j}$, let $D'''\in\cD'(q,u,n,k,\lambda_{\mathrm{run}})$ be a disjoint union of codes in $\cC^{[u]\setminus\{i\}}(q,u,n,k)$ given by \Cref{it:ftsccx} in \Cref{def:ftsuitable}. Specifically, for every $(b_1,j_1),(b_2,j_2),(b_3,j_3)\in[\bar{b}]\times[k]\cong[\bar{b}]\times[k]^{\{i\}}$ where $\bar{Q}_{D\rightarrow D',t}'$ applies transversal $\gCCX^*$ to $K_{b_1}|_{i\rightarrow j_1},K_{b_2}|_{i\rightarrow j_2},K_{b_3}|_{i\rightarrow j_3}$, we choose the respective triples of codes $(C_{b_1,j_1},\Enc_{b_1,j_1}),(C_{b_2,j_2},\Enc_{b_2,j_2}),(C_{b_3,j_3},\Enc_{b_3,j_3})$ to be of the form in \Cref{it:ftsccx} in \Cref{def:ftsuitable}. Then we apply \Cref{lem:switchdec} to switch from $D''$ to $D'''$, apply \Cref{lem:ftall} using \Cref{it:ftaccx} to induce $\bar{Q}_{D\rightarrow D',t}'$ on the logical qudits $K$, and apply \Cref{lem:switchdec} to switch back to $D''$.
      \end{itemize}
    \item\label{it:DDP1loginit2} Apply \Cref{lem:ftall} using \Cref{it:ftainit} to induce logical $\gInit_Z$ on logical qudits $\bigsqcup_{b\in[\bar{b}]\setminus[3]}K_b$ within the decorated code $D''$.
    \item\label{it:DDP1switch} Apply \Cref{lem:switchdec} to switch from $D''$ to $D'$ while applying logical identity.
    \end{enumerate}

    By \Cref{lem:errcorr,lem:ftall,lem:switchdec,lem:compile}, $\cQ_{D\rightarrow D'}$ as defined above uses space $|N'_{D\rightarrow D'}|\leq O(n)^u$ and time $T_{D\rightarrow D'}\leq O((u\log k)^2\cdot u^2n^2)=O(u^4n^2\log^2k)$.
    
    We now prove that $\cQ_{D\rightarrow D'}$ as defined above exhibits the desired mending fault-tolerance. We begin with (non-mending) fault-tolerance. Write $D=(C_{\mathrm{in}},\Enc_{\mathrm{in}},\Gamma_{\mathrm{in}},\cE_{\mathrm{in}})$ and $D'=(C_{\mathrm{out}},\Enc_{\mathrm{out}},\Gamma_{\mathrm{out}},\cE_{\mathrm{out}})$, so that by definition $\cE_{\mathrm{in}}\supseteq 2^{B_{\mathrm{in}}\times[n]^u}|_{\geq\lambda_{\max}}$. For $\ell\in\bN$, let $\rho\in\bC^{\bF_q^{|K(D)|+\ell}\times\bF_q^{|K(D)|+\ell}}$, and let $\sigma_0$ be a Pauli $\cE_{\mathrm{in}}\sqcup [\ell]$-deviation of $\Enc_{\mathrm{in}}\otimes I_\ell(\rho)$. Fix a $\cE_{\mathrm{run}}$-avoiding Pauli fault $\cF$ and a postselection $\zeta$ for $\cQ_{D\rightarrow D'}$. By \Cref{lem:paulift}, it suffices to show that the resulting output $\cQ_{D\rightarrow D'}[\cF,\zeta]\otimes I_\ell(\sigma_0)$ is a $\cE_{\mathrm{out}}\sqcup [\ell]$-deviation of
    \begin{equation*}
      (\Enc_{\mathrm{out}}\circ\cQ_{D\rightarrow D'})\otimes I_\ell(\rho\otimes\ket{0^{K_{\mathrm{out}}'}}\bra{0^{K_{\mathrm{out}}'}}) = \Enc_{\mathrm{out}}\otimes I_\ell\left(\rho\otimes(\ket{0}\bra{0})^{[\bar{b}]\times[k]^u\setminus K(D)}\otimes\ket{0^{K_{\mathrm{out}}'}}\bra{0^{K_{\mathrm{out}}'}}\right).
    \end{equation*}

    By \Cref{lem:errcorr} along with the definition of $D,D'',\lambda_{\min},\lambda_{\max}$, after executing \Cref{it:DDP1start} in the definition of $\cQ_{D\rightarrow D'}$ above on input $\sigma_0$, the resulting state $\sigma_1$ is a $2^{B_{\mathrm{in}}\times[n]^u}|_{\geq\lambda_{\min}}\sqcup [\ell]$-deviation of
    \begin{equation*}
      \Enc''\otimes I_\ell\left(\rho\otimes(\ket{0}\bra{0})^{\otimes[\bar{b}]\times[k]^u\setminus K(D)}\otimes\bC^{\bF_q^{K'_{C''}}\times\bF_q^{K'_{C''}}}\right).
    \end{equation*}
    Therefore by \Cref{lem:errcorr,lem:stateprep}, after executing \Cref{it:DDP1loginit1} above, the resulting state $\sigma_2$ is a $\cE''\sqcup [\ell]=2^{[\bar{b}]\times[n]^u}|_{\geq\lambda_{\min}}\sqcup [\ell]$-deviation of
    \begin{equation*}
      \Enc''\otimes I_\ell\left(\rho\otimes(\ket{0}\bra{0})^{\otimes[\bar{b}]\times[k]^u\setminus K(D)}\otimes\ket{0^{K'_{C''}}}\bra{0^{K'_{C''}}}\right).
    \end{equation*}
    Then by \Cref{lem:ftall,lem:compile}, after applying \Cref{it:DDP1logcirc} above, the resulting state $\sigma_3$ is a $\cE''\sqcup [\ell]$-deviation of
    \begin{equation*}
      \Enc''\otimes I_\ell\left(\rho\otimes(\ket{0}\bra{0})^{\otimes[3]\times[k]^u\setminus K(D)}\otimes\rho'\otimes\ket{0^{K'_{C''}}}\bra{0^{K'_{C''}}}\right)
    \end{equation*}
    for some $\rho'\in\bC^{\bF_q^{([\bar{b}]\setminus[3])\times[k]^u}\times\bF_q^{([\bar{b}]\setminus[3])\times[k]^u}}$. Then by \Cref{lem:ftall}, after applying \Cref{it:DDP1loginit2} above, the resulting state $\sigma_4$ is a $\cE''\sqcup [\ell]$-deviation of
    \begin{equation*}
      \Enc''\otimes I_\ell\left(\rho\otimes(\ket{0}\bra{0})^{\otimes[\bar{b}]\times[k]^u\setminus K(D)}\otimes\ket{0^{K'_{C''}}}\bra{0^{K'_{C''}}}\right).
    \end{equation*}
    Finally, by \Cref{lem:switchdec}, the outputted state $\sigma_5$ following the application of \Cref{it:DDP1switch} above is a $\cE_{\mathrm{out}}\sqcup [\ell]$-deviation of 
    \begin{equation*}
      \Enc_{\mathrm{out}}\otimes I_\ell\left(\rho\otimes(\ket{0}\bra{0})^{\otimes[\bar{b}]\times[k]^u\setminus K(D)}\otimes\ket{0^{K'_{\mathrm{out}}}}\bra{0^{K'_{\mathrm{out}}}}\right).
    \end{equation*}
    Thus $(\cQ_{D\rightarrow D'},\;\cE_{\mathrm{run}},\; D,\; D')$ indeed forms a fault-tolerant gadget for the channel $\bar{O}_{D\rightarrow D'}$.

    To show that this gadget is mending, we use a similar proof as above, except we now assume that $\sigma_0$ is a Pauli $[\ell]$-deviation of $\Enc_{\mathrm{in}}\otimes I_\ell(\rho)$. By the mending property in \Cref{lem:errcorr}, after executing \Cref{it:DDP1start} in the definition of $\cQ_{D\rightarrow D'}$ above on input $\sigma_0$, the resulting state $\sigma_1$ is a linear combination of $2^{B_{\mathrm{in}}\times[n]^u}|_{\geq\lambda_{\min}}\sqcup [\ell]$-deviations of
    \begin{equation*}
      \Enc''\otimes I_\ell\left(L\left(\rho\otimes(\ket{0}\bra{0})^{\otimes B_{\mathrm{in}}\times[k]^u\setminus K(D)}\right)\otimes(\ket{0}\bra{0})^{\otimes([\bar{b}]\setminus B_{\mathrm{in}})\times[k]^u}\otimes\bC^{\bF_q^{K'_{C''}}\times\bF_q^{K'_{C''}}}\right).
    \end{equation*}
    for superoperators $L:\bC^{\bF_q^{K(D)}\times\bF_q^{K(D)}}\rightarrow\bC^{\bF_q^{K(D)}\times\bF_q^{K(D)}}$, which by \Cref{lem:paulidecomp} we may assume are Pauli superoperators. Therefore we may decompose $L=L_1\otimes L_2$ for Pauli superoperators $L_1,L_2$ so that
    \begin{equation*}
      L\left(\rho\otimes(\ket{0}\bra{0})^{\otimes B_{\mathrm{in}}\times[k]^u\setminus K(D)}\right) = L_1(\rho)\otimes L_2\left((\ket{0}\bra{0})^{\otimes B_{\mathrm{in}}\times[k]^u\setminus K(D)}\right).
    \end{equation*}
    Now by \Cref{lem:errcorr,lem:stateprep}, after executing \Cref{it:DDP1loginit1} above, the resulting state $\sigma_2$ is a linear combination of $\cE''\sqcup [\ell]=2^{[\bar{b}]\times[n]^u}|_{\geq\lambda_{\min}}\sqcup [\ell]$-deviations of
    \begin{equation*}
      \Enc''\otimes I_\ell\left(L_1(\rho)\otimes L_2\left((\ket{0}\bra{0})^{\otimes B_{\mathrm{in}}\times[k]^u\setminus K(D)}\right)\otimes(\ket{0}\bra{0})^{\otimes([\bar{b}]\setminus B_{\mathrm{in}})\times[k]^u}\otimes\ket{0^{K'_{C''}}}\bra{0^{K'_{C''}}}\right).
    \end{equation*}
    for Pauli superoperators $L_1,L_2$.
    Then by \Cref{lem:ftall,lem:compile}, after applying \Cref{it:DDP1logcirc} above, the resulting state $\sigma_3$ is a linear combination of $\cE''\sqcup [\ell]$-deviations of
    \begin{equation*}
      \Enc''\otimes I_\ell\left(L_1(\rho)\otimes(\ket{0}\bra{0})^{\otimes([3]\setminus B_{\mathrm{in}})\times[k]^u}\otimes\rho'\otimes\ket{0^{K'_{C''}}}\bra{0^{K'_{C''}}}\right)
    \end{equation*}
    for Pauli superoperators $L_1$ and some $\rho'\in\bC^{\bF_q^{([\bar{b}]\setminus[3])\times[k]^u}\times\bF_q^{([\bar{b}]\setminus[3])\times[k]^u}}$. Then by \Cref{lem:ftall}, after applying \Cref{it:DDP1loginit2} above, the resulting state $\sigma_4$ is a linear combination of $\cE''\sqcup [\ell]$-deviations of
    \begin{equation*}
      \Enc''\otimes I_\ell\left(L_1(\rho)\otimes(\ket{0}\bra{0})^{\otimes[\bar{b}]\times[k]^u\setminus K(D)}\otimes\ket{0^{K'_{C''}}}\bra{0^{K'_{C''}}}\right).
    \end{equation*}
    for Pauli superoperators $L_1$.
    Finally, by \Cref{lem:switchdec}, the outputted state $\sigma_5$ following the application of \Cref{it:DDP1switch} above is a linear combination of $\cE_{\mathrm{out}}\sqcup [\ell]$-deviations of 
    \begin{equation*}
      \Enc_{\mathrm{out}}\otimes I_\ell\left(L_1(\rho)\otimes(\ket{0}\bra{0})^{\otimes[\bar{b}]\times[k]^u\setminus K(D)}\otimes\ket{0^{K'_{\mathrm{out}}}}\bra{0^{K'_{\mathrm{out}}}}\right).
    \end{equation*}
    for Pauli superoperators $L_1$.
    Thus by \Cref{lem:paulift}, $(\cQ_{D\rightarrow D'},\;\cE_{\mathrm{run}},\; D,\; D')$ indeed forms a mending fault-tolerant gadget for the channel $\bar{O}_{D\rightarrow D'}$.
  \item[\Cref{it:DDP2}] Define a circuit
    \begin{equation*}
      \bar{\cQ}_{D'\rightarrow D} = (\bar{Q}_{D'\rightarrow D,1};[3]\times[k]^u,[3]\times[k]^u)
    \end{equation*}
    using space $[3]\times[k]^u$ and times $1$ that simply applies $\gInit_Z$ to all qudits in $[3]\times[k]^u\setminus K(D)$. Let
    \begin{equation*}
      \bar{\cQ}_{D'\rightarrow D}' = (\bar{Q}_{D'\rightarrow D,1},\dots,\bar{Q}_{D'\rightarrow D,\bar{T}_{D'\rightarrow D}'};[3]\times[k]^u,[3]\times[k]^u)
    \end{equation*}
    be the compiled circuit using space $K=\bigsqcup_{b\in[\bar{b}]}K_b$ with each $K_b=\{b\}\times[k]^u$ given by applying \Cref{lem:compile} to $\bar{\cQ}_{D'\rightarrow D}$.

    Let $B_{\mathrm{out}}=\{b\in[3]:D_b\neq\emptyset\}$. We then define $\cQ_{D'\rightarrow D}$ to perform the following:
    \begin{enumerate}
    \item\label{it:DDP2switch} Apply \Cref{lem:switchdec} to switch from $D'$ to $D''$ while applying logical identity.
    \item\label{it:DDP2logcirc} For every $t\in[\bar{T}_{D'\rightarrow D}']$, apply \Cref{lem:ftall} (along with \Cref{lem:switchdec} if necessary) using an appropriate choice in \Cref{it:ftallgad} to induce $\bar{Q}_{D'\rightarrow D,t}'$ on logical qudits $K$. Specifically, we perform the same operation as in \Cref{it:DDP1logcirc} above used to prove \Cref{it:DDP1}, but with $\bar{Q}_{D'\rightarrow D,t}'$ replacing $\bar{Q}_{D\rightarrow D',t}'$.
    \item\label{it:DDP2ret} Return the qudits in $B_{\mathrm{out}}\times[n]^u$.
    \end{enumerate}
    By \Cref{lem:ftall,lem:switchdec}, $\cQ_{D'\rightarrow D}$ as defined above uses space $|N'_{D'\rightarrow D}|\leq O(n)^u$ and time $T_{D'\rightarrow D}\leq O((u\log k)^2\cdot u^2n^2)=O(u^4n^2\log^2k)$.

    We now prove that $\cQ_{D\rightarrow D'}$ as defined above exhibits the desired mending fault-tolerance. We begin with (non-mending) fault-tolerance. Write $D'=(C_{\mathrm{in}},\Enc_{\mathrm{in}},\Gamma_{\mathrm{in}},\cE_{\mathrm{in}})$ and $D=(C_{\mathrm{out}},\Enc_{\mathrm{out}},\Gamma_{\mathrm{out}},\cE_{\mathrm{out}})$. For $\ell\in\bN$, let $\rho\in\bC^{\bF_q^{|K|+\ell}\times\bF_q^{|K|+\ell}}$, and let $\sigma_0$ be a Pauli $\cE_{\mathrm{in}}\sqcup [\ell]$-deviation of $\Enc_{\mathrm{in}}\otimes I_\ell(\rho\otimes\Gamma_{\mathrm{in}})$. Fix a $\cE_{\mathrm{run}}$-avoiding Pauli fault $\cF$ and a postselection $\zeta$ for $\cQ_{D'\rightarrow D}$. By \Cref{lem:paulift}, it suffices to show that the resulting output $\cQ_{D'\rightarrow D}[\cF,\zeta]\otimes I_\ell(\sigma_0)$ is a $\cE_{\mathrm{out}}\sqcup [\ell]$-deviation of
    \begin{equation*}
      (\Enc_{\mathrm{out}}\circ\cQ_{D'\rightarrow D})\otimes I_\ell(\rho) = \Enc_{\mathrm{out}}\otimes I_\ell\left(\tr_{[\bar{b}]\times[k]^u\setminus K(D)}(\rho)\right).
    \end{equation*}

    By \Cref{lem:ftall}, after executing \Cref{it:DDP2switch} in the definition of $\cQ_{D\rightarrow D'}$ above on input $\sigma_0$, the resulting state $\sigma_1$ is a $\cE''\sqcup [\ell]$-deviation of
    \begin{equation*}
      \Enc''\otimes I_\ell\left(\rho\otimes\ket{0^{K'_{C''}}}\bra{0^{K'_{C''}}}\right).
    \end{equation*}
    Therefore by \Cref{lem:compile,lem:ftall,lem:switchdec}, after applying \Cref{it:DDP2logcirc} above, the resulting state $\sigma_2$ is a $\cE''\sqcup [\ell]$-deviation of
    \begin{equation*}
      \Enc''\otimes I_\ell\left(\tr_{[\bar{b}]\times[k]^u\setminus K(D)}(\rho)\otimes(\ket{0}\bra{0})^{\otimes[3]\times[k]^u\setminus K(D)}\otimes\rho'\otimes\ket{0^{K'_{C''}}}\bra{0^{K'_{C''}}}\right).
    \end{equation*}
    for some $\rho'\in\bC^{\bF_q^{([\bar{b}]\setminus[3])\times[k]^u}\times\bF_q^{([\bar{b}]\setminus[3])\times[k]^u}}$. Then because by definition $\cE''=2^N|_{\geq\lambda_{\min}}$ and $\cE_{\mathrm{out}}\subseteq 2^{B_{\mathrm{in}}\times[n]^u}|_{\geq\lambda_{\min}}\subseteq\cE''$, it follows by the definition of $D'',D_{\mathrm{out}}$ that the state $\sigma_3$ returned by \Cref{it:DDP2ret} is a $\cE_{\mathrm{out}}\sqcup [\ell]$-deviation of
    \begin{equation*}
      \Enc_{\mathrm{out}}\otimes I_\ell\left(\tr_{[\bar{b}]\times[k]^u\setminus K(D)}(\rho)\right).
    \end{equation*}
    Thus $(\cQ_{D'\rightarrow D},\; \cE_{\mathrm{run}},\; D,\; D')$ indeed forms a fault-tolerant gadget for the channel $\bar{O}_{D'\rightarrow D}$. The fact that this gadget is mending then follows immediately by \Cref{lem:seqcomp} along with the mending property in \Cref{lem:switchdec}.
  \end{description}

  By \Cref{remark:lacirind,remark:ftacirind,lem:switchdec}, the circuits $\cQ_{D\rightarrow D'},\cQ_{D'\rightarrow D}$ defined above do not depend on the choice of $\lambda_{\mathrm{run}}$.
\end{proof}

We now apply \Cref{lem:ftall,lem:switchdec,lem:switchDDp} to complete the proof of \Cref{thm:laft}. Recall that the proof of \Cref{it:laparam} in \Cref{thm:laft} was given in \Cref{sec:ftlacodes}.

\begin{proof}[Proof of \Cref{it:lascheme} in \Cref{thm:laft}]
  Fix a logical circuit $\bar{\cQ}=(\bar{Q}_1,\dots,\bar{Q}_{\bar{T}};K_{\mathrm{in}},K_{\mathrm{out}})$ of the form described in \Cref{it:lascheme} in \Cref{thm:laft}. Let $\bar{\cQ}'=(\bar{Q}_1',\dots,\bar{Q}_{\bar{T}'}';K_{\mathrm{in}},K_{[3]})$ be the compiled circuit given by applying \Cref{lem:compile} to $\bar{\cQ}$, so that $\bar{\cQ}'$ acts on qudits labeled by $K=K_{[\bar{b}]}=K_1\sqcup\cdots\sqcup K_{\bar{b}}$ with each $K_b=\{b\}\times [k]^u\cong[k]^u$, and uses time
  \begin{equation*}
    \bar{T}' \leq O(\bar{T}\cdot(u\log k)^2).
  \end{equation*}

  For $\alpha\in\{\mathrm{in},\mathrm{out}\}$, fix $D_\alpha=\bigsqcup_{b\in[3]}D_{\alpha,b}$ for some $D_{\alpha,b}\in\cD(q,u,n,k,\lambda_{\mathrm{run}})\cup\{\emptyset\}$ with $K(D_{\alpha,b})=K_\alpha\cap K_b$ for $b\in[3]$. We construct a fault-tolerant circuit $\cQ$ for $\bar{\cQ}'[*](\cdot)\subseteq\spn_{\bC}\{\bar{\cQ}[*](\cdot)\}$ to perform the following, given as input a code state of $D_{\mathrm{in}}$:
  \begin{enumerate}
  \item Apply the gadget in \Cref{it:DDP1} in \Cref{lem:switchDDp} to switch from $D_{\mathrm{in}}$ to some $D''\in\cD'(q,u,n,k,\lambda_{\mathrm{run}})$ while applying logical identity.
  \item For every $t\in[\bar{T}']$, apply \Cref{lem:ftall} (along with \Cref{lem:switchdec} if necessary) using an appropriate choice in \Cref{it:ftallgad} to induce $\bar{Q}_t'$ on logical qudits $K$. Specifically, we perform the same operation as described above in \Cref{it:DDP1logcirc} in the proof of \Cref{it:DDP1} in \Cref{lem:switchDDp}, but with $\bar{Q}_t'$ replacing $\bar{Q}_{D\rightarrow D',t}'$.
  \item Apply the gadget in \Cref{it:DDP2} in \Cref{lem:switchDDp} to switch from $D''$ to $D_{\mathrm{out}}$.
  \end{enumerate}
  By \Cref{remark:ftacirind,lem:switchdec,lem:switchDDp}, $\cQ$ as defined above does not depend on the choice of $\lambda_{\mathrm{run}}$ or on the families of bad sets for $D_{\mathrm{in}},D_{\mathrm{out}}$. By \Cref{lem:ftall,lem:switchdec,lem:switchDDp}, $\cQ$ uses space
  \begin{equation*}
    |N'| \leq O(n)^u,
  \end{equation*}
  time
  \begin{equation*}
    T \leq O(u^4n^2\log^2k)+\bar{T}'\cdot O(u^2n^2) \leq O(\bar{T}\cdot u^4n^2\log^2n),
  \end{equation*}
  gate set $\cG=\{\gH^*,\gX^*,\gZ^*,\gCX^*,\gCCX^*,\gInit_*,\gCO_*\}$. Furthermore, by \Cref{lem:ftall,lem:switchdec,lem:switchDDp} along with \Cref{lem:seqcomp}, $(\cQ,2^{N'}_{\geq\lambda_{\mathrm{run}}},D_{\mathrm{in}},D_{\mathrm{out}})$ is a mending fault-tolerant gadget for $\bar{\cQ}'[*](\cdot)\subseteq\spn_{\bC}\{\bar{\cQ}[*](\cdot)\}$, and hence also for $\bar{\cQ}[*](\cdot)$, as desired.

\end{proof}

\section{Alphabet Reduction via Simulative Composition}
\label{sec:alphred}
In this section, we show how to perform simulative composition on our fault-tolerance scheme in \Cref{thm:laft}. That is, we show how to fault-tolerantly simulate high-dimensional qudits using \Cref{thm:laft} over lower-dimensional qudits. This composition allows us to reduce the qudit dimension of our fault-tolerance scheme, which is exponentially large in \Cref{thm:laft}, down to an arbitrary constant prime power.

\subsection{Main Result Statement}
We state our main result in \Cref{thm:main} below. We will build up the necessary simulative composition machinery in the subsequent subsections, before proving \Cref{thm:main} in \Cref{sec:mainproof}.

\begin{restatable}{theorem}{main}
  \label{thm:main}
  For every prime power $r$, every $0<\epsilon<1/8$, and every $\eta_{\mathrm{gap}}\geq 0$, there exists $\bar{N}_0=\bar{N}_0(r,\epsilon)\geq 0$ such that the following holds. For every set $\bar{N}$ of size $|\bar{N}|\geq\bar{N}_0$ and every subset $K\subseteq\bar{N}$, there is a CSS non-subsystem code $C(r,\epsilon,K,\bar{N})$ with associated encoding map $\Enc_{C(r,\epsilon,K,\bar{N})}$, which satisfy the following for some integer $M=M(r,\epsilon,\bar{N})\leq|\bar{N}|^{4+2^9\epsilon}$:
  \begin{enumerate}
  \item $C(r,\epsilon,K,\bar{N})$ is a $[[n(r,\epsilon,\bar{N}),\; k=|K|,\; d(r,\epsilon,\bar{N})]]_r$ code with logical qudits labeled by the set~$K$, where
    \begin{align*}
      n(r,\epsilon,\bar{N}) &\leq 2^{32\sqrt{\log|\bar{N}|}}\cdot M\cdot|\bar{N}| \\
      d(r,\epsilon,\bar{N}) &\geq 2^{-2^9\sqrt{\log|\bar{N}|}}\cdot M\cdot|\bar{N}|.
    \end{align*}
  \item For every quantum circuit $\bar{\cQ}=(\bar{Q}_1,\dots,\bar{Q}_{\bar{T}};K_{\mathrm{in}},K_{\mathrm{out}})$ acting on a set of $r$-dimensional qudits labeled by $\bar{N}$ using time $\bar{T}$ and gate set $\cG=\{\gH^*,\gX^*,\gZ^*,\gCX^*,\gCCX^*,\gInit_*,\gCO_*\}$, letting
    \begin{align*}
      D_{\mathrm{in}} &= (C_{\mathrm{in}}=C(r,\epsilon,K_{\mathrm{in}},\bar{N}),\; \Enc_{\mathrm{in}}=\Enc_{C_{\mathrm{in}}},\; \cE_{\mathrm{in}}=2^{[n(r,\epsilon,\bar{N})]}|_{\geq\lambda_{\mathrm{in}}}) \\
      D_{\mathrm{out}} &= (C_{\mathrm{out}}=C(r,\epsilon,K_{\mathrm{out}},\bar{N}),\; \Enc_{\mathrm{out}}=\Enc_{C_{\mathrm{out}}},\; \cE_{\mathrm{out}}=2^{[n(r,\epsilon,\bar{N})]}|_{\geq\lambda_{\mathrm{out}}}),
    \end{align*}
    then there exists a fault-tolerant gadget
    \begin{equation*}
      \left(\cQ,\; \cE_{\mathrm{run}}=2^N|_{\geq\lambda_{\mathrm{run}}},\; D_{\mathrm{in}},\; D_{\mathrm{out}}\right)
    \end{equation*}
    for $\bar{\cQ}[*](\cdot)$, where $\cQ$ is a quantum circuit using space $|N|\leq 2^{O(\sqrt{\log|\bar{N}|})}\cdot M\cdot|\bar{N}|$, time $T\leq 2^{2^9\sqrt{\log|\bar{N}|}} \cdot \bar{T}$, and gate set $\cG$, and where
    \begin{align*}
      \lambda_{\mathrm{in}} &= \lambda_{\mathrm{in}}(r,\epsilon,\eta_{\mathrm{gap}},\bar{N}) \geq 2^{-2^{11}\sqrt{\log|\bar{N}|}}\cdot M\cdot|\bar{N}| \\
      \lambda_{\mathrm{run}} &= \lambda_{\mathrm{run}}(r,\epsilon,\eta_{\mathrm{gap}},\bar{N}) \geq 2^{-2^{\eta_{\mathrm{gap}}+O(1)}\sqrt{\log|\bar{N}|}}\cdot M\cdot|\bar{N}| \\
      \lambda_{\mathrm{out}} &= \lambda_{\mathrm{out}}(r,\epsilon,\eta_{\mathrm{gap}},\bar{N}) \leq 2^{-2^{\eta_{\mathrm{gap}}+62}\sqrt{\log|\bar{N}|}}\cdot M\cdot|\bar{N}|.
    \end{align*}
  \end{enumerate}
\end{restatable}

We have not attempted to optimize constants in \Cref{thm:main}, and give some explicit values just for concreteness and clarity in the proof.

We emphasize that while \Cref{thm:main} only fault-tolerantly implements the logical circuit $\bar{\cQ}$ up to arbitrary postselections on the logical classical function gates, we can implement logical circuits consisting of unitary gates and mid-circuit measurements without any classical function gates, and hence without any postselection. In particular, as described in \Cref{sec:faulttolinf}, we can perform mid-circuit logical measurements using $\gInit_*$ and $\gCX^*$ gates, rather than $\gCO_*$ gates.

\subsection{Increasing Logical Alphabet Size}
\label{sec:inclog}
In this section, we show how to perform a quantum computation over a small field $\bF_q$ using qudits over an arbitrary extension field $\bF_{q'}$.

\begin{lemma}
  \label{lem:inclog}
  Let $\bF_q$ be a finite field, let $\bF_{q'}$ be an extension field, and let $\iota:\bF_q\rightarrow\bF_{q'}$ denote the natural inclusion. We also let $\iota:\bC^{\bF_q\times\bF_q}\rightarrow\bC^{\bF_{q'}\times\bF_{q'}}$ denote the associated quantum channel given by $\iota(\ket{x}\bra{y})=\ket{\iota(x)}\bra{\iota(y)}$.
  
  Let $\bar{\cQ}=(\bar{Q}_1,\dots,\bar{Q}_{\bar{T}};K_{\mathrm{in}},K_{\mathrm{out}})$ be a quantum circuit acting on $q$-dimensional qudits labeled by a set $\bar{N}$, using time $\bar{T}$ and gate set $\cG=\{\gH^*,\gX^*,\gZ^*,\gCX^*,\gCCX^*,\gInit_*,\gCO_*\}$ (over $\bF_q$). Then there exists a quantum circuit $\bar{\cQ}'=(\bar{Q}_1',\dots,\bar{Q}_{\bar{T}'}';K_{\mathrm{in}},K_{\mathrm{out}})$ acting on $q'$-dimensional qudits labeled by a set $\bar{N}'\supseteq\bar{N}$ with $|\bar{N}'|\leq|\bar{N}|\cdot 8\log_2q$, using time $\bar{T}'\leq\bar{T}\cdot O(\log q)$ and gate set $\cG$ (over $\bF_{q'}$), such that for every postselection $\zeta'$ for $\bar{\cQ}'$, there exists a postselection $\zeta$ for $\bar{\cQ}$ such that
  \begin{align}
    \label{eq:inclogcirc}
    \bar{\cQ}'[\zeta']\circ\iota^{\sqcup K_{\mathrm{in}}} &\propto \iota^{\sqcup K_{\mathrm{out}}}\circ\bar{\cQ}[\zeta].
  \end{align}
  In the equation above, $\bar{\cQ}[\zeta],\bar{\cQ}'[\zeta']$ denote the superoperators associated to these respective postselected circuits.
\end{lemma}
\begin{proof}
  First, we may replace all $\gZ^*,\gInit_{X,*},\gCO_{X,*}$ gates in $\bar{\cQ}$ with $\gX^*,\gInit_{Z,*},\gCO_{Z,*}$ gates, respectively, conjugated by Hadamard gates. By \Cref{def:gates,fact:Hconj}, such a change does not change the space usage of $\bar{\cQ}$, increases the time usage by at most a factor of $3$, and does not change the associated channel $\bar{\cQ}(\cdot)$. Therefore we may assume that $\bar{\cQ}$ in fact only uses gate set $\cG'=\{\gH^*,\gX^*,\gCX^*,\gCCX^*,\gInit_{Z,*},\gCO_{Z,*}\}$.

  Now for every gate $G\in\cG'$ over $\bF_q$ acting on some number $b$ of qudits, we will construct a circuit $G'$ over $\bF_{q'}$ with $b$ input and output qudits, such that for every postselection $\tilde{\zeta}'$ for $G'$, there exists a postselection $\tilde{\zeta}$ for $G$ with
  \begin{align}
    \label{eq:incloggate}
    G'[\tilde{\zeta}']\circ\iota^{\sqcup b} &\propto \iota^{\sqcup b}\circ G[\tilde{\zeta}].
  \end{align}
  Note that if $G$ or $G'$ does not include $\gCO_*$ gates, the associated postselection is trivial.
  Our final circuit $\bar{\cQ}'$ will then be obtained by replacing each gate $G$ in $\bar{\cQ}$ with the corresponding circuit $G'$ in $\bar{\cQ}'$. $\bar{\cQ}'$ will act on a set of data qudits labeled by $\bar{N}$ that correspond to the respective qudits of $\bar{\cQ}$, along with a set of ancilla qudits used by the circuits $G'$ (specifically for $G\in\gH^*$).


  By definition for every gate $G\in\cG'\setminus\gH^*$ over $\bF_q$, \Cref{eq:incloggate} holds when $G'$ is the circuit that applies a single gate given by the $\bF_{q'}$-version of $G$. Specifically, when $G=\gCO_*$ acts on some number $m$ of qudits, then we take
  \begin{equation}
    \label{eq:tzform}
    \tilde{\zeta}=\tilde{\zeta}'|_{\bF_q^m}
  \end{equation}
  using the natural inclusion $\bF_q^m\subseteq\bF_{q'}^m$, and when $G\in\cG'\setminus\{\gH^*,\gCO_*\}$, then both $\tilde{\zeta},\tilde{\zeta}'$ are trivial. Therefore it only remains to consider $G\in\gH^*$. Below we construct $G'$ for $G=\gH$; the $G=\gH^\dagger$ case is analogous.

  When $G=\gH$, we define $G'=\gH'$ to perform the following steps, given as input a single $q'$-dimensional qudit. Below, we let $p$ denote the characteristic of $\bF_q,\bF_{q'}$, we let $\omega_p=e^{2\pi i/p}$ denote the $p$th root of unity, and we let $\beta\in\bF_{q'}$ be an arbitrary element satisfying $\tr_{\bF_{q'}/\bF_q}(\beta)=1$. We assume the input qudit is in the state $\iota(\rho)$ for some $\rho=\sum_{x,x'\in\bF_q}\rho_{x,x'}\ket{x}\bra{x'}$, where each $\rho_{x,x'}\in\bC$.
  
  \begin{enumerate}
  \item Apply the unitary $\ket{x}\mapsto\ket{\beta{x}}$, so that the resulting state is $\sum_{x,x'\in\bF_q}\rho_{x,x'}\ket{\beta x}\bra{\beta x'}$. To implement this unitary, we can initialize an ancilla qudit to $\ket{0}$ using $\gInit_Z$, then apply a sequence of three $\gCX^*$ gates to map $\ket{x,0}\mapsto\ket{x,x}\mapsto\ket{\beta x,x}\mapsto\ket{\beta x,0}$.
  \item Apply $\gH$ (over $\bF_{q'}$), so that the resulting state is
    \begin{equation*}
      \sum_{x,x'\in\bF_q,\;z,z'\in\bF_{q'}}\rho_{x,x'}\cdot\omega_p^{\tr_{\bF_{q'}/\bF_p}(\beta(xz-x'z'))}\ket{z}\bra{z'}.
    \end{equation*}
  \item\label{it:Hincpower} Apply the isometry $\ket{z}\mapsto\ket{z}\ket{z^q-z}$, so that the resulting state is
    \begin{equation*}
      \sum_{x,x'\in\bF_q,\;z,z'\in\bF_{q'}}\rho_{x,x'}\cdot\omega_p^{\tr_{\bF_{q'}/\bF_p}(\beta(xz-x'z'))}\ket{z}\bra{z'}\otimes\ket{z^q-z}\bra{{z'}^q-z'}.
    \end{equation*}
    To implement this isometry, we use $\gInit_Z$ to initialize\footnote{The factor of $8$ here is not tight, but we will not need to optimize it.} $\leq 8\log_2q$ ancilla qudits to $0$, on which we then apply $O(\log q)$ $\gCCX^{+1}$ gates to perform repeated squaring in order to compute $z^q$. Specifically, using repeated squaring with $\gCCX^{+1}$ gates, we compute $\ket{z^{2^j}}\bra{{z'}^{2^j}}$ in a separate register for every $0\leq j\leq\log_2 q$, and then with $\gCCX^{+1}$ gates compute $\ket{z^q}\bra{{z'}^q}$, using the fact that $z^q$ is the product of a subset of the $z^{2^j}$. We then use a $\gCX^{-1}$ gate to compute $\ket{z^q-z}\bra{{z'}^q-z'}$, before uncomputing all the prior steps from the repeated squaring to return the ancilla qudits to $0$. This procedure by definition uses space $\leq 8\log_2q$ and time $O(\log q)$.
  \item Apply the gate $\gCO_{Z,f}$ to the register containing $z^q-z$, where $f:\bF_{q'}\rightarrow\bF_{q'}$ is the function defined as follows. Given input $w=z^q-z$, then we fix an arbitrary $z_0=z_0(w)\in\bF_{q'}$ satisfying $w=z_0^q-z_0$, and we define $f(w)=z_0-\tr_{\bF_{q'}/\bF_q}(\beta z_0)$. If $w$ is not of the form $z^q-z$ so that no such $z_0$ exists, then $f(w)$ can be arbitrary.
    After this step, assuming we actually apply some $\tilde{\zeta}'$-postselected version $\gCO_{Z,f}^{\tilde{\zeta}'}$ of $\gCO_{Z,f}$, then the resulting state is
    \begin{align}
      \label{eq:Hincret}
      \sum_w\tilde{\zeta}'_w\cdot\sum_{x,x'\in\bF_q,\;z,z'\in\bF_{q'}:z^q-z=w={z'}^q-z'}\rho_{x,x'}\cdot\omega_p^{\tr_{\bF_{q'}/\bF_p}(\beta(xz-x'z'))}\ket{z}\bra{z'}\otimes\ket{f(w)}\bra{f(w)},
    \end{align}
    where the first sum above is over all $w$ of the form $w=z^q-q$ for some $z\in\bF_{q'}$.
  \item Apply $\gCX^{-1}$ to the two registers and return the resulting first register, so that the returned state is
    \begin{align*}
      \sum_w\tilde{\zeta}'_w\cdot\sum_{x,x'\in\bF_q,\;z,z'\in\bF_{q'}:z^q-z=w={z'}^q-z'}\rho_{x,x'}\cdot\omega_p^{\tr_{\bF_{q'}/\bF_p}(\beta(xz-x'z'))}\ket{z-f(w)}\bra{z'-f(w)}
    \end{align*}
    Now by definition the set of $z\in\bF_{q'}$ with $z^q-z=w$ is precisely $z_0+\bF_q=f(w)+\bF_q$, so we may write each such $z=f(w)+y$ for $y\in\bF_q$. Then recalling that by definition $\tr_{\bF_{q'}/\bF_q}(\beta)=1$, we have
    \begin{align*}
      \tr_{\bF_{q'}/\bF_p}(\beta xz)
      &= \tr_{\bF_{q'}/\bF_p}(\beta x(z_0-\tr_{\bF_{q'}/\bF_q}(\beta z_0)+y)) \\
      &= \tr_{\bF_q/\bF_p}\left(x\tr_{\bF_{q'}/\bF_q}(\beta z_0)-x\tr_{\bF_{q'}\bF_q}(\beta z_0)+xy\right) \\
      &= \tr_{\bF_q/\bF_p}(xy).
    \end{align*}
    Applying the same reasoning to $z',x'$ in place of $z,x$, the expression for the returned state in \Cref{eq:Hincret} is proportional to
    \begin{align}
      \label{eq:Hincret2}
      \begin{split}
        \hspace{1em}&\hspace{-1em} \sum_w\tilde{\zeta}'_w\cdot\sum_{x,x'\in\bF_q,\;y,y'\in\bF_q}\rho_{x,x'}\cdot\omega_p^{\tr_{\bF_q/\bF_p}(xy-x'y')}\ket{y}\bra{y'} \\
                    &\propto \left(\sum_w\tilde{\zeta}'_w\right) \cdot \sum_{x,x'\in\bF_q}\rho_{x,x'}\cdot\iota(\gH\ket{x}\bra{x'}\gH^\dagger) \\
                    &= \left(\sum_w\tilde{\zeta}'_w\right) \cdot \iota(\gH\rho\gH^\dagger) \\
        &\propto \iota(\gH\rho\gH^\dagger),
      \end{split}
    \end{align}
    where the $\gH,\gH^\dagger$ above are over $\bF_q$.
  \end{enumerate}
  
  \Cref{eq:Hincret2} shows that our circuit $\gH'$ described above indeed satisfies \Cref{eq:incloggate}; recall that the postselection $\tilde{\zeta}$ here is trivial as $G=\gH$. Furthermore, the circuit described above by definition uses space $8\log_2q$ and time $O(\log q)$ in \Cref{it:Hincpower} and constant space and time in all other steps, for a total space usage of $8\log_qq$ and time usage of $O(\log q)$.

  Therefore replacing each gate $G$ in $\bar{\cQ}$ with the associated circuit $G'$ in $\bar{\cQ}'$ as described above, we obtain a circuit $\bar{\cQ}'$ whose using space $|\bar{N}'|\leq |\bar{N}|\cdot 8\log_2 q$ and time $\bar{T}'=\bar{T}\cdot O(\log q)$. Furthermore, the desired equality \Cref{eq:inclogcirc} follows immediately by applying \Cref{eq:incloggate} to each gate $G$ in $\bar{\cQ}$, where $\zeta$ is given by restricting $\zeta'$ to coordinates given by the field inclusion $\bF_q\subseteq\bF_{q'}$ as in \Cref{eq:tzform}.
\end{proof}

\subsection{Reducing Physical Alphabet Size}
\label{sec:alphabetreduce}
In this section, we show how to reduce the physical alphabet size of a fault-tolerance scheme. 
The main idea is to perform simulative composition to fault-tolerantly simulate each large-alphabet gate using \Cref{thm:laft}. 
We are able to work with non-subsystem codes for simplicity, as \Cref{thm:laft} provides such codes.

In particular, \Cref{lem:simcomp} below is the critical ingredient for proving \Cref{thm:main}.
It shows that from a fault-tolerant gadget over a large field $\bF_{q'}$ we can construct a fault-tolerant gadget over a smaller subfield $\bF_{q}$.
To prove \Cref{thm:main}, we will recursively apply \Cref{lem:simcomp} (or more precisely, the instantiation in \Cref{cor:simcomp}) until the large alphabet size required by the top-level code of the gadget in \Cref{thm:laft} is reduced to a constant size.

Throughout this section, we use the following basic notation, which fixes an isomorphism between an extension field and the corresponding vector space over the base field.

\begin{definition}
  \label{def:fieldiso}
  For every prime power $q$ and every $\kappa\in\bN$, letting $q'=q^\kappa$, then we let $\phi_{q,q'}:\bF_q^\kappa\xrightarrow{\sim}\bF_{q'}$ denote an arbitrary fixed $\bF_q$-linear isomorphism. We also let $\phi_{q,q'}:\bC^{\bF_q^\kappa}\xrightarrow{\sim}\bC^{\bF_{q'}}$ and $\phi_{q,q'}:\bC^{\bF_q^\kappa\times\bF_q^\kappa}\xrightarrow{\sim}\bC^{\bF_{q'}\times\bF_{q'}}$ denote the ismorphisms $\phi\ket{x}=\ket{\phi(x)}$ and $\phi(\ket{x}\bra{x'})=\ket{\phi(x)}\bra{\phi(y)}$ respectively; the use will be clear from context.
\end{definition}

In \Cref{lem:qimp} below, for a prime power $q$ and a larger prime power $q'=q^\kappa$ with $\kappa\in\bN$, we show how to non-fault-tolerantly simulate the gates $\{\gH^*,\gX^*,\gZ^*,\gCX^*,\gCCX^*,\gInit_*\}$ over $q'$-dimensional qudits using the analogous gates over $q$-dimensional qudits. We will subsequently apply \Cref{thm:laft} to these simulation circuits to obtain fault-tolerance. Note that this application will have $\kappa$ growing polynomially in the number of qudits in our overall fault-tolerant circuit. To ensure that our fault-tolerance scheme retains subpolynomial time overhead and almost-linear error resilience, we therefore need the simulation in \Cref{lem:qimp} below to have time usage growing subpolynomially in $\kappa$. To obtain such a low (specifically, logarithmic) time overhead in our simulation, we incur a larger (polynomial) space overhead than would otherwise be needed. Specifically, the key element of our simulation of $\gCX^*,\gCCX^*,\gH^*$ gates is given in \Cref{claim:complin} below, where we show how to implement linear and bilinear maps on $\bF_q^\kappa$ using space $\poly(\kappa)$ and time $O(\log\kappa)$.

\begin{lemma}
  \label{lem:qimp}
  Let $q$ be a prime power, let $\kappa\in\bN$, and let $q'=q^\kappa$. For every gate $G\in\cG:=\{\gH^*,\gX^*,\gZ^*,\gCX^*,\gCCX^*,\gInit_*\}$ over $q'$-dimensional qudits, there exists a circuit $\bar{\cQ}=\bar{\cQ}_G$ using gate set $\cG$ over $q$-dimensional qudits, with $\kappa\cdot\bar{b}$ input and output qudits if $G$ is a $\bar{b}$-qudit gate, using space $|N|\leq 16\kappa^3$ and time $T\leq 32\log(\kappa)+250$, such that
  \begin{align}
    \label{eq:qimp}
    \phi_{q,q'}^{\otimes\bar{b}}\circ\bar{\cQ} &= G\circ\phi_{q,q'}^{\otimes\bar{b}}.
  \end{align}
  In the equation above, $\bar{\cQ},G$ denote the channels associated to the respective circuit/gate.
\end{lemma}
\begin{proof}
  We begin with the following claim that shows how to compute linear and bilinear maps over $\bF_q$ in logarithmic time. Below, all logarithms should be assumed to be base $2$, unless explicitly stated otherwise.
  
  \begin{claim}
    \label{claim:complin}
    For every linear map $A:\bF_q^m\rightarrow\bF_q^n$, there exists a circuit using gate set $\{\gCX^*,\gInit_*\}$, space $\leq 8mn$, and time $\leq 4\log(nm)+32$ over $q$-dimensional qudits that applies the unitary $\ket{x,x'}\mapsto\ket{x,x'+A(x)}$.

    Furthermore, for every bilinear map $B:\bF_q^{m_1}\times\bF_q^{m_2}\rightarrow\bF_q^n$, there exists a circuit using gate set $\{\gCX^*,\gCCX^*,\gInit_*\}$, space $\leq 16m_1m_2n$, and time $\leq 8\log(m_1m_2n)+64$ over $q$-dimensional qudits that applies the unitary $\ket{x_1,x_2,x'}\mapsto\ket{x_1,x_2,x'+B(x_1,x_2)}$.
  \end{claim}
  \begin{proof}
    We first describe the circuit to implement $\ket{x,x'}\mapsto\ket{x,x'+A(x)}$. Let $A\in\bF_q^{n\times m}$ denote the matrix corresponding to $A$, and let $\ket{x}=\ket{x_1,\dots,x_m}$. Our desired circuit begins by implementing the isometry
    \begin{equation*}
      \ket{x} \mapsto \ket{x}\otimes\bigotimes_{i\in[m],j\in[n]}\ket{A_{j,i}x_i}.
    \end{equation*}
    Specifically, for each $i\in[m]$ we define a depth $\leq\log(n)+1$ binary tree on qudits with root given by the qudit $\ket{x_i}$, and with $\leq 2n$ additional nodes given by fresh ancilla qudits initialized to $\ket{0}$ (using $\gInit_Z$), so that there are exactly $n$ leaves. We then perform $\leq\log(n)+1$ rounds of $\gCX^*$ gates, where in the $t$th round we apply $\gCX^a$ from every qudit at distance $t-1$ from the root this tree to each of its children. If the target qudit of the $\gCX^a$ gate is not a leaf of the tree, we choose the coefficient $a=1$. If the target is a leaf (say the $j$th leaf for $j\in[n]$), we choose $a=A_{j,i}$. This procedure by definition takes $\leq 2(\log(n)+1)$ time, as each round takes two timesteps (one for the first child of each node in that round, and one for the second child).

    Next, for each $j\in[n]$, we define a depth $\leq\log(m)+1$ binary tree on qudits with $m$ leaves given by the qudits $\ket{A_{j,i}x_i}$ for $i\in[m]$. The internal nodes of this tree are all fresh ancilla qudits initialized to $\ket{0}$. We then perform $\leq\log(m)+1$ rounds of $\gCX^1$ gates, where in the $t$th round we apply $\gCX^1$ from every qudit at distance $t-1$ from a leaf of this tree to its parent. This procedure by definition takes $\leq 2(\log(m)+1)$ time, and leaves the root of the $j$th tree in the state $\ket{\sum_{i\in[m]}A_{j,i}x_i}=\ket{(Ax)_j}$. Thus the roots of the $n$ trees together form $\ket{Ax}$. We apply $\gCX^1$ gates with these $n$ root qudits as controls and the $n$ qudits in $\ket{x'}$ as targets, to yield $\ket{x'+Ax}$ in the target registers.

    Finally, we apply the inverse of all gates described above (in reverse order as applied) except those involving register forming $\ket{x'+Ax}$, in order to uncompute the ancilla qudits back to their original $\ket{0}$ states. This entire procedure uses space $\leq 8mn$, time $\leq 4(\log(n)+\log(m)+8)=4\log(nm)+32$, and computes $\ket{x}\mapsto\ket{x}\ket{Ax}$, as desired.

    The circuit computing $\ket{x_1,x_2,x'}\mapsto\ket{x_1,x_2,x'+B(x_1,x_2)}$ is similar for the circuit for $A$ described above, except that we first apply the isometry
    \begin{equation*}
      \ket{x_1,x_2} \mapsto \ket{x_1,x_2}\otimes\bigotimes_{(i_1,i_2)\in[m_1]\times[m_2]}\ket{x_{1,i_1}\cdot x_{2,i_2}} = \ket{x_1,x_2}\otimes\ket{x_1\otimes x_2}.
    \end{equation*}
    Specifically, to apply this isometry, we create $m_2$ (resp.~$m_1$) copies of each input qudit in $x_1$ (resp.~$x_2$), again using binary trees of $\gCX^1$ gates, and then we apply $\gCCX^1$ gates to multiply the appropriate leaves of the trees. Then because every bilinear map $B:\bF_q^{m_1}\times\bF_q^{m_2}\rightarrow\bF_q^n$ by definition has the form $B(x_1,x_2)=B'(x_1\otimes x_2)$ for some linear map $B':\bF_q^{m_1}\otimes\bF_q^{m_2}\rightarrow\bF_q^n$, we can apply our procedure above for linear maps with $A=B'$ to compute $\ket{x_1,x_2,x_1\otimes x_2,x'+B'(x_1\otimes x_2)}$, before uncomputing $\ket{x_1\otimes x_2}$ to yield $\ket{x_1,x_2,x'+B(x_1,x_2)}$. This entire procedure uses space $\leq 16m_1m_2n$ and time $\leq 4(\log(m_1)+\log(m_2)+8)+4(\log(m_1m_2)+\log(n)+8)\leq 8\log(m_1m_2n)+64$, as desired.
  \end{proof}

  Let $\phi=\phi_{q,q'}$. If $G=\gX^a$, we simply let $\bar{\cQ}$ apply $\gX^{\phi^{-1}(a)}$. If $G=\gInit_Z$, we simply let $\bar{\cQ}$ apply $\gInit_Z^{\otimes\kappa}$. If $G=\gCX^a$, then as
  \begin{equation*}
    (\phi^{-1})^{\otimes 2}\gCX^a\phi^{\otimes 2}\ket{x_1,x_2} = \ket{x_1,x_2+\phi^{-1}(a\phi(x_1))},
  \end{equation*}
  and $x_1\mapsto\phi^{-1}(a\phi(x_1))$ is a linear map on $\bF_q^\kappa$, \Cref{claim:complin} provides the desired circuit $\bar{\cQ}$ computing the RHS above using space $\leq 8\kappa^2$ and time $8\log(\kappa)+32$. If $G=\gCCX^a$, then as
  \begin{equation*}
    (\phi^{-1})^{\otimes 3}\gCCX^a\phi^{\otimes 3}\ket{x_1,x_2,x_3} = \ket{x_1,x_2,x_3+\phi^{-1}(a\phi(x_1)\phi(x_2))},
  \end{equation*}
  and $(x_1,x_2)\mapsto\phi^{-1}(a\phi(x_1)\phi(x_2))$ is a bilinear map on $\bF_q^\kappa\times\bF_q^\kappa$, \Cref{claim:complin} provides the desired circuit $\bar{\cQ}$ computing the RHS above using space $\leq 16\kappa^3$ and time $\leq 24\log(\kappa)+64$.

  Now consider $G=\gH$ (the $G=\gH^\dagger$ case is analogous). The trace bilinear form $(x,z)\mapsto\tr_{\bF_{q'}/\bF_q}(\phi(x)\phi(z))$ mapping $\bF_q^\kappa\times\bF_q^\kappa\rightarrow\bF_q$ is nondegenerate, so there exists an invertible matrix $A\in\bF_q^{\kappa\times\kappa}$ for which
  \begin{equation*}
    z^\top Ax = \tr_{\bF_{q'}/\bF_q}(\phi(x)\phi(z)).
  \end{equation*}
  We then define $\bar{\cQ}=\bar{\cQ}_{\gH}$ on input $\ket{x}$ to first apply the unitary $\ket{x}\mapsto\ket{Ax}$, and then apply $\gH^{\otimes\kappa}$, to output $\gH^{\otimes\kappa}\ket{Ax}$. The unitary $\ket{x}\mapsto\ket{Ax}$ is implemented by applying $\gInit_Z$ to $\kappa$ ancilla qudits, and then performing
  \begin{equation*}
    \ket{x,0} \mapsto \ket{x,Ax} \mapsto \ket{0,Ax} \mapsto \ket{Ax,Ax} \mapsto \ket{Ax,0},
  \end{equation*}
  where the first step uses \Cref{claim:complin} to apply $\ket{x,x'}\mapsto\ket{x,x'+Ax}$, the second step uses \Cref{claim:complin} to apply $\ket{y',y}\mapsto\ket{y'-A^{-1}y,y}$, the third step uses $\gCX^{+1}$ gates, and the fourth step uses $\gCX^{-1}$ gates. By \Cref{claim:complin}, this entire procedure takes space $\leq 16\kappa^2$ and time $\leq 16\log(\kappa)+70$. This circuit $\bar{\cQ}$ indeed satisfies~\Cref{eq:qimp} because letting $p$ denote the characteristic of $\bF_q,\bF_{q'}$ and letting $\omega_p=e^{2\pi i/p}$, then by definition
  \begin{align}
    \label{eq:hadalphred}
    \begin{split}
      \gH_{q'}\ket{\phi(x)}
      &= \sum_{z\in\bF_q^\kappa}\omega_p^{\tr_{\bF_q/\bF_p}(\tr_{\bF_{q'}/\bF_q}(\phi(x)\phi(z)))}\ket{\phi(z)} \\
      &= \sum_{z\in\bF_q^\kappa}\omega_p^{\tr_{\bF_q/\bF_p}(z^\top Ax)}\ket{\phi(z)} \\
      &= \phi \gH_q^{\otimes\kappa}\ket{Ax}.
    \end{split}
  \end{align}

  It remains to consider the case where $G\in\gZ^*$ or $G=\gInit_X$. For this purpose, by \Cref{def:gates,fact:Hconj}, $\gZ^a=\gH^\dagger X^{-a}\gH$ and $\gInit_X=\gH\circ\gInit_Z$. Therefore if $G$ equals $\gZ^a$ or $\gInit_X$, we let $\bar{\cQ}$ simply apply the circuits described above for $\gH^*$ along with $\gX^{-a}$ or $\gInit_Z$ respectively. In both cases the resulting $\bar{\cQ}$ uses space $\leq 16\kappa^2$ and time $\leq 32\log(\kappa)+250$.
\end{proof}

In \Cref{def:simgad} below, we specify the fault-tolerant circuit $\cQ_{G,D_{\mathrm{in}},D_{\mathrm{out}},T}$ over the small field $\bF_q$ that we use to simulate the action of each gate $G$ over the larger field $\bF_{q'}$, where $D_{\mathrm{in}},D_{\mathrm{out}}$ denote the input and output decorated codes for this circuit, and $T$ denotes the time usage. For all gates except classical function gates, we simply apply \Cref{thm:laft} to the appropriate logical circuit from \Cref{lem:qimp}. Meanwhile, to simulate a large-alphabet classical function gate, we first apply an associated small-alphabet classical function gate. We then initialize a fresh code state (using \Cref{thm:laft}), and apply $\gCX$ from the classical function output to this code state, before applying an error-correction gadget (again using \Cref{thm:laft}) and outputting the resulting state.

\begin{definition}
  \label{def:simgad}
  Let $q$ be a prime power, let $\kappa\in\bN$, and let $q'=q^\kappa$. For $\alpha\in\{\mathrm{in},\mathrm{out}\}$, let $D_\alpha=(C,\Enc,\cE_\alpha)\in\cD(q,u,n,k,\lambda_{\mathrm{run}})$ be decorated codes from the family in \Cref{def:ladec} with the same $C,\Enc$, with $\cE_{\mathrm{in}}=2^{[n]^u}|_{\geq\lambda_{\mathrm{in}}}$ and $\cE_{\mathrm{out}}=2^{[n]^u}|_{\geq\lambda_{\mathrm{out}}}$ for
  \begin{align*}
    \lambda_{\mathrm{in}} &= \lambda_{\max}(u,n) \\
    \lambda_{\mathrm{out}} &= \lambda_{\min}(u,n,\lambda_{\mathrm{run}}).
  \end{align*}
  Also choose $D_{\mathrm{in}},D_{\mathrm{out}}$ such that $K:=K(D_{\mathrm{in}})=K(D_{\mathrm{out}})$ is some set of size $|K|=\kappa$, and fix an isomorphism $K\cong[\kappa]$. We also choose $k=\lfloor n/8\rfloor$ (its maximum allowed value in \Cref{thm:laft}) and we require $\kappa\leq(k^u)^{1/3}/4$.

  For some $T\in\bN$ defined below and for every gate $G\in\cG=\{\gH^*,\gX^*,\gZ^*,\gCX^*,\gCCX^*,\gInit_*,\gCO_{Z,*}\}$ acting on $q'$-dimensional qudits, we define a circuit $\cQ=\cQ_{G,D_{\mathrm{in}},D_{\mathrm{out}},T}$ using time $T$ as follows:
  \begin{enumerate}
  \item\label{it:sgsmall} If $G\in\{\gH^*,\gX^*,\gZ^*,\gCX^*,\gCCX^*,\gInit_*\}$, so that $\bar{b}\in[3]$, let $\bar{\cQ}_G$ be the associated circuit given by \Cref{lem:qimp}, so that $\bar{\cQ}_G$ uses space $\leq 16\kappa^3\leq k^u$ and time $\leq O(\log\kappa)\leq O(u\log n)$, as $\kappa\leq(k^u)^{1/3}/4$. Then let $(\cQ^1,\cE_{\mathrm{run}}^1,D_{\mathrm{in}}^{\sqcup{\bar{b}}},D_{\mathrm{out}}^{\sqcup{\bar{b}}})$ and $(\cQ^2,\cE_{\mathrm{run}}^2,D_{\mathrm{in}}^{\sqcup{\bar{b}}},D_{\mathrm{out}}^{\sqcup{\bar{b}}})$ denote the mending fault-tolerant gadgets for the respective channels $\bar{O}_1=\bar{\cQ}_G$ and $\bar{O}_2=I_{K^{\sqcup\bar{b}}}$ given by \Cref{it:lascheme} in \Cref{thm:laft}. In particular, \Cref{thm:laft} provides such a gadget $\cQ^1$ because $\bar{\cQ}_G$ uses space $\leq k^u$. Let $\cQ^1,\cQ^2$ both use space $\subseteq N'$ and time $T_1,T_2$ respectively, with $|N'|=O(n)^u$, $T_1\leq O(u\log n\cdot u^4n^2\log^2n)\leq O(u^5n^2\log^3n)$, and $T_2\leq O(u^4n^2\log^2n)$. Here we use the fixed isomorphism $K\cong[\kappa]$ to associate the $\bar{b}\cdot|K|$ logical qudits of $D_{\mathrm{in}}^{\sqcup\bar{b}},D_{\mathrm{out}}^{\sqcup\bar{b}}$ with the $\bar{b}\cdot\kappa$ input and output qudits of $\bar{\cQ}_G$.

    For an integer $T\geq T_1+T_2$, we then define $\cQ=\cQ_{G,D_{\mathrm{in}},D_{\mathrm{out}},T}$ as follows. $\cQ$ acts on qudits $N'$, with input and output qudits $([n]^u)^{\sqcup\bar{b}}\subseteq N'$, and performs the following:
    \begin{enumerate}
    \item In timesteps $\{1,\dots,T_1\}$, $\cQ$ applies $\cQ^1$.
    \item In timesteps $\{T_1+1,\dots,T-T_2\}$, $\cQ$ idles (does nothing).
    \item In timesteps $\{T-T_2+1,\dots,T\}$, $\cQ$ applies $\cQ^2$.
    \end{enumerate}
  \item\label{it:sgbig} If $G\in\gCO_{Z,*}$, write $G=\gCO_{Z,f}$ for some function $f:\bF_{q'}^{\bar{b}}\rightarrow\bF_{q'}^{\bar{b}}$.


    Let $(\cQ^1,\cE_{\mathrm{run}}^,D_{\mathrm{in}},D_{\mathrm{out}})$ and $(\cQ^2,\cE_{\mathrm{run}}^,D_{\mathrm{in}},D_{\mathrm{out}})$ denote the mending fault-tolerant gadgets for the respective channels $\bar{O}_1=\gInit_Z^{\otimes K}$ and $\bar{O}_2=I_K$ given by \Cref{it:lascheme} in \Cref{thm:laft}. Let $\cQ^1,\cQ^2$ both use space $\subseteq N_1'$ and time $T_1,T_2$ respectively, with $|N_1'|=O(n)^u$, and $T_1,T_2\leq O(u^4n^2\log^2n)$. We assume that the input and output qudits of $\cQ^1,\cQ^2$ are labeled by\footnote{By inserting swap gates if necessary, we may assume without loss of generality that the input and output qudits of $\cQ^1,\cQ^2$ are labeled by the same subset $N_1$.} a subset $N_1\subseteq N_1'$, so that $N_1\cong[n]^u$.
    
    For an integer $T\geq T_1+T_2+3$, we now define the circuit $\cQ=\cQ_{G,D_{\mathrm{in}},D_{\mathrm{out}},T}$ as follows. $\cQ$ acts on a set of qudits labeled by $N'=(N_0\sqcup N_1')^{\sqcup\bar{b}}$, where $N_0\cong[n]^u$. $\cQ$ has input qudits $N_{\mathrm{in}}=N_0^{\sqcup\bar{b}}$ and output qudits $N_{\mathrm{out}}=N_1^{\sqcup\bar{b}}\subseteq{N_1'}^{\sqcup\bar{b}}$.
    $\cQ$ performs the following:
    \begin{enumerate}
    \item In the first timestep, $\cQ$ applies $\gInit_Z^{\otimes{N_1'}^{\sqcup\bar{b}}}$ to qudits in ${N_1'}^{\sqcup\bar{b}}$, and applies $\gCO_{Z,F}$ to qudits in $N_0^{\sqcup\bar{b}}$, where $F:\bF_q^{N_0^{\sqcup\bar{b}}}\rightarrow\bF_q^{N_0^{\sqcup\bar{b}}}$ is defined as follows. Fix a linear map $\Enc^0:\bF_q^K\rightarrow C_\alpha$ such that $\Enc^0(x)\in\Enc(x)$ for every $x\in\bF_q^K$; such a map $\Enc^0$ exists by the linearity of $\Enc$. For $y_0\in\bF_q^{N_0}$, define $x^{\min}(y_0)\in\bF_q^K$ by\footnote{See \Cref{footnote:setham}.}
      \begin{align}
        \label{eq:sgxmin}
        x^{\min}(y_0) &= \argmin_{x\in\bF_q^K}|y_0-\Enc(x)|,
      \end{align}
      where ties in the $\argmin$ above are broken in some way that is consistent across all $y_0$ within each coset of $C_Z$. That is, for every $y_0\in\bF_q^{N_0}$, every $x_0'\in\bF_q^K$, and every $y_0'\in\Enc(x)\subseteq C_Z$, we require
      \begin{equation}
        \label{eq:sgxmincosets}
        x^{\min}(y_0+y_0')=x^{\min}(y_0)+x_0'.
      \end{equation}
      Then for $y\in\bF_q^{N_0^{\sqcup\bar{b}}}$, we define
      \begin{align}
        \label{eq:sgF}
        F(y) &= (\Enc^0)^{\sqcup b}\circ(\phi_{q,q'}^{-1})^{\sqcup b}\circ f\circ\phi_{q,q'}^{\sqcup b}\circ(x^{\min})^{\sqcup b}(y),
      \end{align}
      where above we use the fixed isomorphism $K\cong[\kappa]$ to associate $\bF_q^K\cong\bF_q^\kappa$.
    \item In timesteps $\{2,\dots,T_1+1\}$, $\cQ$ applies $(\cQ^1)^{\sqcup\bar{b}}$ to qudits in ${N_1'}^{\sqcup\bar{b}}$.
    \item In timestep $T_1+2$, $\cQ$ applies $\gCX^{\otimes([n]^u)^{\sqcup\bar{b}}}$ with control qudits $N_0^{\sqcup\bar{b}}$ and target qudits $N_1^{\sqcup\bar{b}}$.
    \item In timestep $T_1+3$, $\cQ$ applies $\gInit_Z^{\otimes N_0^{\sqcup\bar{b}}}$ to qudits in $N_0^{\sqcup\bar{b}}$. In timesteps $\{T_1+4,\dots,T-T_2\}$, $\cQ$ idles (does nothing).
    \item In timesteps $\{T-T_2+1,\dots,T\}$, $\cQ$ applies $(\cQ^2)^{\sqcup\bar{b}}$ to qudits in ${N_1'}^{\sqcup\bar{b}}$.
    \end{enumerate}
  \end{enumerate}

  We then define $T$ to be the maximum value of $T_1+T_2+3$ over all values of $T_1,T_2$ in \Cref{it:sgsmall,it:sgbig}, so that $T=O(u^5n^2\log^3n)$. In particular, there exists an absolute constant $\eta_T>0$ such that
  \begin{equation}
    \label{eq:etaT}
    T \leq \eta_T\cdot 2^u\cdot n^3.
  \end{equation}
\end{definition}

\begin{remark}
  \label{remark:sgcirind}
  By \Cref{thm:laft}, the circuit $\cQ_{G,D_{\mathrm{in}},D_{\mathrm{out}},T}$ in \Cref{def:simgad} does not depend on the choice of $\lambda_{\mathrm{run}}$.
\end{remark}

We now define a condition called \emph{smooth-mending} for gadgets given by the circuits $\cQ_{G,D_{\mathrm{in}},D_{\mathrm{out}},T}$ defined in \Cref{def:simgad}, where $G\in\gCO_{Z,*}$ is a classical function gate. We cannot simply use the ordinary mending property for classical function gates because such gates may act on arbitrarily many qudits. Therefore in our fault-tolerant circuit $\cQ_{G,D_{\mathrm{in}},D_{\mathrm{out}},T}$ simulating a large-alphabet gate $G$, if a small fraction of the input blocks have large uncorrectable corruptions, we still want to simulate the desired gate $G$ with errors on a small fraction of the inputs. In contrast, the ordinary mending property is bimodal with respect to uncorrectable corruptions: either there is no uncorrectable corruption, in which case the desired logical operation is performed, or else there is an uncorrectable corruption, in which case a logical error may occur on all input qudits.

Specifically, letting $G$ act on $\bar{b}$ qudits, then we let $B_{\mathrm{in}}\subseteq[\bar{b}]$ denote the set of input blocks of $\cQ_{G,D_{\mathrm{in}},D_{\mathrm{out}},T}$ with ``good'' (i.e.~correctable) corruptions, and we let $B_{\mathrm{run}}\subseteq[\bar{b}]$ denote the set of blocks that experience good faults. Roughly speaking, we then define our gadget to be smoothly-mending if it fault-tolerantly simulates the desired gate $G$ up to logical errors on inputs in $B_{\mathrm{in}}$, and bad corruptions on outputs in $B_{\mathrm{run}}$.

Our precise definition of a bad fault for a smoothly-mending gadget does not allow for any errors during the execution of the initialization and error-correction subroutines (see the description of $\cQ_{G,D_{\mathrm{in}},D_{\mathrm{out}},T}$ above), but allows arbitrary errors outside of these subroutines. We will justify this definition by replacing the true fault with an extended fault (see \Cref{def:extfault}) that induces the same corruption, but has no support inside these subroutines. The main idea is to allow the adversary to swap out the relevant qudits to a private register, apply their corrupted subroutine there, and then swap the qudits back; the original subroutine is then only corrupted by the swap gates at the start and end.

\begin{definition}
  \label{def:smoothmend}
  Define all variables as in \Cref{def:simgad}. In particular, let $\cQ=\cQ_{G,D_{\mathrm{in}},D_{\mathrm{out}},T}$ be defined as in \Cref{def:simgad} for some fixed $G\in\gCO_{Z,*}$. Let $\lambda_{\mathrm{run}}\geq 0$.
  We say that
  \begin{equation*}
    (\cQ,\; \lambda_{\mathrm{run}},\; D_{\mathrm{in}}^{\sqcup\bar{b}},\; D_{\mathrm{out}}^{\sqcup\bar{b}})
  \end{equation*}
  forms a \emph{smoothly-mending fault-tolerant gadget} for
  \begin{equation*}
    \bar{O}[*] := \left\{\bar{O}[\bar{\zeta}]:=(\phi_{q,q'}^{-1})^{\sqcup\bar{b}}\circ G[\bar{\zeta}]\circ\phi_{q,q'}^{\sqcup\bar{b}}:\bar{\zeta}\in\bC^{\bF_{q'}^{\bar{b}}}\right\}
  \end{equation*}
  if the following holds for every $B_{\mathrm{in}},B_{\mathrm{run}}\subseteq[\bar{b}]$, every $\lambda_{\mathrm{run},1},\dots,\lambda_{\mathrm{run},\bar{b}}\in[0,\lambda_{\mathrm{run}}]$, and every $\ell\in\bN$.
  For $b\in[\bar{b}]$, let
  \begin{equation*}
    \lambda_{\mathrm{out},b} = \lambda_{\min}(u,n,\lambda_{\mathrm{run},b})=\eta_{\min}(u,n)\cdot\lambda_{\mathrm{run},b}
  \end{equation*}
  for $\lambda_{\min}(\cdot),\eta_{\min}(\cdot)$ defined as in \Cref{thm:laft}. Then let
  \begin{align*}
    \cE_{\mathrm{in},b} &= \begin{cases}
      \cE_{\mathrm{in}},&b\in B_{\mathrm{in}} \\
      \emptyset,&b\notin B_{\mathrm{in}}
    \end{cases} \\
    \cE_{\mathrm{run},b} &= \begin{cases}
      2^{(N_0\sqcup N_1')|_{\geq\lambda_{\mathrm{run},b}}^{\sqcup T}},&b\in B_{\mathrm{run}} \\
      2^{(N_1'\times(\{2,\dots,T_1\}\sqcup\{T-T_2+1,\dots,T-1\}))\sqcup((N_1'\setminus N_1)\times\{T_1+1,T\})},&b\notin B_{\mathrm{run}}
    \end{cases} \\
    \cE_{\mathrm{out},b} &= \begin{cases}
      2^{[n]^u}|_{\geq\lambda_{\mathrm{out},b}},&b\in B_{\mathrm{run}} \\
      \emptyset,&b\notin B_{\mathrm{run}}
    \end{cases},
  \end{align*}
  and let
  \begin{align*}
    \cE_{\mathrm{in}}' &= \bigsqcup_{b\in[\bar{b}]}\cE_{\mathrm{in},b} \sqcup [\ell] \\
    \cE_{\mathrm{run}}' &= \bigsqcup_{b\in[\bar{b}]}\cE_{\mathrm{run},b} \\
    \cE_{\mathrm{out}}' &= \bigsqcup_{b\in[\bar{b}]}\cE_{\mathrm{out},b} \sqcup [\ell]. \\
  \end{align*}
  Then for every $\rho\in\bC^{\bF_q^{\bar{b}\cdot\kappa+\ell}\times\bF_q^{\bar{b}\cdot\kappa+\ell}}$, every $\cE_{\mathrm{in}}'$-deviation $\sigma$ of $\Enc^{\sqcup\bar{b}}\otimes I_\ell(\rho)$, every $\cE_{\mathrm{run}}'$-avoiding fault $\cF$ and postselection $\zeta$ for $\cQ$, the output $\cQ[\cF,\zeta]\otimes I_\ell(\sigma)$ is of the form $\sum_{i\in[m]}\sigma_i$, where each $\sigma_i\in\bC^{\bF_q^{\bar{b}\cdot n^u+\ell}\times\bF_q^{\bar{b}\cdot n^u+\ell}}$ is a $\cE_{\mathrm{out}}'$-deviation of $(\Enc^{\sqcup\bar{b}}\circ \bar{O}[\bar{\zeta}_i]\circ L_i)\otimes I_\ell(\rho)$ for some postselection $\bar{\zeta}_i$ for $\bar{O}$ and some superoperator $L_i:\bC^{\bF_q^{\bar{b}\cdot\kappa}\times\bF_q^{\bar{b}\cdot\kappa}}\rightarrow\bC^{\bF_q^{\bar{b}\cdot\kappa}\times\bF_q^{\bar{b}\cdot\kappa}}$ supported inside $([\bar{b}]\setminus B_{\mathrm{in}})\times[\kappa]$.
\end{definition}

We now prove that our gadgets associated to $\cQ_{G,D_{\mathrm{in}},D_{\mathrm{out}},T}$ are mending for gates $G$ other than classical function gates, and are smoothly-mending for classical function gates $G$.

\begin{lemma}
  \label{lem:simgad}
  Define $q,q'=q^\kappa,\;D_{\mathrm{in}},D_{\mathrm{out}}\in\cD(q,u,n,k,\lambda_{\mathrm{run}}),\;K,T,Q_{G,D_{\mathrm{in}},D_{\mathrm{out}},T}$ as in \Cref{def:simgad}, and assume that
  \begin{equation}
    \label{eq:sglambound}
    \lambda_{\min}(u,n,\lambda_{\mathrm{run}})+T\lambda_{\mathrm{run}} \leq \lambda_{\max}(u,n).
  \end{equation}
  Then the following hold:
  \begin{enumerate}
  \item\label{it:sgnonCO} For every gate $G\in\cG=\{\gH^*,\gX^*,\gZ^*,\gCX^*,\gCCX^*,\gInit_*\}$ acting on some number $\bar{b}\in[3]$ of $q'$-dimensional qudits,
    \begin{align*}
      \left(\cQ_{G,D_{\mathrm{in}},D_{\mathrm{out}},T},\; \cE_{\mathrm{run}}=2^{N'}|_{\geq\lambda_{\mathrm{run}}}^{\sqcup T},\; D_{\mathrm{in}}^{\sqcup\bar{b}},\; D_{\mathrm{out}}^{\sqcup\bar{b}}\right)
    \end{align*}
    forms a mending fault-tolerant gadget for $\bar{O}_G=(\phi_{q,q'}^{-1})^{\sqcup\bar{b}}\circ G\circ\phi_{q,q'}^{\sqcup\bar{b}}$ over $q$-dimensional qudits using space $|N'|\leq O(n)^u$ and time $T\leq O(u^5n^2\log^3n)$.
  \item\label{it:sgCO} For every $G\in\gCO_{Z,*}$ acting on some number $\bar{b}\in\bN$ of $q'$-dimensional qudits,
    \begin{align*}
      \left(\cQ_{G,D_{\mathrm{in}},D_{\mathrm{out}},T},\; \lambda_{\mathrm{run}},\; D_{\mathrm{in}}^{\sqcup\bar{b}},\; D_{\mathrm{out}}^{\sqcup\bar{b}}\right)
    \end{align*}
    forms a smoothly-mending fault-tolerant gadget for $\bar{O}_G[*]=(\phi_{q,q'}^{-1})^{\sqcup\bar{b}}\circ G[*]\circ\phi_{q,q'}^{\sqcup\bar{b}}$ over $q$-dimensional qudits using space $N'=(N_0\sqcup N_1')^{\sqcup\bar{b}}$, so that $|N'|\leq \bar{b}\cdot O(n)^u$, and time $T\leq O(u^5n^2\log^3n)$.
  \end{enumerate}

\end{lemma}
\begin{proof}
  We prove the two items in the lemma statement separately:
  \begin{enumerate}
  \item It follows directly by \Cref{it:lascheme} in \Cref{thm:laft} along with \Cref{eq:sglambound} and \Cref{lem:seqcomp} that $\left(\cQ_{G,D_{\mathrm{in}},D_{\mathrm{out}},T},\; \cE_{\mathrm{run}},\; D_{\mathrm{in}}^{\sqcup\bar{b}},\; D_{\mathrm{out}}^{\sqcup\bar{b}}\right)$ is a mending fault-tolerant gadget for $\bar{O}_G=\bar{\cQ}_G$ defined as in \Cref{it:sgsmall} in \Cref{def:simgad}. By \Cref{lem:qimp}, this channel associated to $\bar{\cQ}_G$ precisely equals $(\phi_{q,q'}^{-1})^{\sqcup\bar{b}}\circ G\circ\phi_{q,q'}^{\sqcup\bar{b}}$, as desired.

    In slightly more detail, $\cQ=\cQ_{G,D_{\mathrm{in}},D_{\mathrm{out}},T}$ first applies $\left(\cQ^1,\; \cE_{\mathrm{run}},\; D_{\mathrm{in}}^{\sqcup\bar{b}},\; D_{\mathrm{out}}^{\sqcup\bar{b}}\right)$, which is a mending-fault-tolerant gadget for $\bar{O}_G$ by \Cref{it:lascheme} in \Cref{thm:laft}. $\cQ$ then idles for $\leq T$ timesteps and then applies the mending fault-tolerant gadget $\left(\cQ^2,\; \cE_{\mathrm{run}},\; D_{\mathrm{in}}^{\sqcup\bar{b}},\; D_{\mathrm{out}}^{\sqcup\bar{b}}\right)$ for the identity channel. By \Cref{lem:paulift}, it suffices to consider a Pauli fault, under which \Cref{eq:sglambound} ensures that a weight $<\lambda_{\min}(u,n,\lambda_{\mathrm{run}})$ error on an outputted code state of $\cQ^1$ will increase to at most a weight $<\lambda_{\max}(u,n)$ error passed as input to $\cQ^2$. Thus the desired result follows by the mending fault-tolerance of $\cQ^2$.
  \item Let $\cQ=\cQ_{G,D_{\mathrm{in}},D_{\mathrm{out}},T}$ and let $\bar{O}=\bar{O}_G$. Let $B_{\mathrm{in}},B_{\mathrm{out}}\subseteq[\bar{b}]$ and $\ell\in\bN$, and for $\alpha\in\{\mathrm{in},\mathrm{run},\mathrm{out}\}$ and $b\in[\bar{b}]$ define $\cE_{\alpha,b}$ and $\cE_\alpha'$ as in \Cref{def:smoothmend}. Let $\rho\in\bC^{\bF_q^{\bar{b}\cdot\kappa+\ell}\times\bF_q^{\bar{b}\cdot\kappa+\ell}}$, let $\sigma$ be a $\cE_{\mathrm{in}}'$-deviation of $\Enc^{\sqcup\bar{b}}\otimes I_\ell(\rho)$, and let $\cF$ be a $\cE_{\mathrm{run}}'$-avoiding fault and $\zeta$ be a postselection for $\cQ$. Our goal is to show that $\cQ[\cF,\zeta]\otimes I_\ell(\sigma)$ is a linear combination of $\cE_{\mathrm{out}}'$-deviations of $(\Enc^{\sqcup\bar{b}}\circ \bar{O}[\bar{\zeta}]\circ L)\otimes I_\ell(\rho)$ for some postselections $\bar{\zeta}$ for $\bar{O}$ and some superoperators $L:\bC^{\bF_q^{\bar{b}\cdot\kappa}\times\bF_q^{\bar{b}\cdot\kappa}}\rightarrow\bC^{\bF_q^{\bar{b}\cdot\kappa}\times\bF_q^{\bar{b}\cdot\kappa}}$ supported inside $([\bar{b}]\setminus B_{\mathrm{in}})\times[\kappa]$. By analogous reasoning as used to prove \Cref{lem:paulift}, we may decompose the error on $\sigma$ into a linear combination of Pauli errors, and decompose $\cF$ into a linear combination of Pauli faults, in order to assume that in fact $\sigma$ is a Pauli $\cE_{\mathrm{in}}'$-deviation of $\Enc^{\sqcup\bar{b}}\otimes I_\ell(\rho)$ and $\cF$ is a $\cE_{\mathrm{run}}'$-avoiding Pauli fault.

    For this purpose, define $F:\bF_q^{N_0^{\sqcup\bar{b}}}\rightarrow\bF_q^{N_0^{\sqcup\bar{b}}}$ as in \Cref{eq:sgF} in \Cref{def:simgad}. We can then write $F=F'\circ F_0$ for $F_0=(x^{\min})^{\sqcup\bar{b}}$ with $x^{\min}:\bF_q^{N_0}\rightarrow\bF_q^{K}$ given \Cref{eq:sgxmin}, and $F':\bF_q^{K^{\sqcup\bar{b}}}\rightarrow\bF_q^{N_0^{\sqcup\bar{b}}}$ given by
    \begin{align}
      \label{eq:Fpdef}
      F' &= (\Enc^0)^{\sqcup b}\circ(\phi_{q,q'}^{-1})^{\sqcup b}\circ f\circ\phi_{q,q'}^{\sqcup b}.
    \end{align}
    Therefore $\gCO_{Z,F}^\zeta=\gCO_{Z,F'}\circ\gCO_{Z,F_0}^\zeta$, where here we implicitly restrict $\zeta$ to the appropriate components $\bF_q^{N_0^{\sqcup\bar{b}}}$ corresponding to the $\gCO_{Z,F}$ gate applied in the first timestep of $\cQ$ (see \Cref{def:simgad}). We therefore begin by analyzing $\gCO_{Z,F_0}\otimes I_\ell(\sigma)=\gCO_{Z,x^{\min}}^{\sqcup\bar{b}}\otimes I_\ell(\sigma)$ in \Cref{claim:sgmeasure} below.

    Just as we decomposed our errors and faults into linear combinations of Paulis above, we may also decompose the (restricted) postselection $\zeta\in\bC^{\bF_q^{N_0^{\sqcup\bar{b}}}}=(\bC^{\bF_q^{N_0}})^{\otimes\bar{b}}$ into a linear combination of postselections of the form $\zeta^1\otimes\cdots\otimes\zeta^{\bar{b}}$ with each $\zeta^b\in\bC^{\bF_q^{N_0}}$, in order to assume that in fact $\zeta$ is of this form $\zeta=\zeta^1\otimes\cdots\otimes\zeta^{\bar{b}}$.
    
    \begin{claim}
      \label{claim:sgmeasure}
      There exists $\bar{\zeta}\in\bC^{\bF_q^{K^{\sqcup\bar{b}}}}$ and a Pauli superoperator $\bar{L}$ acting on qudits $K^{\sqcup\bar{b}}=[\bar{b}]\times K$ with
      \begin{equation}
        \label{eq:barLsupp}
        \supp(\bar{L}) \subseteq ([\bar{b}]\setminus B_{\mathrm{in}})\times K.
      \end{equation}
      such that
      \begin{align*}
        \gCO_{Z,F_0}^\zeta\otimes I_\ell(\sigma)
        &\propto (\gCO_{Z,I_{K^{\sqcup\bar{b}}}}^{\bar{\zeta}}\circ\bar{L})\otimes I_\ell(\rho),
      \end{align*}
      where $I_{K^{\sqcup\bar{b}}}:\bF_q^{K^{\sqcup\bar{b}}}\rightarrow\bF_q^{K^{\sqcup\bar{b}}}$ denotes the identity map.
    \end{claim}
    \begin{proof}
      For $\ell'\in\bN$, consider a state $\rho'\in\bC^{\bF_q^{|K|+\ell'}\times\bF_q^{|K|+\ell'}}$, and a Pauli $[\ell']$-deviation $\sigma'$ of $\Enc\otimes I_{\ell'}(\rho')$, so that we can write
      \begin{align*}
        \rho' &= \sum_{x,x'\in\bF_q^K}\ket{x}\bra{x'}\otimes\rho'_{x,x'}
      \end{align*}
      for some $\rho_{x,x'}\in\bC^{\bF_q^{\ell'}\times\bF_q^{\ell'}}$. Then we can write
      \begin{align*}
        \sigma' &= \sum_{x,x'\in\bF_q^K}Z^{e_Z}\ket{\Enc(x)+e_X}\bra{\Enc(x')+e_X'}(Z^\dagger)^{e_Z'} \otimes \sigma'_{x,x'},
      \end{align*}
      such that $e_X,e_X',e_Z,e_Z'\in\bF_q^{N_0}$ and $\sigma'_{x,x'}=\rho'_{x,x'}$. We will consider both a \emph{high-error case} (corresponding to $b\in[\bar{b}]\setminus B_{\mathrm{in}}$) where $e_X,e_X',e_Z,e_Z'$ may be arbitrarily high-weight, and a \emph{low-error case} (corresponding to $b\in B_{\mathrm{in}}$) where
      \begin{align}
        \label{eq:sgerr}
        |\supp(e_X)\cup\supp(e_X')\cup\supp(e_Z)\cup\supp(e_Z')| &< \lambda_{\max}(u,n) \leq d_u/8,
      \end{align}
      where $C$ has distance $\geq d_u$.

      Now for $\zeta^0\in\bC^{\bF_q^{N_0}}$,
      \begin{align}
        \label{eq:coxmin}
        \begin{split}
          \gCO_{Z,x^{\min}}^{\zeta^0}\otimes I_{\ell'}(\sigma')
          &= \sum_{x,x'\in\bF_q^K,y\in\bF_q^{N_0}} \zeta^0_y\cdot\ket{x^{\min}(y)}\bra{x^{\min}(y)} \\
          &\hspace{5em}\cdot\bra{y}Z^{e_Z}\ket{\Enc(x)+e_X}\bra{\Enc(x')+e_X'}(Z^\dagger)^{e_Z'}\ket{y} \otimes \sigma'_{x,x'}
        \end{split}
      \end{align}
      If $e_x-e_X'\notin C_Z$, then because $\Enc(x),\Enc(x')\subseteq C_Z$, the RHS of \Cref{eq:coxmin} vanishes, so assume that $e_X-e_X'\in C_Z$.

      In the \emph{low-error case} so that \Cref{eq:sgerr} holds, then by the definition of CSS code distance we have $e_X-e_X'\notin C_Z\setminus C_X^\perp$, and hence $e_X-e_X'\in C_X^\perp$. Then the expression inside the sum on the RHS of \Cref{eq:coxmin} is nonvanishing only when $\Enc(x)+e_X=\Enc(x')+e_X'$, or equivalently, when $x=x'$. Therefore letting $p$ denote the characteristic of $\bF_q$ and $\omega_p=e^{2\pi i/p}$, then the RHS of \Cref{eq:coxmin} is proportional to
      \begin{align*}
        \sum_{x\in\bF_q^K}\sum_{y\in\Enc(x)+e_X} \zeta^0_y\cdot\ket{x^{\min}(y)}\bra{x^{\min}(y)} \otimes \sigma'_{x,x} \cdot \omega_p^{\tr_{\bF_q/\bF_p}((e_Z-e_Z')\cdot y)}.
      \end{align*}
      Because $|e_X|<d_u/2$, by the definition of CSS code distance, for every $y\in\Enc(x)+e_X$ and every $x'\in\bF_q^K\setminus\{x\}$ we have $|y-\Enc(x')|>d_u/2$, and hence $x^{\min}(y)=x$. Thus
      \begin{align*}
        \gCO_{Z,x^{\min}}^{\zeta^0}\otimes I_{\ell'}(\sigma')
        &\propto \sum_{x\in\bF_q^K} \ket{x}\bra{x} \otimes \sigma'_{x,x} \cdot \sum_{y\in\Enc(x)+e_X}\zeta^0_y\cdot\omega_p^{\tr_{\bF_q/\bF_p}((e_Z-e_Z')\cdot y)}.
      \end{align*}
      Then defining $\bar{\zeta}^0\in\bF_q^K$ by
      \begin{align}
        \label{eq:barzeta0}
        \bar{\zeta}^0_x &= \sum_{y\in\Enc(x)+e_X}\zeta^0_y\omega_p^{\tr_{\bF_q/\bF_p}((e_Z-e_Z')\cdot y)},
      \end{align}
      then
      \begin{align}
        \label{eq:sgcosmallerr}
        \gCO_{Z,x^{\min}}^{\zeta^0}\otimes I_{\ell'}(\sigma')
        &\propto \sum_{x\in\bF_q^K} \bar{\zeta}^0_x\cdot\ket{x}\bra{x} \otimes \sigma'_{x,x}.
      \end{align}

      In instead the \emph{high-error case}, then \Cref{eq:sgerr} may not hold, so $e_X-e_X'$ may be an arbitrary element of $C_Z$. By \Cref{eq:sgxmincosets}, there exist $\bar{e}_X,\bar{e}_X'\in\bF_q^K$ such that for every $x,x'\in\bF_q^k$ and every $y\in\bF_q^{N_0}$ with $y\in\Enc(x)+e_X=\Enc(x')+e_X'$, then
      \begin{equation}
        \label{eq:barex}
        x^{\min}(y)=x+\bar{e}_X=x'+\bar{e}_X'.
      \end{equation}
      Hence \Cref{eq:coxmin} becomes
      \begin{align*}
        \hspace{1em}&\hspace{-1em}\gCO_{Z,x^{\min}}^{\zeta^0}\otimes I_{\ell'}(\sigma) \\
                    &\propto \sum_{x,x'\in\bF_q^K:x-x'=\bar{e}_X'-\bar{e}_X}\sum_{y\in\Enc(x)+e_X} \zeta^0_y \cdot \ket{x+\bar{e}_X}\bra{x'+\bar{e}_X'} \otimes \sigma'_{x,x'} \cdot \omega_p^{\tr_{\bF_q/\bF_p}((e_Z-e_Z')\cdot y)} \\
                    &= \sum_{x\in\bF_q^K} \ket{x}\bra{x} \otimes \sigma'_{x-\bar{e}_X,x-\bar{e}_X'} \cdot \sum_{y\in\Enc(x-\bar{e}_X)+e_X}\zeta^0_y\cdot\omega_p^{\tr_{\bF_q/\bF_p}((e_Z-e_Z')\cdot y)}
      \end{align*}
      Then defining $\bar{\zeta}^0\in\bF_q^K$ by
      \begin{align}
        \label{eq:barzeta0high}
        \bar{\zeta}^0_x &= \sum_{y\in\Enc(x-\bar{e}_X)+e_X}\zeta^0_y\cdot\omega_p^{\tr_{\bF_q/\bF_p}((e_Z-e_Z')\cdot y)},
      \end{align}
      we have
      \begin{align}
        \label{eq:sgcobigerr}
        \gCO_{Z,x^{\min}}^{\zeta^0}\otimes I_{\ell'}(\sigma')
        &\propto \sum_{x\in\bF_q^K} \bar{\zeta}^0_x \cdot \ket{x}\bra{x} \otimes \sigma'_{x-\bar{e}_X,x-\bar{e}_X'}.
      \end{align}

      Now recall that we may assume the restriction of the postselection $\zeta$ to the $\bar{b}$ respective sets of qudits labeled $N_0$ acted upon by $\gCO_{Z,F}$ in $\cQ$ has the form $\zeta=\zeta^1\otimes\cdots\otimes\zeta^{\bar{b}}$, where each $\zeta^b\in\bC^{\bF_q^{N_0}}$.
      Starting with the state $\rho'=\rho$, we apply $\gCO_{Z,x^{\min}}^{\zeta^b}\otimes I_{N_0^{\sqcup[\bar{b}]\setminus\{b\}}}\otimes I_\ell$ for $b=1,\dots,\bar{b}$ one at a time, where for each $b$ we apply the formula \Cref{eq:sgcosmallerr} if $b\in B_{\mathrm{in}}$ or the formula \Cref{eq:sgcobigerr} if $b\notin B_{\mathrm{in}}$ with $\ell'=|N_0^{\sqcup[\bar{b}]\setminus\{b\}}|+\ell$. We then update $\rho$ accordingly for the next step. In the $b$th step, the Pauli errors $Z^{e_Z}X^{e_X},(Z^{e_Z'}X^{e_X'})^\dagger$ are chosen to match the Pauli error on the $b$th block of $N_0$ qudits in $\sigma$, while we set $\bar{\zeta}^b$ to equal the resulting value of $\bar{\zeta}^0$ in \Cref{eq:barzeta0} or \Cref{eq:barzeta0high}, and we set $\bar{e}_X^b,(\bar{e}_X^b)'\in\bF_q^K$ to equal the resulting values of $\bar{e}_X,\bar{e}_X'$ satisfying \Cref{eq:barex}. Note that if $b\in B_{\mathrm{in}}$ then $\bar{e}_X^b=(\bar{e}_X^b)'=0$. We then let $\bar{\zeta}=\bar{\zeta}^1\otimes\cdots\otimes\bar{\zeta}^{\bar{b}}\in(\bC^{\bF_q^K})^{\otimes\bar{b}}=\bC^{\bF_q^{K^{\sqcup\bar{b}}}}$ and we define a Pauli superoperator $\bar{L}$ acting on qudits $K^{\sqcup\bar{b}}$ by $L(\rho)=X^{(\bar{e}_X^1,\dots,\bar{e}_X^{\bar{b}})}\rho(X^\dagger)^{((\bar{e}_X^1)',\dots,(\bar{e}_X^{\bar{b}})')}$, so that \Cref{eq:barLsupp} holds.
      By \Cref{eq:sgcosmallerr,eq:sgcobigerr}, the resulting state is
      \begin{align*}
        \gCO_{Z,F_0}^\zeta\otimes I_\ell(\sigma)
        &= \gCO_{Z,x^{\min}}^{\sqcup\bar{b}}[\zeta]\otimes I_\ell(\sigma) \\
        &\propto (\gCO_{Z,I_{K^{\sqcup\bar{b}}}}^{\bar{\zeta}}\circ\bar{L})\otimes I_\ell(\rho),
      \end{align*}
      as desired.
    \end{proof}

    We now apply \Cref{claim:sgmeasure} to compute $\gCO_{Z,F}^\zeta\otimes I_\ell(\sigma)$ in the following claim. Below, we define $\bar{\zeta},\bar{L}$ as in \Cref{claim:sgmeasure}. By abuse of notation we also denote the vector $\phi_{q,q'}^{\sqcup\bar{b}}(\bar{\zeta})\in\bF_{q'}^{\bar{b}}$ by $\bar{\zeta}$. Furthermore, we extend the linear map $\Enc^0:\bF_q^K\rightarrow C_Z$ to an isometry $\Enc^0:\bC^{\bF_q^K\times\bF_q^K}\rightarrow\bC^{\bF_q^{N_0}\times\bF_q^{N_0}}$ in the natural way, that is $\Enc^0(\ket{x}\bra{x'})=\ket{\Enc^0(x)}\bra{\Enc^0(x')}$.

    \begin{claim}
      \label{claim:sgcores}
      We have
      \begin{align*}
        \gCO_{Z,F}^\zeta\otimes I_\ell(\sigma)
        &\propto ((\Enc^0)^{\sqcup\bar{b}}\circ\bar{O}[\bar{\zeta}]\circ\bar{L})\otimes I_\ell(\rho).
      \end{align*}
    \end{claim}
    \begin{proof}
      By \Cref{claim:sgmeasure,eq:Fpdef},
      \begin{align*}
        \gCO_{Z,F}^\zeta\otimes I_\ell(\sigma)
        &= (\gCO_{Z,F'}\circ\gCO_{Z,F_0}^\zeta)\otimes I_\ell(\sigma) \\
        &\propto (\gCO_{Z,F'}\circ\gCO_{Z,I_{K^{\sqcup\bar{b}}}}^{\bar{\zeta}}\circ\bar{L})\otimes I_\ell(\rho) \\
        &= ((\Enc^0)^{\sqcup\bar{b}}\circ\gCO_{Z,(\phi_{q,q'}^{-1})^{\sqcup b}\circ f\circ\phi_{q,q'}^{\sqcup b}}^{\bar{\zeta}}\circ\bar{L})\otimes I_\ell(\rho) \\
        &= ((\Enc^0)^{\sqcup\bar{b}}\circ(\phi_{q,q'}^{-1})^{\sqcup\bar{b}}\circ\gCO_{Z,f}^{\bar{\zeta}}\circ\phi_{q,q'}^{\sqcup\bar{b}}\circ\bar{L})\otimes I_\ell(\rho) \\
        &= ((\Enc^0)^{\sqcup\bar{b}}\circ\bar{O}[\bar{\zeta}]\circ\bar{L})\otimes I_\ell(\rho).
      \end{align*}
    \end{proof}

    Let $\sigma_1$ denote the state of the qudits labeled $N_1^{\sqcup\bar{b}}$ following the completion of timestep $T_1+3$ in $\cQ$ on input $\sigma$, with fault $\cF$ and postselection $\zeta$. 

    \begin{claim}
      \label{claim:sgpreec}
      The state $\sigma_1$ is a linear combination of states of the form
      \begin{equation}
        \label{eq:sgpreec}
        (P\circ\Enc^{\sqcup\bar{b}}\circ\bar{O}[\bar{\zeta}']\circ\bar{L})\otimes I_\ell(\rho)
      \end{equation}
      for $\cE_1$-avoiding Paulis $P$ on qudits $N_1^{\sqcup\bar{b}}$ and for postselections $\bar{\zeta}'\in\bC^{\bF_{q'}^{\bar{b}}}$, where
      \begin{equation*}
        \cE_1 = \bigsqcup_{b\in[\bar{b}]}\cE'_{1,b} \hspace{1em}\text{for}\hspace{1em} \cE'_b=\begin{cases}
          2^{N_1}|_{\geq\lambda_{\min}(u,n,\lambda_{\mathrm{run},b})+(T_1+4)\lambda_{\mathrm{run},b}},&b\in B_{\mathrm{run}}\\
          \emptyset,&b\notin B_{\mathrm{run}}.
        \end{cases}
      \end{equation*}
    \end{claim}
    \begin{proof}
      In a slight abuse of notation, for $x\in\bF_q^{K^{\sqcup\bar{b}}}$, we let $f(x)\in\bF_q^{K^{\sqcup\bar{b}}}$ denote the vector $(\phi_{q,q'}^{-1})^{\sqcup\bar{b}}\circ f\circ\phi_{q,q'}^{\sqcup\bar{b}}(x)$.
      Then we may decompose
      \begin{equation}
        \label{eq:sgOLdecomp}
        (\bar{O}[\bar{\zeta}]\circ\bar{L})\otimes I_\ell(\rho) = \sum_{x\in\bF_q^{K^{\sqcup\bar{b}}}}\bar{\zeta}_x\cdot\ket{f(x)}\bra{f(x)}\otimes\rho^1_x
      \end{equation}
      for some $\rho^1_x\in\bC^{\bF_q^\ell\times\bF_q^\ell}$ that only depends on $\rho,\bar{L},x$.
      
      Then by \Cref{claim:sgcores}, the state of qudits in $N_0^{\sqcup\bar{b}}$ just following the $\gCO_{Z,F}^\zeta$ gate in timestep $1$ (but before any errors from the fault are applied) is proportional to
      \begin{align*}
        ((\Enc^0)^{\sqcup\bar{b}}\circ\bar{O}[\bar{\zeta}]\circ\bar{L})\otimes I_\ell(\rho)
        &= \sum_{x\in\bF_q^{K^{\sqcup\bar{b}}}}\bar{\zeta}_x\cdot\ket{(\Enc^0)^{\sqcup\bar{b}}(f(x))}\bra{(\Enc^0)^{\sqcup\bar{b}}(f(x))}\otimes\rho^1_x.
      \end{align*}
      Therefore by the definition of $\cQ^1$ and $\cF$, following timestep $T_1+1$ (now including fault errors as well), the state of qudits in $N_0^{\sqcup\bar{b}}\sqcup N_1^{\sqcup\bar{b}}$ is a linear combination of states of the form
      \begin{align*}
        \sum_{x\in\bF_q^{K^{\sqcup\bar{b}}}} &\bar{\zeta}_x\cdot X^{e_0,X}Z^{e_0,Z}\ket{(\Enc^0)^{\sqcup\bar{b}}(f(x))}\bra{(\Enc^0)^{\sqcup\bar{b}}(f(x))}(Z^\dagger)^{e_{0,Z}'}(X^\dagger)^{e_{0,X}'}\otimes\rho^1_x \\
                                             &\otimes X^{e_{1,X}}Z^{e_{1,Z}}\ket{\Enc^{\sqcup\bar{b}}(0)}\bra{\Enc^{\sqcup\bar{b}}(0)}(Z^\dagger)^{e_{1,Z}'}(X^\dagger)^{e_{1,X}'}
      \end{align*}
      for some $e_{0,X},e_{0,Z},e_{0,X}',e_{0,Z}'\in\bF_q^{N_0^{\sqcup\bar{b}}}$, $e_{1,X},e_{1,Z},e_{1,X}',e_{1,Z}'\in\bF_q^{N_1^{\sqcup\bar{b}}}$. Furthermore, if $b\in B_{\mathrm{run}}$, then writing $e_{i,\alpha}|_b\in\bF_q^{[n]^u}$ to denote the restriction of $e_{i,\alpha}$ to the $b$th set of indices labeled $[n]^u\cong N_0\cong N_1$, we have
      \begin{align*}
        |(\supp(e_{0,X}|_b)\cup\supp(e_{0,Z}|_b)\cup\supp(e_{0,X}'|_b)\cup\supp(e_{0,Z}'|_b))|
        &< (T_1+1)\lambda_{\mathrm{run},b} \\
        |\supp(e_{1,X}|_b)\cup\supp(e_{1,Z}|_b)\cup\supp(e_{1,X}'|_b)\cup\supp(e_{1,Z}'|_b)|
        &< \lambda_{\min}(u,n,\lambda_{\mathrm{run},b})+\lambda_{\mathrm{run},b}.
      \end{align*}
      Here we use the fact that \Cref{it:lascheme} in \Cref{thm:laft} provides the same circuit $\cQ^1$ for all $\lambda_{\mathrm{run},b}\leq\lambda_{\mathrm{run}}$ (see \Cref{remark:sgcirind}).
      
      Then following the $\gCX^1$ gates in timestep $T_1+2$, which use $N_0^{\sqcup\bar{b}}$ as controls and $N_1^{\sqcup\bar{b}}$ as targets, by the definition of $\Enc^0$ and $\Enc$ we obtain a linear combination of states of the form
      \begin{align}
        \label{eq:sgpostcx}
        \begin{split}
          \sum_{x\in\bF_q^{K^{\sqcup\bar{b}}}} &\bar{\zeta}_x\cdot X^{e_2,X}Z^{e_2,Z}\ket{(\Enc^0)^{\sqcup\bar{b}}(f(x))}\bra{(\Enc^0)^{\sqcup\bar{b}}(f(x))}(Z^\dagger)^{e_{2,Z}'}(X^\dagger)^{e_{2,X}'}\otimes\rho^1_x \\
                                               &\otimes X^{e_{3,X}}Z^{e_{3,Z}}\ket{\Enc^{\sqcup\bar{b}}(f(x))}\bra{\Enc^{\sqcup\bar{b}}(f(x))}(Z^\dagger)^{e_{3,Z}'}(X^\dagger)^{e_{3,X}'}
        \end{split}
      \end{align}
      for some $e_{2,X},e_{2,Z},e_{2,X}',e_{2,Z}'\in\bF_q^{N_0^{\sqcup\bar{b}}}$, $e_{3,X},e_{3,Z},e_{3,X}',e_{3,Z}'\in\bF_q^{N_1^{\sqcup\bar{b}}}$. Because $\gCX^1$ gates are Cliffords and hence propagate Pauli errors to Pauli errors, for $b\in B_{\mathrm{run}}$ we have
      \begin{align*}
        |\supp(e_{3,X}|_b)\cup\supp(e_{3,Z}|_b)\cup\supp(e_{3,X}'|_b)\cup\supp(e_{3,Z}'|_b)|
        &< \lambda_{\min}(u,n,\lambda_{\mathrm{run},b})+(T_1+3)\lambda_{\mathrm{run},b}.
      \end{align*}
      
      Now timestep $T_1+3$ simply applies $\gInit_Z$ gates to qudits $N_0^{\sqcup\bar{b}}$ and these qudits are not used again in $\cQ$. Hence this step has the effect of tracing out qudits $N_0^{\sqcup\bar{b}}$, which are given by the register on the first line of \Cref{eq:sgpostcx}. If $e_{2,X}\neq e_{2,X}'$, then tracing out this register collapses the state to $0$. Otherwise, if $e_{2,X}=e_{2,X}'$, then defining $\bar{\zeta}'\in\bC^{\bF_q^{K^{\sqcup\bar{b}}}}$ by
      \begin{align*}
        \bar{\zeta}_x'
        &= \bar{\zeta}_x\cdot\omega_p^{\tr_{\bF_q/\bF_p}(e_{2,Z}-e_{2,Z}'\cdot(\Enc^0)^{\sqcup\bar{b}}(f(x)))},
      \end{align*}
      then tracing out this register collapses the state to
      \begin{align*}
        \sum_{x\in\bF_q^{K^{\sqcup\bar{b}}}}\bar{\zeta}'_x\cdot X^{e_{3,X}}Z^{e_{3,Z}}\ket{\Enc^{\sqcup\bar{b}}(f(x))}\bra{\Enc^{\sqcup\bar{b}}(f(x))}(Z^\dagger)^{e_{3,Z}'}(X^\dagger)^{e_{3,X}'}\otimes\rho^1_x.
      \end{align*}
      Thus if we also apply the Pauli error from the fault in timestep $T_1+3$, we obtain a linear combination of states of the desired form
      \begin{align}
        \label{eq:sgalmost}
        \begin{split}
          \hspace{1em}&\hspace{-1em} \sum_{x\in\bF_q^{K^{\sqcup\bar{b}}}}\bar{\zeta}'_x\cdot P\left(\ket{\Enc^{\sqcup\bar{b}}(f(x))}\bra{\Enc^{\sqcup\bar{b}}(f(x))}\right)\otimes\rho^1_x \\
                      &= (P\circ\Enc^{\sqcup\bar{b}})\otimes I_\ell\left(\sum_{x\in\bF_q^{K^{\sqcup\bar{b}}}}\bar{\zeta}'_x\cdot\ket{f(x)}\bra{f(x)}\otimes\rho^1_x\right) \\
                      &= (P\circ\Enc^{\sqcup\bar{b}}\circ\bar{O}[\bar{\zeta}']\circ\bar{L})\otimes I_\ell(\rho)
        \end{split}
      \end{align}
      for Pauli errors $P$ that for every $b\in B_{\mathrm{run}}$ satisfy
      \begin{align*}
        |P|_b|
        &< \lambda_{\min}(u,n,\lambda_{\mathrm{run},b})+(T_1+4)\lambda_{\mathrm{run},b},
      \end{align*}
      where the final equality in \Cref{eq:sgalmost} holds by \Cref{eq:sgOLdecomp}.
    \end{proof}

    Let $\sigma_2$ denote the state of qudits $N_1^{\sqcup\bar{b}}$ following timestep $T-T_2$, which is passed as input to $\cQ^2$. Then $\sigma_2$ is given by $\sigma_1$ from \Cref{claim:sgpreec} with the additional Pauli errors accumulated from timesteps $T_1+4,\dots,T-T_2$. Therefore $\sigma_2$ is a linear combination of states of the same form \Cref{eq:sgpreec} described in \Cref{claim:sgpreec} for postselections $\bar{\zeta}'\in\bC^{\bF_{q'}^{\bar{b}}}$, but now where the Paulis $P$ are $\cE_2$-avoiding for
    \begin{equation*}
      \cE_2 = \bigsqcup_{b\in[\bar{b}]}\cE'_{2,b} \hspace{1em}\text{for}\hspace{1em} \cE'_b=\begin{cases}
        2^{N_1}|_{\geq\lambda_{\min}(u,n,\lambda_{\mathrm{run},b})+(T-T_2+1)\lambda_{\mathrm{run},b}},&b\in B_{\mathrm{run}}\\
        \emptyset,&b\notin B_{\mathrm{run}}.
      \end{cases}
    \end{equation*}
    Then by the assumption in \Cref{eq:sglambound} that
    \begin{equation*}
      \lambda_{\min}(u,n,\lambda_{\mathrm{run}})+(T-T_2+1)\lambda_{\mathrm{run}} \leq \lambda_{\max}(u,n),
    \end{equation*}
    and because each $\lambda_{\mathrm{run},b}\leq\lambda_{\mathrm{run}}$, the mending fault-tolerance of $(\cQ^2,\cE_{\mathrm{run}}^2,D_{\mathrm{in}}^{\sqcup{\bar{b}}},D_{\mathrm{out}}^{\sqcup{\bar{b}}})$ given by \Cref{it:lascheme} in \Cref{thm:laft} (see \Cref{def:simgad}) ensures that the final output $\sigma_3$ given by applying $\cQ^2$ to $\sigma_2$ with fault $\cF$ and postselection $\bar{\zeta}$ is a linear combination of $\cE_{\mathrm{out}}'$-deviations of states of the form
    \begin{equation*}
      (\Enc^{\sqcup\bar{b}}\circ\bar{L}'\circ\bar{O}[\bar{\zeta}']\circ\bar{L})\otimes I_\ell(\rho).
    \end{equation*}
    for superoperators $\bar{L}'$ supported on qudits $K^{\sqcup([\bar{b}]\setminus B_{\mathrm{run}})}$. Here we use the fact that \Cref{it:lascheme} in \Cref{thm:laft} provides the same circuit $\cQ^2$ for all $\lambda_{\mathrm{run},b}\leq\lambda_{\mathrm{run}}$ (see \Cref{remark:sgcirind}). We may decompose each $\bar{L}'$ into a linear combination of Pauli superoperators by \Cref{lem:paulidecomp}. Then by the definition of $\Enc$, there exists a Pauli superoperator $P'$ supported on qudits $N_1^{\sqcup([\bar{b}]\setminus B_{\mathrm{run}})}$ for which
    \begin{equation*}
      \Enc^{\sqcup\bar{b}}\circ\bar{L}' = P'\circ\Enc^{\sqcup\bar{b}}.
    \end{equation*}
    That is, $P'$ is simply a physical Pauli realization of the logical Pauli $\bar{L}'$ for the CSS code encoded by $\Enc^{\sqcup\bar{b}}$. Therefore our final output $\sigma_3$ is a linear combination of $\cE_{\mathrm{out}}'$-deviations of states of the form
    \begin{equation*}
      (\Enc^{\sqcup\bar{b}}\circ\bar{O}[\bar{\zeta}']\circ\bar{L})\otimes I_\ell(\rho),
    \end{equation*}
    as desired, where the Paulis $P'$ are absorbed into the $\cE_{\mathrm{out}}'$-deviation.
  \end{enumerate}

\end{proof}

In \Cref{lem:simcomp} below, we apply \Cref{lem:simgad} to prove our main result on simulative composition. We first need the following notation regarding concatenated codes.

\begin{definition}
  \label{def:concat}
  Let $q$ be a prime power, let $\kappa\in\bN$, and let $q'=q^\kappa$. Let $C=(C_X,C_Z)$ be a $[[n,\kappa,d]]_q$ CSS non-subsystem code with encoding map $\Enc:\bF_q^\kappa\rightarrow C_Z/C_X^\perp$. Let $C'=(C'_X,C'_Z)$ be a $[[n',k',d']]_{q'}$ CSS non-subsystem code with encoding map $\Enc':\bF_{q'}^{k'}\rightarrow\bF_{q'}^{n'}$. Let $\Enc^\perp,{\Enc'}^\perp$ be the respective dual encoding maps (see \Cref{lem:dualenc}).

  Then the \emph{concatenated code and encoding map}
  \begin{equation*}
    (\tilde{C},\widetilde{\Enc}) = (C,\Enc)\concat(C',\Enc')
  \end{equation*}
  are defined as follows.
  First, for $x\in(\bF_q^\kappa)^{k'}$, we define a subset $\widetilde{\Enc}(x)\subseteq(\bF_q^n)^{n'}$ by
  \begin{equation*}
    \widetilde{\Enc}(x) = \bigcup\Enc^{\sqcup n'}\circ(\phi_{q,q'}^{-1})^{\sqcup n'}\circ\Enc'\circ\phi_{q,q'}^{\sqcup k'}(x).
  \end{equation*}
  As a point of notation above, recall that elements in the image of CSS encoding maps are cosets; therefore for a set $S\subseteq(\bF_q^\kappa)^{\sqcup n'}$ (such as $S=(\phi_{q,q'}^{-1})^{\sqcup n'}\circ\Enc'\circ\phi_{q,q'}^{\sqcup k'}(x)$), then $\bigcup\Enc^{\sqcup n'}(S)$ denotes the union of the cosets $\Enc^{\sqcup n'}(s)$ over all $s\in S$.

  Then we define $\tilde{C}=(\tilde{C}_X,\tilde{C}_Z)$ to be the CSS non-subsystem code given by
  \begin{align*}
    \tilde{C}_X &= \widetilde{\Enc}(0)^\perp \\
    \tilde{C}_Z &= \bigcup\widetilde{\Enc}((\bF_q^\kappa)^{k'}).
  \end{align*}
  By definition, $\tilde{C}_X,\tilde{C}_Z\subseteq(\bF_q^n)^{n'}$ are linear subspaces, and for each $x\in(\bF_q^\kappa)^{k'}$, then $\widetilde{\Enc}(x)$ is a distinct coset in $\tilde{C}_Z/\tilde{C}_X^\perp$. Thus $\tilde{C}$ is a well-defined $[[nn',\kappa k]]_q$ CSS non-subsystem code with encoding map $\widetilde{\Enc}$.
\end{definition}

The following lemma is well known, though we provide a brief proof sketch for intuition.

\begin{lemma}[Well known]
  \label{lem:concatdis}
  The concatenated code $\tilde{C}$ defined in \Cref{def:concat} has distance $dd'$.
\end{lemma}
\begin{proof}[Proof sketch]
  Assume for a contradiction that the physical realization of some nontrivial logical Pauli operator $L$ acting on $\tilde{C}$ has weight $<dd'$. By definition, a code state of $\tilde{C}$ consists of $n'$ (entangled) code states of $C$, each of which in turn contains $n$ qudits. Therefore the restriction of $L$ to each of these $n'$ code states of $C$ must be a logical operator for $C$, and hence either have weight $\geq d$ or else act trivially on the code space. Now if $<d'$ of these $n'$ restrictions of $L$ have weight $\geq d$, then $(\phi_{q,q'}\circ\Enc^{-1})^{\sqcup n'}\circ L\circ(\Enc\circ\phi_{q,q'}^{-1})^{\sqcup n'}$ acts nontrivially on $<d'$ components of $C'$, and hence must act trivially on the code space of $C'$, a contradiction. Therefore we must instead have $\geq d'$ of the $n'$ restrictions of $L$ to code states of $C$ have weight $\geq d$, meaning that $L$ has weight $\geq dd'$.
\end{proof}

\begin{lemma}
  \label{lem:simcomp}
  Let $u,n\in\bN$, and let $q=q_0^{\prod_{i\in[u]}n_i'}$ with $q_0\geq\max_{i\in[u]}n_i$ a prime power and each $n_i'\geq n_i$ an integer. Let $q'=q^\kappa$ for a positive integer $\kappa\leq\lfloor n/8\rfloor^{u/3}/4$. Define $\bar{\lambda}_{\mathrm{run}}=\bar{\lambda}_{\mathrm{run}}(u,n)$ as in \Cref{thm:laft} (see \Cref{eq:lamrun}), define $T,\eta_T$ as in \Cref{def:simgad} (see \Cref{eq:etaT}), and let $\lambda_{\mathrm{run}}\in[0,\bar{\lambda}_{\mathrm{run}}/\eta_T]$. For $\alpha\in\{\mathrm{in},\mathrm{out}\}$, let
  \begin{equation*}
    D_\alpha = (C,\Enc,\cE_\alpha=2^{[n]^u}|_{\geq\lambda_\alpha})
  \end{equation*}
  be the $[[n^u,\;\kappa,\;\geq d_u=(n/8)^u]]_q$ decorated CSS non-subsystem code in \Cref{def:simgad}, and for some $\lambda_\alpha'\geq 0$ let
  \begin{equation*}
    D_\alpha' = (C_\alpha',\; \Enc_\alpha',\; \cE_\alpha'=2^{[n_\alpha']}|_{\geq\lambda_\alpha'})
  \end{equation*}
  be a $[[n_\alpha',k_\alpha',d_\alpha']]_{q'}$ decorated CSS non-subsystem code.
  For some $\lambda_{\mathrm{run}}'\geq 0$, let
  \begin{equation*}
    (\cQ',\; \cE_{\mathrm{run}}'=2^{N'}|_{\geq\lambda_{\mathrm{run}}'}^{\sqcup T'},\; D_{\mathrm{in}}',\; D_{\mathrm{out}}')
  \end{equation*}
  be a fault-tolerant gadget for a set $\bar{\cO}'$ of superoperators $\bar{O}':\bC^{\bF_{q'}^{k_{\mathrm{in}}'}\times\bF_{q'}^{k_{\mathrm{in}}'}}\rightarrow\bC^{\bF_{q'}^{k_{\mathrm{out}}'}\times\bF_{q'}^{k_{\mathrm{out}}'}}$ on $q'$-dimensional qudits, using space $N'$, time $T'$, and gate set $\cG=\{\gH^*,\gX^*,\gZ^*,\gCX^*,\gCCX^*,\gInit_*,\gCO_*\}$.

  Define
  \begin{align}
    \label{eq:lampps}
    \begin{split}
      \lambda_{\mathrm{in}}'' &= \lambda_{\mathrm{in}}'-\lambda_{\mathrm{run}}'/3 \\
      \lambda_{\mathrm{run}}'' &= \lambda_{\mathrm{run}}'/3 \\
      \lambda_{\mathrm{out}}'' &= \lambda_{\mathrm{out}}'.
    \end{split}
  \end{align}
  Then for every
  \begin{align}
    \label{eq:sclamtil}
    \tilde{\lambda}_{\mathrm{run}} &\leq \frac{\lambda_{\mathrm{run}}\cdot\min\{\lambda_{\mathrm{in}}'',\lambda_{\mathrm{run}}''\}}{18T},
  \end{align}
  there exists a fault-tolerant gadget on $q$-dimensional qudits
  \begin{equation*}
    (\tilde{\cQ},\; \tilde{\cE}_{\mathrm{run}}=2^{\tilde{N}}|_{\geq\tilde{\lambda}_{\mathrm{run}}}^{\sqcup\tilde{T}},\; \tilde{D}_{\mathrm{in}},\; \tilde{D}_{\mathrm{out}})
  \end{equation*}
  for $\tilde{\bar{\cO}}=(\phi_{q,q'}^{-1})^{\sqcup k_{\mathrm{out}}'}\circ\bar{\cO}'\circ\phi_{q,q'}^{\sqcup k_{\mathrm{in}}'}$, where:
  \begin{itemize}
  \item $\tilde{\cQ}$ is a quantum circuit using space $|\tilde{N}|\leq |N'|\cdot O(n)^u$ and time $\tilde{T}\leq T'\cdot O(u^5n^2\log^3n)$.
  \item For $\alpha\in\{\mathrm{in},\mathrm{out}\}$, the $[[\tilde{n}_\alpha=n^un_\alpha',\; \tilde{k}_\alpha=\kappa k_\alpha',\; \tilde{d}_\alpha\geq(n/8)^ud_\alpha']]_q$ decorated code $\tilde{D}_\alpha$ is given by
    \begin{equation}
      \label{eq:sctDa}
      \tilde{D}_\alpha = ((\tilde{C},\widetilde{\Enc})=(C,\Enc)\concat(C_\alpha',\Enc_\alpha'),\; \tilde{\cE}_\alpha=2^{[\tilde{n}_\alpha]}|_{\geq\tilde{\lambda}_\alpha}),
    \end{equation}
    (see \Cref{def:concat,lem:concatdis}) such that
    \begin{align}
      \label{eq:sclams}
      \begin{split}
        \tilde{\lambda}_{\mathrm{in}} &= \frac{\lambda_{\mathrm{in}}\cdot\lambda_{\mathrm{in}}''}{6} \\
        \tilde{\lambda}_{\mathrm{out}} &= \left(\frac{9n^u}{\lambda_{\mathrm{run}}}+\eta_{\min}(u,n)\right)\cdot T\cdot\tilde{\lambda}_{\mathrm{run}}+n^u\cdot\lambda_{\mathrm{out}}'',
      \end{split}
    \end{align}
    where $\eta_{\min}(u,n)$ is defined as in \Cref{thm:laft} (see \Cref{eq:lamminmax}).
  \end{itemize}
\end{lemma}

As discussed above, our proof of \Cref{lem:simcomp} below uses swap gates to transform a general Pauli fault on the composed circuit $\tilde{\cQ}$ into an \emph{extended fault} with a nicer structure (see \Cref{def:extfault}). A similar technique was used to deal with general adversarial noise in \cite{he_composable_2025}.

\begin{proof}[Proof of \Cref{lem:simcomp}]
  First, we define $\cQ''$ to be the circuit using gate set $\{\gH^*,\gX^*,\gZ^*,\gCX^*,\gCCX^*,\gInit_*,\gCO_{Z,*}\}$ and time $T''=3T'+1$ as follows. The first $3T'$ timesteps of $\cQ''$ are obtained by replacing each timestep in $\cQ'$ with three timesteps in $\cQ''$. Specifically, for every $\bar{b}$-qudit gate $G$ in $\cQ'$, if $G\in\gCO_{X,*}$ then $G$ is replaced by $\gH^{\otimes\bar{b}}\circ G\circ(\gH^\dagger)^{\otimes\bar{b}}$ in $\cQ''$, while otherwise if $G\in\{\gH^*,\gX^*,\gZ^*,\gCX^*,\gCCX^*,\gInit_*,\gCO_{Z,*}\}$ then is replaced by $I_{\bar{b}}\circ G\circ I_{\bar{b}}$ in $\cQ''$. The final ($T''$th) timestep of $\cQ''$ then applies the identity to all output qudits in $[n_{\mathrm{out}}']$, and applies $\gInit_Z$ to all other qudits in $N'\setminus[n_{\mathrm{out}}']$. Because Hadamard gates are Clifford and hence propagate Pauli errors to Pauli errors, \Cref{lem:paulift} implies that 
  \begin{equation*}
    (\cQ'',\; \cE_{\mathrm{run}}''=2^{N'}|_{\geq\lambda_{\mathrm{run}}''}^{\sqcup T''},\; D_{\mathrm{in}}'',\; D_{\mathrm{out}}'')
  \end{equation*}
  is a fault-tolerant gadget for $\bar{\cO}'$, where for $\alpha\in\{\mathrm{in},\mathrm{out}\}$ we have
  \begin{equation*}
    D_\alpha'' = (C_\alpha',\;\Enc_\alpha',\;\cE_\alpha''=2^{[n_\alpha']}|_{\geq\lambda_\alpha''})
  \end{equation*}
  with $\lambda_{\mathrm{in}}'',\lambda_{\mathrm{run}}'',\lambda_{\mathrm{out}}''$ given by \Cref{eq:lampps}.
  
  We now define $\tilde{\cQ}$ as follows. Define $T\in\bN$ as in \Cref{def:simgad}. Let $N$ be a set of size $|N|=O(n)^u$ such that for every $\bar{b}$-qudit gate $G\in\{\gH^*,\gX^*,\gZ^*,\gCX^*,\gCCX^*,\gInit_*,\gCO_{Z,*}\}$, the circuit $\cQ_{G,D_{\mathrm{in}},D_{\mathrm{out}},T}$ in \Cref{def:simgad} acts on qudits labeled by a subset of $N^{\sqcup\bar{b}}$. Then $\tilde{\cQ}$ uses space $\tilde{N}=N\times N'$ and time $\tilde{T}=T\cdot T''$. For every $t'\in[T'']$, we replace timestep $Q_t''$ in $\cQ''$ with $T$ timesteps $\tilde{Q}_{T(t'-1)+1},\dots,\tilde{Q}_{Tt'}$ in $\tilde{\cQ}$, such that every gate $G$ in $\cQ''$ is replaced by the circuit $\cQ_{G,D_{\mathrm{in}},D_{\mathrm{out}},T}$ in $\tilde{\cQ}$. Note that the identity gate $G=I$ is equal to $\gX^0$.

  By \Cref{def:simgad}, the input and output qudits of $\tilde{\cQ}$ are naturally labeled by $[n]^u\times[n_\alpha']\subseteq N\times N'$ for $\alpha\in\{\mathrm{in},\mathrm{out}\}$ respectively, which is precisely the set of physical qudits of $\tilde{D}_\alpha$. Meanwhile, for each $t'\in[T''-1]$, the qudits that form the output of the gadgets $\cQ_{G,D_{\mathrm{in}},D_{\mathrm{out}},T}$ that terminate at time $Tt'$, and the input of the gadgets that start at time $Tt'+1$, are labeled by $[n]^u\times N'$.

  For $\ell\in\bN$, let $\tilde{\rho}\in\bC^{\bF_q^{\tilde{k}_\alpha+\ell}\times\bF_q^{\tilde{k}_\alpha+\ell}}$. Let $\tilde{\sigma}$ be a Pauli $\tilde{\cE}_{\mathrm{in}}\sqcup [\ell]$-deviation of $\widetilde{\Enc}_{\mathrm{in}}\otimes I_\ell(\tilde{\rho})$. Let $\tilde{\cF}$ be a $\tilde{\cE}_{\mathrm{run}}$-avoiding Pauli fault and $\tilde{\zeta}$ be a postselection for $\tilde{\cQ}$. By \Cref{lem:paulift}, it suffices to show that the resulting output $\tilde{\cQ}[\tilde{\cF},\tilde{\zeta}]\otimes I_\ell(\tilde{\sigma})$ is a $\tilde{\cE}_{\mathrm{out}}\sqcup [\ell]$-deviation of $(\widetilde{\Enc}_{\mathrm{out}}\otimes\tilde{\bar{\cO}})\otimes I_\ell(\tilde{\rho})$.

  For this purpose, we first define an extended fault $\tilde{\cF}'$ (see \Cref{def:extfault}) and a postselection $\tilde{\zeta}'$ for $\tilde{\cQ}$ such that the superoperators $\tilde{\cQ}[\tilde{\cF}',\tilde{\zeta}'](\cdot)=\tilde{\cQ}[\tilde{\cF},\tilde{\zeta}](\cdot)$ are equal. We will ensure that $\supp(\tilde{\cF}')$ has a structure amenable to applying \Cref{lem:simgad}. In a slight departure from our standard notation, we also allow $\tilde{\cF}'=(\tilde{F}'_0,\dots,\tilde{F}'_{\tilde{T}})$ to include a superoperator $\tilde{F}'_0$ applied before the first timestep of $\tilde{\cQ}$.

  Specifically, we let the extended fault $\tilde{\cF}'$ have access to another set of qudits $\tilde{A}\cong\tilde{N}$. For every $t'\in[T'']$, let $G_{t',1},\dots,G_{t',m_{t'}}$ denote the gates making up $Q''_{t'}$, acting on qudits $B_{t',1},\dots,B_{t',m_{t'}}$ respectively with $B_{t',1}\sqcup\cdots\sqcup B_{t',m_{t'}}= N'$. For every $i\in[m_{t'}]$, we define the action of $\tilde{\cF}'$ along with the postselection $\tilde{\zeta}'$ on qudits $N\times B_{t',i}$ over timesteps $T(t'-1),\dots,Tt'$ as follows:
  \begin{enumerate}
  \item If $G=G_{t',i}\notin\gCO_{Z,*}$:
    \begin{enumerate}
    \item If it holds for every $\tilde{t}\in\{T(t'-1)+1,\dots,Tt'\}$ that
      \begin{equation}
        \label{eq:scGgood}
        |\supp(\tilde{F}_{\tilde{t}})\cap(N\times B_{t',i})|<\lambda_{\mathrm{run}},
      \end{equation}
      then for $\tilde{t}\in\{T(t'-1)+1,\dots,Tt'\}$, on qudits $N\times B_{t',i}\subseteq\tilde{N}$ we let $\tilde{F}'_{\tilde{t}}$ apply the same Paulis as $\tilde{F}_{\tilde{t}}$, and we let $\tilde{\zeta}'$ apply the same postselection as $\tilde{\zeta}$.
    \item If \Cref{eq:scGgood} is violated for some $\tilde{t}\in\{T(t'-1)+1,\dots,Tt'\}$, we let $\tilde{F}'_{T(t'-1)}$ apply $\gInit_Z$ to the qudits labeled $N\times B_{t',i}$ in $\tilde{A}$, and then swap the contents of the qudits labeled $[n]^u\times B_{t',i}\subseteq N\times B_{t',i}$ in $\tilde{A}$ with their counterparts (i.e.~the qudits with the same labels $[n]^u\times B_{t',i}$) in $\tilde{N}$, and subsequently apply the entire noisy gadget\footnote{\label{footnote:screstrict} In this expression $\cQ_{G,D_{\mathrm{in}},D_{\mathrm{out}},T}[\tilde{\cF},\tilde{\zeta}]$, we implicitly restrict $\tilde{\cF},\tilde{\zeta}$ to gates acting on qudits $N\times B_{t',i}$ in timesteps $T(t'-1)+1,\dots,Tt'$.} $\cQ_{G,D_{\mathrm{in}},D_{\mathrm{out}},T}[\tilde{\cF},\tilde{\zeta}]$ to these resulting qudits in $\tilde{A}$. The superoperators $\tilde{F}'_{T(t'-1)+1},\dots,\tilde{F}'_{Tt'-1}$ then act as the identity on qudits labeled $N\times B_{t',i}$. The postselection $\tilde{\zeta}'$ is trivial, i.e.~has all values equal to $1$, on gates performed on qudits $N\times B_{t',i}\subseteq\tilde{N}$ during timesteps $\{T(t'-1)+1,\dots,Tt'\}$. Finally, the superoperator $\tilde{F}'_{Tt'}$ swaps back the qudits labeled $[n]^u\times B_{t',i}$ in $\tilde{A}$ with their counterparts in $\tilde{N}$, and then applies $\gInit_Z$ to the resulting qudits labeled $N\times B_{t',i}$ in $\tilde{A}$.
    \end{enumerate}
  \item If $G=G_{t',i}\in\gCO_{Z,*}$, recall from \Cref{def:simgad} that $\cQ_{G,D_{\mathrm{in}},D_{\mathrm{out}},T}$ acts on qudits $(N_0\sqcup N_1')^{\sqcup\bar{b}}\cong(N_0\sqcup N_1')\times B_{t',i}$ with $[n]^u\cong N_0\cong N_1\subseteq N_1'$. (Qudits in $(N\setminus(N_0\sqcup N_1'))^{\sqcup\bar{b}}$ are acted upon trivially by $G$ and can be ignored.) Also define $T_1,T_2,\cQ^1,\cQ^2$ as in \Cref{it:sgbig} in \Cref{def:simgad}. Then for each $b\in B_{t',i}$:
    \begin{enumerate}
    \item If it holds for every $\tilde{t}\in\{T(t'-1)+1,\dots,Tt'\}$ that
      \begin{equation}
        \label{eq:scCOgood}
        |\supp(\tilde{F}_{\tilde{t}})\cap(N\times\{b\})| < \lambda_{\mathrm{run}},
      \end{equation}
      then for $\tilde{t}\in\{T(t'-1)+1,\dots,Tt'\}$, on qudits $(N_0\sqcup N_1')\times\{b\}\subseteq\tilde{N}$ we let $\tilde{F}'_{\tilde{t}}$ apply the same Paulis as $\tilde{F}_{\tilde{t}}$, and we let $\tilde{\zeta}'$ apply the same postselection as $\tilde{\zeta}$.
    \item If \Cref{eq:scCOgood} is violated for some $\tilde{t}\in\{T(t'-1)+1,\dots,Tt'\}$, then for $\tilde{t}\in\{T(t'-1)+1,\dots,Tt'\}$, on qudits $N_0\times\{b\}\subseteq\tilde{N}$ we let $\tilde{F}'_{\tilde{t}}$ apply the same Paulis as $\tilde{F}_{\tilde{t}}$, and we let $\tilde{\zeta}'$ apply the same postselection as $\tilde{\zeta}$. Furthermore, for
      \begin{equation*}
        \tilde{t}\in T(t'-1) + ([T]\setminus(\{2,\dots,T_1+1\}\sqcup\{T-T_2+1,\dots,T\})),
      \end{equation*}
      we let $\tilde{F}'_{\tilde{t}}$ apply the same Paulis as $\tilde{F}_{\tilde{t}}$ to qudits $N_1'\times\{b\}\subseteq\tilde{N}$; there are no $\gCO_*$ gates (and hence no postselection to specify) acting on these qudits $N_1'\times\{b\}$ during these timesteps $\tilde{t}$.

      However, following this Pauli error from $\tilde{F}_{T(t'-1)+1}$, we let $\tilde{F}'_{T(t'-1)+1}$ apply $\gInit_Z$ to the qudits labeled $N_1'\times\{b\}$ in $\tilde{A}$, and then swap the contents of the qudits labeled $N_1\times\{b\}\subseteq N_1'\times\{b\}$ in $\tilde{A}$ with their counterparts (i.e.~the qudits with the same labels $N_1\times\{b\}$) in $\tilde{N}$, and subsequently apply the entire noisy gadget\footnote{As in \Cref{footnote:screstrict}, the expression $\cQ^1[\tilde{\cF},\tilde{\zeta}]$ implicitly restricts $\tilde{\cF},\tilde{\zeta}$ to the appropriate gates corresponding to $\cQ^1$.} $\cQ^1[\tilde{\cF},\tilde{\zeta}]$ to these resulting qudits in $\tilde{A}$. The superoperators $\tilde{F}'_{T(t'-1)+2},\dots,\tilde{F}'_{T(t'-1)+T_1}$ then act as the identity on qudits $N_1'\times\{b\}$. The postselection $\tilde{\zeta}'$ is trivial, i.e.~has all values equal to $1$, on gates performed on qudits $N_1'\times\{b\}\subseteq\tilde{N}$ during timesteps $\{T(t'-1)+2,\dots,T(t'-1)+T_1+1\}$. The superoperator $\tilde{F}_{T(t'-1)+T_1+1}$ swaps back the qudits labeled $N_1\times\{b\}$ in $\tilde{A}$ with their counterparts in $\tilde{N}$, and then applies $\gInit_Z$ to the resulting qudits labeled $N_1'\times\{b\}$ in $\tilde{A}$.

      Similarly, following the Pauli error from $\tilde{F}_{Tt'-T_2}$, we let $\tilde{F}'_{Tt'-T_2}$ apply $\gInit_Z$ to the qudits labeled $N_1'\times\{b\}$ in $\tilde{A}$, then swap the contents of qudits labeled $N_1\times\{b\}$ in $\tilde{A}$ with their counterparts in $\tilde{N}$, and subsequently apply the entire noisy gadget $\cQ^2[\tilde{\cF},\tilde{\zeta}]$ to these resulting qudits in $\tilde{A}$. The superoperators $\tilde{F}'_{Tt'-T_2+1},\dots,\tilde{F}'_{Tt'-1}$ then act as the identity on qudits $N_1'\times\{b\}$. The postselection $\tilde{\zeta}'$ is trivial on qudits $N_1'\times\{b\}\subseteq\tilde{N}$ during timesteps $\{Tt'-T_2+1,\dots,Tt'\}$. The superoperator $\tilde{F}_{Tt'}$ swaps back the qudits labeled $N_1\times\{b\}$ in $\tilde{A}$ with their counterparts in $\tilde{N}$, and then applies $\gInit_Z$ to the resulting qudits labeled $N_1'\times\{b\}$ in $\tilde{A}$.
    \end{enumerate}
  \end{enumerate}
  We then have an equality of superoperators $\tilde{\cQ}[\tilde{\cF}',\tilde{\zeta}'](\cdot)=\tilde{\cQ}[\tilde{\cF},\tilde{\zeta}](\cdot)$, as $\tilde{\cF}'$ implements the exact same Pauli corruption as $\tilde{\cF}$, just sometimes in the roundabout fashion of swapping out some qudits in $\tilde{N}$ to $\tilde{A}$, applying the corrupted gates on $\tilde{A}$, and then swapping the qudits back. Specifically, each of these pairs of swaps is sandwiched by $\gInit_Z$ gates on the involved qudits in $\tilde{A}$, and the postselection $\tilde{\zeta}'$ on the associated region in $\tilde{N}$ is trivial. Hence up to relabeling of some qudits from the swap gates, $\tilde{\cQ}[\tilde{\cF}',\tilde{\zeta}'](\cdot)$ simply runs $\tilde{\cQ}[\tilde{\cF},\tilde{\zeta}](\cdot)$, while sometimes initializing an additional block of qudits in the $\ket{0}$ state, applying $\cQ_{G,D_{\mathrm{in}},D_{\mathrm{out}},T}$, $\cQ^1$, or $\cQ^2$ (all of which are channels due to the trivial postselection), and then tracing out the channel's output. 
  These additional initializations, channels, and tracing-outs have no effect on the execution of $\tilde{\cQ}[\tilde{\cF},\tilde{\zeta}](\cdot)$, as desired.

  Therefore it suffices to show that the resulting output $\tilde{\cQ}[\tilde{\cF}',\tilde{\zeta}']\otimes I_\ell(\tilde{\sigma})$ is a $\tilde{\cE}_{\mathrm{out}}\sqcup [\ell]$-deviation of $(\widetilde{\Enc}_{\mathrm{out}}\otimes\tilde{\bar{\cO}})\otimes I_\ell(\tilde{\rho})$. For this purpose, similarly as in the proof of \Cref{lem:extft}, for each $\tilde{t}\in[\tilde{T}]$, we may decompose $\tilde{F}'_{\tilde{t}}$ into a linear combination of Pauli superoperators $\tilde{F}''_{\tilde{t}}$ acting on qudits $\tilde{N}\sqcup\tilde{A}$, with $\supp(\tilde{F}'')_{\tilde{t}}\subseteq\supp(\tilde{F}')_{\tilde{t}}$. Letting each $\tilde{\cF}''=(\tilde{F}''_1,\dots,\tilde{F}''_{\tilde{T}})$, then $\tilde{\cQ}[\tilde{\cF}',\tilde{\zeta}']\otimes I_\ell(\tilde{\sigma})$ is a linear combination of states of the form $\tilde{\cQ}[\tilde{\cF}'',\tilde{\zeta}']\otimes I_\ell(\tilde{\sigma})$ for Pauli faults $\tilde{\cF}''$ with $\supp(\tilde{\cF}'')\subseteq\supp(\tilde{\cF}')$. As in the proof of \Cref{lem:extft}, because each Pauli superoperator $\tilde{F}''_{\tilde{t}}$ does not entangle qudits $\tilde{N}$ and $\tilde{A}$, these two sets of qudits never become entangled, and hence the output $\tilde{\cQ}[\tilde{\cF}'',\tilde{\zeta}']\otimes I_\ell(\tilde{\sigma})$ is preserved (up to a global phase) if we restrict attention to $\tilde{N}$, and replace the Paulis gates in $\tilde{\cF}''$ acting on $\tilde{A}$ with identity gates (or equivalently, simply remove qudits $\tilde{A}$).

  Thus we have reduced our problem to showing the following claim:

  \begin{claim}
    \label{claim:scnicefault}
    For
    every (non-extended) Pauli fault $\tilde{\cF}''$ with $\supp(\tilde{\cF}'')\subseteq\supp(\tilde{\cF}')$ and every postselection $\tilde{\zeta}'$ for $\tilde{\cQ}$, the output $\tilde{\cQ}[\tilde{\cF}'',\tilde{\zeta}']\otimes I_\ell(\tilde{\sigma})$ is a $\tilde{\cE}_{\mathrm{out}}\sqcup [\ell]$-deviation of $(\widetilde{\Enc}_{\mathrm{out}}\otimes\tilde{\bar{\cO}})\otimes I_\ell(\tilde{\rho})$.
  \end{claim}

  We now turn to proving \Cref{claim:scnicefault}. We begin by introducing some additional notation. Below, for a circuit $\bar{\cQ}$ using space $\bar{N}$, we let $\bar{\cQ}_{\leq\bar{t}}[\bar{\cF},\bar{\zeta}](\bar{\sigma})$ denote the state of qudits $\bar{N}$ after running the first $\bar{t}$ timesteps of $\bar{\cQ}$ on input $\bar{\sigma}$, under the corresponding restrictions of $\bar{\cF},\bar{\zeta}$ to the first $t$ timesteps.

  We also fix $\tilde{\cF}',\tilde{\cF}'',\tilde{\zeta}'$ as in \Cref{claim:scnicefault}. Recalling that $\tilde{\sigma}$ is a Pauli $\tilde{\cE}_{\mathrm{in}}\sqcup [\ell]$-deviation of $\widetilde{\Enc}_{\mathrm{in}}\otimes I_\ell(\tilde{\rho})$, we can write
  \begin{align}
    \label{eq:scinput}
    \tilde{\sigma} &= (\tilde{P_0}\circ\widetilde{\Enc}_{\mathrm{in}})\otimes I_\ell(\tilde{\rho}) = (\tilde{P}_0\circ\Enc^{\sqcup n_{\mathrm{in}}'}\circ(\phi_{q,q'}^{-1})^{\sqcup n_{\mathrm{in}}'}\circ\Enc'\circ\phi_{q,q'}^{\sqcup k_{\mathrm{in}}'})\otimes I_\ell(\tilde{\rho})
  \end{align}
  for some Pauli superoperator $\tilde{P}_0$ of weight $|\tilde{P}_0|<\tilde{\lambda}_{\mathrm{in}}$ acting on qudits $[n]^u\times[n_{\mathrm{in}}']$.
  We then define $A_0\subseteq[n_{\mathrm{in}}']$ by
  \begin{align*}
    A_0 &= \{b\in[n_{\mathrm{in}}']:|\supp(\tilde{P}_0)\cap([n]^u\times\{b\})|\geq\lambda_{\mathrm{in}}/3\},
  \end{align*}
  so that
  \begin{align}
    \label{eq:scA0}
    |A_0| &< \frac{3\tilde{\lambda}_{\mathrm{in}}}{\lambda_{\mathrm{in}}} = \frac{\lambda_{\mathrm{in}}''}{2}.
  \end{align}
  For $t'\in[T'']$, we define $A_{t'}\subseteq N'$ to contain every $b\in N'$ such that the following holds: letting $i\in[m_{t'}]$ denote the unique index for which $B_{t',i}\subseteq N'$ defined above contains $b$, then either $G_{t',i}\notin\gCO_{Z,*}$ and \Cref{eq:scGgood} fails to hold, or $G_{t',i}\in\gCO_{Z,*}$ and \Cref{eq:scCOgood} fails to hold. In other words, $A_{t'}$ contains every $b\in N'$ for which we insert swap gates on qudits $N\times\{b\}$ during timesteps $\{T(t'-1)+1,\dots,Tt'\}$ in the definition of the extended fault $\tilde{\cF}'$ above. Then
  \begin{align}
    \label{eq:scAtp}
    |A_{t'}| &< \frac{9T\cdot\tilde{\lambda}_{\mathrm{run}}}{\lambda_{\mathrm{run}}},
  \end{align}
  as because $\tilde{\cF}$ is $\tilde{\cE}_{\mathrm{run}}$-avoiding, there are $<T\cdot\tilde{\lambda}_{\mathrm{run}}/(\lambda_{\mathrm{run}}/3)$ distinct $b'\in N'$ for which $\supp(\tilde{\cF})$ contains $\geq\lambda_{\mathrm{run}}/3$ elements in the set $(N\times\{b'\})\times\{T(t'-1)+1,\dots,Tt'\}$, and by definition every $b\in A_{t'}$ either equals such a $b'$, or else shares a $\gCX^*$ or $\gCCX^*$ gate $G_{t',i}$ with such a $b'$.

  We will use \Cref{claim:scinduct} below, which we will prove by induction, to prove \Cref{claim:scnicefault}. Below, we use the fact that by \Cref{eq:lamrun,eq:lamminmax}, we have $\lambda_{\mathrm{out}}=\lambda_{\mathrm{in}}(u,n,\lambda_{\mathrm{run}})\leq\lambda_{\max}(u,n)/3=\lambda_{\mathrm{in}}/3$.
  
  \begin{claim}
    \label{claim:scinduct}
    For every $t'\in[T'']$, the state $\tilde{\cQ}_{\leq Tt'}[\tilde{\cF}'',\tilde{\zeta}']\otimes I_\ell(\tilde{\sigma})$ is a linear combination of states of the form
    \begin{equation*}
      (\tilde{P}_{t'}\circ\Enc^{\sqcup N'}\circ(\phi_{q,q'}^{-1})^{\sqcup N'}\circ\cQ''_{\leq t'}[\cF',\zeta'])\otimes I_\ell(\sigma'),
    \end{equation*}
    where:
    \begin{itemize}
    \item $\sigma'$ is a $\cE_{\mathrm{in}}''\sqcup [\ell]$-deviation of $(\Enc'\circ\phi_{q,q'}^{\sqcup k_{\mathrm{in}}})\otimes I_\ell(\tilde{\rho})$,
    \item $\cF'$ is a $\cE_{\mathrm{run},<t'}'':=\cE_{\mathrm{run}}''\sqcup 2^{N'\times\{t',\dots,T''\}}$-avoiding Pauli fault for $\cQ''_{\leq t'}$,
    \item $\zeta'$ is a postselection for $\cQ''_{\leq t'}$,
    \item $\tilde{P}_{t'}$ is a Pauli superoperator acting on qudits $[n]^u\times N'$ satisfying
      \begin{align}
        \label{eq:scPt}
        \begin{split}
          |\supp(\tilde{P}_{t'})\setminus([n]^u\times A_{t'})| &< \eta_{\min}(u,n)\cdot T\cdot\tilde{\lambda}_{\mathrm{run}} \\
          |\supp(\tilde{P}_{t'})\cap([n]^u\times\{b\})| &< \lambda_{\mathrm{out}} \leq \lambda_{\mathrm{in}}/3 \hspace{1em} \forall b\in N'\setminus A_{t'}.
        \end{split}
      \end{align}
    \end{itemize}
  \end{claim}
  \begin{proof}
    We prove the claim by induction on $t'\in[T'']$. For the base case, we will show that the claim holds for $t'=1$. The input state $\tilde{\sigma}$ is given by \Cref{eq:scinput}. By definition, for each $i\in[m_1]$, $\tilde{\cQ}_{\leq T}[\tilde{\cF}'',\tilde{\zeta}']$ applies $\cQ_{G_{1,i},D_{\mathrm{in}},D_{\mathrm{out}},T}[\tilde{\cF}'',\tilde{\zeta}']$ to qudits $N\times B_{1,i}$, where $\tilde{\cF}'',\tilde{\zeta}'$ here are implicity restricted to these qudits and the appropriate timesteps $\{1,\dots,T\}$.

    By definition $\lambda_{\mathrm{run}}\leq\bar{\lambda}_{\mathrm{run}}/\eta_T$, and therefore \Cref{eq:sglambound} holds. Then \Cref{lem:simgad} implies that $\tilde{\cQ}_{\leq T}[\tilde{\cF}'',\tilde{\zeta}']\otimes I_\ell(\tilde{\sigma})$ is a linear combination of states of the form
    \begin{align*}
      (\tilde{P}_1\circ\Enc^{\sqcup N'}\circ(\phi_{q,q'}^{-1})^{\sqcup N'}\circ\cQ''_{\leq 1}[\zeta'])\otimes I_\ell(\sigma'),
    \end{align*}
    where:
    \begin{itemize}
    \item $\tilde{P}_1$ is a Pauli superoperator acting on qudits $[n]^u\times N'$ satisfying \Cref{eq:scPt}. Specifically, for $B_{1,i}\subseteq N'\setminus A_1$ with $G_{1,i}\notin\gCO_{Z,*}$, then let
      \begin{equation*}
        \lambda_{\mathrm{run},B_{1,i}} = \max_{\tilde{t}\in\{1,\dots,T\}} \supp(\tilde{F}''_{\tilde{t}})\cap(N\times B_{1,i}).
      \end{equation*}
      By the definition of $A_1$ (see \Cref{eq:scGgood}), we have $\lambda_{\mathrm{run},B_{1,i}}\leq\lambda_{\mathrm{run}}$. Then by \Cref{it:sgnonCO} in \Cref{lem:simgad}, the output error $\tilde{P}_1|_{[n]^u\times B_{1,i}}$ from running $\cQ_{G_{1,i},D_{\mathrm{in}},D_{\mathrm{out}},T}[\tilde{\cF}'',\tilde{\zeta}']$ on qudits $N\times B_{1,i}$ has weight
      \begin{equation}
        \label{eq:scPt1G}
        |\supp(\tilde{P}_1)\cap([n]^u\times B_{1,i})| \leq \eta_{\min}(u,n)\cdot\lambda_{\mathrm{run},B_{1,i}} < \lambda_{\mathrm{out}}.
      \end{equation}
      Note that here we may use $\lambda_{\mathrm{run},B_{1,i}}$ in place of $\lambda_{\mathrm{run}}$ in \Cref{lem:simgad} by \Cref{remark:sgcirind}. Similarly, for $B_{1,i}\subseteq N'$ with $G_{1,i}\in\gCO_{Z,*}$ and for $b\in B_{1,i}\setminus A_{t'}$, let
      \begin{equation*}
        \lambda_{\mathrm{run},b} = \max_{\tilde{t}\in\{1,\dots,T\}} \supp(\tilde{F}''_{\tilde{t}})\cap(N\times\{b\}).
      \end{equation*}
      By the definition of $A_1$ (see \Cref{eq:scCOgood}), we have $\lambda_{\mathrm{run},b}\leq\lambda_{\mathrm{run}}$. Then applying \Cref{it:sgCO} in \Cref{lem:simgad} to the execution of $\cQ_{G_{1,i},D_{\mathrm{in}},D_{\mathrm{out}},T}[\tilde{\cF}'',\tilde{\zeta}']$ on qudits $N\times\{b\}\subseteq N\times B_{1,i}$ gives that
      \begin{equation}
        \label{eq:scPt1CO}
        |\supp(\tilde{P}_1)\cap([n]^u\times\{b\})| \leq \eta_{\min}(u,n)\cdot\lambda_{\mathrm{run},b} < \lambda_{\mathrm{out}}.
      \end{equation}
      Here we use the fact that $\supp(\tilde{\cF}'')\subseteq\supp(\tilde{\cF}')$ is defined precisely to allow for the application of smoothly-mending fault-tolerance, specifically with regard to the definition of the families $\cE_{\mathrm{run},b}$ in \Cref{def:smoothmend}.
      Thus for a given $b\in N'\setminus A_1$, the second inequality in \Cref{eq:scPt} with $t'=1$ holds by \Cref{eq:scPt1G,eq:scPt1CO}. Meanwhile, the first inequality in \Cref{eq:scPt} holds by summing \Cref{eq:scPt1G,eq:scPt1CO} over all $B_{1,i}\subseteq N'\setminus A_1$ for $B_{1,i}\notin\gCO_{Z,*}$ and over all $b\in B_{1,i}\setminus A_1$ for $B_{1,i}\in\gCO_{Z,*}$, and using the fact that the associated sum of all $\lambda_{\mathrm{run},B_{1,i}}$ and $\lambda_{\mathrm{run},b}$ is $\leq T\cdot\tilde{\lambda}_{\mathrm{run}}$ by the definition of $A_1$ and $\tilde{\cF}''$.
    \item $\sigma'=(F'_0\circ\Enc'\circ\phi_{q,q'}^{\sqcup k_{\mathrm{in}}'})\otimes I_\ell(\tilde{\rho})$ for some superoperator $F'_0$ acting on ($q'$-dimensional) qudits $N'$ with $\supp(F'_0)\subseteq A_0\cup A_1$. (Here similarly as described above for $\tilde{P}_1$, we apply \Cref{lem:simgad} to the execution of each $\cQ_{G_{1,i},D_{\mathrm{in}},D_{\mathrm{out}},T}[\tilde{\cF}'',\tilde{\zeta}']$; we omit the details to avoid redundancy.) Therefore by \Cref{eq:sclamtil,eq:scA0,eq:scAtp}, we have $|F'_0|\leq|A_0\cup A_1|<\lambda_{\mathrm{in}}''$, so $\sigma'$ is a $\cE_{\mathrm{in}}''\sqcup [\ell]$-deviation of $(\Enc'\circ\phi_{q,q'}^{\sqcup k_{\mathrm{in}}'})\otimes I_\ell(\tilde{\rho})$.
    \item $\zeta'$ is a postselection for $\cQ''_{\leq 1}$ (again by \Cref{lem:simgad}).
    \end{itemize}

    Thus we have proven the desired claim statement for $t'=1$, completing the base case of our induction. For the inductive step, assume that the claim holds for some $t'-1\in\{1,\dots,T''-1\}$. Then \Cref{lem:simgad} implies that $\tilde{\cQ}_{\leq Tt'}[\tilde{\cF}'',\tilde{\zeta}']\otimes I_\ell(\tilde{\sigma})$ is a linear combination of states of the form
    \begin{equation*}
      (\tilde{P}_{t'}\circ\Enc^{\sqcup N'}\circ(\phi_{q,q'}^{-1})^{\sqcup N'}\circ\cQ''_{\leq 1}[\cF',\zeta'])\otimes I_\ell(\sigma'),
    \end{equation*}
    where:
    \begin{itemize}
    \item $\tilde{P}_{t'}$ is a Pauli superoperator acting on qudits $[n]^u\times N'$ satisfying \Cref{eq:scPt}, by analogous reasoning as in the proof of the base case above.
    \item $\cF'=(F'_1,\dots,F'_{t'-1},I)$ is a $\cE''_{\mathrm{run},<t'}$-avoiding Pauli fault for $\cQ''_{\leq t'}$. Specifically, $(F'_1,\dots,F'_{t'-2},I)$ is the Pauli fault given by the $t'-1$ case of the inductive hypothesis, and $F'_{t'-1}$ is some superoperator acting on ($q'$-dimensional) qudits $N'$ with $\supp(F'_{t'-1})\subseteq A_{t'-1}\cup A_{t'}$. Therefore indeed $|F'_{t'-1}|\leq|A_{t'-1}\cup A_{t'}|\leq\lambda_{\mathrm{run}}''$ by \Cref{eq:sclamtil,eq:scAtp}.
    \item $\tilde{\zeta}'$ is a postselection for $\cQ''_{\leq t'}$.
    \end{itemize}
    Thus the $t'$ case of the claim holds, completing the inductive step.
  \end{proof}

  We now apply \Cref{claim:scinduct} to prove \Cref{claim:scnicefault}, thereby completing the proof of \Cref{lem:simcomp}.

  \begin{proof}[Proof of \Cref{claim:scnicefault}]
    By definition
    \begin{align}
      \label{eq:scoutput}
      \tilde{\cQ}[\tilde{\cF}'',\tilde{\zeta}']\otimes I_\ell(\tilde{\sigma})
      &= \tr_{N\times(N'\setminus[n'_{\mathrm{out}}])}(\tilde{\cQ}_{\leq TT''}[\tilde{\cF}'',\tilde{\zeta}']\otimes I_\ell(\tilde{\sigma})),
    \end{align}
    where $\tilde{\cQ}_{\leq TT''}[\tilde{\cF}'',\tilde{\zeta}']\otimes I_\ell(\tilde{\sigma})$ is of the form given in \Cref{claim:scinduct}. In particular, because we defined $\cQ''$ to apply $\gInit_Z$ gates to all qudits in $N'\setminus[n'_{\mathrm{out}}]$ in its final timestep $T''$, \Cref{claim:scinduct} implies that qudits $N\times(N'\setminus[n'_{\mathrm{out}}])$ are unentanged from qudits $N\times[n'_{\mathrm{out}}]$ in the state $\tilde{\cQ}_{\leq TT''}[\tilde{\cF}'',\tilde{\zeta}']\otimes I_\ell(\tilde{\sigma})$. Therefore by \Cref{eq:scoutput,claim:scinduct}, $\tilde{\cQ}[\tilde{\cF}'',\tilde{\zeta}']\otimes I_\ell(\tilde{\sigma})$ is proportional to a linear combination of states of the form
    \begin{align}
      \label{eq:sctroutput}
      (\tilde{P}_{T''}|_{[n]^u\times[n_{\mathrm{out}}']}\circ\Enc^{\sqcup n_{\mathrm{out}}'}\circ(\phi_{q,q'}^{-1})^{\sqcup n_{\mathrm{out}}'}\circ\cQ''[\cF',\zeta'])\otimes I_\ell(\sigma'), 
    \end{align}
    where:
    \begin{itemize}
    \item $\sigma'$ is a $\cE_{\mathrm{in}}''\sqcup [\ell]$-deviation of $(\Enc'\circ\phi_{q,q'}^{\sqcup k_{\mathrm{in}}'})\otimes I_\ell(\tilde{\rho})$,
    \item $\cF'$ is a $\cE_{\mathrm{run},<T''}'':=\cE_{\mathrm{run}}''\sqcup 2^{N'\times\{T''\}}$-avoiding Pauli fault for $\cQ''$,
    \item $\zeta'$ is a postselection for $\cQ''$,
    \item $\tilde{P}_{T''}|_{[n]^u\times[n_{\mathrm{out}}']}$ is the restriction to qudits $[n]^u\times[n_{\mathrm{out}}']$ of a Pauli superoperator $\tilde{P}_{T''}$ acting on qudits $[n]^u\times N'$ satisfying \Cref{eq:scPt}.
    \end{itemize}

    By the fault-tolerance of $(\cQ'',\; \cE_{\mathrm{run}}'',\; D_{\mathrm{in}}'',\; D_{\mathrm{out}}'')$, the state $\cQ''[\cF',\zeta']\otimes I_\ell(\sigma')$ is a $\cE_{\mathrm{out}}''\sqcup [\ell]$-deviation of $(\Enc_{\mathrm{out}}'\circ\bar{\cO}'\circ\phi_{q,q'}^{\sqcup k_{\mathrm{in}}'})\otimes I_\ell(\tilde{\rho})$. Therefore we can write the state in \Cref{eq:sctroutput} as a linear combination of states of the form
    \begin{align*}
      \hspace{1em}&\hspace{-1em} (\tilde{P}_{\mathrm{out}}\circ\Enc^{\sqcup n_{\mathrm{out}}'}\circ(\phi_{q,q'}^{-1})^{\sqcup n_{\mathrm{out}}'}\circ\Enc_{\mathrm{out}}'\circ\bar{O}'\circ\phi_{q,q'}^{\sqcup k_{\mathrm{in}}'})\otimes I_\ell(\tilde{\rho}) \\
                  &= (\tilde{P}_{\mathrm{out}}\circ\widetilde{\Enc}_{\mathrm{out}}\circ(\phi_{q,q'}^{-1})^{\sqcup k_{\mathrm{out}}'}\circ\bar{O}'\circ\phi_{q,q'}^{\sqcup k_{\mathrm{in}}'})\otimes I_\ell(\tilde{\rho}) \\
                  &= (\tilde{P}_{\mathrm{out}}\circ\widetilde{\Enc}_{\mathrm{out}}\circ\tilde{\bar{O}})\otimes I_\ell(\tilde{\rho}),
    \end{align*}
    where:
    \begin{itemize}
    \item $\bar{O}'\in\bar{\cO}'$ and hence $\tilde{\bar{O}}:=(\phi_{q,q'}^{-1})^{\sqcup k_{\mathrm{out}}'}\circ\bar{O}'\circ\phi_{q,q'}^{\sqcup k_{\mathrm{in}}'}\in\tilde{\bar{\cO}}$.
    \item $\tilde{P}_{\mathrm{out}}$ is a Pauli superoperator acting on qudits $[n]^u\times[n_{\mathrm{out}}']$ with
      \begin{align*}
        |\tilde{P}_{\mathrm{out}}|
        &< |\tilde{P}_{T''}|+n^u\cdot\lambda_{\mathrm{out}}'' \\
        &\leq (n^u\cdot|A_{T''}| + \eta_{\min}(u,n)\cdot T\cdot\tilde{\lambda}_{\mathrm{run}}) + n^u\cdot\lambda_{\mathrm{out}}'' \\
        &\leq \tilde{\lambda}_{\mathrm{out}}.
      \end{align*}
      Specifically, the first inequality above holds because we define $\tilde{P}_{\mathrm{out}}$ to be the product of the Pauli error $\tilde{P}_{T''}$ and the $\cE_{\mathrm{out}}''\sqcup [\ell]$-deviation on $(\Enc_{\mathrm{out}}'\circ\bar{\cO}'\circ\phi_{q,q'}^{\sqcup k_{\mathrm{in}}'})\otimes I_\ell(\tilde{\rho})$, the latter of which we decompose into a linear combination of Pauli errors (over $\bF_{q'}$) that we propagate through $\Enc^{\sqcup n_{\mathrm{out}}'}\circ(\phi_{q,q'}^{-1})^{\sqcup n_{\mathrm{out}}'}$, to obtain a linear combination of Pauli errors (over $\bF_q$) of weight $<n^u\cdot\lambda_{\mathrm{out}}''$. The second inequality above holds by the first inequality in \Cref{eq:scPt}. The third inequality above holds by \Cref{eq:scAtp,eq:sclams}. Thus we have shown that $\tilde{\cQ}[\tilde{\cF}'',\tilde{\zeta}']\otimes I_\ell(\tilde{\sigma})$ is a $\tilde{\cE}_{\mathrm{out}}\sqcup [\ell]$-deviation of $(\widetilde{\Enc}_{\mathrm{out}}\circ\tilde{\bar{\cO}})\otimes I_\ell(\tilde{\rho})$, completing the proof of the claim.
    \end{itemize}
  \end{proof}

\end{proof}

The following corollary instantiates \Cref{lem:simcomp} with parameters that will be useful for proving our main result.

\begin{corollary}
  \label{cor:simcomp}
  For every $0<\epsilon<1/4$, there exists $v_0=v_0(\epsilon)\geq 1$, and there exists an absolute constant $\xi\geq 16$, such that the following holds. Fix a prime power $r$. Let $q'=r^{2^{v'}}$ for some $v'\geq v_0$. For $\alpha\in\{\mathrm{in},\mathrm{out}\}$ for some $\lambda_\alpha'\geq 0$ let
  \begin{equation*}
    D_\alpha' = (C_\alpha',\; \Enc_\alpha',\; \cE_\alpha'=2^{[n_\alpha']}|_{\geq\lambda_\alpha'})
  \end{equation*}
  be a $[[n_\alpha',k_\alpha',d_\alpha']]_{q'}$ decorated CSS non-subsystem code.
  For some
  \begin{equation}
    \label{eq:sclamrunsmaller}
    0 \leq \lambda_{\mathrm{run}} ' \leq\min\{\lambda_{\mathrm{in}}',\lambda_{\mathrm{out}}'\},
  \end{equation}
  let
  \begin{equation*}
    (\cQ',\; \cE_{\mathrm{run}}'=2^{N'}|_{\geq\lambda_{\mathrm{run}}'}^{\sqcup T'},\; D_{\mathrm{in}}',\; D_{\mathrm{out}}')
  \end{equation*}
  be a fault-tolerant gadget for a set $\bar{\cO}'$ of superoperators $\bar{O}':\bC^{\bF_{q'}^{k_{\mathrm{in}}'}\times\bF_{q'}^{k_{\mathrm{in}}'}}\rightarrow\bC^{\bF_{q'}^{k_{\mathrm{out}}'}\times\bF_{q'}^{k_{\mathrm{out}}'}}$ on $q'$-dimensional qudits, using space $N'$, time $T'$, and gate set $\cG=\{\gH^*,\gX^*,\gZ^*,\gCX^*,\gCCX^*,\gInit_*,\gCO_*\}$.

  Let $v=\lfloor(3/4+\epsilon) v'\rfloor$, and let $q=r^{2^v}$. Then there exists a code with encoding map $(C,\Enc)$, which depends only on $q'$ and has alphabet size $q$, such that the following holds. Letting
  \begin{align}
    \label{eq:tlamrundef}
    \tilde{\lambda}_{\mathrm{run}} &= 2^{v-\xi\sqrt{v}}\cdot\lambda_{\mathrm{run}}',
  \end{align}
  there exists a fault-tolerant gadget on $q$-dimensional qudits
  \begin{equation*}
    (\tilde{\cQ},\; \tilde{\cE}_{\mathrm{run}}=2^{\tilde{N}}|_{\geq\tilde{\lambda}_{\mathrm{run}}}^{\sqcup\tilde{T}},\; \tilde{D}_{\mathrm{in}},\; \tilde{D}_{\mathrm{out}})
  \end{equation*}
  for $\tilde{\bar{\cO}}=(\phi_{q,q'}^{-1})^{\sqcup k_{\mathrm{out}}'}\circ\bar{\cO}'\circ\phi_{q,q'}^{\sqcup k_{\mathrm{in}}'}$, where:
  \begin{itemize}
  \item $\tilde{\cQ}$ is a quantum circuit using space $|\tilde{N}|\leq 2^{v+O(\sqrt{v})}\cdot|N'|$ and time $\tilde{T}\leq 2^{4\sqrt{v}}\cdot T'$.
  \item For $\alpha\in\{\mathrm{in},\mathrm{out}\}$, the
    \begin{equation*}
      [[\tilde{n}_\alpha\leq 2^v\cdot n_\alpha',\; \tilde{k}_\alpha=2^{v'-v}\cdot k_\alpha'\geq 2^{(1/3-4\epsilon)v}\cdot k_\alpha',\; \tilde{d}_\alpha\geq 2^{v-8\sqrt{v}}\cdot d_\alpha']]_q
    \end{equation*}
    decorated code $\tilde{D}_\alpha$ is given by
    \begin{equation*}
      \tilde{D}_\alpha = ((\tilde{C},\widetilde{\Enc})=(C,\Enc)\concat(C_\alpha',\Enc_\alpha'),\; \tilde{\cE}_\alpha=2^{[\tilde{n}_\alpha]}|_{\geq\tilde{\lambda}_\alpha}),
    \end{equation*}
    (see \Cref{def:concat,lem:concatdis}) such that
    \begin{align}
      \label{eq:sclamscor}
      \begin{split}
        \tilde{\lambda}_{\mathrm{in}} &\geq 2^{v-16\sqrt{v}}\cdot\lambda_{\mathrm{in}}' \\
        \tilde{\lambda}_{\mathrm{out}} &= 2^v\cdot\lambda_{\mathrm{out}}'.
      \end{split}
    \end{align}
  \end{itemize}
\end{corollary}
\begin{proof}
  We will simply instantiate \Cref{lem:simcomp} with appropriate parameters. We will specify $\xi$ explicitly below; we take $v_0$ to be sufficiently large so that various inequalities that hold as $v'\rightarrow\infty$, specified below, in fact hold for every $v'\geq v_0$. All logarithms below are taken base $2$, unless explicitly specified otherwise.

  To begin, we define $n\in\bN$ to be the greatest integer for which $v\geq\log^2n+2\log n$, and we set $u=\lfloor\log n\rfloor$. Assuming $v_0$ is sufficiently large, and because $v'\geq v_0$, we have
  \begin{align}
    \label{eq:vbounds}
    \begin{split}
      v &\in [\log^2n+2\log n,\; \log^2n+3\log n] \\
      v &\geq \lfloor\log n\rfloor\cdot\lceil\log n\rceil + \lceil\log\lceil\log(n)/\log(r)\rceil\rceil.
    \end{split}
  \end{align}
  Let $q_0=r^{2^{\lceil\log\lceil\log(n)/\log(r)\rceil\rceil}}$, let $n'=2^{\lceil\log n\rceil}$, and let $n''=2^{v''}$ for
  \begin{equation*}
    v''=v-(\lceil\log\lceil\log(n)/\log(r)\rceil\rceil+(u-1)\lceil\log n\rceil).
  \end{equation*}
  Then $q_0\geq n$, and by \Cref{eq:vbounds} we have $n''\geq 2^{\lceil\log n\rceil}$, so $n',n''\geq n$.

  Let $\kappa=2^{v'-v}$. Assuming $v_0=v_0(\epsilon)$ is sufficiently large relative to $1/\epsilon$, then \Cref{eq:vbounds} implies that
  \begin{align*}
    v'-v
    &\leq \frac{1/4-\epsilon}{3/4+\epsilon}v+O(1) \\
    &\leq \frac{1/4-\epsilon}{3/4+\epsilon}(\log^2n+3\log n)+O(1) \\
    &\leq \frac13\cdot\lfloor\log n\rfloor\cdot(\log(n)-4),
  \end{align*}
  so again assuming $v_0$ (and hence $u,n$) is sufficiently large,
  \begin{align*}
    \kappa &= 2^{v'-v} \leq (n/16)^{u/3} \leq \lfloor n/8\rfloor^{u/3}/4.
  \end{align*}

  Define $\bar{\lambda}_{\mathrm{run}}=\bar{\lambda}_{\mathrm{run}}(u,n)$ as in \Cref{thm:laft} (see \Cref{eq:lamrun}), and define $T,\eta_T$ as in \Cref{def:simgad} (see \Cref{eq:etaT}). By this definition, there exists an absolute constant $\xi_{\mathrm{run}}>0$ such that when $u=\lfloor\log n\rfloor$ and $v_0$ (and hence $u,n$) is sufficiently large, then $n^{u-\xi_{\mathrm{run}}}\leq\bar{\lambda}_{\mathrm{run}}(u,n)/\eta_T$. Let $\lambda_{\mathrm{run}}= n^{u-\xi_{\mathrm{run}}}$.

  Now by construction $q=q_0^{{n'}^{u-1}n''}$ with $q_0,n',n''\geq n$ and $q'=q^\kappa$ with $\kappa\leq\lfloor n/8\rfloor^{u/3}/4$. Also, $\lambda_{\mathrm{run}}\leq\bar{\lambda}_{\mathrm{run}}(u,n)/\eta_T$. Thus $q,q',u,n$ satisfy the conditions in \Cref{lem:simcomp}. Furthermore, defining $\lambda_\alpha''$ for $\alpha\in\{\mathrm{in},\mathrm{run},\mathrm{out}\}$ as in \Cref{eq:lampps}, the assumption that $\lambda_{\mathrm{run}}'\leq\min\{\lambda_{\mathrm{in}}',\lambda_{\mathrm{out}}'\}$ implies that $\lambda_{\mathrm{in}}''\geq\lambda_{\mathrm{in}}'/2$ and that $\min\{\lambda_{\mathrm{in}}'',\lambda_{\mathrm{run}}''\}=\lambda_{\mathrm{run}}''=\lambda_{\mathrm{run}}'/3$.

  Define $\eta_{\min}(u,n)$ as in \Cref{thm:laft} (see \Cref{eq:lamminmax}). Then there exists some absolute constant $\xi_{\min}>0$ such that when $u=\lfloor\log n\rfloor$ and $v_0$ (and hence $u,n$) is sufficiently large, then $\eta_{\min}(u,n)\leq n^{\xi_{\min}}$. Define the absolute constant $\xi=\max\{\xi_{\mathrm{run}},\xi_{\min}\}+16$, and recall that $\tilde{\lambda}_{\mathrm{run}}=2^{v-\xi\sqrt{v}}\cdot\lambda_{\mathrm{run}}'$ by \Cref{eq:tlamrundef}. Then assuming $v_0$ is sufficiently large,
  \begin{align*}
    \tilde{\lambda}_{\mathrm{run}}
    &\leq 2^{\log^2n+3\log n-(\xi_{\mathrm{run}}+16)\log n}\cdot\lambda_{\mathrm{run}}' \\
    &= n^{\log n-(\xi_{\mathrm{run}}+13)}\cdot\lambda_{\mathrm{run}}' \\
    &\leq n^{(\log n-1)-\xi_{\mathrm{run}}-5}\cdot\lambda_{\mathrm{run}}' \\
    &\leq \frac{n^{u-\xi_{\mathrm{run}}}\cdot\lambda_{\mathrm{run}}'}{n^5} \\
    &\leq \frac{n^{u-\xi_{\mathrm{run}}}\cdot\lambda_{\mathrm{run}}'/3}{18T} \\
    &= \frac{\lambda_{\mathrm{run}}\cdot\min\{\lambda_{\mathrm{in}}'',\lambda_{\mathrm{run}}''\}}{18T}.
  \end{align*}
  The first inequality above holds by \Cref{eq:vbounds}. The fourth inequality above holds because $T\leq\eta_T\cdot n^4$ by \Cref{eq:etaT}, so when $v_0$ (and hence $n$) is sufficiently large we have $54T\leq n^5$. The final equality above holds by the definition of $\lambda_{\mathrm{run}}$, and because $\min\{\lambda_{\mathrm{in}}'',\lambda_{\mathrm{run}}''\}=\lambda_{\mathrm{run}}'/3$ as shown above. Thus \Cref{eq:sclamtil} holds.

  It follows that \Cref{lem:simcomp} gives a fault-tolerant gadget $(\tilde{\cQ},\; \tilde{\cE}_{\mathrm{run}}=2^{\tilde{N}}|_{\geq\tilde{\lambda}_{\mathrm{run}}}^{\sqcup\tilde{T}},\; \tilde{D}_{\mathrm{in}},\; \tilde{D}_{\mathrm{out}})$ for $\tilde{\bar{\cO}}=(\phi_{q,q'}^{-1})^{\sqcup k_{\mathrm{out}}'}\circ\bar{\cO}'\circ\phi_{q,q'}^{\sqcup k_{\mathrm{in}}'}$ using the desired space
  \begin{align*}
    |\tilde{N}|
    &\leq |N'|\cdot O(n)^u \\
    &\leq |N'|\cdot n^u\cdot 2^{O(u)} \\
    &\leq |N'|\cdot 2^{v+O(\sqrt{v})}
  \end{align*}
  and time
  \begin{align*}
    \tilde{T}
    &\leq T'\cdot O(u^5n^2\log^3n) \\
    &\leq T'\cdot 2^{4\sqrt{v}},
  \end{align*}
  where the final inequality above holds by \Cref{eq:vbounds} when $v_0$ (and hence $u,n$) is sufficiently large.

  By \Cref{lem:simcomp}, for $\alpha\in\{\mathrm{in},\mathrm{out}\}$ the decorated code $\tilde{D}_\alpha$ given by \Cref{eq:sctDa} has the desired parameters
  \begin{align*}
    \tilde{n}_\alpha &= n^u\cdot n_\alpha' \leq 2^v\cdot n_\alpha' \\
    \tilde{k}_\alpha &= \kappa\cdot k_\alpha' = 2^{v'-v}\cdot k_\alpha' \geq 2^{(1/3-4\epsilon)v}\cdot k_\alpha' \\
    \tilde{d}_\alpha &\geq (n/8)^u\cdot d_\alpha' \geq 2^{(\log(n)-1)(\log(n)-3)}\cdot d_\alpha' \geq 2^{v-8\sqrt{v}}\cdot d_\alpha',
  \end{align*}
  where the bound above on $\tilde{k}_\alpha$ holds by the definition of $v=\lfloor(3/4+\epsilon) v'\rfloor$, and the bound on $\tilde{d}_\alpha$ holds by \Cref{eq:vbounds}.

  By \Cref{eq:sclams} in \Cref{lem:simcomp} we have
  \begin{align*}
    \tilde{\lambda}_{\mathrm{in}}
    &= \frac{\lambda_{\mathrm{in}}\cdot\lambda_{\mathrm{in}}''}{6} \\
    &\geq \frac{\lambda_{\max}(u,n)\cdot\lambda_{\mathrm{in}}'}{12} \\
    &= \frac{(n/8)^{u-2}/4\cdot\lambda_{\mathrm{in}}'}{12} \\
    &\geq (n/16)^{u-2}\cdot\lambda_{\mathrm{in}}' \\
    &\geq 2^{(\log(n)-3)(\log(n)-4)}\cdot\lambda_{\mathrm{in}}' \\
    &\geq 2^{v-16\sqrt{v}}\cdot\lambda_{\mathrm{in}}'.
  \end{align*}
  The first inequality above holds by the definition of $\lambda_{\mathrm{in}}=\lambda_{\max}(u,n)$ in \Cref{def:simgad,lem:simcomp} and by the assumption that $\lambda_{\mathrm{in}}'\geq\lambda_{\mathrm{run}}'$ so that $\lambda_{\mathrm{in}}''\geq\lambda_{\mathrm{in}}'/2$. The second inequality holds assuming $v_0$ (and hence $u,n$) is sufficiently large. The third inequality holds because $u=\lfloor\log n\rfloor$, and the fourth inequality holds by \Cref{eq:vbounds}. Also by \Cref{eq:sclams},
  \begin{align}
    \label{eq:sclamout}
    \begin{split}
      \tilde{\lambda}_{\mathrm{out}}
      &= \left(\frac{9n^u}{\lambda_{\mathrm{run}}}+\eta_{\min}(u,n)\right)\cdot T\cdot\tilde{\lambda}_{\mathrm{run}}+n^u\cdot\lambda_{\mathrm{out}}'' \\
      &\leq (9n^{\xi_{\mathrm{run}}}+n^{\xi_{\min}})\cdot (n^5/18)\cdot\tilde{\lambda}_{\mathrm{run}}+n^u\cdot\lambda_{\mathrm{out}}' \\
      &\leq n^{\max\{\xi_{\mathrm{run}},\xi_{\min}\}+5}\cdot\tilde{\lambda}_{\mathrm{run}}+n^u\cdot\lambda_{\mathrm{out}}' \\
      &\leq 2^{(\max\{\xi_{\mathrm{run}},\xi_{\min}\}+5)\sqrt{v}}\cdot 2^{v-\xi\sqrt{v}}\cdot\lambda_{\mathrm{run}}'+2^{v-1}\cdot\lambda_{\mathrm{out}}' \\
      &\leq 2^{v-1}\cdot\lambda_{\mathrm{run}}'+2^{v-1}\cdot\lambda_{\mathrm{out}}' \\
      &\leq 2^v\cdot\lambda_{\mathrm{out}}',
    \end{split}
  \end{align}
  The first inequality above holds by the definition of $\lambda_{\mathrm{run}},\xi_{\min},\lambda_{\mathrm{out}}''$, and because $18T\leq n^5$ assuming $v_0$ is sufficiently large as described previously. The third inequality holds by \Cref{eq:vbounds} and by the definition of $\tilde{\lambda}_{\mathrm{run}}$. The fourth inequality holds by the definition of $\xi$. The fifth inequality holds by the assumption that $\lambda_{\mathrm{run}}'\leq\lambda_{\mathrm{out}}'$. As $(\tilde{\cQ},\; \tilde{\cE}_{\mathrm{run}},\; \tilde{D}_{\mathrm{in}},\; \tilde{D}_{\mathrm{out}})$ remains a fault-tolerant gadget if we increase $\tilde{\lambda}_{\mathrm{out}}$, by \Cref{eq:sclamout} we may simply take
  \begin{equation*}
    \tilde{\lambda}_{\mathrm{out}} = 2^v\cdot\lambda_{\mathrm{out}}'.
  \end{equation*}
  Thus we have shown that \Cref{eq:sclamscor} holds, completing the proof of the lemma.
\end{proof}




\Cref{cor:simcomp} reduces the alphabet size of a fault-tolerance scheme with just a mild decay in the fault-tolerance properties. However, the assumption that $v'\geq v_0(\epsilon)$ means that we cannot apply \Cref{cor:simcomp} when the alphabet size is too small (i.e.~below some constant assuming $r,\;1/\epsilon=O(1)$). We therefore also show the more basic alphabet reduction result in \Cref{lem:basiccomp} below, which works for arbitrary alphabet sizes, at the cost of a more severe decay in the fault-tolerance properties. This significant decay in parameters means that \Cref{lem:basiccomp} will be unsuitable when the starting alphabet size is exponentially large in the number of physical qudits, in which case we will need to first apply \Cref{cor:simcomp}.

\begin{lemma}
  \label{lem:basiccomp}
  Fix a prime power $r$. Let $q'=r^\kappa$ for some $\kappa\in\bN$. For $\alpha\in\{\mathrm{in},\mathrm{out}\}$ for some $\lambda_\alpha'\geq 0$ let
  \begin{equation*}
    D_\alpha' = (C_\alpha',\; \Enc_\alpha',\; \cE_\alpha'=2^{[n_\alpha']}|_{\geq\lambda_\alpha'})
  \end{equation*}
  be a $[[n_\alpha',k_\alpha',d_\alpha']]_{q'}$ decorated CSS non-subsystem code. For some $\lambda_{\mathrm{run}}'\geq 0$, let
  \begin{equation*}
    (\cQ',\; \cE_{\mathrm{run}}'=2^{N'}|_{\geq\lambda_{\mathrm{run}}'}^{\sqcup T'},\; D_{\mathrm{in}}',\; D_{\mathrm{out}}')
  \end{equation*}
  be a fault-tolerant gadget for a set $\bar{\cO}'$ of superoperators $\bar{O}':\bC^{\bF_{q'}^{k_{\mathrm{in}}'}\times\bF_{q'}^{k_{\mathrm{in}}'}}\rightarrow\bC^{\bF_{q'}^{k_{\mathrm{out}}'}\times\bF_{q'}^{k_{\mathrm{out}}'}}$ on $q'$-dimensional qudits, using space $N'$, time $T'$, and gate set $\cG=\{\gH^*,\gX^*,\gZ^*,\gCX^*,\gCCX^*,\gInit_*,\gCO_*\}$.

  For $\alpha\in\{\mathrm{in},\mathrm{out}\}$, define the $[[\tilde{n}_\alpha=\kappa\cdot n_\alpha',\; \tilde{k}_\alpha=\kappa\cdot k_\alpha',\; \tilde{d}_\alpha\geq d_\alpha']]_r$ code $\tilde{C}_\alpha$ with associated encoding map $\widetilde{\Enc}_\alpha$ by
  \begin{align*}
    \tilde{C}_\alpha &= (\phi_{r,q'}^{-1})^{\sqcup n_\alpha}(C_\alpha') \\
    \widetilde{\Enc}_\alpha &= (\phi_{r,q'}^{-1})^{\sqcup n_\alpha'}\circ\Enc_\alpha'\circ\phi_{r,q'}^{\sqcup k_\alpha'}.
  \end{align*}
  Define an associated decorated code $\tilde{D}_\alpha=(\tilde{C}_\alpha,\; \widetilde{\Enc}_\alpha,\; \tilde{\cE}_\alpha=2^{[\kappa]\times[n_\alpha']}|_{\geq\tilde{\lambda}_\alpha})$ for
  \begin{align*}
    \tilde{\lambda}_{\mathrm{in}} &= \lambda_{\mathrm{in}}' - \lambda_{\mathrm{run}}' \\
    \tilde{\lambda}_{\mathrm{out}} &= \kappa\cdot\lambda_{\mathrm{out}}',
  \end{align*}
  and let
  \begin{align*}
    \tilde{\lambda}_{\mathrm{run}} &= \frac{1}{6(32\log(\kappa)+256)}\cdot\lambda_{\mathrm{run}}'
  \end{align*}

  Then there exists a fault-tolerant gadget on $r$-dimensional qudits
  \begin{equation*}
    (\tilde{\cQ},\; \tilde{\cE}_{\mathrm{run}}=2^{\tilde{N}}|_{\geq\tilde{\lambda}_{\mathrm{run}}}^{\sqcup\tilde{T}},\; \tilde{D}_{\mathrm{in}},\; \tilde{D}_{\mathrm{out}})
  \end{equation*}
  for $\tilde{\bar{\cO}}=(\phi_{r,q'}^{-1})^{\sqcup k_{\mathrm{out}}'}\circ\bar{\cO}'\circ\phi_{r,q'}^{\sqcup k_{\mathrm{in}}'}$, where $\tilde{\cQ}$ is a quantum circuit using space $|\tilde{N}|\leq 16\kappa^3\cdot|N'|$ and time $\tilde{T}\leq(32\log(\kappa)+256)\cdot T'$.
\end{lemma}
\begin{proof}
  Define $N=[16\kappa^3]$ and $T=32\log(\kappa)+256$. We define $\tilde{\cQ}$ to act on physical qudits $\tilde{N}=N\times N'$ and run in time $\tilde{T}=T\cdot T'$. For simplicity we let the input and output qudits of $\cQ'$ be labeled by $[n_{\mathrm{in}}'],\;[n_{\mathrm{out}}']\subseteq N'$ respectively. Then $\tilde{\cQ}$ has input and output qudits $[\kappa]\times[n_{\mathrm{in}}'],\;[\kappa]\times[n_{\mathrm{out}}']\subseteq\tilde{N}$ respectively. We replace each gate $G_{B,t}$ in $\cQ'$ acting on some set of qudits $B\subseteq N'$ in timestep $t\in[T']$ with a subcircuit $\tilde{\cQ}_{B,t}$ of $\tilde{\cQ}$ acting on qudits $N\times B\subseteq\tilde{N}$ in timesteps $T(t-1)+1,\dots,Tt$. In timesteps $T(t-1)+1$ and $Tt$, we let $\tilde{\cQ}_{B,t}$ call $\gInit_Z$ on qudits $(N\setminus[\kappa])\times B$.

  If $G_{B,t}\in\{\gH^*,\gX^*,\gZ^*,\gCX^*,\gCCX^*,\gInit_*\}$, then in timesteps $T(t-1)+2,\dots,Tt-1$, we let $\tilde{\cQ}_{B,t}$ call the circuit $\bar{\cQ}_{G_{B,t}}$ in \Cref{lem:qimp}, and then idle for any remaining timesteps before $Tt$. Thus circuit $\tilde{\cQ}_{B,t}$ fits within our alloted qudits $N\times B$ and timesteps $T(t-1)+2,\dots,Tt-1$ by the definition of $N,T$ along with the space and time bounds in \Cref{lem:qimp}. \Cref{lem:qimp} also implies that
  \begin{equation}
    \label{eq:bcgateconj}
    \tilde{\cQ}_{B,t} = (\phi_{r,q'}^{-1})^{\sqcup B}\circ G_{B,t}\circ\phi_{r,q'}^{\sqcup B},
  \end{equation}
  where in the above equation $\tilde{\cQ}_{B,t}$ denotes the channel associated to this circuit with $r$-dimensional input and output qudits labeled by $[\kappa]\times B$.

  If instead $G_{B,t}=\gCO_{\alpha,f}$ for some $\alpha\in\{X,Z\}$ and some $f:\bF_{q'}^B\rightarrow\bF_{q'}^B$, then we let $\tilde{\cQ}_{B,t}$ apply the gate $\gCO_{\alpha,F}$ to qudits $[\kappa]\times B$ in timestep $T(t-1)+1$, where $F:\bF_r^{N\times B}\rightarrow\bF_r^{N\times B}$ is defined as follows. If $\alpha=Z$, we let
  \begin{equation*}
    F = (\phi_{r,q'}^{-1})^{\sqcup B}\circ f\circ\phi_{r,q'}^{\sqcup B}.
  \end{equation*}
  If instead $\alpha=X$, we first let $A\in\bF_r^{\kappa\times\kappa}$ be the unique invertible matrix satisfying
  \begin{equation*}
    z^\top Ax = \tr_{\bF_{q'}/\bF_r}(\phi_{r,q'}(x)\phi_{r,q'}(z))
  \end{equation*}
  for every $x,z\in\bF_r^\kappa$. The nondegeneracy of the trace bilinear form ensures that $A$ is well defined. Also letting $A$ denote the associated unitary in $\bC^{\bF_r^\kappa\times\bF_r^\kappa}$ given by $A\ket{x}=\ket{Ax}$, then as shown in \Cref{eq:hadalphred} in the proof of \Cref{lem:qimp}, we have
  \begin{equation}
    \label{eq:bchad}
    \gH_{q'}\phi_{r,q'} = \phi_{r,q'} \gH_r^{\otimes\kappa}A,
  \end{equation}
  where $\gH_r,\gH_{q'}$ denote the Hadamard gates over the respective fields $\bF_r,\bF_{q'}$.
  We then define
  \begin{equation*}
    F = (A\circ\phi_{r,q'}^{-1})^{\sqcup B}\circ f\circ(\phi_{r,q'}\circ A^{-1})^{\sqcup B}.
  \end{equation*}
  We then let $\tilde{\cQ}_{B,t}$ idle for the remaining timesteps $T(t-1)+2,\dots,Tt-1$. This definition ensures that \Cref{eq:bcgateconj} also holds for $G_{B,t}\in\gCO_{\alpha,f}$, where in particular the $\alpha=X$ case follows by \Cref{eq:bchad} along with the fact that by \Cref{def:gates} we have
  \begin{equation*}
    \gCO_{X,f} = \gH^{\sqcup B}\circ\gCO_{Z,f}\circ(\gH^\dagger)^{\sqcup B}.
  \end{equation*}

  The isomorphisms $\phi_{r,q'}$ and $A$ together provide an isomorphism between the space of postselections for $\tilde{\cQ}$ and for $\cQ'$. Specifically, because \Cref{lem:qimp} does not use any gates in $\gCO_*$, every $\gCO_*$ gate in $\tilde{\cQ}$ corresponds to the $\gCO_*$ gate in $\tilde{\cQ}_{B,t}$ for some $B,t$ such that $G_{B,t}\in\gCO_*$. If $\zeta_{B,t}'\in\bC^{\bF_{q'}^B}$ is a postselection for $G_{B,t}\in\gCO_{Z,*}$ (resp.~$G_{B,t}\in\gCO_{X,*}$), then applying the isomorphism $(\phi_{r,q'}^{-1})^{\sqcup B}$ (resp.~$(A\circ\phi_{r,q'}^{-1})^{\sqcup B}$) yields a corresponding postselection $\tilde{\zeta}_{B,t}\in\bC^{(\bF_r^\kappa)^B}$ for $\tilde{\cQ}_{B,t}$, such that
  \begin{equation}
    \label{eq:bcgateconjpost}
    \tilde{\cQ}_{B,t}[\tilde{\zeta}_{B,t}] = (\phi_{r,q'}^{-1})^{\sqcup B}\circ G_{B,t}[\zeta_{B,t}']\circ\phi_{r,q'}^{\sqcup B}.
  \end{equation}
  The isomorphism $\phi_{r,q'}^{\sqcup B}$ (resp.~$\phi_{r,q'}^{\sqcup B}\circ A^{-1}$) similarly maps every $\tilde{\zeta}_{B,t}$ back to the corresponding $\zeta_{B,t}'$.
  Because \Cref{lem:qimp} does not use any gates in $\gCO_*$, \Cref{eq:bcgateconjpost} is vacuously true for $B,t$ with $G_{B,t}\notin\gCO_*$. Therefore in a slight abuse of notation, we may view a postselection $\tilde{\zeta}_{B,t}$ for $\tilde{\cQ}_{B,t}$ also as a postselection for $G_{B,t}$, and vice versa. We may also combine these postselections across all $B,t$, to obtain an isomorphism between postselections for $\tilde{\cQ}$ and for $\cQ'$. Therefore we also may view a postselection $\tilde{\zeta}$ for $\tilde{\cQ}$ as a postselection for $\cQ'$, and vice versa.
  
  We will now show that $(\tilde{\cQ},\; \tilde{\cE}_{\mathrm{run}},\; \tilde{D}_{\mathrm{in}},\; \tilde{D}_{\mathrm{out}})$ is a fault-tolerant gadget for $\tilde{\bar{\cO}}=(\phi_{r,q'}^{-1})^{\sqcup k_{\mathrm{out}}'}\circ\bar{\cO}'\circ\phi_{r,q'}^{\sqcup k_{\mathrm{in}}'}$. Let $\ell\in\bN$, let $\tilde{\rho}\in\bC^{\bF_r^{\kappa k_{\mathrm{in}}'+\ell}\times\bF_r^{\kappa k_{\mathrm{in}}'+\ell}}$, and let $\tilde{\sigma}$ be a $\tilde{\cE}_{\mathrm{in}}\sqcup [\ell]$-deviation of $\widetilde{\Enc}_{\mathrm{in}}\otimes I_\ell(\tilde{\rho})$. Let $\tilde{\cF}$ be a $\cE_{\mathrm{run}}$-avoiding Pauli fault for $\tilde{\cQ}$, and let $\tilde{\zeta}$ be a postselection for $\tilde{\cQ}$. By \Cref{lem:paulift}, it suffices to show that $\tilde{\cQ}[\tilde{\cF},\tilde{\zeta}]\otimes I_\ell(\tilde{\sigma})$ is a $\tilde{\cE}_{\mathrm{out}}\sqcup [\ell]$-deviation of $(\widetilde{\Enc}_{\mathrm{out}}\circ\tilde{\bar{\cO}})\otimes I_\ell(\rho)$.

  For this purpose, similarly as in the proof of \Cref{lem:simcomp}, we first define an extended fault $\tilde{\cF}'$ (see \Cref{def:extfault})
  such that the superoperators $\tilde{\cQ}[\tilde{\cF}',\tilde{\zeta}](\cdot)=\tilde{\cQ}[\tilde{\cF},\tilde{\zeta}](\cdot)$ are equal. We will ensure that $\supp(\tilde{\cF}')$ has a structure amenable to applying \Cref{eq:bcgateconjpost}. In a slight departure from our standard notation, we also allow $\tilde{\cF}'=(\tilde{F}'_0,\dots,\tilde{F}'_{\tilde{T}})$ to include a superoperator $\tilde{F}'_0$ applied before the first timestep of $\tilde{\cQ}$.
  
  Specifically, we let the extended fault $\tilde{\cF}'$ have access to another set of qudits $\tilde{A}\cong\tilde{N}$. For every $t\in[T']$, let $G_{t,1},\dots,G_{t,m_t}$ denote the gates making up $Q_t'$, acting on qudits $B_{t,1},\dots,B_{t,m_t}$ respectively with $B_{t,1}\sqcup\cdots\sqcup B_{t,m_t}=N'$, so that $G_{t,i}=G_{B_{m_i},t}$. Similarly let $\tilde{\cQ}_{t,i}=\tilde{\cQ}_{B_{m_i},t}$. For every $i\in[m_t]$, we define the action of $\tilde{\cF}'$
  on qudits $N\times B_{t,i}$ over timesteps $T(t-1),\dots,Tt$ as follows:
  \begin{enumerate}
  \item If $G_{t,i}\notin\gCO_*$ (so that the postselection $\tilde{\zeta}$ are trivial as $\tilde{\cQ}_{t,i}$ has no $\gCO_*$ gates by \Cref{lem:qimp}):
    \begin{enumerate}
    \item If it holds for every $\tilde{t}\in\{T(t-1)+1,\dots,Tt\}$ that
      \begin{equation}
        \label{eq:bcgood}
        \supp(\tilde{F}_{\tilde{t}})\cap(N\times B_{t,i})| = \emptyset,
      \end{equation}
      then for $\tilde{t}\in\{T(t-1)+1,\dots,Tt\}$, on qudits $N\times B_{t,i}\subseteq\tilde{N}$ we let $\tilde{F}_{\tilde{t}}'$ apply the identity (just like $\tilde{F}_{\tilde{t}}$ does).
    \item If \Cref{eq:bcgood} is violated for some $\tilde{t}\in\{T(t-1)+1,\dots,Tt\}$, we let $\tilde{F}_{T(t-1)}'$ apply $\gInit_Z$ to the qudits labeled $N\times B_{t,i}$ in $\tilde{A}$, and then swap the contents of the qudits labeled $[\kappa]\times B_{t,i}\subseteq N\times B_{t,i}$ in $\tilde{A}$ with their counterparts (i.e.~the qudits with the same labels $[\kappa]\times B_{t,i}$) in $\tilde{N}$, and subsequently apply the entire noisy circuit\footnote{In this expression $\tilde{\cQ}_{t,i}[\tilde{\cF},\tilde{\zeta}]$, we implicitly restrict $\tilde{\cF},\tilde{\zeta}$ to gates acting on qudits $N\times B_{t,i}$, in timesteps $T(t-1)+1,\dots,Tt$.} $\tilde{\cQ}_{t,i}[\tilde{\cF},\tilde{\zeta}]$ to these resulting qudits in $\tilde{A}$. The superoperators $\tilde{F}_{T(t-1)+1}',\dots,\tilde{F}_{Tt'-1}'$ then act as the identity on qudits labeled $N\times B_{t,i}$. Finally, the superoperator $\tilde{F}_{Tt'}'$ swaps back the qudits labeled $[\kappa]\times B_{t,i}$ in $\tilde{A}$ with their counterparts in $\tilde{N}$, and then applies $\gInit_Z$ to the resulting qudits labeled $N\times B_{t,i}$ in $\tilde{A}$.
    \end{enumerate}
  \item If $G_{t,i}\in\gCO_*$, recall from above that on qudits $[\kappa]\times B_{t,i}$, the circuit $\tilde{\cQ}_{t,i}$ applies a $\gCO_*$ gate in timestep $T(t-1)+1$, and then idles (i.e.~applies identity gates) for the remaining timesteps $T(t-1)+2,\dots,Tt$. Therefore on qudits $[\kappa]\times B_{t,i}$, $\tilde{\cQ}_{t,i}[\tilde{\cF},\tilde{\zeta}]$ simply applies a $\gCO_*$ gate followed by a Pauli error, given by the product (i.e.~sequential composition) $F_{t,i}$ of the restrictions of $\tilde{F}_{T(t-1)+1},\dots,\tilde{F}_{Tt}$ to qudits $[\kappa]\times B_{t,i}$. Meanwhile, on qudits $N\setminus[\kappa]\times B_{t,i}$, the circuit $\tilde{\cQ}_{t,i}$ simply applies $\gInit_Z$ gates in timesteps $T(t-1)+1$ and $Tt$, and idles for the intermediate timesteps $T(t-1)+2,\dots,Tt-1$. Therefore on qudits $N\setminus[\kappa]\times B_{t,i}$, $\tilde{\cQ}_{t,i}[\tilde{\cF},\tilde{\zeta}]$ has the effect of applying $\gInit_Z$ while rescaling the state by a phase $\beta_{t,i}$ (which may be $0$) that arises when tracing out states given by applying Pauli superoperators to $\ket{0}\bra{0}$ states. Thus we define $\tilde{F}_{T(t-1)+1}',\dots,\tilde{F}_{Tt-1}$ to idle on qudits $N\times B_{t,i}$, and we let $\tilde{F}_{Tt}$ apply $\beta_{t,i}F_{t,i}$ to qudits $[\kappa]\times B_{t,i}$.
  \end{enumerate}
  We then have an equality of superoperators $\tilde{\cQ}[\tilde{\cF}',\tilde{\zeta}](\cdot)=\tilde{\cQ}[\tilde{\cF},\tilde{\zeta}](\cdot)$, as $\tilde{\cF}'$ implements the exact same Pauli corruption as $\tilde{\cF}$, just sometimes in the roundabout fashions of:
  \begin{enumerate}
  \item swapping out some qudits in $\tilde{N}$ to $\tilde{A}$, applying the corrupted gates on $\tilde{A}$, and then swapping the qudits back, and
  \item accumulating Pauli errors across idling space-time regions of the circuit, and applying them together in the last timestep of the region.
  \end{enumerate}
  Specifically, each of the pairs of swaps described above is sandwiched by $\gInit_Z$ gates on the involved qudits in $\tilde{A}$, and the postselection $\tilde{\zeta}$ on the associated region in $\tilde{N}$ is trivial. Hence up to relabeling of some qudits from the swap gates, $\tilde{\cQ}[\tilde{\cF}',\tilde{\zeta}](\cdot)$ simply runs $\tilde{\cQ}[\tilde{\cF},\tilde{\zeta}](\cdot)$, while sometimes initializing an additional block of qudits in the $\ket{0}$ state, applying $\cQ_{t,i}=\cQ_{B_{t,i},t}$ (which is a channel due to the trivial postselection), and then tracing out this channel's output. These additional initializations, channels, and tracing-outs have no effect on the execution of $\tilde{\cQ}[\tilde{\cF},\tilde{\zeta}](\cdot)$, as desired.

  Therefore it suffices to show that the resulting output $\tilde{\cQ}[\tilde{\cF}',\tilde{\zeta}]\otimes I_\ell(\tilde{\sigma})$ is a $\tilde{\cE}_{\mathrm{out}}\sqcup [\ell]$-deviation of $(\widetilde{\Enc}_{\mathrm{out}}\otimes\tilde{\bar{\cO}})\otimes I_\ell(\tilde{\rho})$. For this purpose, similarly as in the proof of \Cref{lem:extft}, for each $\tilde{t}\in[\tilde{T}]$, we may decompose $\tilde{F}'_{\tilde{t}}$ into a linear combination of Pauli superoperators $\tilde{F}''_{\tilde{t}}$ acting on qudits $\tilde{N}\sqcup\tilde{A}$, with $\supp(\tilde{F}'')_{\tilde{t}}\subseteq\supp(\tilde{F}')_{\tilde{t}}$. Letting each $\tilde{\cF}''=(\tilde{F}''_1,\dots,\tilde{F}''_{\tilde{T}})$, then $\tilde{\cQ}[\tilde{\cF}',\tilde{\zeta}]\otimes I_\ell(\tilde{\sigma})$ is a linear combination of states of the form $\tilde{\cQ}[\tilde{\cF}'',\tilde{\zeta}]\otimes I_\ell(\tilde{\sigma})$ for Pauli faults $\tilde{\cF}''$ with $\supp(\tilde{\cF}'')\subseteq\supp(\tilde{\cF}')$. As in the proof of \Cref{lem:extft}, because each Pauli superoperator $\tilde{F}''_{\tilde{t}}$ does not entangle qudits $\tilde{N}$ and $\tilde{A}$, these two sets of qudits never become entangled, and hence the output $\tilde{\cQ}[\tilde{\cF}'',\tilde{\zeta}]\otimes I_\ell(\tilde{\sigma})$ is preserved (up to a global phase) if we restrict attention to $\tilde{N}$, and replace the Paulis gates in $\tilde{\cF}''$ acting on $\tilde{A}$ with identity gates (or equivalently, simply remove qudits $\tilde{A}$).

  Thus we have reduced our problem to showing the following claim:

  \begin{claim}
    \label{claim:bcnicefault}
    For
    every (non-extended) Pauli fault $\tilde{\cF}''$ with $\supp(\tilde{\cF}'')\subseteq\supp(\tilde{\cF}')$
    for $\tilde{\cQ}$, the output $\tilde{\cQ}[\tilde{\cF}'',\tilde{\zeta}]\otimes I_\ell(\tilde{\sigma})$ is a $\tilde{\cE}_{\mathrm{out}}\sqcup [\ell]$-deviation of $(\widetilde{\Enc}_{\mathrm{out}}\otimes\tilde{\bar{\cO}})\otimes I_\ell(\tilde{\rho})$.
  \end{claim}

  We now turn to proving \Cref{claim:bcnicefault}. The key idea is that by construction, $\tilde{\cF}'$ and hence $\tilde{\cF}''$ has support inside timesteps $\{Tt:t\in[T']\}$, and hence we obtain a precise equivalence between the execution of $\tilde{\cQ}[\tilde{\cF}'',\tilde{\zeta}]$ and that of $\cQ'[\cF',\tilde{\zeta}]$ for appropriate $\cF'$.

  We begin by introducing some additional notation.
  Below, for $t\in[T']$, we let $\tilde{\cQ}_{\leq Tt}$ denote the circuit given by the first $Tt$ timesteps of $\tilde{\cQ}$, with input qudits $[\kappa]\times[n_{\mathrm{in}}']$ and output qudits $[\kappa]\times N'$. We write $\tilde{\cQ}_{\leq Tt}[\tilde{\cF},\tilde{\zeta}]$ to denote $\tilde{\cQ}_{\leq Tt}$ with fault $\tilde{\cF}$ and postselection $\tilde{\zeta}$ restricted to the first $Tt$ timesteps. We similarly let $\cQ'_{\leq t}$ denote the circuit given by the first $t$ timesteps of $\cQ'$, with input qudits $[n_{\mathrm{in}}']$ and output qudits $N'$. We also write $\cQ'_{\leq t}[\cF',\zeta']$ to denote $\cQ'_{\leq t}$ with fault $\cF'$ and postselection $\zeta'$ restricted to the first $t$ timesteps.

  We also fix $\tilde{\cF}',\tilde{\cF}''$ as in \Cref{claim:bcnicefault}. Recalling that $\tilde{\sigma}$ is a Pauli $\tilde{\cE}_{\mathrm{in}}\sqcup [\ell]$-deviation of $\widetilde{\Enc}_{\mathrm{in}}\otimes I_\ell(\tilde{\rho})$, we can write
  \begin{equation*}
    \tilde{\sigma} = (\tilde{P}_0\circ\widetilde{\Enc}_{\mathrm{in}})\otimes I_\ell(\tilde{\rho}) = (\tilde{P}_0\circ(\phi_{r,q'}^{-1})^{\sqcup n_{\mathrm{in}}'}\circ\Enc'\circ\phi_{r,q'}^{\sqcup k_{\mathrm{in}}'})\otimes I_\ell(\tilde{\rho}).
  \end{equation*}
  for some Pauli superoperator $\tilde{P}_0$ of weight $|\tilde{P}_0|<\tilde{\lambda}_{\mathrm{in}}$ acting on qudits $[\kappa]\times[n_{\mathrm{in}}']$.
  For $t\in[T']$, we define $A_t\subseteq N'$ to contain every $b\in N'$ such that the following holds: letting $i\in[m_t]$ denote the unique index for which $B_{t,i}\subseteq N'$ defined above contains $b$, then there exists some $\tilde{t}\in\{T(t-1)+1,\dots,Tt\}$ such that either $G_{t,i}\notin\gCO_*$ and $\supp(\tilde{F}_{\tilde{t}})\cap(N\times B_{t,i})\neq\emptyset$, or $G_{t,i}\in\gCO_*$ and $\supp(\tilde{F}_{\tilde{t}})\cap(N\times\{b\})\neq\emptyset$. In other words, $A_t$ contains every $b\in N'$ for which we either insert swap gates, or accumulate a nonzero number of Pauli errors, on qudits $N\times\{b\}$ during timesteps $T(t'-1)+1,\dots,Tt'$ in the definition of the extended fault $\tilde{\cF}'$ above. Then letting $A_0=A_{T'+1}=\emptyset$, by construction we have
  \begin{align}
    \label{eq:bcsuppF}
    \begin{split}
      \supp(\tilde{F}_{Tt}'') \subseteq \supp(\tilde{F}_{Tt}') &\subseteq A_t\cup A_{t+1} \hspace{4em} \forall\; t\in\{0,\dots,T'\} \\
      \supp(\tilde{F}_{\tilde{t}}'') \subseteq \supp(\tilde{F}_{\tilde{t}}') &= \emptyset \hspace{4em} \forall\; \tilde{t}\in[\tilde{T}]\setminus T\cdot[T'].
    \end{split}
  \end{align}
  Furthermore, for $t\in[T']$,
  \begin{equation}
    \label{eq:bcAtbound}
    |A_t| < 3T\cdot\tilde{\lambda}_{\mathrm{run}} = \frac12\cdot\lambda_{\mathrm{run}}',
  \end{equation}
  as because $\tilde{\cF}$ is $\tilde{\cE}_{\mathrm{run}}$-avoiding, there are $<T\cdot\tilde{\lambda}_{\mathrm{run}}$ distinct $b'\in N'$ for which $\supp(\tilde{F})$ has nonempty support within the set $(N\times\{b'\})\times\{T(t-1)+1,\dots,Tt\}$, and by definition every $b\in A_t$ either equals such a $b'$, or else shares a $\gCX^*$ or $\gCCX^*$ gates $G_{t,i}$ with such a $b'$.

  We will use \Cref{claim:bcinduct} below, which we will prove by induction, to prove \Cref{claim:bcnicefault}.

  \begin{claim}
    \label{claim:bcinduct}
    For every $t\in[T']$, there exists a $\cE_{\mathrm{in}}'\sqcup [\ell]$-deviation $\sigma'$ of $(\Enc'\circ\phi_{r,q'}^{\sqcup k_{\mathrm{in}}})\otimes I_\ell(\tilde{\rho})$ and a $\cE_{\mathrm{run}}'$-avoiding Pauli $\cF'$ fault for $\cQ'_{\leq t}$ such that
    \begin{align}
      \label{eq:bcinduct}
      \tilde{\cQ}_{\leq Tt}[\tilde{\cF}'',\tilde{\zeta}]\otimes I_\ell(\tilde{\sigma})
      &= ((\phi_{r,q'}^{-1})^{\sqcup N'}\circ\cQ'_{\leq t}[\cF',\tilde{\zeta}])\otimes I_\ell(\sigma').
    \end{align}
  \end{claim}
  \begin{proof}
    We prove the claim by induction on $t\in[T']$. For the base case, we will show that the claim holds for $t=1$. Recall that the input state $\tilde{\sigma}$ is a $\tilde{\cE}_{\mathrm{in}}\sqcup [\ell]$-deviation of
    \begin{equation*}
      \widetilde{\Enc}_{\mathrm{in}}\otimes I_\ell(\tilde{\rho}) = ((\phi_{r,q'}^{-1})^{\sqcup n_{\mathrm{in}}'}\circ\Enc_{\mathrm{in}}'\circ\phi_{r,q'}^{\sqcup k_{\mathrm{in}}'})\otimes I_\ell(\tilde{\rho}),
    \end{equation*}
    which means that the state
    \begin{equation*}
      \sigma_0' := \phi_{r,q'}^{\sqcup n_{\mathrm{in}}'}(\tilde{\sigma})
    \end{equation*}
    is a $2^{[n_{\mathrm{in}}']}|_{\geq\tilde{\lambda}_{\mathrm{in}}}\sqcup [\ell]$-deviation of
    \begin{equation*}
      (\Enc_{\mathrm{in}}'\circ\phi_{r,q'}^{\sqcup k_{\mathrm{in}}'})\otimes I_\ell(\tilde{\rho}).
    \end{equation*}
    For each $i\in[m_1]$, by definition $\tilde{\cQ}_{\leq T}[\tilde{\cF}'',\tilde{\zeta}]$ applies $\tilde{\cQ}_{1,i}[\tilde{\cF}'',\tilde{\zeta}]=\tilde{\cQ}_{B_{1,i},1}[\tilde{\cF}'',\tilde{\zeta}]$ to qudits $N\times B_{1,i}$, where $\tilde{\cF}'',\tilde{\zeta}$ here are implicitly restricted to these qudits and the appropriate timesteps $\{1,\dots,T\}$. By \Cref{eq:bcgateconjpost,eq:bcsuppF}, it follows that
    \begin{align*}
      \tilde{\cQ}_{\leq T}[\tilde{\cF}'',\tilde{\zeta}]\otimes I_\ell(\tilde{\sigma})
      &= \cQ'_{\leq 1}[\cF',\tilde{\zeta}]\otimes I_\ell(\sigma')
    \end{align*}
    where $\sigma'=P(\sigma_0')$ for some Pauli superoperator $P$ supported inside $A_1$, and $F_1'=\phi_{r,q'}^{\sqcup N'}\circ\tilde{F}_T''\circ(\phi_{r,q'}^{-1})^{\sqcup N'}$ is supported inside $A_1\cup A_2$. Therefore by \Cref{eq:bcAtbound}, $\sigma'$ differs from $(\Enc_{\mathrm{in}}'\circ\phi_{r,q'}^{\sqcup k_{\mathrm{in}}'})\otimes I_\ell(\tilde{\rho})$ by a Pauli error of weight less than
    \begin{equation*}
      \tilde{\lambda}_{\mathrm{in}}+|A|_1 \leq (\lambda_{\mathrm{in}}'-\lambda_{\mathrm{run}}')+\frac12\cdot\lambda_{\mathrm{run}}' \leq \lambda_{\mathrm{in}}',
    \end{equation*}
    while $F_1'$ is a Pauli error of weight less than
    \begin{equation*}
      |A_1\cup A_2| \leq |A_1|+|A_2| \leq \lambda_{\mathrm{run}}'.
    \end{equation*}
    Thus we have shown that \Cref{eq:bcinduct} holds for $t=1$, completing the base case of our induction.

    For the inductive step, assume that \Cref{eq:bcinduct} holds for some $t-1\in\{1,\dots,T'-1\}$. Then by similar reasoning as for the base case above, \Cref{eq:bcgateconjpost,eq:bcsuppF} imply that
    \begin{align*}
      \tilde{\cQ}_{\leq Tt}[\tilde{\cF}'',\tilde{\zeta}]\otimes I_\ell(\tilde{\sigma})
      &= \cQ'_{\leq t}[\cF',\tilde{\zeta}]\otimes I_\ell(\sigma'),
    \end{align*}
    where $\sigma'$ and $F_1',\dots,F_{t-1}'$ are as defined above, and $F_t'=\phi_{r,q'}^{\sqcup N'}\circ\tilde{F}_{Tt}''\circ(\phi_{r,q'}^{-1})^{\sqcup N'}$ . Therefore by \Cref{eq:bcAtbound}, $F_t'$ is a Pauli error of weight less than
    \begin{equation*}
      |A_t\cup A_{t+1}| \leq |A_t|+|A_{t+1}| \leq \lambda_{\mathrm{run}}'.
    \end{equation*}
    Thus we have shown that \Cref{eq:bcinduct} holds for $t$, completing the inductive step.
  \end{proof}

  We now apply \Cref{claim:bcinduct} to prove \Cref{claim:bcnicefault}, thereby completing the proof of \Cref{lem:basiccomp}.

  \begin{proof}[Proof of \Cref{claim:bcnicefault}]
    let $\sigma$ be the $\cE_{\mathrm{in}}'\sqcup [\ell]$-deviation of $(\Enc'\circ\phi_{r,q'}^{\sqcup k_{\mathrm{in}}})\otimes I_\ell(\tilde{\rho})$ and let $\cF'$ be the $\cE_{\mathrm{run}}'$-avoiding Pauli fault for $\cQ'_{\leq t}$ given in \Cref{claim:bcinduct}. Then
    \begin{align*}
      \tilde{\cQ}[\tilde{\cF}'',\tilde{\zeta}]\otimes I_\ell(\tilde{\sigma})
      &= \tr_{[\kappa]\times(N'\setminus[n'_{\mathrm{out}}])}(\tilde{\cQ}_{\leq TT'}[\tilde{\cF}'',\tilde{\zeta}]\otimes I_\ell(\tilde{\sigma})) \\
      &= \tr_{[\kappa]\times(N'\setminus[n'_{\mathrm{out}}])}(((\phi_{r,q'}^{-1})^{\sqcup N'}\circ\cQ'_{\leq T'}[\cF',\tilde{\zeta}])\otimes I_\ell(\sigma')) \\
      &= ((\phi_{r,q'}^{-1})^{\sqcup n_{\mathrm{out}}'}\circ\cQ'[\cF',\tilde{\zeta}])\otimes I_\ell(\sigma'),
    \end{align*}
    where the second equality above holds by \Cref{claim:bcinduct}. Because $(\cQ',\cE_{\mathrm{run}}',D_{\mathrm{in}}',D_{\mathrm{out}})$ is a fault-tolerant gadget for $\bar{\cO}'$, the state $\cQ'[\cF',\tilde{\zeta}]\otimes I_\ell(\sigma')$ is a $\cE_{\mathrm{out}}'\sqcup [\ell]$-deviation of $(\Enc_{\mathrm{out}}'\circ\bar{\cO}'\circ\phi_{r,q'}^{\sqcup k_{\mathrm{in}}})\otimes I_\ell(\tilde{\rho})$. Applying the map $(\phi_{r,q'}^{-1})^{\sqcup n_{\mathrm{out}}'}$ then transforms a Pauli error on a $q'$-dimensional qudit in into $\leq\kappa$ Pauli errors on $r$-dimensional qudits, so because $\tilde{\lambda}_{\mathrm{out}}=\kappa\cdot\lambda_{\mathrm{out}}'$, the RHS above is a $\tilde{\cE}_{\mathrm{out}}\sqcup [\ell]$-deviation of
    \begin{align*}
      &((\phi_{r,q'}^{-1})^{\sqcup n_{\mathrm{out}}'}\circ\Enc_{\mathrm{out}}'\circ\bar{\cO}'\circ\phi_{r,q'}^{\sqcup k_{\mathrm{in}}})\otimes I_\ell(\tilde{\rho}) \\
      &= (\widetilde{\Enc}_{\mathrm{out}}\circ(\phi_{r,q'}^{-1})^{\sqcup k_{\mathrm{out}}'}\circ\bar{\cO}'\circ\phi_{r,q'}^{\sqcup k_{\mathrm{in}}'})\otimes I_\ell(\tilde{\rho}) \\
      &= (\widetilde{\Enc}_{\mathrm{out}}\circ\tilde{\bar{\cO}})\otimes I_\ell(\tilde{\rho})
    \end{align*}
    where the first equality above holds by the definition of $\widetilde{\Enc}_{\mathrm{out}}$, and the second equality holds by the definition of $\tilde{\bar{\cO}}$. Thus we have shown that $\tilde{\cQ}[\tilde{\cF}'',\tilde{\zeta}]$ is a $\tilde{\cE}_{\mathrm{out}}\sqcup [\ell]$-deviation of $(\widetilde{\Enc}_{\mathrm{out}}\circ\tilde{\bar{\cO}})\otimes I_\ell(\tilde{\rho})$, as desired.
  \end{proof}
\end{proof}


\subsection{Main Result Proof}
\label{sec:mainproof}
In this section, we combine \Cref{thm:laft,lem:inclog,cor:simcomp,lem:basiccomp} to prove our main result \Cref{thm:main}, which we restate below for convenience.

\main*

\begin{proof}
  At a high level, the proof outline is as follows. We will first apply \Cref{lem:inclog} to $\bar{\cQ}$ to obtain a logical circuit $\bar{\cQ}'$ over qudits of large dimension $q'\gg r$, which implements the same superoperator as $\bar{\cQ}$ when restricting each qudit to the subspace $\bC^{\bF_r}\subseteq\bC^{\bF_{q'}}$. We will choose $q'$ to be sufficiently large so that we can apply \Cref{thm:laft} to obtain a physical circuit $\cQ'$ that provides a fault-tolerant gadget for $\bar{\cQ}'[*](\cdot)$. We will then repeatedly apply \Cref{cor:simcomp} to reduce the qudit dimension from $q'$ to some much smaller $\tilde{q}$, which will be bounded above by a value depending only on $r,\epsilon$. Finally, we will apply \Cref{lem:basiccomp} to reduce the qudit dimension from $\tilde{q}$ to $r$.
  
  We now present the details. We choose $\bar{N}_0=\bar{N}_0(r,\epsilon)$ to be sufficiently large such that various inequalities specified below that hold asymptotically as $|\bar{N}|\rightarrow\infty$ in fact hold for every $|\bar{N}|\geq\bar{N}_0$. Below, all logarithms are assumed to be base~$2$ unless explicitly stated otherwise. Let $\bar{N}'\supseteq\bar{N}$ be a superset of $\bar{N}$ of size $|\bar{N}'|=8\log(r)\cdot|\bar{N}|$ Define variables\footnote{Note that the variable $n$ we define here is different from the variable $n(r,\epsilon,\bar{N})$ in the statement of \Cref{thm:main}.}
  \begin{align}
    \label{eq:lavars}
    \begin{split}
      n &= 2^{\lfloor\sqrt{\log|\bar{N}'|}\rfloor+5} \\
      k &= \lfloor n/8\rfloor \\
      u &= \log n = \lfloor\sqrt{\log|\bar{N}'|}\rfloor+5 \\
      v_0' &= \lceil\log\lceil\log(n)/\log(r)\rceil\rceil \\
      v' &= v_0'+u^2 \\
      q' &= (r^{2^{v_0'}})^{n^u} = r^{2^{v'}}.
    \end{split}
  \end{align}
  By construction,
  \begin{equation}
    \label{eq:vpbound}
    v' \leq (\log n)^2+\log\log n+2 \leq \log|\bar{N}'|+16\sqrt{\log|\bar{N}'|},
  \end{equation}
  where the second inequality above holds by \Cref{eq:lavars} assuming $\bar{N}_0$, and therefore $|\bar{N}'|$, is sufficiently large.

  Let $\iota:\bF_r\rightarrow\bF_{q'}$ be the natural inclusion, and also let $\iota:\bC^{\bF_r\times\bF_r}\rightarrow\bC^{\bF_{q'}\times\bF_{q'}}$ denote the associated quantum channel. By \Cref{lem:inclog}, there exists a quantum circuit $\bar{\cQ}'=(\bar{\cQ}_1',\dots,\bar{\cQ}_{\bar{T}}';K_{\mathrm{in}},K_{\mathrm{out}})$ acting on $q'$-dimensional qudits labeled by $\bar{N}'$, using time $\bar{T}'\leq O(\log r)\cdot\bar{T}$ and gate set $\cG$ (over~$\bF_{q'}$), such that
  \begin{equation}
    \label{eq:mainQi}
    \bar{\cQ}'[*]\circ\iota^{\sqcup K_{\mathrm{in}}} \subseteq \iota^{\sqcup K_{\mathrm{out}}}\circ\bar{\cQ}[*].
  \end{equation}
  In the equation above, $\bar{\cQ}[*],\bar{\cQ}'[*]$ denote the sets of superoperators associated to all possible postselections of the respective circuits $\bar{\cQ},\bar{\cQ}'$.

  By definition, the expression $\bar{\lambda}_{\mathrm{run}}=\bar{\lambda}_{\mathrm{run}}(u,n)$ in \Cref{eq:lamrun} in \Cref{thm:laft} satisfies $\bar{\lambda}_{\mathrm{run}}\geq n^u/(2^u\cdot n)^{O(1)}$. Therefore there exists an absolute constant $\xi_{\mathrm{run}}>0$ and a sufficiently large choice of $\bar{N}_0$ such that for every $|\bar{N}|\geq\bar{N}_0$, then it holds under our choice of $u=\log n$ that $\bar{\lambda}_{\mathrm{run}}\geq n^{u-\xi_{\mathrm{run}}}$. Similarly, the expression $\eta_{\min}=\eta_{\min}(u,n)$ in \Cref{eq:lamminmax} in \Cref{thm:laft} satisfies $\eta_{\min}\leq(2^u\cdot n)^{O(1)}$. Therefore we assume $\bar{N}_0$ is sufficiently large such that for every $|\bar{N}|\geq\bar{N}_0$, then for an absolute constant $\xi_{\min}>0$, it holds with $u=\log n$ that $\eta_{\min}\leq n^{\xi_{\min}}$. We then define
  \begin{equation}
    \label{eq:xirunp}
    \xi_{\mathrm{run}}' := \max\{\xi_{\mathrm{run}},\xi_{\min}\}+2^{\eta_{\mathrm{gap}}+64},
  \end{equation}
  and we let
  \begin{align}
    \label{eq:mainlamrunp}
    \lambda_{\mathrm{run}}' &= 2^{\log|\bar{N}'|-\xi_{\mathrm{run}}'\sqrt{\log|\bar{N}'|}}.
  \end{align}
  Then define
  \begin{align*}
    \lambda_{\min} &= \lambda_{\min}(u,n,\lambda_{\mathrm{run}}') = \eta_{\min}\cdot\lambda_{\mathrm{run}}' \\
    \lambda_{\max} &= \lambda_{\max}(u,n) = \frac{(n/8)^{u-2}}{4}
  \end{align*}
  as in \Cref{eq:lamminmax} in \Cref{thm:laft}.

  By definition $r^{2^{v_0'}}\geq n$. Assuming $\bar{N}_0$ and therefore $|\bar{N}|$ is sufficiently large, then we also have $k^u\geq(n/16)^{\log n}=2^{\log(n)\cdot(\log(n)-4)}\geq|\bar{N}'|$. Hence we may fix an arbitrary isomorphism between $\bar{N}'$ and some size-$|\bar{N}'|$ subset of $[k]^u$, so that we may write $\bar{N}\subseteq\bar{N}'\subseteq[k]^u$.
  Then for every $K\subseteq\bar{N}$, we fix the code and encoding map
  \begin{equation*}
    (C'(r,K,\bar{N}),\Enc_{C'(r,K,\bar{N})}) := (C(q',u,n,k,K),\Enc_{C(q',u,n,k,K)}),
  \end{equation*}
  where the RHS above consists of the code and encoding map defined in \Cref{thm:laft}, with logical qudits labeled by $K$.
  By \Cref{thm:laft}, $C'(r,K,\bar{N})$ has parameters $[[n^u,|K|,(n/8)^u]]_{q'}$.
  We then define decorated codes $D_{\mathrm{in}}',D_{\mathrm{out}}'\in\cD(q',u,n,k,\lambda_{\mathrm{run}}')$ (see \Cref{def:ladec}) by
  \begin{align*}
    D_{\mathrm{in}}' &= (C'(r,K_{\mathrm{in}},\bar{N}),\; \Enc_{C'(r,K_{\mathrm{in}},\bar{N})},\; 2^{[n]^u}|_{\geq\lambda_{\mathrm{in}}'}) \\
    D_{\mathrm{out}}' &= (C'(r,K_{\mathrm{out}},\bar{N}),\; \Enc_{C'(r,K_{\mathrm{out}},\bar{N})},\; 2^{[n]^u}|_{\geq\lambda_{\mathrm{out}}'})
  \end{align*}
  for $\lambda_{\mathrm{in}}'=\lambda_{\max}$ and $\lambda_{\mathrm{out}}'=\lambda_{\min}$, so that when $\bar{N}_0$ and therefore $|\bar{N}|$ is sufficiently large then
  \begin{align}
    \label{eq:mainlamp}
    \begin{split}
      \lambda_{\mathrm{in}}' &= \frac{(n/8)^{u-2}}{4} \geq (n/16)^{u-2} = 2^{(\log(n)-2)(\log(n)-4)} \geq 2^{\log|\bar{N}'|} = |\bar{N}'| \\
      \lambda_{\mathrm{out}}' &\leq n^{\xi_{\min}}\cdot\lambda_{\mathrm{run}}' = 2^{\xi_{\min}\cdot\log n}\cdot 2^{\log|\bar{N}'|-\xi_{\mathrm{run}}'\cdot\sqrt{\log|\bar{N}'|}} \leq 2^{\log|\bar{N}'|-(\xi_{\mathrm{run}}'-\xi_{\min}-1)\sqrt{\log|\bar{N}'|}}.
    \end{split}
  \end{align}
  Because by definition $\xi_{\mathrm{run}}'\geq 0$ and $\xi_{\min}\geq 0$, we have
  \begin{equation}
    \label{eq:mainlamrunsmaller}
    0 \leq \lambda_{\mathrm{run}}' \leq \min\{\lambda_{\mathrm{in}}',\lambda_{\mathrm{out}}'\}.
  \end{equation}
  Furthermore, for $\alpha\in\{\mathrm{in},\mathrm{out}\}$, letting $[[n_\alpha',k_\alpha',d_\alpha']]_{q'}=[[n^u,|K_\alpha|,(n/8)^u]]_{q'}$ denote the parameters of $D_{\mathrm{in}}'$ (i.e.~of $C'(r,K_\alpha,\bar{N})$), then
  \begin{align}
    \label{eq:primeparams}
    \begin{split}
      n_\alpha' &= n^u = 2^{(\log n)^2} \leq 2^{\log|\bar{N}'|+16\sqrt{\log|\bar{N}'|}} \\
      k_\alpha' &= |K_\alpha| \\
      d_\alpha' &\geq (n/8)^u = 2^{\log(n)\cdot(\log(n)-3)} \geq 2^{\log|\bar{N}'|} = |\bar{N}'|.
    \end{split}
  \end{align}
  
  Now applying \Cref{it:lascheme} in \Cref{thm:laft} to the circuit $\bar{\cQ}'$, we obtain a fault-tolerant gadget
  \begin{equation*}
    (\cQ',\; \cE_{\mathrm{run}}'=2^{N'}|_{\geq\lambda_{\mathrm{run}}'}^{\sqcup T'},\; D_{\mathrm{in}}',\; D_{\mathrm{out}}')
  \end{equation*}
  for $\bar{\cO}'(\cdot)=\bar{\cQ}'[*](\cdot)$, where $\cQ'$ is a quantum circuit using space
  \begin{equation}
    \label{eq:mainQpspace}
    |N'| \leq O(n)^u = 2^{(\log n)^2+O(\log n)} = 2^{\log|\bar{N}'|+O(\sqrt{\log|\bar{N}'|})},
  \end{equation}
  time
  \begin{equation}
    \label{eq:mainQptime}
    T' \leq O(u^4n^2(\log n)^2)\cdot\bar{T}' \leq 2^{4\sqrt{\log|\bar{N}'|}}\cdot\bar{T}'
  \end{equation}
  (assuming $\bar{N}_0$ and therefore $|\bar{N}|$ is sufficiently large), and gate set $\cG$.

  We will next repeatedly apply \Cref{cor:simcomp} to reduce the physical alphabet size of this gadget. First, for $v_1\leq v_2\in\bZ_{\geq 0}$, we choose the isomorphisms $\phi_{r^{2^{v_1}},r^{2^{v_2}}}$ in \Cref{def:fieldiso} to respect composition as follows. First, we fix an arbitrary choice of this isomorphism for every $(v_1,v_2)$ with $v_2=v_1+1$. Then for general $v_1\leq v_2$, we define
  \begin{equation*}
    \phi_{r^{2^{v_1}},r^{2^{v_2}}} = \phi_{r^{2^{v_2-1}},r^{2^{v_2}}}^{\sqcup 1} \circ \cdots \circ \phi_{r^{2^{v_1+1}},r^{2^{v_1+2}}}^{\sqcup 2^{v_2-v_1-2}} \circ \phi_{r^{2^{v_1}},r^{2^{v_1+1}}}^{\sqcup 2^{v_2-v_1-1}}.
  \end{equation*}
  Therefore for every $v_1\leq v_2\leq v_3$, we have
  \begin{equation}
    \label{eq:phicomp}
    \phi_{r^{2^{v_1}},r^{2^{v_3}}} = \phi_{r^{2^{v_2}},r^{2^{v_3}}} \circ \phi_{r^{2^{v_1}},r^{2^{v_2}}}^{\sqcup 2^{v_3-v_2}}.
  \end{equation}

  Define $v_0(\epsilon)$ as in \Cref{cor:simcomp}. Let $v^{(0)}=v'$, let
  \begin{equation*}
    (\cQ^{(0)},\; \cE_{\mathrm{run}}^{(0)},\; D_{\mathrm{in}}^{(0)},\; D_{\mathrm{out}}^{(0)}) = (\cQ',\; \cE_{\mathrm{run}}',\; D_{\mathrm{in}}',\; D_{\mathrm{out}}').
  \end{equation*}
  Iterating $i=0,1,2,\dots$ while $v^{(i)}\geq v_0(\epsilon)$, we let $v^{(i+1)}=\lfloor(3/4+\epsilon)v^{(i)}\rfloor$, and we let
  \begin{equation*}
    (\cQ^{(i+1)},\; \cE_{\mathrm{run}}^{(i+1)},\; D_{\mathrm{in}}^{(i+1)},\; D_{\mathrm{out}}^{(i+1)})
  \end{equation*}
  be the fault-tolerant gadget (labeled by tildes in \Cref{cor:simcomp})
  given by applying \Cref{cor:simcomp} to $(\cQ^{(i)},\; \cE_{\mathrm{run}}^{(i)},\; D_{\mathrm{in}}^{(i)},\; D_{\mathrm{out}}^{(i)})$. Let $m\in\bN$ be the least integer for which $v^{(m)}<v_0(\epsilon)$. Such an $m$ must exist by the assumption that $\epsilon<1/8$, so that if $v^{(i)}\geq v_0(\epsilon)\geq 1$ then $v^{(i+1)}\leq v^{(i)}$. This repeated application of \Cref{cor:simcomp} does not violate the required condition \Cref{eq:sclamrunsmaller} at each step because this inequality is satisfied in the first step by \Cref{eq:mainlamrunsmaller}, and then it is satisfied in each subsequent step by \Cref{eq:tlamrundef,eq:sclamscor} (recalling that \Cref{cor:simcomp} defines the absolute constant $\xi\geq 16$).

  Let $\tilde{v}=v^{(m)}$, $\tilde{q}=r^{2^{\tilde{v}}}$ and let
  \begin{equation*}
    (\tilde{\cQ},\; \tilde{\cE}_{\mathrm{run}},\; \tilde{D}_{\mathrm{in}},\; \tilde{D}_{\mathrm{out}}) = (\cQ^{(m)},\; \cE_{\mathrm{run}}^{(m)},\; D_{\mathrm{in}}^{(m)},\; D_{\mathrm{out}}^{(m)}).
  \end{equation*}
  For $\alpha\in\{\mathrm{in},\mathrm{out}\}$, write
  \begin{equation*}
    \tilde{D}_\alpha = (\tilde{C}_\alpha,\; \widetilde{\Enc}_\alpha,\; \tilde{\cE}_\alpha=2^{[\tilde{n}_\alpha]}|_{\geq\tilde{\lambda}_\alpha}),
  \end{equation*}
  and let $[[\tilde{n}_\alpha,\tilde{k}_\alpha,\tilde{d}_\alpha]]_{\tilde{q}}$ denote the parameters of $\tilde{C}$. Letting
  \begin{align*}
    M &= 2^{\sum_{i=1}^m v^{(i)}},
  \end{align*}
  then recalling that $0<\epsilon<1/8$ and applying \Cref{eq:vpbound}, we have
  \begin{align}
    \label{eq:maingeoser}
    \begin{split}
      M &\leq 2^{\sum_{i=1}^\infty(3/4+\epsilon)^iv'} = 2^{v'/(1/4-\epsilon)} \leq 2^{(\log|\bar{N}'|+16\sqrt{\log|\bar{N}'|})\cdot 4/(1-4\epsilon)} \leq |\bar{N}'|^{4+2^8\epsilon}\cdot 2^{2^8\sqrt{\log|\bar{N}'|}} \\
      \sum_{i=1}^m\sqrt{v^{(i)}} &\leq \sum_{i=1}^\infty\sqrt{(3/4+\epsilon)^iv'} = \frac{\sqrt{v'}}{1-\sqrt{3/4+\epsilon}} \leq \frac{\sqrt{v'}}{1-\sqrt{7/8}} \leq 16\sqrt{v'} \leq 32\sqrt{\log|\bar{N}'|}.
    \end{split}
  \end{align}
  The final inequality in the bound for $\sum_{i=1}^m\sqrt{v^{(i)}}$ above assumes that $\bar{N}_0$ and therefore $|\bar{N}|$ is sufficiently large so that $\sqrt{v'}\leq 2\sqrt{\log|\bar{N}'|}$ by \Cref{eq:vpbound}.
  Therefore by \Cref{cor:simcomp},
  \begin{align}
    \label{eq:tildeparams}
    \begin{split}
      \tilde{n}_\alpha &\leq 2^{\sum_{i=1}^m v^{(i)}}\cdot n_\alpha' = M\cdot n_\alpha' \\
      \tilde{k}_\alpha &= 2^{v^{(1)}-v^{(m)}}\cdot k_\alpha' = 2^{v'-\tilde{v}}\cdot k_\alpha' \\
      \tilde{d}_\alpha &\geq 2^{\sum_{i=1}^m(v^{(i)}-8\sqrt{v^{(i)}})}\cdot d_\alpha' \geq 2^{-2^8\sqrt{\log|\bar{N}'|}}\cdot M\cdot d_\alpha'.
    \end{split}
  \end{align}
  By \Cref{cor:simcomp}, the pair $(\tilde{C}_\alpha,\widetilde{\Enc}_\alpha)$ is obtained by repeated concatenation of $(C'(r,K_\alpha,\bar{N}),\Enc_{C'(r,K_\alpha,\bar{N})})$ with codes that only depend on $v^{(0)},\dots,v^{(m)}$, which in turn only depend on $v',\epsilon$, and therefore only on $r,\epsilon,\bar{N}$. Thus $(\tilde{C}_\alpha,\widetilde{\Enc}_\alpha)$ only depend on $r,\epsilon,K_\alpha,\bar{N}$, so for every $K\subseteq\bar{N}$ there exists a code $\tilde{C}(r,\epsilon,K,\bar{N})$ with encoding map $\Enc_{\tilde{C}(r,\epsilon,K,\bar{N})}$ such that $(\tilde{C}_\alpha,\widetilde{\Enc}_\alpha)=(\tilde{C}(r,\epsilon,K_\alpha,\bar{N})),\Enc_{\tilde{C}(r,\epsilon,K_\alpha,\bar{N})}$. By \Cref{eq:primeparams,eq:tildeparams}, the parameters $[[\tilde{n},\tilde{k},\tilde{d}]]_{\tilde{q}}$ of $\tilde{C}(r,\epsilon,K,\bar{N})$ satisfy
  \begin{align}
    \label{eq:tildeparams2}
    \begin{split}
      \tilde{n} &\leq 2^{16\sqrt{\log|\bar{N}'|}}\cdot M\cdot|\bar{N}'| \\
      \tilde{k} &= 2^{v'-\tilde{v}}\cdot|K| \\
      \tilde{d} &\geq 2^{-2^8\sqrt{\log|\bar{N}'|}}\cdot M\cdot|\bar{N}'|.
    \end{split}
  \end{align}

  We now describe the fault-tolerance properties of
  \begin{equation*}
    (\tilde{\cQ},\; \tilde{\cE}_{\mathrm{run}}=2^{\tilde{N}}|_{\geq\tilde{\lambda}_{\mathrm{run}}}^{\sqcup\tilde{T}},\; \tilde{D}_{\mathrm{in}},\; \tilde{D}_{\mathrm{out}}).
  \end{equation*}
  By \Cref{cor:simcomp}, the above data provides a fault-tolerant gadget for the set $\tilde{\bar{\cO}}$ of superoperators obtained by precomposing and postcomposing $\bar{\cO}'$ by the composition of $\phi_{r^{2^{v^{(i)}}},r^{2^{v^{(i+1)}}}}$ for $i=0,\dots,m-1$ and their inverses, respectively. By \Cref{eq:phicomp}, and recalling that $\tilde{q}=r^{2^{\tilde{v}}}$, $q'=r^{2^{v'}}$, it follows that
  \begin{equation}
    \label{eq:maintbO}
    \tilde{\bar{\cO}} = (\phi_{\tilde{q},q'}^{-1})^{\sqcup k_{\mathrm{out}}'} \circ \bar{\cO}' \circ \phi_{\tilde{q},q'}^{\sqcup k_{\mathrm{in}}'}.
  \end{equation}
  Defining the absolute constant $\xi$ as in \Cref{cor:simcomp}, then by \Cref{cor:simcomp,eq:maingeoser,eq:mainlamrunp} we have
  \begin{align*}
    \tilde{\lambda}_{\mathrm{run}}
    &= 2^{\sum_{i=1}^m(v^{(i)}-\xi\sqrt{v^{(i)}})}\cdot \lambda_{\mathrm{run}}' \\
    &\geq 2^{-32\xi\sqrt{\log|\bar{N}'|}}\cdot M\cdot\lambda_{\mathrm{run}}' \\
    &\geq 2^{-(\xi_{\mathrm{run}}'+32\xi)\sqrt{\log|\bar{N}'|}}\cdot M\cdot|\bar{N}'|,
  \end{align*}
  by \Cref{cor:simcomp,eq:maingeoser,eq:mainlamp} we have
  \begin{align*}
    \tilde{\lambda}_{\mathrm{in}}
    &\geq 2^{\sum_{i=1}^m(v^{(i)}-16\sqrt{v^{(i)}})}\cdot\lambda_{\mathrm{in}}' \\
    &\geq 2^{-2^9\sqrt{\log|\bar{N}'|}}\cdot M\cdot\lambda_{\mathrm{in}}' \\
    &\geq 2^{-2^9\sqrt{\log|\bar{N}'|}}\cdot M\cdot|\bar{N}'|,
  \end{align*}
  and by \Cref{cor:simcomp,eq:maingeoser,eq:mainlamp,eq:xirunp} we have
  \begin{align*}
    \tilde{\lambda}_{\mathrm{out}}
    &= 2^{\sum_{i=1}^mv^{(i)}}\cdot\lambda_{\mathrm{out}}' \\
    &= M\cdot\lambda_{\mathrm{out}}' \\
    &\leq 2^{-(\xi_{\mathrm{run}}'-\xi_{\min}-1)\sqrt{\log|\bar{N}'|}}\cdot M\cdot|\bar{N}'| \\
    &\leq 2^{-2^{\eta_{\mathrm{gap}}+63}\sqrt{\log|\bar{N}'|}}\cdot M\cdot|\bar{N}'|.
  \end{align*}
  By \Cref{cor:simcomp,eq:maingeoser,eq:mainQpspace}, $\tilde{\cQ}$ uses space
  \begin{align}
    \label{eq:maintQspace}
    \begin{split}
      |\tilde{N}|
      &\leq 2^{\sum_{i=1}^m(v^{(i)}+O(\sqrt{v^{(i)}}))} \cdot |N'| \\
      &\leq 2^{O(\sqrt{\log|\bar{N}'|})}\cdot M \cdot 2^{\log|\bar{N}'|+O(\sqrt{\log|\bar{N}'|})} \\
      &\leq 2^{O(\sqrt{\log|\bar{N}'|})}\cdot M\cdot|\bar{N}'|,
    \end{split}
  \end{align}
  and by \Cref{cor:simcomp,eq:maingeoser,eq:mainQptime}, $\tilde{\cQ}$ uses time
  \begin{align}
    \label{eq:maintQtime}
    \begin{split}
      \tilde{T}
      &\leq 2^{4\sum_{i=1}^m\sqrt{v^{(i)}}}\cdot T' \\
      &\leq 2^{2^7\sqrt{\log|\bar{N}'|}}\cdot 2^{4\sqrt{\log|\bar{N}'|}} \cdot \bar{T}' \\
      &\leq 2^{2^8\sqrt{\log|\bar{N}'|}} \cdot \bar{T}'.
    \end{split}
  \end{align}

  Let $\kappa=2^{\tilde{v}}$. We now apply \Cref{lem:basiccomp} to the gadget\footnote{Note that there is an unfortunate clash of notation here: tilded variables $\tilde{\cQ},\dots$ in this proof correspond to primed variables $\cQ',\dots$ in our application of \Cref{lem:basiccomp}, while plain (non-primed and non-tilded) variables $\cQ,\dots$ in this proof correspond to tilded variables $\tilde{\cQ},\dots$ in \Cref{lem:basiccomp}.} $(\tilde{\cQ},\; \tilde{\cE}_{\mathrm{run}},\; \tilde{D}_{\mathrm{in}},\; \tilde{D}_{\mathrm{out}})$ over $\tilde{q}=r^\kappa$-dimensional qudits to obtain a fault-tolerant gadget
  \begin{equation*}
    (\cQ,\; \cE_{\mathrm{run}}=2^N|_{\geq\lambda_{\mathrm{run}}}^{\sqcup T},\; D_{\mathrm{in}}^0,\; D_{\mathrm{out}}^0)
  \end{equation*}
  over $r$-dimensional qudits for
  \begin{align*}
    \bar{\cO}^0
    &:= (\phi_{r,\tilde{q}}^{-1})^{\sqcup\tilde{k}_{\mathrm{out}}}\circ\tilde{\bar{\cO}}\circ\phi_{r,\tilde{q}}^{\sqcup\tilde{k}_{\mathrm{in}}} \\
    &= (\phi_{r,q'}^{-1})^{\sqcup k_{\mathrm{out}}'} \circ \bar{\cO}' \circ \phi_{r,q'}^{\sqcup k_{\mathrm{in}}'},
  \end{align*}
  where the second equality above holds by \Cref{eq:maintbO,eq:phicomp}.

  For $K\subseteq\bar{N}$, defining $\tilde{n},\tilde{k},\tilde{d}$ as in \Cref{eq:tildeparams}, then let
  \begin{align*}
    C^0(r,\epsilon,K,\bar{N}) &= (\phi_{r,\tilde{q}}^{-1})^{\sqcup\tilde{n}}(\tilde{C}(r,\epsilon,K,\bar{N})) \\
    \Enc_{C^0(r,\epsilon,K,\bar{N})} &= (\phi_{r,\tilde{q}}^{-1})^{\sqcup\tilde{n}} \circ \Enc_{\tilde{C}(r,\epsilon,K,\bar{N})} \circ \phi_{r,\tilde{q}}^{\sqcup\tilde{k}}.
  \end{align*}
  Recall above that $\tilde{k}=2^{v'-\tilde{v}}\cdot|K|$ depends on $K$, but $\tilde{n},\tilde{d}$ do not.
  Then by \Cref{lem:basiccomp}, for $\alpha\in\{\mathrm{in},\mathrm{out}\}$ we have
  \begin{align*}
    D_\alpha^0 &= (C^0(r,\epsilon,K_\alpha,\bar{N}),\; \Enc_{C^0(r,\epsilon,K_\alpha,\bar{N})},\; \cE_\alpha=2^{[\kappa]\times[\tilde{n}]}|_{\geq\lambda_\alpha})
  \end{align*}
  where
  \begin{align}
    \label{eq:mainlamfinal}
    \begin{split}
      \lambda_{\mathrm{run}}
      &= \frac{1}{6(32\log(\kappa)+256)}\cdot\tilde{\lambda}_{\mathrm{run}} \\
      &\geq \frac{1}{6(32v_0(\epsilon)+256)} \cdot 2^{-(\xi_{\mathrm{run}}'+32\xi)\sqrt{\log|\bar{N}'|}}\cdot M\cdot|\bar{N}'| \\
      &\geq 2^{-2(\xi_{\mathrm{run}}'+32\xi+1)\sqrt{\log|\bar{N}|}}\cdot M\cdot|\bar{N}| \\
      &= 2^{-2(\max\{\xi_{\mathrm{run}},\xi_{\min}\}+2^{\eta_{\mathrm{gap}}+64}+32\xi+1)\sqrt{\log|\bar{N}|}}\cdot M\cdot|\bar{N}| \\
      &\geq 2^{-2^{\eta_{\mathrm{gap}}+O(1)}\sqrt{\log|\bar{N}|}}\cdot M\cdot|\bar{N}| \\
      \lambda_{\mathrm{in}}
      &= \tilde{\lambda}_{\mathrm{in}}-\tilde{\lambda}_{\mathrm{out}} \\
      &\geq 2^{-2^{10}\sqrt{\log|\bar{N}'|}}\cdot M\cdot|\bar{N}'| \\
      &\geq 2^{-2^{11}\sqrt{\log|\bar{N}|}}\cdot M\cdot|\bar{N}| \\
      \lambda_{\mathrm{out}}
      &= \kappa\cdot\tilde{\lambda}_{\mathrm{out}} \\
      &\leq 2^{v_0(\epsilon)}\cdot 2^{-2^{\eta_{\mathrm{gap}}+63}\sqrt{\log|\bar{N}'|}}\cdot M\cdot|\bar{N}'| \\
      &\leq 8\log(r)\cdot 2^{v_0(\epsilon)}\cdot 2^{-2^{\eta_{\mathrm{gap}}+63}\sqrt{\log|\bar{N}|}}\cdot M\cdot|\bar{N}| \\
      &\leq 2^{-2^{\eta_{\mathrm{gap}}+62}\sqrt{\log|\bar{N}|}}\cdot M\cdot|\bar{N}|.
    \end{split}
  \end{align}
  Above we apply the fact that by definition $|\bar{N}'|=8\log(r)\cdot|\bar{N}|$, $\kappa=2^{\tilde{v}}$, and $\tilde{v}\leq v_0(\epsilon)$, and we assume that $|\bar{N}|\geq\bar{N}_0=\bar{N}_0(r,\epsilon)$ is sufficiently large so that $6(32v_0(\epsilon)+256)\leq 2^{\sqrt{\log|\bar{N}|}}$, so that $|\bar{N}|\geq 8\log(r)$ and therefore $|\bar{N}|^2\geq|\bar{N}'|$, and so that $8\log(r)\cdot 2^{v_0(\epsilon)}\leq 2^{\sqrt{\log|\bar{N}|}}$.

  Furthermore, again assuming $|\bar{N}|\geq\bar{N}_0=\bar{N}_0(r,\epsilon)$ is sufficiently large, then by \Cref{eq:tildeparams2} the parameters $[[n^0,k^0,d^0]]_r$ of $C^0(r,\epsilon,K,\bar{N})$ satisfy
  \begin{align*}
    n^0
    &= \kappa\cdot\tilde{n} \\
    &\leq 2^{v_0(\epsilon)} \cdot 2^{16\sqrt{\log|\bar{N}'|}}\cdot M\cdot|\bar{N}'| \\
    &\leq 2^{32\sqrt{\log|\bar{N}|}}\cdot M\cdot|\bar{N}| \\
    k^0 &= \kappa\cdot\tilde{k} = 2^{v'}\cdot|K| \\
    d^0 &\geq \tilde{d} \geq 2^{-2^9\sqrt{\log|\bar{N}|}}\cdot M\cdot|\bar{N}|,
  \end{align*}
  where we emphasize that $\tilde{k}$ and therefore $k^0$ depends on $|K|$, whereas $n^0,d^0$ do not depend on $K$.

  Also, by \Cref{eq:maingeoser}, assuming $|\bar{N}|\geq\bar{N}_0=\bar{N}_0(r,\epsilon)$ is sufficiently large, we have
  \begin{align}
    \label{eq:mainMbound}
    M
    &\leq |\bar{N}'|^{4+2^8\epsilon}\cdot 2^{2^8\sqrt{\log|\bar{N}'|}} \leq |\bar{N}|^{4+(2^8+1)\epsilon}\cdot 2^{2^9\sqrt{\log|\bar{N}|}} \leq |\bar{N}|^{4+2^9\epsilon}.
  \end{align}
  
  Meanwhile, by \Cref{lem:basiccomp,eq:maintQspace}, assuming $|\bar{N}|\geq\bar{N}_0=\bar{N}_0(r,\epsilon)$ is sufficiently large, then $\cQ$ uses space
  \begin{align}
    \label{eq:mainQspace}
    \begin{split}
      |N|
      &\leq 16\kappa^3\cdot|\tilde{N}| \\
      &\leq 2^{3v_0(\epsilon)+4} \cdot 2^{O(\sqrt{\log|\bar{N}'|})}\cdot M\cdot|\bar{N}'| \\
      &\leq 2^{O(\sqrt{\log|\bar{N}|})}\cdot M\cdot|\bar{N}|.
    \end{split}
  \end{align}
  Similarly, by \Cref{lem:basiccomp,eq:maintQtime}, and recalling that $\bar{T}'\leq O(\log r)\cdot\bar{T}$, then $\cQ$ uses time
  \begin{align}
    \label{eq:mainQtime}
    \begin{split}
      T
      &\leq (32\log(\kappa)+256)\cdot \tilde{T} \\
      &\leq (32v_0(\epsilon)+256) \cdot 2^{2^8\sqrt{\log|\bar{N}'|}} \cdot \bar{T}' \\
      &\leq (32v_0(\epsilon)+256) \cdot 2^{2^8\sqrt{\log|\bar{N}'|}} \cdot O(\log r)\cdot\bar{T} \\
      &\leq 2^{2^9\sqrt{\log|\bar{N}|}} \cdot \bar{T}.
    \end{split}
  \end{align}

  Now recall that $(\cQ,\; \cE_{\mathrm{run}},\; D_{\mathrm{in}}^0,\; D_{\mathrm{out}}^0)$ is a fault-tolerant gadget for $\bar{\cO}^0=(\phi_{r,q'}^{-1})^{\sqcup k_{\mathrm{out}}'} \circ \bar{\cO}' \circ \phi_{r,q'}^{\sqcup k_{\mathrm{in}}'}$, where $\bar{\cO}'(\cdot)=\bar{\cQ}'[*](\cdot)$. Recall that our goal is to obtain a fault-tolerant gadget for $\bar{\cO}$. For this purpose, we define a CSS non-subsystem code $C(r,\epsilon,K,\bar{N})$ with encoding map $\Enc_{C(r,\epsilon,K,\bar{N})}$ by first defining the encoding map
  \begin{align*}
    \Enc_{C(r,\epsilon,K,\bar{N})} = \Enc_{C^0(r,\epsilon,K,\bar{N})} \circ (\phi_{r,q'}^{-1}\circ\iota)^{\sqcup K},
  \end{align*}
  and then defining the code $C(r,\epsilon,K,\bar{N})=(C_X(r,\epsilon,K,\bar{N}),C_Z(r,\epsilon,K,\bar{N}))$ by
  \begin{align*}
    C_X^\perp(r,\epsilon,K,\bar{N}) &= \Enc_{C(r,\epsilon,K,\bar{N})}(0) \\
    C_Z(r,\epsilon,K,\bar{N}) &= \bigsqcup\Enc_{C(r,\epsilon,K,\bar{N})}(\bF_r^{K}),
  \end{align*}
  where above we recall that $\Enc_{C(r,\epsilon,K,\bar{N})}(\bF_r^{K})$ is a set of cosets in $\bF_r^{n^0}/C_X^\perp(r,\epsilon,K,\bar{N})$, so that $C_Z(r,\epsilon,K,\bar{N})$ is defined to be the union of these cosets.
  In words, $C(r,\epsilon,K,\bar{N})$ is simply a subcode of $C^0(r,\epsilon,K,\bar{N})$, meaning that the former simply is given by the latter restricted to a subspace of the entire logical (message) space. Similarly, $\Enc_{C(r,\epsilon,K,\bar{N})}$ is simply the restriction of $\Enc_{C^0(r,\epsilon,K,\bar{N})}$ to inputs in $(\phi_{r,q'}^{-1}\circ\iota)^{\sqcup K}(\bF_r^{K})\subseteq\bF_r^{[2^{v'}]\times K}=\bF_r^{k^0}$. By definition, $C(r,\epsilon,K,\bar{N})$ has parameters $[[n(r,\epsilon,\bar{N}),\; |K|,\; d(r,\epsilon,\bar{N})]]_r$ with
  \begin{align}
    \label{eq:mainCfinal}
    \begin{split}
      n(r,\epsilon,\bar{N}) &= n^0 \leq 2^{32\sqrt{\log|\bar{N}|}}\cdot M\cdot|\bar{N}| \\
      d(r,\epsilon,\bar{N}) &\geq d^0 \geq 2^{-2^9\sqrt{\log|\bar{N}|}}\cdot M\cdot|\bar{N}|,
    \end{split}
  \end{align}
  and the logical qudits are naturally labeled by $K$ (inherited from the labeling of the logical qudits of $C'(r,K,\bar{N})$ by $K$).
  
  By \Cref{eq:mainQi},
  \begin{align}
    \label{eq:mainlog}
    \begin{split}
      \bar{\cO}^0\circ(\phi_{r,q'}^{-1}\circ\iota)^{\sqcup K_{\mathrm{in}}}
      &= (\phi_{r,q'}^{-1})^{\sqcup K_{\mathrm{out}}} \circ \bar{\cO}' \circ \iota^{\sqcup K_{\mathrm{in}}} \\
      &= (\phi_{r,q'}^{-1})^{\sqcup K_{\mathrm{out}}} \circ \bar{\cQ}'[*] \circ \iota^{\sqcup K_{\mathrm{in}}} \\
      &\subseteq (\phi_{r,q'}^{-1}\circ\iota)^{\sqcup K_{\mathrm{out}}} \circ \bar{\cQ}[*].
    \end{split}
  \end{align}
  Therefore for $\alpha\in\{\mathrm{in},\mathrm{out}\}$ defining
  \begin{align*}
    D_\alpha &= (C(r,\epsilon,K_\alpha,\bar{N}),\; \Enc_{C(r,\epsilon,K_\alpha,\bar{N})},\; \cE_\alpha),
  \end{align*}
  then
  \begin{equation*}
    (\cQ,\; \cE_{\mathrm{run}}=2^N|_{\geq\lambda_{\mathrm{run}}}^{\sqcup T},\; D_{\mathrm{in}},\; D_{\mathrm{out}})
  \end{equation*}
  is a fault-tolerant gadget for $\bar{\cO}(\cdot)=\bar{\cQ}[*](\cdot)$. That is, we have simply taken the fault-tolerant gadget $(\cQ,\; \cE_{\mathrm{run}},\; D_{\mathrm{in}}^0,\; D_{\mathrm{out}}^0)$, and restricted attention to inputs $\sigma$ that are $\cE_{\mathrm{in}}\sqcup [\ell]$-deviations of states of the form
  \begin{equation*}
    (\Enc_{C^0(r,\epsilon,K_{\mathrm{in}},\bar{N})}\circ(\phi_{r,q'}^{-1}\circ\iota)^{\sqcup K_{\mathrm{in}}})\otimes I_\ell(\rho) = \Enc_{C(r,\epsilon,K_{\mathrm{in}},\bar{N})}\otimes I_\ell(\rho)
  \end{equation*}
  for $\rho\in\bC^{\bF_r^{|K_{\mathrm{in}}|+\ell}\times\bF_r^{|K_{\mathrm{in}}|+\ell}}$. By the fault-tolerance of the gadget $(\cQ,\; \cE_{\mathrm{run}},\; D_{\mathrm{in}}^0,\; D_{\mathrm{out}}^0)$ along with \Cref{eq:mainlog}, the output $\cQ[\cF,\zeta](\sigma)$ for a $\cE_{\mathrm{run}}$-avoiding fault $\cF$ and postselection $\zeta$ for $\cQ$ then must be a $\cE_{\mathrm{out}}\sqcup [\ell]$-deviation of
  \begin{align*}
    (\Enc_{C^0(r,\epsilon,K_{\mathrm{out}},\bar{N})} \circ \bar{\cO}^0 \circ (\phi_{r,q'}^{-1}\circ\iota)^{\sqcup K_{\mathrm{in}}})\otimes I_\ell(\rho)
    &\subseteq (\Enc_{C^0(r,\epsilon,K_{\mathrm{out}},\bar{N})} \circ (\phi_{r,q'}^{-1}\circ\iota)^{\sqcup K_{\mathrm{out}}} \circ \bar{\cQ}[*])\otimes I_\ell(\rho) \\
    &= (\Enc_{C(r,\epsilon,K_{\mathrm{out}},\bar{N})} \circ \bar{\cQ}[*])\otimes I_\ell(\rho).
  \end{align*}

  Thus indeed $(\cQ,\; \cE_{\mathrm{run}},\; D_{\mathrm{in}},\; D_{\mathrm{out}})$ is a fault-tolerant gadget for $\bar{\cQ}[*](\cdot)$. The values $\lambda_{\mathrm{run}},\lambda_{\mathrm{in}},\lambda_{\mathrm{out}}$ and $M$ satisfy the desired inequalities in \Cref{eq:mainlamfinal,eq:mainMbound}, while the codes $C(r,\epsilon,K,\bar{N})$ have the desired parameters given by \Cref{eq:mainCfinal}. The circuit $\cQ$ has the desired space and time usage in \Cref{eq:mainQspace,eq:mainQtime}. Thus the desired result holds.
\end{proof}

\section*{Acknowledgments}
We thank Anurag Anshu, Venkatesan Guruswami, Ting-Chun Lin, Quynh T.~Nguyen, Christopher A.~Pattison, and Noga Ron-Zewi for helpful discussions.

N.P.B.~acknowledges support through the Quantum Research Pod, CIQC and the Simons Institute for the Theory of Computing.
This work was done in part while N.P.B.~was visiting the Simons Institute for the Theory of Computing, supported by NSF QLCI Award OMA-2016245.
L.G.~acknowledges support from a Google PhD Fellowship, a National Science Foundation Graduate Research Fellowship under Grant No.~DGE 2146752, ONR grant N00014-24-1-2491, and a UC Noyce initiative award.
U.V.~acknowledges support from NSF grant 2311733, DOE grant DE-SC0024124 and NSF QLCI Award OMA-2016245.

\textbf{AI use disclosure:} All proofs and writing are the work of the human authors. ChatGPT was used to check the final draft for typos and minor presentation issues.

\bibliographystyle{alpha}
\bibliography{bibliography.bib,library.bib}

\appendix



\section{Non-Subsystem Codes via Homological Products}
\label{sec:homprod}
In this section, we show how to construct CSS non-subsystem codes with low-weight stabilizers by appropriately gauge-fixing the codes in \Cref{def:pevprod}. While we will not directly use these non-subsystem codes in our fault-tolerance scheme, we believe them to be of independent interest. The main theorem of this section is stated below.

\begin{theorem}
  \label{thm:homconstruct}
  Let $u,k,n,q_0,n_1',\dots,n_u'\in\bN$ such that $k<n$ with $k\equiv n\pmod{2}$, and such that $q_0\geq n$ is a prime power and $n_1',\dots,n_u'\geq n$, and let $q=q_0^{\prod_{i\in[u]}n_i'}$. Then there exists an explicit family of
  \begin{equation*}
    \left[\left[N=n^u,\; K=k^u,\; D\geq\left(\frac{n-k}{2}+1\right)^u\right]\right]_q
  \end{equation*}
  codes with stabilizer weight $W=un$. These codes are CSS non-subsystem codes constructed from homological products (see \Cref{def:homprod}).
\end{theorem}

In \Cref{thm:homconstruct}, if we for instance set $k=(1-1/u)n$, and for arbitrary constant $\epsilon>0$ set $u=\lceil 1/2\epsilon\rceil$, then letting $n\rightarrow\infty$, we obtain a family of explicit CSS non-subsystem codes with parameters $[[N,\Theta(N),\Theta(N)]]_q$ and stabilizer weight $W\leq N^\epsilon$. Instead letting $u$ grow slowly with $n$, such as $u=\log\log\log n$, gives parameters $[[N,\Theta(N),\tilde{\Omega}(N)]]_q$ and stabilizer weight $W\leq N^{1-o(1)}$. These codes do require an exponentially large alphabet $q=2^{\tilde{\Theta}(N)}$. See \Cref{sec:codeimpinf} for a comparison to prior code constructions.


We now turn to proving \Cref{thm:homconstruct}. We will use the following notion of homological product of quantum codes, which is similar to the subsystem product in \Cref{def:subsystemcode}, but outputs a non-subsystem product code.

\begin{definition}
  \label{def:homprod}
  For $u\in\bN$, for $i\in[u]$ let $C^i=(C^i_X=\ker(H^i_X),\;C^i_Z=\ker(H^i_Z))$ be a $[[n_i,k_i]]_q$ CSS non-subsystem code over some finite field $\bF_q$ of characteristic $2$ with chosen full-rank parity-check matrices $H^i_X,H^i_Z\in\bF_q^{\ell_i\times n_i}$, so that $\dim(C^i_X)=\ell_i=\dim(C^i_Z)$. Let $\partial^i=(H^i_X)^\top H^i_Z\in\bF_q^{n_i\times n_i}$. Then the \emph{single-sector homological product} $C$ of $C^1,\dots,C^u$ is the CSS non-subsystem code $C=(C_X,C_Z)$ given by $C_X=\ker(\partial^\top)$ and $C_Z=\ker(\partial)$ for the matrix $\partial\in\bigotimes_{i\in[u]}\bF_q^{n_i\times n_i}\cong\bF_q^{n_1\cdots n_u\times n_1\cdots n_u}$ given by
  \begin{align*}
    \partial &:= \sum_{i\in[u]}I^{\otimes i-1}\otimes\partial^i\otimes I^{\otimes u-i}.
  \end{align*}
\end{definition}

Single-sector homological products were first studied in the context of quantum codes by \cite{bravyi_homological_2014}, and more recently by \cite{zeng_minimal_2020,golowich_near-asymptotically-good_2025}. Similarly as in these prior works, we restrict attention to such products over fields of characteristic $2$ to avoid signing issues. These products are a ``single-sector'' variant of the more standard tensor product (sometimes called a homological product) of chain complexes. These products were introduced for creating codes with low-weight stabilizers. Indeed, the \emph{stabilizer weight}, meaning the maximum weight of any row or column of the matrix $\partial$ in \Cref{def:homprod}, is at most $\sum_{i\in[u]}n_i$.

The homological product code in \Cref{def:homprod} has length $n_1\cdots n_u$; the following result of~\cite{bravyi_homological_2014} shows that the dimension is $k_1\cdots k_u$, as for the subsystem product code in \Cref{def:subsystemcode}. While \cite{bravyi_homological_2014} only proved the $u=2$ case of \Cref{prop:sskunneth} below, the result holds for general $u\in\bN$ as stated below because the homological product over fields of characteristic $2$ is by definition associative.

\begin{proposition}[Single-sector K\"{u}nneth formula \cite{bravyi_homological_2014}]
  \label{prop:sskunneth}
  Let $(C^i=(C^i_X=\ker(H^i_X),\;C^i_Z=\ker(H^i_Z)))_{i\in[u]}$ be a $[[n_i,k_i]]_q$ CSS non-subsystem codes over some $\bF_q$ of characteristic $2$. Then the associated homological product $C=(C_X,C_Z)$ has
  \begin{equation*}
    C_Z/C_X^\perp \cong \bigotimes_{i\in[u]}C_Z^i/{C_X^i}^\perp,
  \end{equation*}
  where for $(c_i\in C_Z^i)_{i\in[u]}$, the above isomorphism maps
  \begin{equation*}
    \bigotimes_{i\in[u]}c_i+C_X^\perp \mapsfrom \bigotimes_{i\in[u]}(c_i+{C_X^i}^\perp).
  \end{equation*}
\end{proposition}

Note that while we have stated \Cref{prop:sskunneth} for $C_Z/C_X^\perp$, by the symmetry of swapping $X$ and $Z$, an analogous result holds for $C_X/C_Z^\perp$.

The following corollary of \Cref{prop:sskunneth} bounds the distance of homological products in terms of the distance of the associated subsystem product.

\begin{corollary}
  \label{cor:subtohom}
  Let $(C^i=(C^i_X=\ker(H^i_X),\;C^i_Z=\ker(H^i_Z)))_{i\in[u]}$ be a $[[n_i,k_i]]_q$ CSS non-subsystem codes over some $\bF_q$ of characteristic $2$. Let $C=(C_X,C_Z)$ be the homological product, and let $C'=(C_X',C_Z')$ be the subsystem  product. Then we have inclusions $C_X\subseteq C_X'+{C_Z'}^\perp$ and $C_Z\subseteq C_Z'+{C_X'}^\perp$ that induce isomorphisms
  \begin{align*}
    C_X/C_Z^\perp &\xrightarrow{\sim} (C_X'+{C_Z'}^\perp)/{C_Z'}^\perp \\
    C_Z/C_X^\perp &\xrightarrow{\sim} (C_Z'+{C_X'}^\perp)/{C_X'}^\perp.
  \end{align*}
  In particular, the $X$- and $Z$-distances $d_X,d_Z$ are at least the respective distances $d_X',d_Z'$ of $C'$, that is, $d_X\geq d_X'$ and $d_Z\geq d_Z'$.
\end{corollary}
\begin{proof}
  Recall that the subsystem product $C'$ is defined by $C_X'=\bigotimes_{i\in[u]}C_X^i$ and $C_Z'=\bigotimes_{i\in[u]}C_Z^i$. We then have
  \begin{align*}
    C_X &= C_X'+C_Z^\perp \subseteq C_X'+{C_Z'}^\perp \\
    C_Z &= C_Z'+C_X^\perp \subseteq C_Z'+{C_X'}^\perp,
  \end{align*}
  where the equalities above hold by \Cref{prop:sskunneth}, and these equalities imply that $C_X'\subseteq C_X$ and $C_Z'\subseteq C_Z$, which in turn imply the inclusions above. These inclusions then induce well-defined linear maps
  \begin{align*}
    C_X/C_Z^\perp &\rightarrow(C_X'+{C_Z'}^\perp)/{C_Z'}^\perp \\
    C_Z/C_X^\perp &\rightarrow(C_Z'+{C_X'}^\perp)/{C_X'}^\perp.
  \end{align*}
  By \Cref{prop:sskunneth} and \Cref{prop:spkunneth}, we may respectively prepend and postpend isomorphisms to obtain
  \begin{align*}
    \bigotimes_{i\in[u]}(C_X^i/{C_Z^i}^\perp) \xrightarrow{\sim} C_X/C_Z^\perp &\rightarrow(C_X'+{C_Z'}^\perp)/{C_Z'}^\perp \xrightarrow{\sim} \bigotimes_{i\in[u]}(C_X^i/{C_Z^i}^\perp) \\
    \bigotimes_{i\in[u]}(C_Z^i/{C_X^i}^\perp) \xrightarrow{\sim} C_Z/C_X^\perp &\rightarrow(C_Z'+{C_X'}^\perp)/{C_X'}^\perp \xrightarrow{\sim} \bigotimes_{i\in[u]}(C_Z^i/{C_X^i}^\perp),
  \end{align*}
  such that the compositions of both sequences of maps above gives the identity, as every map above sends the coset containing a given tensor $\bigotimes_{i\in[u]}c_i$ to a coset containing the same tensor. Thus the middle map in each sequence above is an isomorphism. Furthermore, because every coset $c+C_Z^\perp\in C_X/C_Z^\perp$ is contained in the respective coset $c+{C_Z'}^\perp\in(C_X'+{C_Z'}^\perp)/{C_Z'}^\perp$, it follows that $d_X\geq d_X'$. By analogous reasoning, $d_Z\geq d_Z'$.
\end{proof}

\Cref{cor:subtohom} was implicit in the proof of \cite[Theorem 4.1]{golowich_near-asymptotically-good_2025}, which used a bound on subsystem product distance to bound homological product distance. We have included the general proof for completeness.

In the language of stabilizer codes, \Cref{cor:subtohom} can be interpreted as saying that the homological product $C$ is obtained from the subsystem product $C'$ by fixing the gauge qudits to an appropriate state. Applying this result with \Cref{thm:pevmain} immediately yields \Cref{thm:homconstruct}:


\begin{proof}[Proof of \Cref{thm:homconstruct}]
  The result follows immediately by applying \Cref{cor:subtohom} to the subsystem product codes in \Cref{thm:pevmain}, instantiated with each $n_i=n$, $k_i=k$, and $\ell_i=(n+k)/2$.
\end{proof}

A similar result as in \Cref{thm:homconstruct} was proven in \cite[Corollary 4.7]{golowich_near-asymptotically-good_2025}. However, in \Cref{thm:homconstruct} we use explicit Reed-Solomon codes from \Cref{thm:pevmain}, whereas the construction of \cite{golowich_near-asymptotically-good_2025} was non-explicit, and in fact used uniformly random codes. Whereas our distance proof in \Cref{thm:pevmain} (which implies that in \Cref{thm:homconstruct}) is a fairly concise algebraic argument, \cite{golowich_near-asymptotically-good_2025} bounded distance using the notion of \emph{higher-order product-expansion}, which was shown in \cite{kalachev_maximally_2025} to be satisfied by random codes.

\end{document}